\documentclass[11pt]{report}
\usepackage[a4paper, left=2.5cm, right=2.5cm, top=2cm, bottom=3cm]{geometry}
\usepackage[utf8]{inputenc}
\usepackage{graphicx}
\usepackage{float}
\usepackage{booktabs}
\usepackage{hyperref}
\usepackage{bm}
\usepackage{paralist}
\usepackage{hypcap}
\hypersetup{
  colorlinks=true,
  citecolor=black,
  linkcolor=black,
  urlcolor=black,
  allbordercolors={0 0 0},
  pdfborderstyle={/S/U/W 1}
}

\newcommand{\norm}[1]{\ensuremath{\lVert#1\rVert}}

\def \pycbc     {{\textsc{PyCBC}}}

\def \param     {\vec \Lambda}

\def \ngauss    {N_{\text{Gauss}}}
\def \nunif     {N_{\text{Unif}}}
\def \relbin    {{\textsc{RelBin}}}

\usepackage[section]{placeins}

\usepackage{enumitem}
\setlist[enumerate]{leftmargin=14pt, labelsep=0.20em, itemsep=0.em}
\usepackage{subcaption}
\usepackage{siunitx}
\usepackage{lipsum, multirow, microtype, amsmath, amssymb, newfloat, bm, color}
\usepackage{graphicx} 
\usepackage{dcolumn}  

\usepackage[dvipsnames]{xcolor}

\usepackage{orcidlink}
\newtheorem{theorem}{Theorem}

\usepackage{hyperref}
\usepackage{hypcap}
\usepackage{mathtools}
\hypersetup{
  colorlinks=true,
  citecolor=black,
  linkcolor=black,
  urlcolor=black,
  allbordercolors={0 0 0},
  pdfborderstyle={/S/U/W 1}
}

\begin{document}
\begin{titlepage}
    \begin{center}
        \vspace*{0.4cm}
        \huge
        \textbf{A meshfree approach for rapid reconstruction of compact binary sources
detected in the network of international gravitational wave detectors}
        
        \vspace*{0.8cm}
        \Large
        A Thesis submitted\\
        in partial fulfillment of the requirements for the Degree of\\
        \vspace{0.3cm}
        \huge
        \textbf{Doctor of Philosophy}\\
        \vspace{0.3cm}
        \Large
        by
        \vspace{0.3 cm}

        \Large
       \textbf{Lalit Pathak} \\
        \textbf{Roll No: 18310022}

        \vspace{0.2cm}

        \Large
         \textbf{Indian Institute of Technology Gandhinagar}
    
         \vspace{0.7cm}
    
         \includegraphics[width=0.25\textwidth]{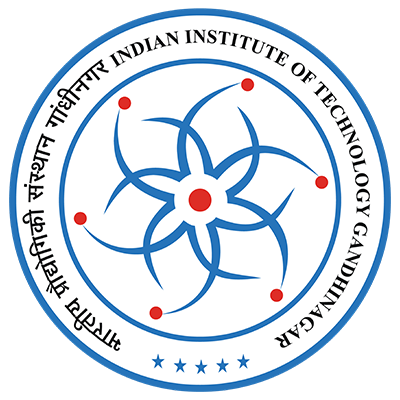} 
     
        \vspace{0.3cm}
        \Large
      Under the guidance of \\
\vspace{0.3cm}
\textbf{Prof. Anand Sengupta}\\
 \textit{Associate Professor} \\
        \vspace{0.2cm}
         \Large
         \textbf{Department of Physics}\\
         \vspace{0.1cm}
         \textbf{Indian Institute of Technology Gandhinagar}\\ 
         \vspace{0.1cm}
         \textbf{Gandhinagar - 382055, Gujarat, India}\\  
        \vspace{0.3cm}
        
\textbf{June 2023}
            
    \end{center}
\end{titlepage}  

\chapter*{}
\begin{center}
    \vspace{50mm}
    \textit{This page is intentionally left blank.}
\end{center}

\chapter*{Certificate}
It is certified that the work contained in the thesis titled \textbf{``A meshfree approach for rapid reconstruction of compact binary sources detected in the network of international gravitational wave detectors"} by \textbf{``Mr. Lalit Pathak"} (Roll no. 18310022), has been carried out under my supervision and this work has not been submitted elsewhere for a degree.

\vspace{4cm}
\begin{flushright}
    (Thesis Supervisor) \\
    \textbf{Prof. Anand Sengupta} \\
    Associate Professor \\
    Department of Physics \\
    Indian Institute of Technology Gandhinagar, \\
    Gujarat, India
\end{flushright}



\chapter*{Acknowledgements}
I extend my heartfelt appreciation to Professor Anand Sengupta for his unwavering guidance and thoughtful support. I am deeply grateful to my family for their constant love, care, and enduring support, demonstrating immense patience throughout my life's journey. Additionally, I am indebted to my friends at IIT Gandhinagar, whose continuous support and timely assistance have been invaluable. Special thanks go to Soumen Roy, Amit Reza, Rahul Shastri, Abhishek Sharma, Sanket Munishwar, and Sachin Shukla for their consistent encouragement and engaging discussions. I am grateful to the TCS Research Scholar program for providing the funding for my Ph.D. studies. I express my gratitude to IIT Gandhinagar for offering the awareness and resources that have significantly contributed to my personal and professional development throughout the duration of my five-and-a-half-year Ph.D. program.

\chapter*{}
\begin{center}
    \vspace{50mm}
    \textit{Dedicated to my family and my love ``Rashmi"}
\end{center}

\tableofcontents

\addcontentsline{toc}{chapter}{Publications}
\chapter*{Publications}

The chapters of this thesis are based on the following publications: 

\begin{enumerate}
    \item Fast and faithful interpolation of numerical relativity surrogate waveforms using meshfree approximation, \textbf{Lalit Pathak}, Amit Reza, Anand Sengupta, \href{https://arxiv.org/abs/2403.19162}{arXiv:2403.19162 (2024)}
    
    \item Fast likelihood evaluation using meshfree approximations for reconstructing compact binary sources, \textbf{Lalit Pathak}, Amit Reza, and Anand S. Sengupta, \href{https://journals.aps.org/prd/abstract/10.1103/PhysRevD.108.064055}{Phys. Rev. D \textbf{108}, 064055 (2023)}

    \item Prompt sky localization of compact binary coalescences using meshfree approximation, \textbf{Lalit Pathak}, Sanket Munishwar, Amit Reza, and Anand S. Sengupta, \href{https://journals.aps.org/prd/abstract/10.1103/PhysRevD.109.024053}{Phys. Rev. D \textbf{109}, 024053 (2024)}
    
    \item Pinpointing Binary Neutron Star sources with a global network of GW Detectors, including LIGO-Aundha, Sachin Shukla, \textbf{Lalit Pathak}, Anand Sengupta, \href{https://link.aps.org/doi/10.1103/PhysRevD.109.044051}{Phys. Rev. D \textbf{109}, 044051 (2024)}
    
\end{enumerate}
    
\paragraph{\textbf{Other publications not included in this thesis}}
\begin{enumerate}
    \item Does the speed of gravitational waves depend on the source velocity?, Rajes Ghosh, Sreejith Nair, \textbf{Lalit Pathak}, Sudipta Sarkar, Anand S. Sengupta, \href{https://journals.aps.org/prd/abstract/10.1103/PhysRevD.108.124017}{Phys. Rev. D \textbf{108}, 124017 (2023)}
\end{enumerate}
%












\chapter{Introduction}
\label{chap:chapter_1}
\section{Introduction to Gravitational waves (GWs)}

\subsection{Einstein's equation and plane wave solution}
\label{sec:1.1}
The detection of the first Gravitational wave signal in September 2015~\cite{PhysRevLett.116.061102, PhysRevLett.116.241102} ushered in a new era of gravitational wave astronomy. Albert Einstein, in his seminal work on General relativity (GR)~\cite{Einstein:1916vd}, predicted the existence of gravitational waves~\cite{Einstein:1916cc, Einstein:1918btx} but also lamented our possible inability to detect extremely faint signals using the technology available during his time. This raises the question of why Einstein held such reservations and what precisely we measure in a gravitational wave (GW) detector.  To address such questions, we need to establish a theoretical foundation for these detections, beginning with the fundamental principles. Let's start with Einstein's field equations.
\begin{equation}
    R_{\mu\nu} - \frac{1}{2}g_{\mu\nu} R = \frac{8\pi G}{c^4}T_{\mu\nu}
    \label{eq:einstein_field_equation}
\end{equation}
where $R_{\mu\nu}$, $R$, $g_{\mu\nu}$, and $T_{\mu\nu}$ are Ricci tensor, Ricci scalar, metric, and Energy-momentum tensor, respectively (For definitions, refer to Appendix~\ref{appendix:definitions_GR_chap1}). To see how GWs appear from Einstein's equations, let's linearize the field equations by considering a perturbation $h_{\mu\nu}$ on a flat Minkowski background $\eta_{\mu\nu}$
\begin{equation}
    g_{\mu\nu} = \eta_{\mu\nu} + h_{\mu\nu}, \:\:\: |h_{\mu\nu}| \ll 1, 
    \label{eq:linearized_metric}
\end{equation}
Now, let's perform a finite, global Lorentz transformation defined by
\begin{equation}
    x^{\mu} \rightarrow \Lambda^{\mu}_{\nu} x^{\nu}
    \label{eq:lorentz_transformn}
\end{equation}
Under Lorentz transformation, the matrix $\Lambda^{\mu}_{\nu}$ transforms as
\begin{equation}
    \eta_{\mu\nu} = \Lambda_{\mu}^{\:\:\:\rho} \Lambda_{\nu}^{\:\:\:\sigma}\eta_{\rho\sigma},
    \label{eq:transf_rule_lambda}
\end{equation}
which leads to the following transformation for $g_{\mu\nu}$:
\begin{equation}
\begin{split}
 g_{\mu\nu}(x) \rightarrow g^{'}_{\mu\nu}(x^{'}) 
    &= \Lambda_{\mu}^{\:\:\:\rho} \Lambda_{\nu}^{\:\:\:\sigma}\eta_{\rho\sigma}g_{\rho\sigma}(x) \\
    &= \Lambda_{\mu}^{\:\:\:\rho} \Lambda_{\nu}^{\:\:\:\sigma}\eta_{\rho\sigma}[\eta_{\rho\sigma} + h_{\rho\sigma}(x)]\\
    &= \eta_{\mu\nu} + \Lambda_{\mu}^{\:\:\:\rho} \Lambda_{\nu}^{\:\:\:\sigma}\eta_{\rho\sigma}h_{\rho\sigma}(x) = \eta_{\mu\nu} + h^{'}_{\mu\nu}(x^{'}),
\end{split}
\label{eq:transf_rule_metric}
\end{equation}
The above equation implies that $h_{\mu\nu}$ transforms as a tensor under Lorentz transformations. To linearize the Eq.~\eqref{eq:einstein_field_equation}, we need to linearize both $R_{\mu\nu}$ and $R$. First we express $R_{\mu\nu}$ in terms of Christoffel symbol $\Gamma^{\alpha}_{\mu\nu}$ as the following:
\begin{equation}
    R_{\mu\nu} = R^{\alpha}_{\mu\alpha\nu} = \partial_{\alpha}\Gamma^{\alpha}_{\mu\nu} - \partial_{\nu}\Gamma^{\alpha}_{\mu\alpha} + \Gamma^{\alpha}_{\beta\alpha}\Gamma^{\beta}_{\mu\nu} - \Gamma^{\alpha}_{\beta\nu}\Gamma^{\beta}_{\mu\alpha}
    \label{eq:ricci_tensor}
\end{equation} 
where the Christoffel symbol is given by
\begin{equation}
    \Gamma^{\alpha}_{\mu\nu} = \frac{1}{2}g^{\alpha\sigma} (\partial_{\mu}g_{\sigma\nu} + \partial_{\nu}g_{\sigma\mu} - \partial_{\sigma}g_{\mu\nu})
    \label{eq:christoffel_symb}
\end{equation}
The inverse metric $g^{\mu\nu}$ can be written as $\eta^{\mu\nu} - h^{\mu\alpha}$. Pluggin it in Eq.~\eqref{eq:christoffel_symb} along with Eq.~\eqref{eq:linearized_metric}, to $\mathcal{O}(h)$, we find 
\begin{equation}
    \Gamma^{\alpha}_{\mu\nu} = \frac{1}{2}\eta^{\alpha\sigma} (\partial_{\mu}h_{\sigma\nu} + \partial_{\nu}h_{\sigma\mu} - \partial_{\sigma}h_{\mu\nu})
    \label{eq:christoffel_symb_linear}
\end{equation}
Pluggin the above expression in Eq.~\eqref{eq:ricci_tensor}, the linearized Riemann tensor to $\mathcal{O}(h)$ becomes
\begin{equation}
    R_{\mu\nu\rho\sigma} = \frac{1}{2}(\partial_{\nu}\partial_{\rho}h_{\mu\rho} + \partial_{\mu}\partial_{\sigma}h_{\nu\rho}-\partial_{\nu}\partial_{\sigma}h_{\mu\rho})
    \label{eq:linearlized_riemann_tensor}
\end{equation}
To compactify the linearized equations, we define

\begin{equation}
    h = \eta^{\mu\nu}h_{\mu\nu}
    \label{eq:contracted_h_mu_nu}
\end{equation}

and
\begin{equation}
    \Bar{h}_{\mu\nu} = h_{\mu\nu} - \frac{1}{2}\eta_{\mu\nu}h
    \label{eq:redefined_h}
\end{equation}
From Eq.~\eqref{eq:redefined_h}, using $\Bar{h} = -h$, and inverting the above equation, we find

\begin{equation}
    h_{\mu\nu} = \Bar{h}_{\mu\nu} - \frac{1}{2}\eta_{\mu\nu}\Bar{h}
    \label{eq:inverted_hbar}
\end{equation}
Using Eq.~\eqref{eq:linearlized_riemann_tensor} and Eq.~\eqref{eq:inverted_hbar}, the linearization of Einstein's equations, Eq.~\eqref{eq:einstein_field_equation} leads to 
\begin{equation}
    \Box \Bar{h}_{\mu\nu} + \eta_{\mu\nu}\partial^{\rho}\partial^{\sigma}\Bar{h}_{\rho\sigma} - \partial^{\rho}\partial_{\nu}\Bar{h}_{\mu\rho} - \partial^{\rho}\partial_{\mu}\Bar{h}_{\nu\rho} = -\frac{16\pi G}{c^4}T_{\mu\nu}
    \label{eq:linearized_einstein_eqn}
\end{equation}

where $\Box \equiv -\frac{1}{c^2}\partial^2_0 + \nabla^2$.

Now, recalling that we chose a frame where Eq.~\eqref{eq:linearized_metric} is valid, there is still a residual gauge symmetry that can be exploited. Under a coordinate transformation defined by $x^{\mu} \rightarrow x^{'\:\mu} = x^{\mu} + \psi^{\mu}(x)$ and using transformation law (Eq.~\eqref{eq:transf_rule_metric}, we find
\begin{equation}
    h^{'}_{\mu\nu}(x') = h_{\mu\nu}(x) - (\partial_{\sigma}\psi_{\rho} + \partial_{\rho}\psi_{\sigma}) - (\partial_{\sigma}\psi^{\mu}h^{'}_{\mu\sigma}(x') + \partial_{\sigma}\psi^{\nu}h^{'}_{\rho\nu}(x'))
    \label{eq:tranform_res_gauge}
\end{equation}
Assuming $|\partial_{\mu}\psi_{\nu}| \sim \mathcal{O}(|h_{\mu\nu}|)$, the above expression in linear order becomes
\begin{equation}
    h^{'}_{\mu\nu}(x') = h_{\mu\nu}(x) - (\partial_{\sigma}\psi_{\rho} + \partial_{\rho}\psi_{\sigma})
    \label{eq:linear_gauge_tranformn}
\end{equation}
Using the above transformation rule, we find
\begin{equation}
    \begin{split}
    \Bar{h}^{'}_{\mu\nu} 
     &= h_{\mu\nu} - (\partial_{\mu}\psi_{\nu} + \partial_{\nu}\psi_{\mu}) - \frac{1}{2}\eta_{\mu\nu}h + \eta_{\mu\nu}\eta^{\alpha\beta}\partial_{\alpha}\psi_{\beta} \\
     &= \Bar{h}_{\mu\nu} - (\partial_{\mu}\psi_{\nu} + \partial_{\nu}\psi_{\mu} + \eta_{\mu\nu}\partial_{\rho}\psi^{\rho})
     \label{eq:redefined_h_tranform_gauge_symm}
    \end{split}
\end{equation}
which implies
\begin{equation}
    (\partial^{\nu}\Bar{h}_{\mu\nu})^{'} = \partial^{\nu}\Bar{h}_{\mu\nu} - \Box \psi_{\mu}
    \label{eq:redefined_h_deriv_tranformn}
\end{equation}
Now, we can always choose $\Box \psi_{\mu}$ such that $(\partial^{\nu}\Bar{h}_{\mu\nu})^{'} = 0$. It is known by the names of Lorentz/Hilbert/harmonic/De Donder gauge. In this gauge, the last three terms of the Eq.~\eqref{eq:linearized_einstein_eqn} vanishes, giving us the following wave equation:
\begin{equation}
    \Box \Bar{h}_{\mu\nu} = -\frac{16\pi G}{c^4}T_{\mu\nu}
    \label{eq:wave_equation1}
\end{equation}
Note that we started with a symmetric $h_{\mu\nu}$ with \textbf{ten} independent components. With the Lorentz gauge, we have $4$ conditions, which leads $h_{\mu\nu}$ to \textbf{six} independent components. To understand the propagation of the GWs and their effect on a GW detector, which is essentially happening outside the source, we can put $T_{\mu\nu} = 0$. It implies that 
\begin{equation}
    \Box \Bar{h}_{\mu\nu} = 0
    \label{eq:wave_eqn_outside_source}
\end{equation}
Since the Lorentz gauge didn't fix the gauge completely, let's perform another coordinate transformation such that $\Box \psi_{\mu} = 0$. It also implies $\Box \psi_{\mu\nu} = 0$, where 
\begin{equation}
    \psi_{\mu\nu} \equiv \partial_{\mu}\psi_{\nu} + \partial_{\nu}\psi_{\mu} - \eta_{\mu\nu}\partial_{\rho}\psi^{\rho},
    \label{eq:furthur_gauge_tranform}
\end{equation}
So we now have four extra conditions, coming from $\Box \psi_{\mu\nu} = 0$, since it depends on four functions $\psi_{\mu}$. These conditions can be imposed on $\Bar{h}_{\mu\nu}$, which already have six independent components. Furthurmore, $\psi^0$ can be chosen such that $\Bar{h} = 0$ which also implies that $\Bar{h}_{\mu\nu} = h_{\mu\nu}$. The rest of the functions $\psi^{i}(x)$ are chosen such that $h^{0i}(x) = 0$. This is what is known as the transverse-traceless gauge (TT gauge). So, finally, we have reduced the \textbf{ten} degrees of freedom associated with $h_{\mu\nu}$ to \textbf{two} degrees of freedom by exploiting the Lorentz gauge and a residual gauge associated with four functions $\psi^{\mu}$. Unless otherwise stated, we will refer to metric in TT gauge as $h_{\mu\nu}$.

The Eq.~\eqref{eq:wave_eqn_outside_source} have plane-wave solutions, $h_{ij}(x) = e_{ij}(\vec k) e^{jkx}$ with four vector $k^{\mu} = (\omega/c, \vec k)$ and $\omega/c = |\vec k|$, where $k^{\mu}$ has dimensions of inverse length, and $e_{ij}(\vec k)$ is the polarization tensor. From the above discussion, it is clear that the non-zero components of $h_{ij}$ lie in a plane perpendicular to $\hat{n}$ and the condition $\partial^jh_{ij} = 0$ is equivalent to $n^ih_{ij} = 0$ for a plane wave. With $\hat{n} = \vec{k}/|\vec{k}|$ (direction of wave propagation) along the $z$-axis and a symmetric and traceless $h_{ij}$, we get
\begin{equation}
    h_{ij}(t,z) =  \begin{pmatrix}
h_{+} &  h_{\times}&  0\\
h_{\times}&  -h_{+}&  0\\
 0&  0&  0\\
\end{pmatrix}_{ij}\cos[\omega(t - z/c)],
\label{eq:metric_perturbation1}
\end{equation}
In particular, we can simplify the above expression as the following:
\begin{equation}
    h_{ab}(t,z) =  \begin{pmatrix}
h_{+} &  h_{\times}\\
h_{\times}&  -h_{+}
\end{pmatrix}_{ab}\cos[\omega(t - z/c)],
\label{eq:metric_perturbation2}
\end{equation}
where $h_{+}$ and $h_{\times}$ are called the amplitudes of the ``plus'' and ``cross'' polarization of the wave, and indices $a, b = 1, 2$ represent the non-zero components of $h_{ij}$. The equation of motion  Eq.~\eqref{eq:wave_eqn_outside_source} implies that we can expand $h_{ij}$ as
\begin{equation}
    h_{ij}(x) = \int \frac{d^3k}{(2\pi)^3} (\mathcal{B}_{ij}(\vec k)e^{jkx} + \mathcal{B}^{*}_{ij}(\vec k)e^{-jkx})
    \label{eq:plane_waveform_expansion1}
\end{equation}
where $d^3k = (2\pi / c)^3f^2dfd\Omega$ with $f > 0$. Denoting $d^2\hat{n} = d\cos \theta d\theta d\phi$ as the integration over solid-angle, the above equation reads,
\begin{equation}
    h_{ij}(x) = \frac{1}{c^3}\int_0^{\infty} df f^2 \int d^2\hat{n} (\mathcal{B}_{ij}(f, \hat{n}) e^{-2\pi j f(t - \hat{n}\cdot x/c)} + c.c.)
    \label{eq:plane_waveform_expansion2}
\end{equation}
The TT gauge conditions implies that $\mathcal{B}^{i}_{i}(\vec k)$ and $\mathcal{B}_{ij}(\vec k) = 0$. Note that for superimposition of waves with different $\hat{n}$, $h_{ij}$ will no longer be a $2\times 2$ matrix. However, for the GWs detected on the Earth, $\hat{n}$ is very well defined and therefore, can be written as
\begin{equation}
    \mathcal{B}_{ij}(\vec k) = B_{ij}(f) \delta^{(2)}(\hat{n} - \hat{n}_0)
\end{equation}
where $\hat{n}_0$ is the direction of wave propagation.

Using Eq.~\eqref{eq:plane_waveform_expansion2}, we can write 

\begin{equation}
    h_{ab}(t, \vec x) = \int_{0}^{\infty} df \left(\tilde{h}_{ab}(f, \vec x) e^{-2\pi j f t} + \tilde{h}^{*}_{ab}(f, \vec x) e^{2\pi j f t}\right),
    \label{eq:transverse_plane_wave_soln1}
\end{equation}
where
\begin{equation}
    \begin{split}
       \tilde{h}_{ab}(f, \vec x) 
       &= \frac{f^2}{c^3}\int d^2\hat{n} \mathcal{A}_{ab}(f, \hat{n}) e^{2\pi j f\hat{n}\cdot x/c} \\
       &= \frac{f^2}{c^3}A_{ab}(f)e^{2\pi j f\hat{n}_0\cdot x/c}
    \end{split}
\end{equation}
From Eq.~\eqref{eq:metric_perturbation2} and the above equation, it follows that

\begin{equation}
    \tilde{h}_{ab}(f) = \begin{pmatrix}
\tilde{h}_{+}(f) &  \tilde{h}_{\times}(f)\\
\tilde{h}_{\times}(f) &  -\tilde{h}_{+}(f)\\
\end{pmatrix}_{ab}
\end{equation}
where the ``$+$'' and ``$\times$'' polarizations are defined with respect to a given choice of axes in the transverse plane. As evident from Eq.~\eqref{eq:transverse_plane_wave_soln1}, only physical frequencies ($f > 0$) are present in the equations. We can extend it to negative frequencies by defining 
\begin{equation}
    \tilde{h}_{ab}(-f, \vec x) = \tilde{h}^{*}_{ab}(f, \vec x),
\end{equation}
plugging in the Eq.~\eqref{eq:transverse_plane_wave_soln1}, we get

\begin{equation}
    h_{ab}(t) = \int_{-\infty}^{\infty} df\tilde{h}_{ab}(f)e^{-2\pi j f t}.
\end{equation}
Apart from the above forms, we can express the $h_{ab}(t, \vec x)$ in terms of so called polarization tensors $e^{\alpha}_{ij}(\hat{n})$ (with $\alpha$ defining labels $+$ and $\times$) which are defined as
\begin{equation}
\begin{aligned}
  e^{+}_{ij}(\hat{n}) &= \hat{u}_i \hat{u}_j - \hat{v}_i \hat{v}_j \\
  e^{\times}_{ij}(\hat{n}) &= \hat{u}_i \hat{v}_j + \hat{v}_i \hat{u}_j,
  \label{eq:polarization_tensors2}
\end{aligned}
\end{equation}
where $\hat{u}$ and $\hat{v}$ are unit vectors orthogonal to $\hat{n}$ and to each other. With $\hat{n}$ along $z$-axis in a given frame, one can choose $\hat{u}=\hat{x}$ and $\hat{v}=\hat{y}$ and express polarization tensors in the following form:
\begin{equation}
\begin{aligned}
  e^{+}_{ab} &=  \begin{pmatrix}
 1&  0\\
 0&  -1\\
\end{pmatrix}_{ab}\\
  e^{\times}_{ab} &= \begin{pmatrix}
 0&  1\\
 1&  0\\
\end{pmatrix}_{ab}
  \label{eq:polarization_tensors1}
\end{aligned}
\end{equation}
In a generic frame, the amplitdues $\tilde{h}_A(f, \hat{n})$ can be read from the following:
\begin{equation}
    \frac{f^2}{c^3}\mathcal{A}_{ij}(f, \hat{n}) = \sum_{\alpha = +, \times} \hat{h}_{\alpha}(f, \hat{n}) e^{\alpha}_{ij}(\hat{n})
\end{equation}
The Eq.~\eqref{eq:plane_waveform_expansion2} now be written as 
\begin{equation}
    h_{ab}(t, \vec x) = \sum_{\alpha = +, \times} \int_{-\infty}^{\infty} df \int d^2\hat{n} \hat{h}_{\alpha}(f, \hat{n})e^{\alpha}_{ab}(\hat{n}) e^{-2\pi j f (t - \hat{n}\cdot x/c)},
\end{equation}
with $\hat{h}_{\alpha}(-f, \hat{n}) = \tilde{h}^{*}_{\alpha}(f, \hat{n})$. Now, we will discuss how the GWs interact with test masses. 
\subsection{Test masses interaction with GWs}
\subsubsection{Geodesic equation and Local inertial frame}
In GR, the choice of a particular gauge is synonymous with the selection of a specific observer. As discussed in the preceding section, GWs exhibit a simplified expression in the TT gauge. This prompts the question of what reference frame corresponds to the TT gauge. In order to understand the effects of GWs on test masses, let's start with a geodesic equation which is given by
\begin{equation}
    \frac{d^2x^{\mu}}{d\tau^2} + \Gamma^{\mu}_{\nu\rho}(x)\frac{dx^{\nu}}{d\tau} \frac{dx^{\rho}}{d\tau} = 0
    \label{eq:geodesic_eqn}
\end{equation}
which describes the motion of a test mass in curved background metric $g_{\mu\nu}$ in the absence of non-gravitational forces, and where $\tau$ is the proper time. The above equation can rewritten in terms of four-velocity $u^{\mu}$ as
\begin{equation}
    \frac{du^{\mu}}{d\tau} + \Gamma^{\mu}_{\nu\rho}u^{\nu}u^{\rho} = 0
    \label{eq:geodesic_eqn_4vel}
\end{equation}
To find the equation of geodesic deviation, let's consider two nearby geodesics represented by $x^{\mu}(\tau)$ and the $x^{\mu}(\tau) + \zeta^{\mu}(\tau)$. The geodesic equation for the latter is given by
\begin{equation}
    \frac{d^2(x^{\mu} + \zeta^{\mu})}{d\tau^2} + \Gamma^{\mu}_{\nu\rho}(x + \zeta) \frac{d(x^{\nu} + \zeta^{\nu})}{d\tau} \frac{d(x^{\rho} + \zeta^{\rho})}{d\tau} = 0 
\end{equation}
If $|\zeta^{\mu}|$ is much smaller than the typical scale of variation of the gravitational field, the above equation can be expanded to first order in $\zeta$ as
\begin{equation}
    \frac{d^2\zeta^{\mu}}{d\tau^2} + 2\,\Gamma^{\mu}_{\nu\rho}(x)\frac{dx^{\nu}}{d\tau}\frac{d\zeta^{\rho}}{d\tau} + \zeta^{\sigma}\partial_{\sigma}\Gamma^{\mu}_{\nu\rho}(x)\frac{dx^{\nu}}{d\tau}\frac{dx^{\rho}}{d\tau}
    \label{eq:eqn_of_geodesic_dev}
\end{equation}
It can be written in a more elegant way in terms of the covariant derivative (for definition, refer to Eq.~\eqref{eq:cov_der_def} in Appendix~\ref{appendix:definitions_GR_chap1}) of a vector field $\zeta^{\mu}$ along the curve $x^{\mu}(\tau)$ as
\begin{equation}
    \frac{D^2\zeta^{\mu}}{d\tau^2} = -R^{\mu}_{\:\:\nu\rho\sigma}\zeta^{\rho}\frac{dx^{\nu}}{d\tau}\frac{x^{\sigma}}{d\tau},
    \label{eq:eq:eqn_of_geodesic_dev_cov_dev}
\end{equation}
The implication of the equation mentioned above is that a tidal gravitational force exists between nearby time-like geodesics, and its determination relies on the Riemann tensor. This understanding aids in elucidating the behavior of test masses influenced by gravitational waves for a specific observer. Before we delve further into a crucial discussion on reference frames, let us briefly discuss the construction of a local inertial frame (LIF). Within the framework of General Relativity (GR), it is always possible to perform coordinate transformations in such a manner that, at a specific spacetime point P, all components of the Christoffel symbol are zero identically ($\Gamma^{\mu}_{\nu\rho}(\text{P}) = 0$). This implies that the test mass is in a state of free fall in this reference frame, albeit only at a particular point in space and time.
\begin{equation}
    \frac{d^2x^{\mu}}{d\tau^2}\bigg|_{\text{P}} = 0
    \label{eq:local_inertial_frame_condition}
\end{equation}
As previously explained, within the Local Inertial Frame (LIF), the test mass moves freely at a specific point in spacetime. Alternatively, we can construct a reference frame in which the test mass remains in free fall throughout its journey along the geodesic. This can be achieved by employing a freely spinning gyroscope in motion along a time-like geodesic $x^{\mu}(\tau)$ obeying the following equation:
\begin{equation}
    \frac{ds^{\mu}}{d\tau} + \Gamma^{\mu}_{\nu\rho} s^{\nu}\frac{dx^{\rho}}{d\tau} = 0
    \label{eq:gyroscope_along_geodesic}
\end{equation}
where $s^{\mu}$ is the spin four-vector. We can start with an LIF at P with three gyroscopes defining the directions of the spatial axes. During the movement along the geodesic, the spatial axes continuously orient themselves with the directions indicated by gyroscopes. Note that since gyroscopes define the orientation of the axes, they do not rotate relative to the axes within this frame. This results in $ds^{\mu}/d\tau = 0$ throughout the geodesic. In line with Eq.~\eqref{eq:gyroscope_along_geodesic}, this condition implies that $\Gamma^{\mu}_{\nu\rho}$ vanishes along the entire geodesic. Such a reference frame is commonly referred to as a freely-falling frame. For a more comprehensive and detailed discussion of the construction of the Local Inertial Frame, please refer to Section $8.4$ of Hartle~\cite{hartle2021gravity}.

\subsubsection{TT frame and proper detector frame}
As found in the earlier sections, in the TT gauge, the GWs have a simple form. We now seek to determine the reference frame (or observer) associated with this gauge. Writing down the Eq.~\eqref{eq:geodesic_eqn} for a test mass at rest at $\tau = 0$
\begin{equation}
    \begin{split}
        \frac{d^2x^{i}}{d\tau^2}\bigg|_{\tau=0} 
        & = -\left[\Gamma^{i}_{\nu\rho}(x)\frac{dx^{\nu}}{d\tau}\frac{dx^{\rho}}{d\tau}\right]\bigg|_{\tau=0}\\
        &= -\left[\Gamma^{i}_{00}\left(\frac{dx^{0}}{d\tau}\right)^2\right]_{\tau=0},
    \end{split}
\end{equation}

From Eq.~\eqref{eq:christoffel_symb_linear}, we find that 
\begin{equation}
    \Gamma^{i}_{00} = \frac{1}{2}(2\partial_0h_{0i} - \partial_ih_{00}).
\end{equation}

Since $h_00 = h_{0i} = 0$ in the TT gauge, the above equation implies $d^2x^{i}/d\tau^2 = 0$ at $\tau=0$. It means that the test mass remains at rest at all times, implying that the GWs have no effect on the motion of the test mass. To see what happens to the separation of two test masses which are initial at test, the spatial component of the Eq.~\eqref{eq:eqn_of_geodesic_dev} is written as
\begin{equation}
    \frac{d^2\xi^{i}}{d\tau^2}\bigg|_{\tau=0} = -\left[2c\Gamma^{i}_{0\rho} \frac{d\xi^{\rho}}{d\tau} + c^2 \xi^{\sigma} \partial_{\sigma} \Gamma^{i}_{00} \right]_{\tau = 0}
\end{equation}
The first term in the above equation is non-zero only when $\rho$ represents a spatial index, whereas the second term is zero identically. From Eq.~\eqref{eq:christoffel_symb_linear}, we find $\Gamma^{i}_{0j} = (\frac{1}{2})\partial_0h_{ij}$. Therefore, the above equation is reduced to
\begin{equation}
    \frac{d^2\zeta^{i}}{d\tau^2}\bigg|_{\tau=0} = -\left[\dot{h}_{ij}\frac{d\zeta^{i}}{d\tau}\right]_{\tau = 0},
    \label{eq:eqn_of_geodesic_eqn_TTframe}
\end{equation}
which implies that if at $\tau = 0$, the $d\zeta^{i}/d\tau = 0$, then $\frac{d^2\zeta^{i}}{d\tau^2} = 0$ and hence the separation $\zeta^{i}$ remains constant at all times.

From the above discussion, it seems as if there is no physical effect of GWs passing by test masses. However, the physical effects of GWs indeed exist and can be found by monitoring the proper distances instead of coordinate distances. To see this, let us write the proper distance between two events (say $(t, x_1, 0, 0)$ and $(t, x_2, 0, 0)$) as
\begin{equation}
    \begin{split}
        s
        &= (x_2 - x_1)[1 + h_{+}\cos (\omega t)]^{1/2}\\
        &\simeq L\, [1 + \frac{1}{2} h_{+}\cos (\omega t)],
    \end{split}
\end{equation}

which implies that the proper distance changes periodically in time due to GWs. Generally, for two events separated by a vector $\vec L$, the proper distance is given by $s^2 = L^2 + h_{ij}(t)L_iL_j$, where $h_{ij}(t)$ is the GW tensor at time $t$. To linear order in $h$, this equation can be approximated as $s \simeq L + h_{ij}(L_iL_j/2L)$, giving
\begin{equation}
    \begin{split}
        \ddot{s} &\simeq \frac{1}{2}\ddot{h}_{ij}\frac{L_i}{L}L_j\\
        &\simeq \frac{1}{2}\ddot{h}_{ij}s_j,
    \end{split}
\end{equation}

where $n_i = L_i/L$ and $s = n$. This is the geodesic equation expressed in terms of proper distances. The fact that GWs affect proper distances implies that any quantity dependent on proper distances can be measured, facilitating the detection of GWs. In fact, this forms the basis of interferometer-based GW detectors, where a laser beam travels back and forth between the mirrors (akin to test masses) whose round trip time is dependent on the proper distance. We will study this in the later sections while discussing the GW detectors. Now, let's briefly introduce the proper detector frame. As mentioned earlier, the TT frame simplifies the form of GWs. However, it may not be a suitable frame for experimentalists describing GW detectors since, in a detector, the positions are not represented by freely falling particles. It can be demonstrated that in the proper detector frame, the equation of geodesic deviation~\footnote{For derivation, refer~\cite{maggiore2008gravitational}} can be expressed as follows: 
\begin{equation}
    \ddot{\xi}^{i} = \frac{1}{2}\ddot{h}^{\text{TT}}_{ij}\xi^j
    \label{eq:eqn_of_geodesic_dev_proper_det_frame}
\end{equation}
\begin{figure*}[!hbt]
    \centering
    \includegraphics[width=0.7\linewidth]{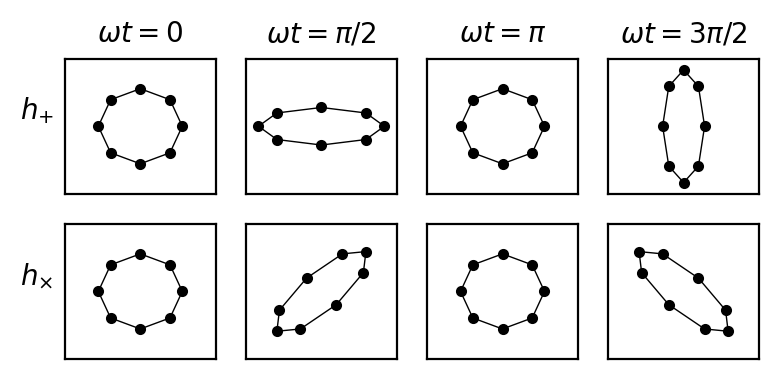}
    \caption{The ring of test masses shows deformation due $+$ and $\times$ polarizations of GWs as they enter into the plane of the paper. Notice that the pattern ring follows as the phase changes. $h_{+}$ polarization follows a ``$+$'' pattern while $h_{\times}$ follows a ``$\times$'' pattern and hence the denominations ``plus'' and ``cross'' polarizations.}
    \label{fig:deformation_ring_of_test_masses_GW}
\end{figure*}
Unlike Eq.~\eqref{eq:eqn_of_geodesic_eqn_TTframe}, the above equation implies that the effects of GW on a point particle of mass $m$ can be expressed in terms of a Newtonian force in the proper detector frame as
\begin{equation}
    F_i = \frac{m}{2}\ddot{h}^{\text{TT}}_{ij}\xi^j,
    \label{eq:newtonian_force_GW}
\end{equation}

which is a great simplification over the TT frame. Naturally, there are other slowly varying Newtonian forces, the influence of which can be minimized by carefully selecting a frequency window $[f_{min}, f_{max}]$. Here, the choice of $f_{min}$ ensures that the slowly varying Newtonian and seismic noise remains significantly low, while $f_{max}$ is determined to prevent other sources of noise from dominating. Consequently, the proper detector frame offers two key advantages: (i) in this frame, the description of the detector can be done in Newtonian language, which is more intuitive, and (ii) The Newtonian force induced by gravitational waves can be expressed in terms of $h_{ij}$, which is calculated in the TT gauge and has a highly simplified form.
Now, let us examine the effect of GWs on a ring of test masses which is placed in the $x$-$y$ plane and a GW is propagating along the $z$-direction. In such a scenario, the motion of the test particles remains confined to the $x$-$y$ plane in accordance with Eq.~\eqref{eq:eqn_of_geodesic_dev_proper_det_frame}. To begin, let's consider the plus $h_{+}$ polarization, which at $z=0$ can be represented as:
\begin{equation}
    h_{ab} = h_{+}\sin(\omega t) \begin{pmatrix}
1 &  0\\
0 &  -1\\
\end{pmatrix},
\end{equation}
We choose $\zeta_a(t)$ (see Eq.~\eqref{eq:eqn_of_geodesic_dev_proper_det_frame}) to be $(x_0 + \delta x(t), y_0 + \delta y(t))$, where $x_0$ and $y_0$ are the unperturbed positions and $\delta x(t)$ and $\delta y(t)$ induced by the GW. From Eq.~\eqref{eq:eqn_of_geodesic_dev_proper_det_frame}, we find
\begin{subequations}
    \begin{align}
        \delta \ddot{x} &= -\frac{h_{+}}{2} (x_0 + \delta x)\omega^2 \sin (\omega t),\\
        \delta \ddot{y} &= \frac{h_{+}}{2} (y_0 + \delta y)\omega^2 \sin (\omega t).
    \end{align}
\end{subequations}
Neglecting the $O(h)$ in the right-hand side of the above equations and integrating, we get
\begin{subequations}
    \begin{align}
        \delta x &= \frac{h{+}}{2} x_0 \sin (\omega t),\\
        \delta y &= -\frac{h{+}}{2} y_0 \sin (\omega t).
    \end{align}
\end{subequations}
Similarly for cross $h_{\times}$ polarizations, we find 
\begin{subequations}
    \begin{align}
        \delta x &= \frac{h_{\times}}{2} y_0 \sin (\omega t),\\
        \delta y &= \frac{h_{\times}}{2} x_0 \sin (\omega t).
    \end{align}
\end{subequations}
Fig.~\ref{fig:deformation_ring_of_test_masses_GW} shows how a ring of test masses is deformed by a passing gravitational wave (GW). We now briefly discuss how the detector responds to GWs.
\subsection{GW interaction with detector}
\label{sec:interaction_with_det}
We again revisit the end of Sec.~\ref{sec:1.1}, and we saw that we could describe the GWs in the TT gauge as a linear combination of polarization tensors $e^{\alpha}_{ij}$.
\begin{subequations}
    \begin{align}
        h_{ij} &= h_{+}e^{+}_{ij} + h_{\times}e^{\times}_{ij}\\
        \boldsymbol{h} &= h_{+}\boldsymbol{e}^{+} + h_{\times}\boldsymbol{e}^{\times}
    \end{align}
    \label{h_comb_pol_tensors}
\end{subequations}
where $\left\{e^{+}_{ij}\right\}$ and $\left\{e^{\times}_{ij}\right\}$ can be written as
\begin{equation}
   \left\{e^{+}_{ij}\right\} \equiv \boldsymbol{e}^{+} = \begin{pmatrix}
1 &  0&  0\\
0 &  -1&  0\\
0 &  0&  0\\
\end{pmatrix}
   \quad\text{and}\quad 
 \left\{e^{\times}_{ij}\right\} \equiv \boldsymbol{e}^{\times} = \begin{pmatrix}
0 &  1&  0\\
1 &  0&  0\\
0 &  0&  0\\
\end{pmatrix}
\end{equation}

However, it is more useful to express these quantities in a coordinate-independent manner. For example, we can represent a vector $\vec A$ in different bases, e.g., $\left\{\vec e_{i}\right\}$ and $\left\{\vec e^{\:'}_{i}\right\}$ and its components in these coordinate systems are defined by
\begin{equation}
   A_i = \vec e_i \cdot \vec A
   \quad\text{or}\quad 
   A^{'}_i = \vec e^{\:'}_i \cdot \vec A
\end{equation}

We can represent the same physical quantity in various manners without altering the underlying physics. To achieve this, we will introduce an abstract index notation, where $A_a$ is equivalent to the familiar vector $\vec A$, referring to an object with components $\left\{A_{i}\right\}$ when analyzed within the basis $\vec e_i$. Returning to Eq.~\eqref{h_comb_pol_tensors}, we would like to formulate it in a manner that is independent of a specific set of basis vectors. This can be accomplished by constructing matrices $\boldsymbol{e}^{+}$ and $\boldsymbol{e}^{\times}$ using the components of the unit vectors $\vec \ell$ and $\vec m$. These unit vectors, in combination with $\vec k-$representing the direction of gravitational wave propagation—form an orthonormal triple. In the coordinate system we have been working with, these vectors have the following components:
\begin{equation}
   \left\{\ell_{i}\right\} = \boldsymbol{\ell} = \begin{pmatrix} 1\\ 0\\ 0 \end{pmatrix}
   \quad\text{and}\quad \left\{m_{i}\right\} = \boldsymbol{m} = \begin{pmatrix} 0\\ 1\\ 0 \end{pmatrix}
\end{equation}
and we can express $\boldsymbol{e}^{+}$ and $\boldsymbol{e}^{\times}$ in terms of the above vectors as
\begin{subequations}
    \begin{align}
        \boldsymbol{e}^{+} &= \boldsymbol{\ell}\boldsymbol{\ell}^T - \boldsymbol{m}\boldsymbol{m}^T\\
        \boldsymbol{e}^{\times} &= \boldsymbol{\ell}\boldsymbol{m}^T + \boldsymbol{m}\boldsymbol{\ell}^T
    \end{align}
\end{subequations}
Or, in terms of components, 
\begin{subequations}
    \begin{align}
        \boldsymbol{e}^{+}_{ij} &= \ell_i \ell_j - m_im_j\\
        \boldsymbol{e}^{\times}_{ij} &= \ell_im_j + m_i\ell_j
    \end{align}
\end{subequations}
In the abstract index notation, we can describe the above polarization matrices in the following way:
\begin{subequations}
    \begin{align}
        e^{+}_{ab} &= \ell_a \ell_b - m_am_b\\
        e^{\times}_{ab} &= \ell_am_b + m_a\ell_b
    \end{align}
\end{subequations}
Or, these basis vectors can be expressed in terms of the tensor (dyad) products as 
\begin{subequations}
    \begin{align}
        \stackrel{\leftrightarrow}{e}^{\: +} &= \vec{\ell} \otimes \vec{\ell} - \vec{m} \otimes \vec{m}\\
        \stackrel{\leftrightarrow}{e}^{\: \times} &= \vec{\ell} \otimes \vec{m} + \vec{m} \otimes \vec{\ell}
    \end{align}
    \label{eq:natural_pol_tensor}
\end{subequations}
using which we can write the general plane wave propagating along $\vec k$ in covariant tensor notation as
\begin{equation}
    \stackrel{\leftrightarrow}{h} = h_{+}\stackrel{\leftrightarrow}{e}^{\: +} + h_{\times}\stackrel{\leftrightarrow}{e}^{\: \times}
\end{equation}
where $h_{+}$ and $h_{\times}$ are the functions of $t - \vec k \cdot \vec r / c$.
\begin{figure*}[!hbt]
    \centering
    \includegraphics[width=0.5\linewidth]{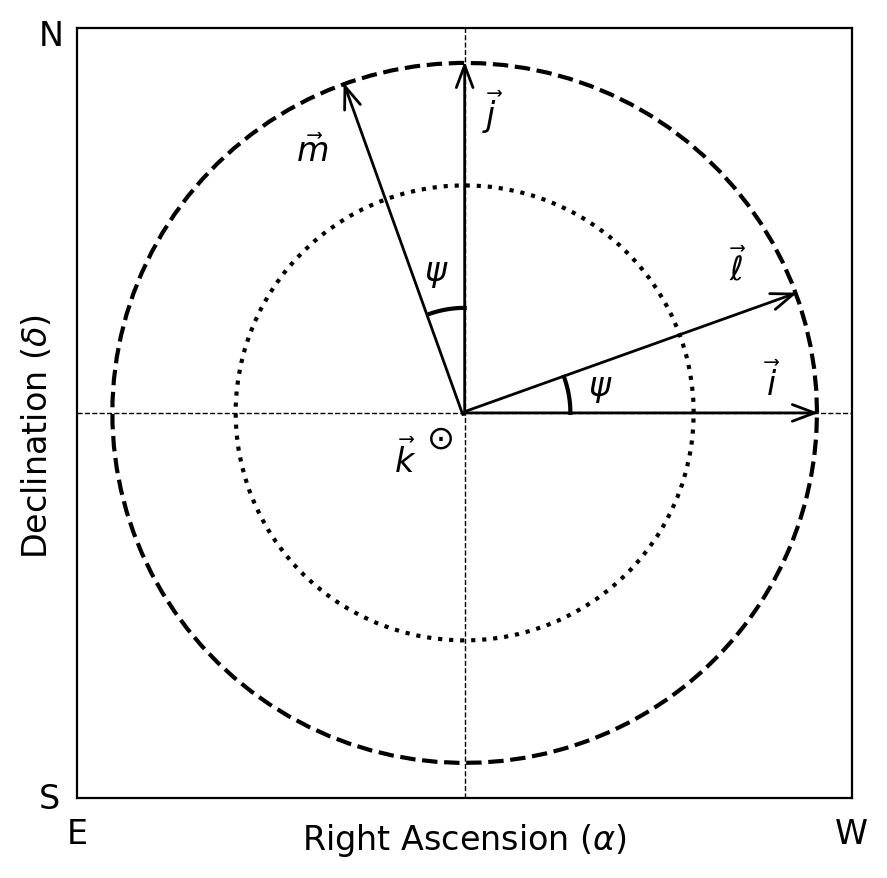}
    \caption{Natural polarization basis is obtained by rotating the reference basis around the direction of propagation of GW. $\vec i$ is chosen in the direction of decreasing $\alpha$, so that $\vec j$ points into the Northern celestial hemisphere. (reproduced from~\cite{whelan_notes})}
    \label{fig:nat_basis_vs_ref_basis}
\end{figure*}
Note that the propagation direction alone is insufficient to uniquely determine the construction of $\stackrel{\leftrightarrow}{e}^{+}$ and $\stackrel{\leftrightarrow}{e}^{\times}$. To establish a complete orthonormal basis, the selection of a vector $\vec \ell$ within the plane perpendicular to $\vec k$ is also necessary. This choice finalizes the definition of the orthonormal triple comprising $\vec \ell$, $\vec m$, and $\vec k$.
The specific polarization basis adopted may be influenced by the nature of the sources being studied and the requirements of the analysis. In some cases, a ``convenient'' polarization basis may be preferred. It might also be desirable to transition from this convenient polarization basis to a reference basis constructed solely based on the direction of GW propagation and some absolute reference directions. For a clearer illustration, let's consider an example where the source's position in the sky is specified using its right ascension $\alpha$ and declination $\delta$, which is equivalent to defining the propagation direction $\vec k$. At each sky location, a reference basis can be assigned by introducing two additional unit vectors, $\vec i$ and $\vec j$, which together form an orthonormal set alongside $\vec k$. From these unit vectors, a reference polarization basis can be constructed for traceless symmetric tensors oriented transversely to $\vec k$. This basis can be expressed as follows:
\begin{subequations}
    \begin{align}
        \stackrel{\leftrightarrow}{\varepsilon}^{\: +} &= \vec i \otimes \vec i - \vec j \otimes \vec j \\
        \stackrel{\leftrightarrow}{\varepsilon}^{\: \times} &= \vec i \otimes \vec j + \vec j \otimes \vec i
    \end{align}
    \label{eq:refernce_pol_tensors}
\end{subequations}
Moreover, $\vec \ell$, $\vec m$ forming the natural polarization basis, lie in the same plane as $\vec i$ and $\vec j$ and hence perpendicular to $\vec k$. It implies that we can go from the reference basis to the natural basis by rotating the reference basis counter-clockwise ($\vec i$ to $\vec \ell$) by some angle $\psi$ and can be written in the following form:
\begin{subequations}
\begin{align}
        \vec \ell &= \vec i \cos \psi + \vec j \sin \psi \\  
        \vec m &= -\vec i \sin \psi + \vec j \cos \psi 
    \end{align}
\end{subequations}

Substituting the above equations in Eq.~\eqref{eq:natural_pol_tensor}, we find, 
\begin{subequations}
    \begin{align}
        \stackrel{\leftrightarrow}{e}^{\: +} &= \vec i \otimes \vec i \cos 2\psi + \vec i \otimes \vec j \sin 2\psi + \vec j \otimes \vec i \sin 2\psi - \vec j \otimes \vec j \cos 2\psi\\
        \stackrel{\leftrightarrow}{e}^{\: \times} &= -\vec i \otimes \vec i \sin 2\psi + \vec i \otimes \vec j \cos 2\psi + \vec j \otimes \vec i \cos 2\psi + \vec j \otimes \vec j \sin 2\psi
    \end{align}
\end{subequations}
which can be expressed in terms of $\stackrel{\leftrightarrow}{\varepsilon}^{\: +}$ and $\stackrel{\leftrightarrow}{\varepsilon}^{\: \times}$ using the Eq.~\eqref{eq:refernce_pol_tensors} as
\begin{subequations}
\begin{align}
        \stackrel{\leftrightarrow}{e}^{\: +} &= \stackrel{\leftrightarrow}{\varepsilon}^{\: +} \cos 2\psi \: + \stackrel{\leftrightarrow}{\varepsilon}^{\: \times} \sin 2\psi \\  
        \stackrel{\leftrightarrow}{e}^{\: \times} &= -\stackrel{\leftrightarrow}{\varepsilon}^{\: +} \sin 2\psi \: + \stackrel{\leftrightarrow}{\varepsilon}^{\: \times} \cos 2\psi 
    \end{align}
\end{subequations}

which shows that three angles are needed to specify a polarization basis associated with a particular source: (i) $\alpha$, (ii) $\delta$, and (iii) an additional angle $\psi$, which is also known as the polarization angle defining the orientation of the preferred polarization basis $\left\{ \stackrel{\leftrightarrow}{e}^{\: +}, \stackrel{\leftrightarrow}{e}^{\: \times}\right\}$ relative to some reference basis $\left\{ \stackrel{\leftrightarrow}{\varepsilon}^{\: +}, \stackrel{\leftrightarrow}{\varepsilon}^{\: \times}\right\}$.

Now, we discuss how GWs affect a GW detector. For the sake of simplicity, we will focus on an ``L'' shaped GW interferometer, where each arm has a fixed length $L_0$. A natural question that arises is how we can make measurements when both the spacetime and the detector itself are subject to stretching and squeezing due to GWs. To clarify, we actually measure the phase difference between the light beams returning from the two arms of the interferometer after reflecting off the mirrors. This phase difference is essentially equivalent to determining the roundtrip travel time of photons traveling down both arms. The beauty of this measurement lies in the fact that it remains unaffected by the presence of gravitational waves, mainly because the time component of the spacetime metric remains constant in the TT gauge. Additionally, in the TT gauge, points with constant coordinates experience free fall. Assuming the beam splitter is located at the origin and the end mirror of one of the arms is positioned at coordinates $(x^1, x^2, x^3) = (L_0, 0, 0)$, these coordinates remain unchanged despite the passage of GWs. Considering the path of the photon traveling from the origin $(0, 0, 0)$ to the end mirror at $(L_0, 0, 0)$ and back, its trajectory follows the equation:
\begin{equation}
    ds^2 = -c^2dt^2 + (1 + h_{11})(dx^1)^2 = 0
\end{equation}
which implies, 
\begin{equation}
    dt = \frac{\sqrt{1 + h_{11}}}{c}|dx^1| \approx \left(1 + \frac{1}{2}h_{11}\right)\frac{|dx^1|}{c}
\end{equation}

In general the GW amplitude $h_{11}$ is a function of $\left(t - \frac{\vec k \cdot \vec r}{c}\right)$. However, in the ``long-wavelength'' limit where the wavelength of the GW is much longer than the arm length of the detector, we can approximate $h_{11}$ to be a constant along the photon's travel path. This approximation is not valid for space-borne GW detectors, such as LISA~\cite{amaroseoane2017laser, Robson_2019, Cornish_2020}. Under this approximation, the time taken by a photon to travel from $(0,0,0)$ to $(L_0, 0, 0)$ and back is
\begin{equation}
    T_1 = \left(1 + \frac{1}{2}h_{11}\right)\frac{2L_0}{c}
\end{equation}

We now describe the detector arm in a coordinate-independent manner and define a unit vector $\vec u$ along one of the arms whose components (in a specialized coordinate system) are
\begin{equation}
    \left\{ u^i\right\} \equiv \boldsymbol{u} = \begin{pmatrix}
1 \\
0 \\
0
\end{pmatrix}
\end{equation}
The metric perturbation $h_{11}$ above can be expressed as
\begin{equation}
    h_{11} = u^ih_{ij}u^j = \boldsymbol{u}^T\boldsymbol{h}\boldsymbol{u}
\end{equation}
which can be written in the abstract index notations as
\begin{equation}
    u^ah_{ab}u^b = \vec u \:\cdot \stackrel{\leftrightarrow}{h} \cdot \:\vec u
\end{equation}
so that 
\begin{equation}
    L_{\vec u} = L_0\left(1 + \frac{1}{2}h_{ab}u^au^b\right) = L_0\left(1 + \frac{1}{2}\vec u \:\cdot \stackrel{\leftrightarrow}{h} \cdot \:\vec u\right)
\end{equation}

Choosing a unit vector $\vec v$ along the other arm, the difference in the roundtrip times down the two arms is given by
\begin{figure*}[!htp]
\begin{subfigure}{0.5\linewidth}
    \includegraphics[width=\linewidth]{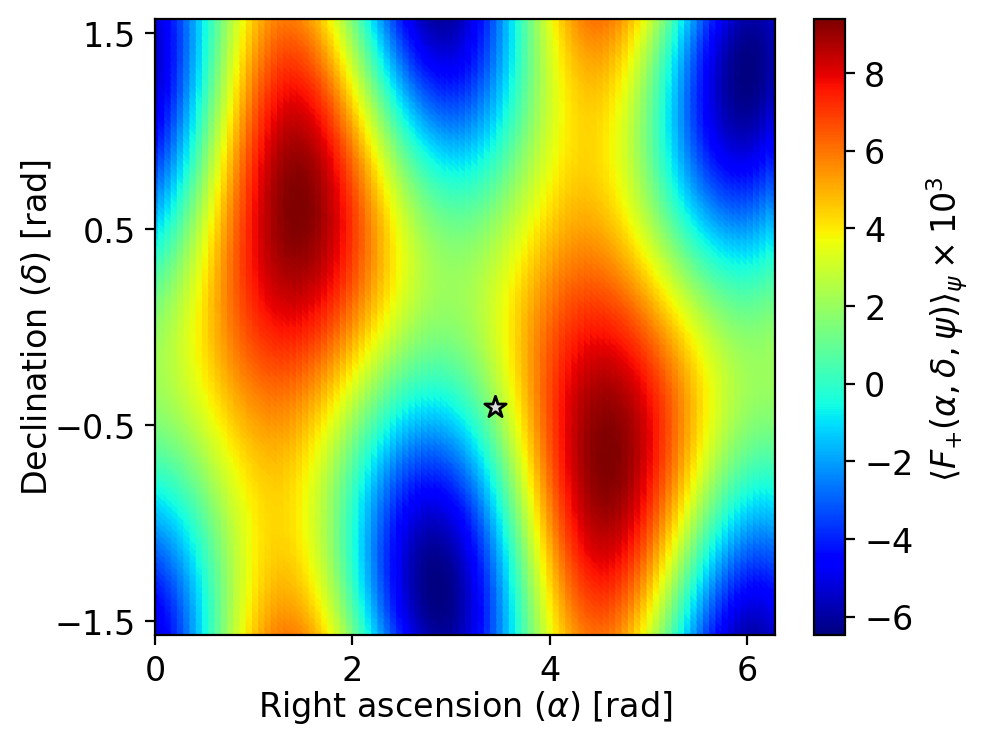}
\end{subfigure}\hfill
\begin{subfigure}{0.5\linewidth}
    \includegraphics[width=\linewidth]{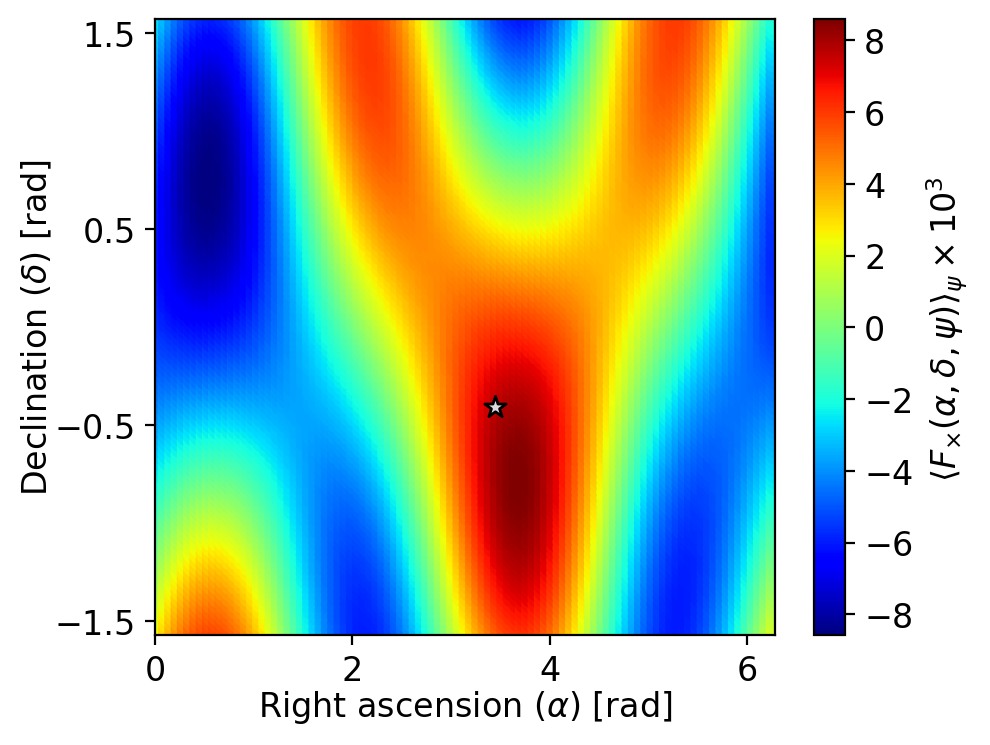}
\end{subfigure}
\caption{Antenna pattern functions $F_{+}$ and $F_{\times}$ averaged over the polarization angle $\psi$, for a given detector (in this case LIGO Livingston ``L1''~\cite{adv_ligo_2015}) at a given sidereal time (in this case the trigger time of GW170817~\cite{abbott2017gw170817, PhysRevX.9.011001} event). The star represents the sky location of the GW170817 BNS event. We used \textsc{PyCBC}~\cite{usman2016pycbc, biwer2019pycbc} (a Python package to analyze CBC events) for generating these plots.} 
\label{fig:antenna_pattern_funcs_plots}
\end{figure*}

\begin{equation}
    \frac{2(L_{\vec u} - L_{\vec v})}{c} = \frac{1}{c}L_0\left(\vec u \:\cdot \stackrel{\leftrightarrow}{h} \cdot \:\vec u - \vec v \:\cdot \stackrel{\leftrightarrow}{h} \cdot \:\vec v\right)
\end{equation}
Dividing the above equation by $2L_0/c$, what we find is the Gravitational wave strain $h(t)$:
\begin{equation}
    \begin{split}
        h &= \frac{1}{2}\left(\vec u \:\cdot \stackrel{\leftrightarrow}{h} \cdot \:\vec u - \vec v \:\cdot \stackrel{\leftrightarrow}{h} \cdot \:\vec v\right)\\
        &= \: h_{ab}\frac{u^au^b - v^av^b}{2} = h_{ab}D^{ab} = \stackrel{\leftrightarrow}{h}\: : \:\stackrel{\leftrightarrow}{D}
    \end{split}
\end{equation}
where $\stackrel{\leftrightarrow}{d}$ is detector tensor, defined as
\begin{equation}
    D^{ab} = \frac{u^au^b - v^av^b}{2} \quad \text{or} \quad \frac{\vec u \otimes \vec u - \vec v \otimes \vec v}{2}
\end{equation}

The detector tensor gives the response of the detector to a GW tensor $\stackrel{\leftrightarrow}{h}$. Using it, we can write down the strain measured by a detector as
\begin{equation}
    \begin{split}
        h &= \: \stackrel{\leftrightarrow}{h}\: : \: \stackrel{\leftrightarrow}{D}\\
        &= \: (h_{+}\stackrel{\leftrightarrow}{e}^{+} + \: h_{\times}\stackrel{\leftrightarrow}{e}^{\times}) \: : \: \stackrel{\leftrightarrow}{D}\\
        &= h_{+}F_{+} + h_{\times}F_{\times}
    \end{split}
    \label{eq:ht_as_lin_comb}
\end{equation}
where $F_+$ and $F_{\times}$ are the antenna pattern functions which are given by
\begin{subequations}
\begin{align}
        F_{+} &= \stackrel{\leftrightarrow}{D}\: : \: \stackrel{\leftrightarrow}{e}^{+} = D^{ab}e_{ab}^{+}  \\  
        F_{\times} &= \stackrel{\leftrightarrow}{D}\: : \: \stackrel{\leftrightarrow}{e}^{\times} = D^{ab}e_{ab}^{\times}
    \end{align}
\end{subequations}
%

For a given detector at a given time (i.e., for a fixed detector tensor $\stackrel{\leftrightarrow}{D}$), $F_{+}$ and $F_{\times}$ depends on the three angles that define the sky location and polarization basis of the GW source with respect to some reference system. For example, in the equatorial coordinate system, they will depend on $\alpha$, $\delta$, and $\psi$. Given the sky location of the source, we can always construct a reference polarization basis $\left\{ \stackrel{\leftrightarrow}{\varepsilon}^{\: +}, \stackrel{\leftrightarrow}{\varepsilon}^{\: \times}\right\}$ and then construct the following combinations:
\begin{subequations}
\begin{align}
        a &= \stackrel{\leftrightarrow}{D}\: : \: \stackrel{\leftrightarrow}{\varepsilon}^{+} = D^{ab}\varepsilon_{ab}^{+}\\  
        b &= \stackrel{\leftrightarrow}{D}\: : \: \stackrel{\leftrightarrow}{\varepsilon}^{\times} = D^{ab}\varepsilon_{ab}^{\times}
    \end{align}
\end{subequations}
using which we can calculate the antenna pattern functions:
\begin{subequations}
\begin{align}
        F_{+}(\alpha, \delta, \psi) &= a(\alpha, \delta)\cos 2\psi + b(\alpha, \delta)\sin 2\psi\\  
        F_{\times}(\alpha, \delta, \psi) &= -a(\alpha, \delta)\sin 2\psi + b(\alpha, \delta)\cos 2\psi
    \end{align}
    \label{eq:antenna_pattern_funcs}
\end{subequations}

From the observations in Fig.~\ref{fig:antenna_pattern_funcs_plots}, it's clear that the antenna pattern functions exhibit a continuous variation across different sky locations. Furthermore, GW detectors have a wide sky coverage spanning $\sim 4\pi$, with only a few specific directions where they are blind. This is in contrast to conventional astronomy, where telescopes must be pointed very precisely towards a source for detection. However, it's important to note that a single GW detector cannot uniquely determine $h_{+}$, $h_{\times}$, $\alpha$, and $\delta$. When using two detectors, we have access to two GW strains, $h_1(t)$ and $h_2(t)$, as well as a time delay $\tau_{12}$. However, when employing three or more detectors, we can acquire at least three or more GW strains and two-time delays. With this additional information, we can confidently and uniquely determine all four unknowns.
\subsection{GW generation in linearized theory}
In the context of linearized theory, as discussed in Section~\ref{sec:1.1}, we assume that the source produces a sufficiently weak gravitational field. This allows us to perform an expansion around the flat space-time. This assumption further implies that within a system primarily influenced by gravitational forces, the typical velocities involved are relatively low. For instance, in a two-body system held together by gravitational interactions and characterized by a reduced mass $\mu$ and total mass $M$,
\begin{equation}
    \frac{1}{2}\mu v^2 = \frac{G\mu M}{r} \quad \Rightarrow \quad v^2/c^2 = R_S/2r
\end{equation}
where $R_s = 2GM/c^2$ is the Schwarzschild radius corresponding to a mass $M$, from the above equation, we can deduce that in the presence of a weak gravitational field, the condition $R_s/r \ll 1$ holds, indicating that $v \ll c$. However, when dealing with systems primarily influenced by non-gravitational forces, we can explore weak field sources with velocities of arbitrary magnitudes. This simplifying assumption allows us to comprehend how gravitational waves (GWs) manifest in flat background space-time, which encompasses both Newtonian and special relativistic dynamics. Note that for many of the astrophysical sources of interest, such as neutron stars, black holes, and compact binary systems, which are inherently self-gravitating, the Newtonian approximation may not be applicable. In such cases, the Post-Newtonian (PN)~\cite{PhysRevD.51.5360, Blanchet_2008, Arun_2009} formalism becomes necessary for a comprehensive understanding of GW generation. However, we do not discuss the PN formalism here. Interested readers seeking a deeper exploration of this topic can refer to Michele Maggiore's book~\cite{maggiore2008gravitational} and Luc Blanchet's Living Review in Relativity~\cite{Blanchet_2014}. Now, let's turn to weak field sources characterized by arbitrary velocities while recalling the Eq.~\eqref{eq:wave_equation1}, 
\begin{equation}
    \Box \Bar{h}_{\mu\nu} = -\frac{16\pi G}{c^4}T_{\mu\nu}
    \label{eq:wave_equation2}
\end{equation}
we found that in the Lorentz gauge, $\partial^{\mu}\bar{h}_{\mu} = 0$ and $T_{\mu\nu}$ satisfies that flat-space conservation law $\partial^{\mu}T_{\mu\nu} = 0$. Since the equation presented above is linear with respect to $h_{\mu\nu}$, we can employ Green's function method (see Eq.~\eqref{eq:green_fn_def} in Appendix~\ref{appendix:definitions_GR_chap1}) to find a solution. The solution to this equation in regions outside the source is as follows:
\begin{equation}
    h_{ij}(t, \vec x) = \frac{4G}{c^4}\Lambda_{ij,kl}(\boldsymbol{\hat{n}})\int d^3x'\frac{1}{|\vec x - \vec x'|}T_{kl}\left(t - \frac{|\vec x - \vec x^{\: '}|}{c}, \vec x^{\: '}\right).
    \label{eq:greens_function_soln}
\end{equation}
where $\Lambda_{ij, kl}$ is the Lambda tensor, which can be written in terms of Kronecker deltas as 
\begin{equation}
    \begin{split}
        \Lambda_{ij, kl}(\boldsymbol{\hat{n}}) &= \delta_{ik}\delta_{jl} - \frac{1}{2}\delta_{ij}\delta_{kl} - n_jn_l\delta_{ik} - n_in_k\delta_{jl} \\
        &+ \frac{1}{2}n_kn_l\delta_{ij} + \frac{1}{2}n_in_j\delta_{kl} + \frac{1}{2}n_in_jn_kn_l
    \end{split}
\end{equation}
Here, we can expand $|\vec x - \vec x^{\: '}|$ at $r \gg d$, where $d$ is the typical radius of the source 
\begin{equation}
    |\vec x - \vec x^{\: '}| = r - \vec x^{\: '} \cdot \boldsymbol{\hat{n}} + \mathcal{O}\left(\frac{d^2}{r}\right),
\end{equation}
To find the gravitational wave amplitude $h_{ij}$ at the detector, which is far away from the source, we can take the limit $r\rightarrow \infty$ at fixed time $t$ and keep only the leading order terms in the Eq.~\eqref{eq:greens_function_soln}. Setting $|\vec x - \vec x^{\: '}| = r$, at large distances, the Eq.~\eqref{eq:greens_function_soln} becomes 
\begin{equation}
    h_{ij}(t, \vec x) = \frac{4G}{rc^4}\Lambda_{ij,kl}(\boldsymbol{\hat{n}})\int d^3x'\frac{1}{|\vec x - \vec x'|}T_{kl}\left(t - \frac{r}{c} + \frac{\vec x^{\: '}\cdot \boldsymbol{\hat{n}}}{c}, \: \vec x^{\: '}\right),
    \label{eq:gw_tensor_at_large_distance}
\end{equation}
and we write $T_{kl}$ in terms of its Fourier tranforms, we get
\begin{equation}
    T_{kl}\left(t - \frac{r}{c} + \frac{\vec x^{\: '}\cdot \boldsymbol{\hat{n}}}{c}, \: \vec x^{\: '}\right) = \int\frac{d^4k}{(2\pi)^4}\tilde{T}_{kl}(\omega, \vec k)e^{-i\omega (t - r/c + \vec x^{\: '}\cdot \boldsymbol{\hat{n}}/c) + i\vec k \cdot \vec x^{\: '}}.
    \label{eq:energy_moment_FT}
\end{equation}
For non-relativistic sources, the Fourier transform of the stress-energy tensor, 
$\tilde{T}_{kl}(\omega, \vec k)$ has a peak around a characteristic frequency $\omega_s$, where $\omega_s d \ll c$. Additionally, inside the source, $\tilde{T}_{kl}$ is non-vanishing, which means that the integral 
in the above equation is restricted to the region $\vec x^{\: '} \leq d$. Therefore, the frequencies $\omega$ which contributes dominantly to $h_{ij}$ satisfy
\begin{equation}
    \frac{\omega}{c}\vec x^{\: '}\cdot \boldsymbol{\hat{n}} \lesssim \frac{\omega_s d}{c} \ll 1,    
\end{equation}
Using the above condition in the Eq.~\eqref{eq:energy_moment_FT}, we get
\begin{equation}
    T_{kl}\left(t - \frac{r}{c} + \frac{\vec x^{\: '}\cdot \boldsymbol{\hat{n}}}{c}, \: \vec x^{\: '}\right) \lesssim T_{kl}(t - \frac{r}{c}, \vec x^{\: '}) + \frac{x^{'i}n^{i}}{c} \partial_0T_{kl} + \frac{1}{2c^2}x^{'i}x^{'j}n^in^j\partial^2_0T_{kl} + ....
\end{equation}
where all the derivatives are evaluated at the point $(t - r/c, \vec x^{\: '})$. Putting the above expansion in Eq.~\eqref{eq:gw_tensor_at_large_distance}, we obtain
\begin{equation}
    h_{ij}(t, \vec x) = \frac{4G}{rc^4}\Lambda_{ij,kl}(\boldsymbol{\hat{n}}) \left[S^{kl} + \frac{1}{c}n_{m}\dot{S}^{kl,m} + \frac{1}{2c^2}n_{m}n_p\ddot{S}^{kl,mp} + ...\right]_{\text{ret}},
    \label{eq:gw_tensor_multipol_basis}
\end{equation}
Here $S$ represents various momentas of stress-energy tensor $T^{ij}$ (see Eq.~\eqref{eq:momentas_of_stres_energy} in  Appendix~\ref{appendix:definitions_GR_chap1}). The subscript ``ret'' indicates that all the quantities are evaluated at retarded time $t - r/c$. We also define second-order momentas of $T^{00}/c^2$ (which will be used later) as
\begin{equation}
    M^{ij} = \frac{1}{c^2}\int d^x T^{00}(t, \vec x) x^i x^j,
    \label{eq:third_momenta_of_T00}
\end{equation}
It turns out that $S^{ij}$ and $M^{ij}$ are related by the following relation,
\begin{equation}
    S^{ij} = \frac{1}{2}\ddot{M}^{ij}.
\end{equation}
Now, using the above equation in Eq.~\eqref{eq:gw_tensor_multipol_basis} and keeping the leading term, we get 
\begin{equation}
    [h_{ij}(t, \vec x)]_{\text{quad}} = \frac{2G}{rc^4}\Lambda_{ij, kl}(\boldsymbol{\hat{n}})\ddot{M}^{kl}(t - r/c).
    \label{eq:gw_quad_eq1}
\end{equation}
where ``quad'' refers to the quadrupolar term. Denoting ${\rho = T^{00}/c^2}$ (in the lowest order in $v/c$, $\rho$ becomes mass density) and defining the ``quadrupole moment'' as
\begin{equation}
    \begin{split}
        Q^{ij} &\equiv M^{ij} - \frac{1}{3}\delta^{ij}M_{kk}\\
        &= \int d^3x \rho(t, \vec x)(x^i x^j -\frac{1}{3}r^2 \delta^{ij}),
    \end{split}
    \label{eq:quadrupole_moment}
\end{equation}
and using the above equation in Eq.~\eqref{eq:gw_quad_eq1}, we obtain

\begin{equation}
    \begin{split}
        [h_{ij}(t, \vec x)]_{\text{quad}} &= \frac{2G}{rc^4}\Lambda_{ij, kl}(\boldsymbol{\hat{n}})\ddot{Q}_{kl}(t - r/c)\\
        &\equiv \frac{2G}{rc^4}\ddot{Q}_{ij}(t - r/c)
    \end{split}
    \label{eq:gw_tensor_quad_exp}
\end{equation}
For a case when the direction of propagation $\boldsymbol{\hat{n}}$ of GW to be $\boldsymbol{\hat{z}}$, the two polarization amplitudes can be written as
\begin{subequations}
    \begin{align}
        h_{+} &= \frac{G}{rc^4}(\ddot{M}_{11} - \ddot{M}_{22})\\
        h_{\times} &= \frac{G}{rc^4}\ddot{M}_{12},
    \end{align}
\end{subequations}
It's important to highlight that the dominant term in Eq.~\eqref{eq:gw_tensor_quad_exp} corresponds to the mass quadrupole and excludes monopole or dipole terms. To introduce the monopole and dipole terms, they would need to depend on mass density and momentum density, which are conserved quantities in linearized theory. Given that $h_{ij}$ depends on the derivatives of the multipole moments, their contribution would vanish. Notably, the absence of these terms remains consistent even in the context of more generalized non-linear theory. Now, let's proceed to formulate the energy flux of GWs, which quantifies the energy carried by these waves per unit time through a unit surface located at a substantial distance from the source.
\begin{equation}
    \frac{dE}{dt} = \frac{c^3r^2}{32\pi G}\int d\Omega \langle \dot{h}_{ij} \dot{h}_{ij}\rangle
\end{equation}
Substituting the Eq.~\eqref{eq:gw_tensor_quad_exp} in the above equation, we find that in the quadrupole approximation, the radiated power per unit solid angle is given by
\begin{equation}
    \begin{split}
        \left(\frac{dP}{d\Omega}\right)_{\text{quad}} &= \frac{r^2c^3}{32\pi G}\langle \dot{h}_{ij} \dot{h}_{ij}\rangle\\
        &= \frac{G}{8\pi c^5}\Lambda_{ij, kl}(\boldsymbol{\hat{n}}) \langle \dddot{Q}_{ij} \dddot{Q}_{ij}\rangle
        \label{eq:pow_rad_per_unit_solid_angle}
    \end{split}
\end{equation}
where the average is taken over over several characteristic periods of the GW, and $\dddot{Q}_{ij}$ is evaluated at the retarded time $t - r/c$. Using Eq.~\eqref{eq:direction_ids} in Appendix~\ref{appendix:definitions_GR_chap1}, we find the total radiative power  
\begin{equation}
    P_{\text{quad}} = \frac{G}{5c^5} \langle \dddot{Q}_{ij} \dddot{Q}_{ij}\rangle, 
\end{equation}
which is the Einstein's famous quadrupole formula~\cite{Einstein:1918btx}. Note that, in the quadrupole approximation, linear momentum is conserved. This can be seen by writing the rate of change of linear momentum as
\begin{equation}
    \frac{dP^i}{dt} = -\frac{G}{8\pi c^5}\int d\Omega \dddot{Q}_{ab} \partial^{i}\ddot{Q}_{ab}
\end{equation}
where the integrand is odd under reflection ($\vec x \rightarrow \vec x^{:'}$), owing to the fact that $Q$ remains invariant under reflection while $\partial^{i} \rightarrow -\partial^{i}$. Consequently, the derivative $dP^i/dt$ becomes zero. We conclude this section by showing quadrupole radiation from a mass in a circular orbit. Consider a binary system comprised of masses $m_1$ and $m_2$ where a circular motion is performed by the relative coordinate. Assuming the lowest order of $v/c$, which allows us to work in a flat space-time for describing a self-gravitating system. We choose a reference frame denoted as $(x, y, z)$, aligning with the binary orbit lying in the $(x, y)$ plane, which is described by the following equation:"
\begin{equation}
    \begin{split}
        x_0(t) &= R\cos(\omega_s t + \pi/2)\\
        y_0(t) &= R\sin(\omega_s t + \pi/2)\\
        z_0(t) &= 0
    \end{split}
\end{equation}
The second mass moment for this binary system is $M^{ij} = \mu x^{i}_{0}(t)x^{j}_{0}(t)$, where $\mu = m_1m_2/(m_1 + m_2)$ is the reduced mass of the system, so the non-vanishing components are 
\begin{equation}
    \begin{split}
        M_{11} &= \mu R^2 \frac{1 - \cos 2\omega_s t}{2}, \\
        M_{22} &= \mu R^2 \frac{1 + \cos 2\omega_s t}{2},\\
        M_{12} &= -\frac{1}{2}\mu R^2\sin(2\omega_s t),
    \end{split}
\end{equation}
\begin{figure*}[!htp]
\centering
\begin{subfigure}{0.5\linewidth}
    \centering
    \includegraphics[width=0.8\linewidth, height=0.8\linewidth]{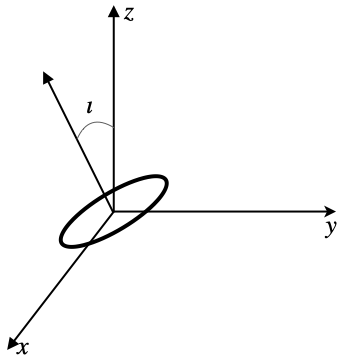}
\end{subfigure}\hfill
\begin{subfigure}{0.4\linewidth}
    \includegraphics[width=0.5\linewidth, height=1\linewidth]{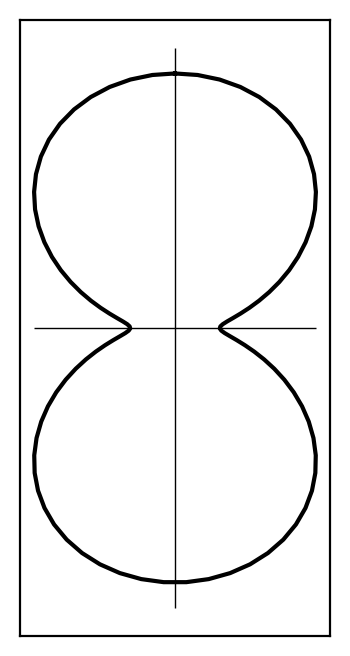}
\end{subfigure}
\caption{The left panel shows the coordinate system chosen for the binary system we have considered here. The right panel shows $g(\iota)$ in polar coordinates where $\iota$ is measured from the verticle axis (reproduced from~\cite{maggiore2008gravitational}).} 
\label{fig:coordinate_sys_g_theta}
\end{figure*}
Using the above equations, we get $\ddot{M}_{11} = 2\mu R^2\omega_s^2 \cos(2\omega_s t)$ and $\ddot{M}_{12} = 2\mu R^2\omega_s^2 \sin(2\omega_s t)$ while $\ddot{M}_{22} = - \ddot{M}_{11}$. Using these expressions (see Eq. $(3.72)$ in~\cite{maggiore2008gravitational}), we find 
\begin{subequations}
    \begin{align}
        h_{+}(t; \theta, \phi) &= \frac{4G\mu \omega_s^2 R^2}{rc^4}\left(\frac{1 + \cos^2 \theta}{2}\right) \cos(2\omega_s t_{\text{ret}} + 2\phi),\\
        h_{\times}(t; \theta, \phi) &= \frac{4G\mu \omega_s^2 R^2}{rc^4}\cos \theta \sin(2\omega_s t_{\text{ret}} + 2\phi)
    \end{align}
    \label{eq:plus_cross_pols_point_binary_with_phi}
\end{subequations}
Notice that the quadrupole radiation occurs at a frequency that is precisely ``twice" the source frequency $\omega_s$. When observed, we only detect the radiation that is emitted in our direction from the binary system. Therefore, the angle $\theta$ corresponds to the inclination angle $\iota$ between the normal to the orbital plane and our line of sight. To simplify the equations, we can fix $\phi$ at zero. Consequently, the equations above can be expressed as follows (see~\cite{maggiore2008gravitational} for detail):
\begin{subequations}
    \begin{align}
        h_{+}(t; \theta, \phi) &= \frac{4G\mu \omega_s^2 R^2}{rc^4}\left(\frac{1 + \cos^2 \iota}{2}\right) \cos(2\omega_s t_{\text{ret}}),\\
        h_{\times}(t; \theta, \phi) &= \frac{4G\mu \omega_s^2 R^2}{rc^4}\cos \iota \sin(2\omega_s t_{\text{ret}})
    \end{align}
    \label{eq:plus_cross_pols_point_binary}
\end{subequations}

For $\iota =  0$ (corresponding to a face-on orbit), both $h_{+}$ and $h_{\times}$ have identical amplitude, and GW becomes circularly polarized. In contrast, In contrast, when $\iota$ equals $\pi/2$ (representing an edge-on orbit), $h_{\times}$ becomes zero, and the gravitational wave radiation becomes linearly polarized. For $0\geq \iota \leq \pi/2$, the $h_{+}$ and $h_{\times}$ amplitudes are different, causing the gravitational waves to exhibit elliptical polarization. It's worth noting that by measuring the relative amplitudes of these two polarizations, the inclination angle $\iota$ of the orbit can be estimated. Using Eq.~\eqref{eq:pow_rad_per_unit_solid_angle} and insert Eq.~\eqref{eq:plus_cross_pols_point_binary}, the angular distribution of the radiative power is 
\begin{equation}
    \left(\frac{dP}{d\Omega}\right)_{\text{quad}} = \frac{2G\mu^2R^4\omega_s^6}{\pi c^5}g(\iota),
    \label{eq:power_per_solid_angle}
\end{equation}
where $g(\iota) = (1 + \cos^2\iota)^2/2 \: + \: \cos^2\iota$. It is worth mentioning that $g(\theta)$ never vanishes irrespective of the inclination angle $\iota$, implying the non-vanishing component of the source's motion perpendicular to the light of sight of the observer as evident from Fig.~\ref{fig:coordinate_sys_g_theta}. In the next section, we will discuss the inspiral of compact binaries, which are the most interesting astrophysical sources of GWs.

\begin{figure*}[!hbt]
    \centering
    \includegraphics[width=\linewidth]{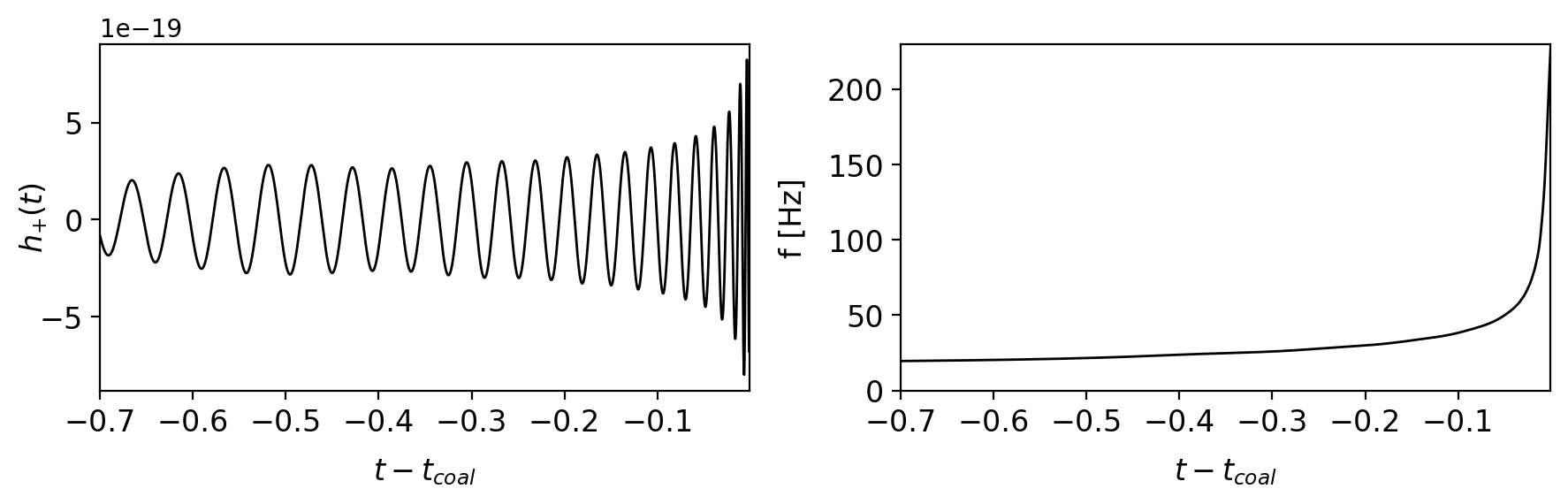}
    \caption{\textit{Left panel}: The time evolution of the GW amplitude $h_{+}(t)$ in the inspiral phase of compact binary of equal mass of $35 \: M_{\odot}$. The binary spends $\sim 0.7$ seconds in the frequency band starting from $20$ Hz before merging. \textit{Right panel}: The corresponding frequency evolution of the GW signal exhibiting a typical chirp-like behavior. We used \textsc{PyCBC}~\cite{usman2016pycbc, biwer2019pycbc} (a Python package to analyze CBC events) for generating these plots.}
    \label{fig:gw_inspiral}
\end{figure*}

\subsection{Inspiraling compact binaries}
In the preceding section, we considered a scenario involving a binary system comprising two point masses, denoted as $m_1$ and $m_2$, which were in a circular orbit. By invoking Kepler's law, we can establish a relation between the orbital frequency $\omega_s$ and the orbital radius $R$ as follows:
\begin{equation}
    \omega^2_s = \frac{GM}{R^3} \quad \Rightarrow \quad R = \left(\frac{GM}{\omega^2_s}\right)^{1/3}
\end{equation}
We substitue $R$ from the above equation in Eq.\eqref{eq:plus_cross_pols_point_binary_with_phi} and write $\omega_{\text{gw}} = 2\omega_s = 2\pi f_{\text{gw}}$ and find
\begin{subequations}
    \begin{align}
        h_{+}(t) &= \frac{4}{r}\left(\frac{G\mathcal{M}}{c^2}\right)^{5/3}\left(\frac{\pi f_{\text{gw}}}{c}\right)^{2/3}\frac{1 + \cos^2\iota}{2}\cos(2\pi f_{\text{gw} t_{\text{ret}}} + 2\phi)\\
        h_{\times}(t) &= \frac{4}{r}\left(\frac{G\mathcal{M}}{c^2}\right)^{5/3}\left(\frac{\pi f_{\text{gw}}}{c}\right)^{2/3}\cos \iota \sin(2\pi f_{\text{gw} t_{\text{ret}}} + 2\phi),
    \end{align}
    \label{eq:hp_hc_inspiraling}
\end{subequations}
where $\mathcal{M}$ is the chirp mass ${\mathcal{M} = \mu^{3/5}M^{2/5} = (m_1m_2)^{3/5}/(m_1 + m2_2)^{1/5}}$. Note that in the quadrupole approximation, the amplitudes $h_{+}$ and $h_{\times}$ (see Fig.~\ref{fig:gw_inspiral}) of the GWs emitted depend on the masses $m_1$ and $m_2$ only through the combination $\mathcal{M}$.
Similarly, we write the Eq.~\eqref{eq:power_per_solid_angle} in terms of $\mathcal{M}$ and $\omega_{\text{gw}}$
\begin{equation}
    \frac{dP}{d\Omega} = \frac{2}{\pi}\frac{c^5}{G}\left(\frac{G\mathcal{M}\omega_{\text{gw}}}{2c^3}\right)^{10/3}g(\iota),
    \label{eq:pow_rad_mc}
\end{equation}
The total radiated power $P$ is therefore
\begin{equation}
    P = \frac{32}{5}\frac{c^5}{G}\left(\frac{G\mathcal{M}\omega_{\text{gw}}}{2c^3}\right)^{10/3}.
    \label{eq:total_radiated_power}
\end{equation}
where we have used the fact that the angular average of $g(\iota)$ is $4/5$. As the GWs are emitted, the orbital energy decreases, implying the $R$ must decrease over time. This reduction in $R$ results in an increase in $\omega_s$, which leads to a further increase in the radiated power. It starts a runaway process which, on a sufficiently long time scale, results in a coalescence of the binary system. As long as $\dot{\omega}_s \ll \omega_s^2$, the binary is in the regime called ``quasi-circular'' which can be translated to $\dot{R} = -2/3 (\omega_sR)\dot{\omega}_s/\omega_s^2$ where $|\dot{R}|\ll \omega_sR$ as long as the quasi-circular condition holds. To calculate the rate of change of the GW frequency, we can equate the rate of change of orbit energy $\dot{E}_{\text{orbit}}$ to the total power radiated given in the Eq.~\eqref{eq:total_radiated_power} 
\begin{equation}
    \dot{f}_{\text{gw}} = \frac{96}{5}\pi^{8/3} \left(\frac{G\mathcal{M}}{c^3}\right)^{5/3}f^{11/3}_{\text{gw}}.
\end{equation}
which, when written in terms of the time to coalescence $\tau \equiv t_{\text{coal}} - t$ with $t$ as the observer time and $t_{\text{coal}}$ as the time of coalescence, is given by
\begin{equation}
    f_{\text{gw}}(\tau) = \frac{1}{\pi}\left(\frac{5}{256\:\tau}\right)^{3/8}\left(\frac{G\mathcal{M}}{c^3}\right)^{-5/8}.
\end{equation}
where $f_{\text{gw}}$ diverges at the time of coalescence $f_{\text{gw}}$ beyond which the quadrupole approximation breaks down. Using the above equation, we can calculate the time to coalescence $\tau$
\begin{equation}
    \tau \simeq 2.18s \left(\frac{1.21\: M_{\odot}}{\mathcal{M}}\right)^{5/3}\left(\frac{100\:\text{Hz}}{f_{\text{gw}}}\right)^{8/3}.
    \label{eq:time_to_coalscence}
\end{equation}
To estimate the typical time to coalescence, $\tau$, we consider a binary system of two neutron stars with equal masses $m_1 = m_2 = 1.4 \: M_{\odot}$ at the $f_{\text{gw}} = 10 \: \text{Hz}$, we get around $\tau = 17 \:\text{min}$ to coalescence. We can also calculate the number of cycles $\mathcal{N}_{\text{cyc}}$ spent within a frequency band accessible to a detector as
\begin{subequations}
    \begin{align}
        \mathcal{N}_{\text{cyc}} &= \int_{f_{\text{min}}}^{f_{\text{max}}} df_{\text{gw}}\frac{f_{\text{gw}}}{\dot{f}_{\text{gw}}}\\
        &\simeq 1.6 \times 10^4 \left(\frac{10 \: \text{Hz}}{f_{\text{min}}}\right)^{5/3}\left(\frac{1.2\: M_{\odot}}{\mathcal{M}}\right)^{5/3},
    \end{align}
    \label{eq:number_of_GW_cycles}
\end{subequations}
From the equation above, we can infer that ground-based GW detectors can access thousands of cycles throughout the GW signal's evolution, whereas space-borne detectors can record millions of cycles, highlighting the importance of precisely modeling GW waveforms. When considering the GW's evolution, we can reasonably assume that $\dot{R}$ is negligible, as long as $f_{\text{gw}} \ll 13\text{kHz} (1.2\,M_{\odot}/\mathcal{M})$. Consequently, we can reframe Eq.~\eqref{eq:hp_hc_inspiraling} in terms of the observer's measurement of the time to coalescence $\tau$.
\begin{subequations}
    \begin{align}
        h_{+}(t) &= \frac{1}{r}\left(\frac{G\mathcal{M}}{c^2}\right)^{5/4}\left(\frac{5}{c\tau}\right)^{1/4}\frac{1 + \cos^2\iota}{2}\cos[\Phi(\tau)] = A_{+}\cos[\Phi(\tau)] \\
        h_{\times}(t) &= \frac{1}{r}\left(\frac{G\mathcal{M}}{c^2}\right)^{5/4}\left(\frac{5}{c\tau}\right)^{1/4}\cos \iota \sin[\Phi(\tau)] = A_{\times}\sin[\Phi(\tau)],
    \end{align}
    \label{eq:hp_hc_inspiraling_with_tau}
\end{subequations}
where $\Phi(\tau) = -2\left(\frac{5G\mathcal{M}}{c^3}\right)^{-5/8}\tau^{5/8} + \Phi_{0}$, and $\Phi_{0}$ is the phase at coalescence. As evident from the above equations, it is clear that both the frequency and amplitude of the signal increase as the binary system approaches its merger. Up to this point, all calculations have been conducted under the assumption of a flat background. In a realistic scenario involving compact binaries like binary neutron stars or binary black holes, the gravitational field is significantly influenced by the proximity of these massive objects, impacting the binary's dynamics. In accordance with the Schwarzschild geometry, there exist a minimum radial distance beyond which a stable circular orbit is not possible. This critical boundary is known as the Innermost Stable Circular Orbit (ISCO) and is defined as:
\begin{equation}
    r_{\text{ISCO}} = \frac{6GM}{c^2}
    \label{eq:schwarchild_radius}
\end{equation}
where $M$ is the total mass of the binary. The corresponding frequency beyond which the condition of ``quasi-circular'' orbits no longer holds can be written as
\begin{equation}
    \begin{split}
        f_{\text{ISCO}} &= \frac{1}{6\sqrt{6}(2\pi)}\frac{c^3}{GM}\\
        &\simeq 2.2 \text{kHz}\left(\frac{M_{\odot}}{M}\right)
    \end{split}
    \label{eq:fISCO}
\end{equation}
For a binary neutron star system with both masses being approximately $1.4\, M_{\odot}$, the ISCO frequency, denoted as $f_{\text{ISCO}}$, is estimated to be $\sim 800$ Hz. In the case of a binary black hole system where both masses are roughly $15\, M_{\odot}$, the ISCO frequency is $\sim 73$ Hz. In conclusion, we will now write down the Fourier transform of the chirp signal presented in Eq.~\eqref{eq:hp_hc_inspiraling_with_tau}. The corresponding Fourier transforms are as follows (see~\cite{maggiore2008gravitational} for derivation):
\begin{subequations}
    \begin{align}
        \tilde{h}_{+}(f) &= Ae^{i\Psi_{+}(f)}\frac{c}{r}\left(\frac{G\mathcal{M}}{c^3}\right)^{5/6}\frac{1}{f^{7/6}}\left(\frac{1 + \cos^2\iota}{2}\right), \\
        \tilde{h}_{\times}(f) &= Ae^{i\Psi_{\times}(f)}\frac{c}{r}\left(\frac{G\mathcal{M}}{c^3}\right)^{5/6}\frac{1}{f^{7/6}}\cos\iota,
    \end{align}
\end{subequations}
where $A = 1/\pi^{2/3} (5/24)^{1/2}$ and phases are given by 
\begin{equation}
    \Psi_{+}(f) = 2\pi f(t_c + r/c) - \Phi_0 - \frac{\pi}{4} + \frac{3}{4}\left(\frac{8\pi f G\mathcal{M}}{c^3}\right)^{-5/3},
\end{equation}
where $\Phi_0$ is phase of coalescence, with phase for cross-polarization is $\Psi_{\times} = \Psi_{+} + \pi/2$. The next section will discuss the first indirect and direct evidence of GWs.
\subsection{Indirect evidence and direct evidence of GWs}
\subsubsection{Indirect evidence: Hulse-Taylor binary pulsar}
\label{par:indirect_evidence_HT_pulsar}
In July 1974, Hulse and Taylor made a breakthrough discovery when they detected a pulsar called PSR B1913+16, also known as the Hulse-Taylor binary pulsar~\cite{1975ApJ...195L..51H, RevModPhys.66.699, RevModPhys.66.711}. This pulsar, characterized by a period of approximately $59$ ms, was observed using the Arecibo Observatory in Puerto Rico. Pulsars are rapidly rotating neutron stars with intense magnetic fields that emit electromagnetic radiation in the form of a beam through their magnetic poles. This radiation becomes observable to us only when the beam is oriented toward Earth, much like a lighthouse effect. 
\begin{figure*}[!hbt]
\centering
\begin{subfigure}{0.45\linewidth}
    \centering
    \includegraphics[width=\linewidth]{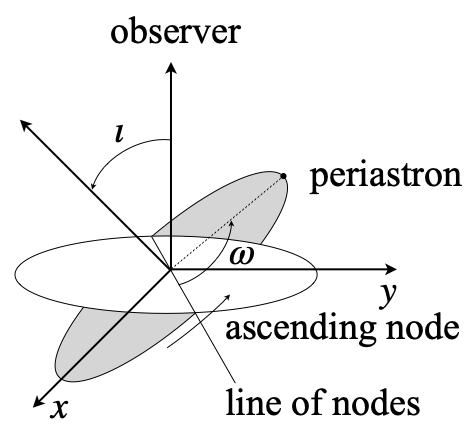}
\end{subfigure}\hfill
\begin{subfigure}{0.49\linewidth}
    \includegraphics[width=\linewidth, height=1\linewidth]{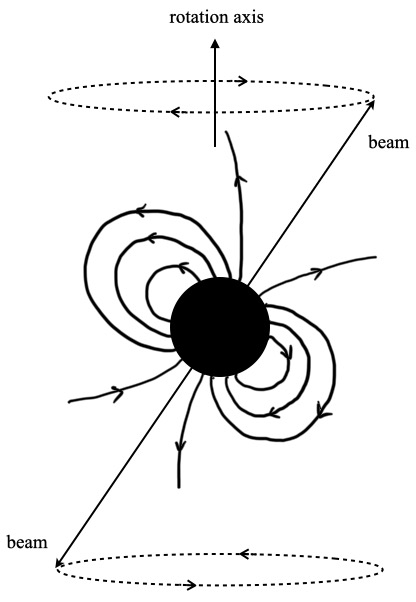}
\end{subfigure}
\caption{\textit{Left panel}: The pulsar's orbit geometry where the orbital plane is shown in grey. \textit{Right panel}: A schematic of the magnetosphere of the pulsar and the beams of outgoing radiation (reproduced from~\cite{maggiore2008gravitational}).} 
\label{fig:pulsar_orbit_geom_and_em_jet}
\end{figure*}
The Hulse-Taylor binary pulsar exhibits an orbital period of around $7.6$ hours and an orbital velocity of approximately $10^{-3}$ times the speed of light ($c$), indicating its relativistic nature. An intriguing aspect of this observation is the rate of periastron advance, denoted as $\dot{\omega}$ which is $\sim 4.22$ degrees per year, significantly larger than the perihelion advance of Mercury, which is $\sim 43$ arcseconds per century~\cite{1986SvA....30..365G, Will_2018}. This highlights the substantial impact of GR effects in the Hulse-Taylor pulsar system. Through the determination of $\langle \dot{\omega}\rangle$, where $\omega$ represents the angular position of the periastron, measured from the ascending node and the Einstein parameter $\gamma$, which characterizes second-order Doppler-gravitational-redshift timing, the masses of the pulsar ($m_p$) and its companion ($m_c$) can be measured and found to be approximately $1.4412 \: M_{\odot}$ and $1.3867 \: M_{\odot}$, respectively. Utilizing these values of masses along with the orbital period, the semimajor axis ($a$) is calculated using Kepler's formula, yielding an estimate of roughly $3 \: R_{\odot}$. The compactness of the orbit and the absence of eclipses during observations strongly suggest that the companion is a compact star, possibly a binary black hole (BBH) or binary neutron star (BNS). However, the value of $m_c$ indicates that it is more likely a BNS. Finally, $\dot{P}$, the rate of change of the orbital period, can be measured by combining information about the masses and eccentricity, using the predictions of GR regarding the decrease in the orbital period due to the emission of GW. The ratio of the experimentally determined value of $\dot{P_b}$ ($(\dot{P}_{b})_{\text{exp}}$) to the value predicted by GR ($(\dot{P}_{b})_{\text{GR}}$) comes out to be
\begin{equation}
    \frac{(\dot{P}_{b})_{\text{exp}}}{(\dot{P}_{b})_{\text{GR}}} \sim 1.0013
    \label{eq:agreement_bw_exp_gr}
\end{equation}
This discovery provided a remarkable validation of GR and the existence of GWs. Next, we will discuss the process of detecting the signals emitted by the pulsar and explore the various factors influencing the detection.
\paragraph{Pulsar as stable clocks}
As previously mentioned, pulsars are rapidly rotating, magnetized neutron stars with short rotational periods on the order of milliseconds, a characteristic resulting from the conservation of angular momentum. Additionally, the conservation of magnetic flux leads to the presence of extremely strong magnetic fields, reaching magnitudes as high as $10^{12}$ gauss. These powerful magnetic fields influence the behavior of charged particles in the vicinity of the pulsar. Within a critical distance from the rotation axis, denoted as $\rho_c = c/\Omega$ (where $\rho$ represents the distance from the rotation axis in cylindrical coordinates), magnetic field lines are closed. Beyond this critical radius, at $\rho > \rho_c$, magnetic field lines are open, allowing charged particles to escape into space. This region, known as the magnetosphere, encompasses a cylindrical volume with a radius of $\rho_c$ and is filled with ionized, high-energy plasma that co-rotates with the neutron star. In the magnetosphere, the high-energy particles follow a constrained path along the open magnetic field lines.
This motion gives rise to EM radiation, which is primarily observed in the form of narrow radio waves emitted from the magnetic poles of the pulsar. Each of these radiation beams traces out a circle in the sky, and if one of these beams intersects the observer's line of sight on Earth, it results in the detection of pulses. To study and characterize pulsars, a coherent addition of these pulses is performed, creating a single, stable profile that acts as a unique fingerprint for each pulsar. The resulting averaged profile becomes a template for further analysis. A least square fit is employed to obtain the time-of-arrival (TOAs) of the individual pulses. The data consisting of TOAs, collected over time, offers remarkable precision and allows researchers to determine the exact number of pulses that occurred between the closure of the Arecibo telescope in the mid-1990s and the present day.

\begin{figure*}[!hbt]
    \centering
    \includegraphics[width=0.5\linewidth, height=0.6\linewidth]{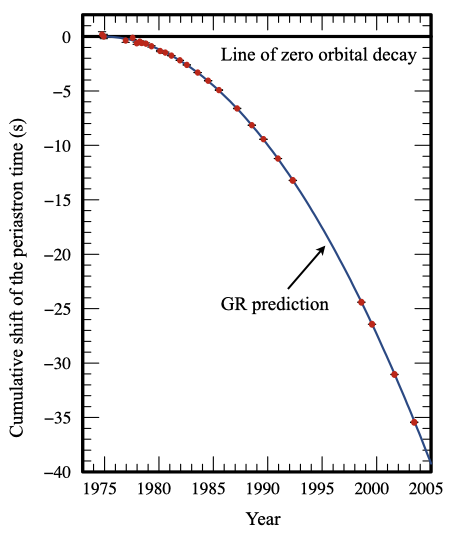}
    \caption{The plot depicts the orbital decay of the Hulse-Taylor pulsar PSRB1913+16~\cite{1975ApJ...195L..51H, weisberg2004relativistic}. The red dots represent the observed variations in the epoch of periastron with time. Meanwhile, the parabolic curve, illustrated by the solid blue line, represents the GR prediction for a system emitting gravitational radiation.}
    \label{fig:HT_pulsar_theory_vs_exp}
\end{figure*}

While the times of arrival (TOAs) of these pulses from pulsars demonstrate remarkable stability, their precision can be influenced by various time-dependent factors. These factors include the Earth's motion around the Sun and the gravitational effects of the solar system as predicted by general relativity. Additionally, in cases where the pulsar is part of a binary system, there are further modulations to consider when calculating the TOAs of the pulses. Since TOAs are dependent on the specific parameters of the binary system, such as the component masses, their measurements can be used to estimate these parameters. Now, we will briefly discuss various factors affecting the TOAs of the pulses.
\paragraph{Time-delays and Dispersion}
There are three kinds of delays that can affect the TOAs of the pulses arriving from the pulsars: (i) Roemer time delay, (ii) Shapiro time delay, and (iii) Einstein time delay. The Roemer time delay is given by 
\begin{equation}
    \Delta_{R, \odot} = -\vec r_{\text{ob}} \cdot \boldsymbol{\hat{s}}/c,
\end{equation}
where $r_{\text{ob}} = r_{\text{oe}} + r_{\text{es}} + r_{\text{sb}}$. Here $r_{\text{oe}}$ is the vector pointing from the observer to the Earth's center, $r_{\text{es}}$ from the Earth's center to the Sun's center, $r_{\text{sb}}$ from the Sun's center to the solar system barycenter (SSB), and $\boldsymbol{\hat{s}}$ is the unit vector from the SSB to the pulsar. Taking into account that light's propagation is also affected by the general relativistic effects due to the solar system's gravitational field, the arrival time $t_{\text{obs}}$ at the observer is given by
\begin{equation}
    t_{\text{SSB}} = t_{\text{obs}} + \Delta_{R, \odot} + \frac{2}{c}\int_{\vec r_{\text{obs}}}^{\vec r_{\text{p}}} |d\vec x| \phi(\vec x).
\end{equation}
In this equation, $\phi(\vec x)$ represents the metric perturbation, and $t_{\text{SSB}} = t_e + |\vec r_p - r_b|/c$ corresponds to the time when the pulse would reach the Solar System Barycenter (SSB) without the influence of the solar system's gravitational effects. Here, $\vec r_p$ and $\vec r_b$ denote the positions of the pulsar and the SSB, respectively. The final term, preceded by a negative sign, denotes the Shapiro time delay~\cite{PhysRevLett.13.789} due to the solar system, represented as $\Delta_{S, \odot}$. It's important to note that a laboratory clock measures its proper time $\tau$ rather than the coordinate time $t$ utilized in the calculations of the previous two delays. Using the relationship between coordinate time $t$ and proper time $\tau$, we can express this as:
\begin{equation}
    t \simeq \tau + \int^{t}dt' \left[\frac{v_{\text{obs}}^2(t')}{2c^2} - \phi (\vec x_{\text{obs}}(t'))\right]
\end{equation}
where second term represents the Einstein time delay $\Delta_{E, \odot}$. In addition to the previously mentioned delays, there is another correction applied to the TOAs due to the radio waves traveling through ionized interstellar gas, causing the waves to disperse. This dispersion effect correction is expressed as $-D/\nu^2$, with $D = (e^2/2\pi m_e c)$ representing the Dispersion Measure (DM) and $\nu$ indicating the frequency of the radio pulse. The Dispersion Measure, often referred to as DM, is calculated as DM $\equiv \int_0^L n_e dl$, where $n_e$ signifies the electron number density, and $L$ is the distance traveled. To account for all these factors, we can combine the various delays and dispersion effects as follows:
\begin{equation}
    t_{\text{SSB}} = \tau_{\text{obs}} - \frac{D}{\nu^2} + \Delta_{E, \odot} + \Delta_{R, \odot} - \Delta_{S, \odot}.
\end{equation}
where $t_{\text{SSB}}$ is the coordinate time at which laboratory clock records the signal, and $\tau_{\text{obs}}$ is observer's proper time. Up to this point, we have exclusively focused on the case of a single pulsar emitting electromagnetic pulses. In scenarios where the pulsar is part of a binary system, there exist additional delays and corrections that must be integrated into the final timing formula. However, the detailed treatment of such complexities falls outside the scope of this thesis. Readers interested in exploring these intricacies further are encouraged to consult Section $6.2.3$ of Michele Maggiore's Volume 1~\cite{maggiore2008gravitational}. To conclude this section, we highlight several crucial observations stemming from the Hulse-Taylor binary system. Through a least-squares fitting of Time of Arrivals (TOAs), we can extract all the required parameters of the binary system, accounting for delays, dispersion, and necessary corrections. Referring to Equation \eqref{eq:agreement_bw_exp_gr}, we identify a remarkable alignment between the observed rate of change in the orbital period and the value predicted by General Relativity. The corresponding accumulated phase $\phi_b$ is 
\begin{equation}
    \frac{\phi_b(T)}{2\pi} = \nu_b T + \frac{1}{2} \dot{\nu}_b T^2 + ...,
\end{equation}
where $\nu_b$ is the orbital frequency, and $T$ is the proper time in the pulsar frame. The $n$-$\text{th}$ time of periastron passage, $T_n$ is defined by the condition, $\phi_{b}(T_n) = 2\pi n$
\begin{equation}
    \nu_bT_n + \frac{1}{2}\dot{\nu}_bT_n^2 = n.
\end{equation}
The cumulative difference between the periastron passages $T_n$ and the values $nP_b$ can be written as follows:
\begin{equation}
    T_n - nP_b = \frac{\dot{P}_b}{2P_b}T_n^2.
\end{equation}
which is equation of a parabola with $\dot{P}_b/(2P_b) < 0$ as the coefficient. As depicted in Fig.~\ref{fig:HT_pulsar_theory_vs_exp}, the agreement between the theoretical predictions and the experimental data is exceptional and has demonstrated remarkable consistency over the years. Recent assessments of the Hulse-Taylor pulsar suggest that it will cease to be observable after the year 2025~\cite{Kramer_1998, Weisberg_2002}. This is due to its geodetic precession, which arises from the effects of general relativity associated with the spin-orbit coupling.
\subsubsection{Direct evidence of GWs}
\paragraph{GW150914 BBH event}
On September 14, 2015, approximately a century after Einstein's initial prediction, a significant milestone was achieved when the two ground-based LIGO detectors, LIGO-Livingston and LIGO-Hanford~\cite{adv_ligo_2015}, jointly observed a simultaneous gravitational wave event referred to as ``GW150914"~\cite{abbott2016observation, PhysRevLett.116.241102}. This event originated from a binary black hole system with source component masses of $\sim 36\:M_{\odot}$ for one black hole and $\sim 29:M_{\odot}$ for the other, resulting in a final black hole with a mass of approximately $62\:M_{\odot}$ and a final black hole spin of $a_f\sim 0.67$. The source was estimated to be situated at a distance of approximately $410$ million parsecs, corresponding to a redshift of around $z\sim 0.099$. The detection of GW150914 marked a historic moment, offering the first direct confirmation of GW's existence and reaffirming the validity of Einstein's theory of General Relativity. This breakthrough ushered in a new era of gravitational wave astronomy.
\begin{figure*}[!hbt]
    \centering
    \includegraphics[width=1\linewidth, height=0.68\linewidth]{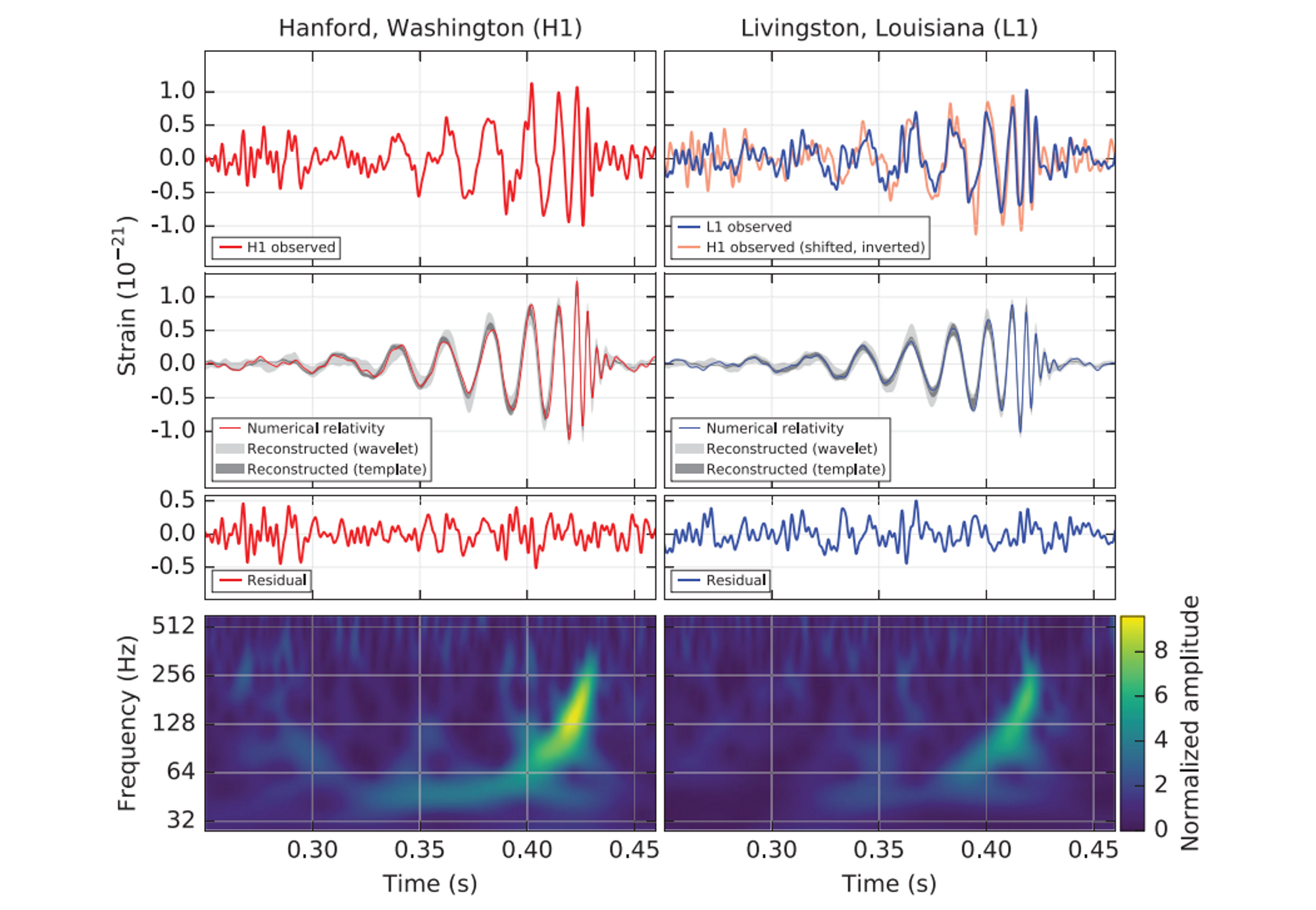}
    \caption{\textit{Top row}: The GW strain (bandpased) as observed by H1 (left) and L1 (right). \textit{Second row}: Gravitational-wave strain projected onto each
    detector in the 35–350 Hz band. The solid red line represents the numerical relativity waveform generated for a system with parameters consistent with those recovered from GW150914. Dark (light) grey areas show $90\%$ credible region of the signal reconstructed using BBH template waveforms (a linear combination of sine-Gaussian wavelets). \textit{Third row}: Residuals after subtracting the filtered numerical relativity waveform from the filtered detector time series. \textit{Bottom row}: A time-frequency map of the strain data, showing a typical chirp behavior of the BBH signal (taken from~\cite{abbott2016observation}).}
    \label{fig:GW150914_imag1}
\end{figure*}
\paragraph{GW170817 BNS event: First multimessenger event} 
The GW170817 binary neutron star (BNS) event marked a significant milestone as the first multimessenger event detected by LIGO detectors~\cite{adv_ligo_2015, AdVirgo}, during the second observation run (O2)~\cite{abbott2017gw170817}. Its electromagnetic (EM) counterpart was also observed by various EM facilities worldwide~\cite{2017ApJ...848L..12A, doi:10.1126/science.aap9580}. This event had a combined Signal-to-Noise Ratio (SNR) of $32.4$, featuring component masses of approximately $1.4\, M_{\odot}$ and $1.2\, M_{\odot}$ and component spins that were nearly zero. It originated from a host galaxy known as ``NGC4993"~\cite{x_ray_counterpart_gw170817} positioned at a distance of roughly $40$ million parsecs from Earth. Approximately $1.7$ seconds after the GW detection, a gamma-ray burst was also observed, which played a crucial role in constraining the propagation speed of gravitational waves. These multimessenger events provide valuable insights into constraining the physical models governing the emission processes. Consequently, early detection of such events becomes imperative, necessitating the development of an accurate and rapid parameter estimation technique to estimate the source's sky location and component masses. One such technique will be elaborated upon in Chapter~\ref{chap:chapter_4}, where we discuss the rapid parameter estimation for BNS/NSBH sources.
\begin{figure}[!hbt]
    \centering
    \includegraphics[width=0.75\linewidth]{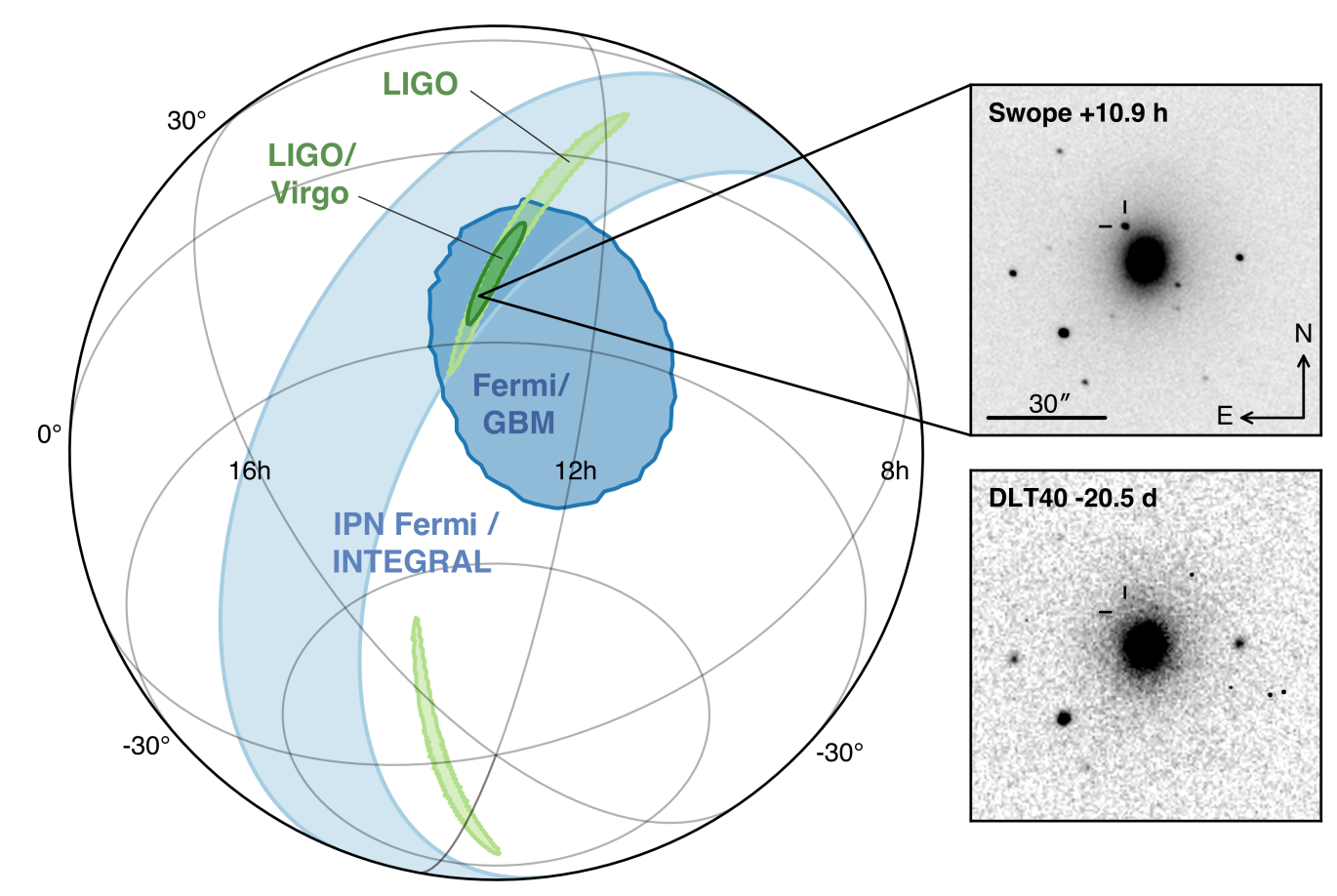}
    \caption{The localization of the GW170817 event, along with the gamma-ray and optical signals, is illustrated in the figure. In the left panel, skymaps indicating the $90\%$ credible regions are displayed. These regions are represented by light green for LIGO (dark green for LIGO-Virgo), IPN triangulation based on the time delay between Fermi and INTEGRAL (light blue), and Fermi-GBM. Additionally (dark blue), the inset provides a view of the apparent location of the host galaxy NGC $4993$. It is also depicted using the Swope optical discovery image captured approximately $10.9$ hours after the merger (top right). The bottom right image shows the DLT40 per-discovery image taken roughly $20.5$ days prior to the merger (taken from~\cite{2017ApJ...848L..12A}).}
    \label{fig:GW170817_LIGO_sky_map}
\end{figure}
\subsection{Detection strategies}
In the preceding sections, we examined the generation and sources of GWs. In this section, we will explore methods for detecting these signals. Given that GW signals are exceedingly faint and often obscured by various types of significantly stronger noise sources (as discussed in Section~\ref{subsec:noise_sources}), it is imperative to develop optimal techniques to extract these signals from the noise. The output of a GW detector typically takes the form of a time series that describes various parameters, such as the motion of test masses or phase shifts in light. Naturally, we anticipate that this output contains the real signal along with the noise. A GW detector can be viewed as a linear system, where the input represents the GW signal $h(t) = D^{ij}h_{ij}$, with $h_{ij}$ characterizing the GW and $D^{ij}$ denoting the detector's sensor (as detailed in Section~\ref{sec:interaction_with_det}), which is dependent upon the detector's geometry. The resulting output is a time series expressed as follows:
\begin{equation}
    d(t) = h(t) + n(t)
\end{equation}
Assuming the noise is stationary and the corresponding Fourier components uncorrelated, their ensemble average is given by
\begin{equation}
    \langle \tilde{n}^{*}(f)\tilde{n}(f')\rangle = \delta(f - f')\frac{1}{2}S_n(f)
    \label{eq:deftn_psd}
\end{equation}
where $S_n(f)$ is called the ``single-sided'' power spectral density and it has dimensions $\text{Hz}^{-1}$. The term single-sided comes from the fact that only positive frequencies are used to calculate the $\langle n^2(t)\rangle$. From the above equation, we can write 
\begin{equation}
    \langle|\tilde{n}(f)|^2\rangle = \frac{1}{2}S_n(f)T
\end{equation}
where we are restricting ourselves to a time interval $ -T/2 < t < T/2$ for calculating $\tilde{n}(f)$ which implies that the frequency resolution is given by is $\Delta f = 1/T$. Now, we will discuss the technique of matched filtering in the context of GWs.
\subsubsection{Matched filtering}
As discussed in the previous section, we can express the detector's output as a combination of gravitational wave strain, denoted as $h(t)$, and noise, represented as $n(t)$. Typically, the noise component exhibits significantly greater magnitude compared to $h(t)$, especially considering Earth-based detections. This raises a crucial question: how can we detect such faint signals that are deeply embedded within the noise? Surprisingly, we can indeed detect these signals without requiring that $|h(t)| \ll |n(t)|$. It is sufficient for $h_0$ to satisfy the condition $h_0 > (\tau_0/T)^{1/2}n_0$, where $h_0$ and $n_0$ are the characteristic amplitudes of $h(t)$ and $n(t),$ $\tau_0$ is the typical characteristic time of $h(t),$ and $T$ is the duration of the observation. Defining 
\begin{equation}
    \hat{d} = \int_{-\infty}^{\infty} dt \: d(t)K(t), 
    \label{eq:define_d}
\end{equation}
where $K(t)$ is the filter function. Now, we need to estimate optimal $F(t)$, which maximizes the signal-to-noise ratio (SNR), denoted as $\rho_m$. It is the ratio of the expected value of $\hat{d}$ in the presence of a signal and the expected value of the $\hat{d}$ when the signal is not present. Assuming $\langle n(t)\rangle = 0$, the $\rho_m$ is given by,
\begin{equation}
    \rho_m = \frac{\int_{-\infty}^{\infty} df \langle \tilde{h}(f)\rangle \tilde{K}^{*}(f)}{\left[\int_{-\infty}^{\infty} df (1/2) S_n(f) |\tilde{K}(f)|^2\right]^{1/2}}
    \label{eq:matched_filter_snr}
\end{equation}
The above equation can be succinctly written as
\begin{equation}
    \rho_m = \frac{\langle u \mid h \rangle}{\langle u \mid u \rangle^{1/2}}.
\end{equation}
where $u(t)$ is a function whose Fourier transform is given by $\tilde{u}(f) = S_n(f)\tilde{K}(f)$ and $\langle \cdot \mid \cdot \rangle$ denotes the inner product
\footnote{The inner product is defined as $\langle x\mid y\rangle = 4\:\text{Re}\int_{0}^{\infty}df \: \frac{\tilde{x}^{*}(f)\tilde{y}(f)}{S_n(f)}$}. It is clear that $u$, when chosen parallel to the $h$, gives the maximum value of $\rho_m$. It implies that the optimal filter (also known as the matched filter~\cite{Allen_2012, matched_filter_1960}) is given by
\begin{equation}
    \tilde{K}(f) \propto \frac{\tilde{h}(f)}{S_n(f)}
\end{equation}
Using the above equations, we find the optimal SNR denoted as $\rho_{\text{opt}}$ as 
\begin{equation}
    \rho_{\text{opt}} = \sqrt{4\int_{0}^{\infty}df \: \frac{|\tilde{h}(f)|^2}{S_n(f)}},
    \label{eq:optimal_snr}
\end{equation}
\subsubsection{Source reconstruction}
\paragraph{Maximum likelihood estimate}
\label{subsec:max_like_estm}
There are primarily two approaches to source reconstruction: (i) the frequentist approach and (ii) the Bayesian approach. In the context of parameter estimation for GW sources, we adopt the Bayesian approach because we lack an ensemble of identical GW events, unlike particle physics experiments where experimenters can manipulate various experiment parameters. The Bayesian approach is advantageous as it can provide answers to questions such as the most probable values of the source parameters given the GW data. Now, let's consider a scenario in which a GW detection has been made. This implies that for a particular template $h(t, \vec \theta)$, where $\vec \theta$ represents various source parameters like the masses and spins of the binary, etc., the signal-to-noise ratio, denoted as $\rho_m$, has surpassed a predetermined threshold. At this point, we aim to estimate the source parameters. We can achieve this by estimating the posterior distribution over the source parameters $\vec \theta$. Utilizing Eq.~\eqref{eq:deftn_psd}, we can express the probability distribution of noise as:
\begin{equation}
    \begin{split}
    P(n_0) &= \mathcal{N} \exp\left[-\frac{1}{2}\int_{-\infty}^{\infty} df \frac{|\tilde{n}_0(f)|^2}{(1/2)S_n(f)}\right]\\
    &= \mathcal{N}\exp\left[-\langle n_0\mid n_0\rangle/2\right].
    \end{split}
    \label{eq:noise_distrbn}
\end{equation}
where we assumed a stationary and Gaussian noise $n(t)$ and $\mathcal{N}$ is the normalization constant. Since $d(t) = h(t, \vec \theta_t) + n_0(t)$, plugging $n_0(t)$ back in the above equation, we get
\begin{equation}
    \begin{split}
    P(d\mid \theta_t) &= \mathcal{N} \exp\left[-\frac{1}{2}\langle d - h(\theta_t)\rangle \mid \langle d - h(\theta_t)\rangle\right],\\
    &= \mathcal{N}\exp\left[\langle h(\theta_t)\mid d\rangle - \frac{1}{2}\langle h(\theta_t)\mid h(\theta_t)\rangle - \langle d\mid d\rangle\right]
    \end{split}
    \label{eq:likelihood_source_reconst}
\end{equation}
Together with a prior distribution over $\vec \theta_t$, $P(\vec \theta_t)$ and using Bayes' theorem, we can express the posterior distribution over $\vec \theta_t$ given a data $d$ as 
\begin{equation}
    P(\vec \theta_t\mid d) = \mathcal{N}P(\vec \theta_t)\exp\left[\langle h(\theta_t)\mid d\rangle - \frac{1}{2}\langle h(\theta_t)\mid h(\theta_t)\rangle\right]
    \label{eq:log_likelihood_basic}
\end{equation}
where ${\langle d \mid d \rangle}$ is a constant for a given data $d$ and is reabsorbed in $\mathcal{N}$. In principle, you can evaluate the posterior ${P(\vec \theta_t\mid d)}$ at various values of $\vec \theta_t$. However, for a typical binary black hole (BBH) system, $\vec \theta_t$ belongs to a fifteen-dimensional parameter space, which presents significant computational challenge when evaluating the posterior. In Section [parameter estimation], we have explored several efficient techniques for estimating posterior distributions. This thesis primarily focuses on a rapid parameter estimation technique developed to address the computational challenges, especially for low-mass binaries like binary neutron stars (BNS), where the in-band signal has an extended duration. For now, let's focus on extracting essential information (an estimator) from the posterior, such as the maximum probable value denoted as $\hat{\vec \theta}$, and the corresponding estimation uncertainties or errors. An ideal estimator should adhere to fundamental criteria, including consistency, unbiasedness, efficiency, and robustness. One reasonable choice is the maximum likelihood estimator ($\hat{ \theta}_{\text{ML}}$), which is a value of $\vec \theta$ that maximizes the likelihood. When assuming a flat prior distribution, maximizing the posterior ${P(\vec \theta_t\mid d)}$ is equivalent to maximizing the likelihood $p(d\mid \vec \theta_t)$. By taking the natural logarithm of Eq.~\eqref{eq:likelihood_source_reconst} and omitting the constant terms, we obtain:
\begin{equation}
    \ln P(d\mid \vec \theta_t) = \langle h(\theta_t) \mid d \rangle - \frac{1}{2}\langle h(\vec \theta_t) \mid h(\vec \theta_t) \rangle
    \label{eq:likelihood_source_reconst_log}
\end{equation}
Maximizing the above equation amounts to taking its derivative and setting it to zero, 
\begin{equation}
    \langle \frac{\partial h(\vec \theta_t)}{\partial \vec \theta_i}  \mid d \rangle - \langle \frac{\partial h(\vec \theta_t)}{\partial \vec \theta_i} \mid h(\vec \theta_t) \rangle = 0
\end{equation}
and the corresponding errors, denoted as $\Delta \vec \theta_i$, can be estimated by using the width of the posterior distribution $P(\vec \theta_t \mid d)$. Interestingly, it is found that $\hat{\theta}_{\text{ML}}$ is identical to the $\vec \theta$ that maximizes $\rho_m$ in the matched-filtering process. To illustrate this, let's represent a generic template as $h(t;\vec \theta) = Ah_0(t; \vec \lambda)$, where the amplitude $A$ depends on certain parameters, while the remaining parameters are represented by $\vec \lambda$. Substituting it in the Eq.~\eqref{eq:likelihood_source_reconst_log} and maximizing it with respect to $A$, we find
\begin{equation}
    \hat{A}_{\text{ML}}(d) = \frac{\langle h_0(\vec \lambda)\mid d\rangle}{\langle h_0(\vec \lambda)\mid h_0(\vec \lambda)\rangle}.
\end{equation}
and the corresponding maximum likelihood is
\begin{equation}
    \ln P(d\mid \vec \lambda) = \frac{1}{2}\frac{\langle h_0(\vec \lambda)\mid d\rangle^2}{\langle h_0(\vec \lambda)\mid h_0(\vec \lambda)\rangle}
    \label{eq:max_like_wrt_amp}
\end{equation}
It can be further maximized with respect to $\vec \lambda$ and is equivalent to maximizing the overlap between $d$ with $h_0(\vec \lambda)/\langle h_0(\vec \lambda)\mid h_0(\vec \lambda)\rangle^{1/2}$ which is the matched-filtering procedure we discussed earlier. For the signal with very high SNRs, the corresponding errors of the estimated parameters are expected to be small so that we can expand $h(\vec \theta_i)$ around some estimate, say $\hat{\theta}_{\text{ML}}$ and the Eq.~\eqref{eq:log_likelihood_basic}, takes a simple form, 
\begin{equation}
    P(\vec \theta \mid d) = \mathcal{N} \exp\left[-\frac{1}{2}\Gamma^{ij}\Delta \vec \theta_{i}\Delta \vec \theta_j\right],
\end{equation}
where 
\begin{subequations}
    \begin{align}
        \Gamma^{ij} &= \langle \frac{\partial}{\partial \vec \theta_i}\frac{\partial h}{\partial \vec \theta_j}\mid h - d\rangle + \langle \frac{\partial h}{\partial \vec \theta_i}\mid \frac{\partial h}{\partial \vec \theta_j}\rangle, \label{eq:fish_no_approx}\\
        &\simeq \langle \frac{\partial h}{\partial \vec \theta_i}\mid \frac{\partial h}{\partial \vec \theta_j}\rangle\Bigr|_{\substack{_{\vec \theta_i = \hat{\theta}_{\text{ML}}}}}\label{eq:fish_high_snr_approx}
    \end{align}
\end{subequations}
where we arrive at the Eq.~\eqref{eq:fish_high_snr_approx} from Eq.~\eqref{eq:fish_no_approx} by using a large SNR limit and ignoring the first term in Eq.~\eqref{eq:fish_no_approx}. $\Gamma^{ij}$ in the above equation is known as the Fisher information matrix. The inverse of the Fisher information matrix gives the estimate of the average errors over the estimated parameters, $(\Gamma^{-1})_{ij} = \langle \Delta \vec \theta_i \Delta \vec \theta_j\rangle$.
\begin{figure}[!hbt]
    \centering
    \includegraphics[width=0.75\linewidth, height=0.55\linewidth]{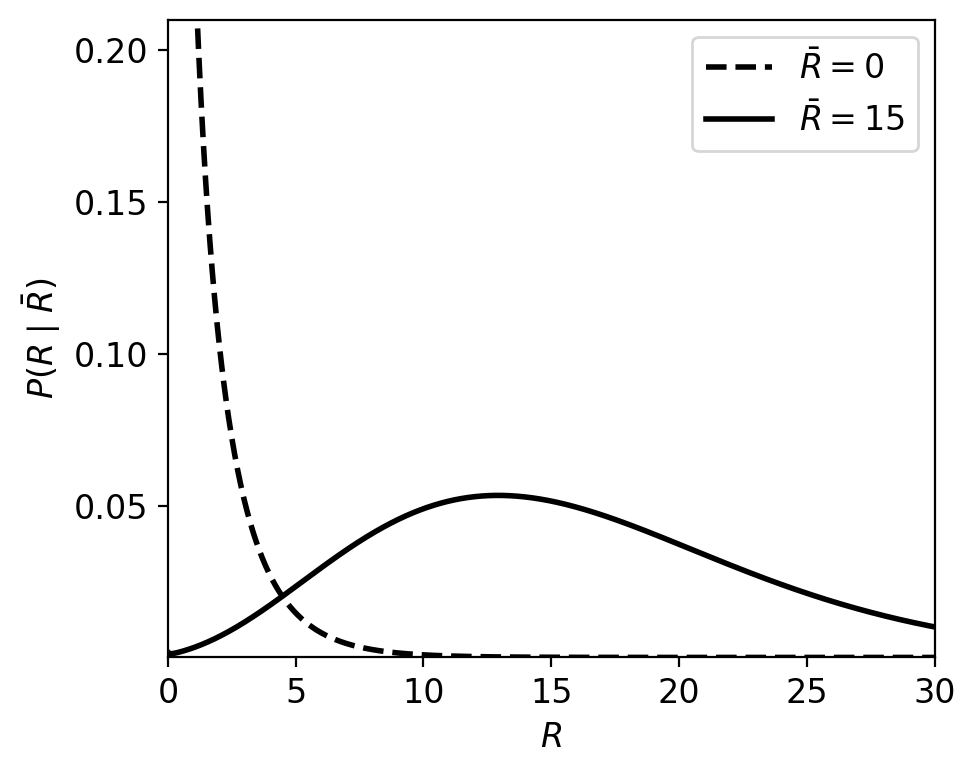}
    \caption{Probability distribution of $R$ for $\bar{R} = 0$ and $\bar{R} = 15$ (reproduced from~\cite{maggiore2008gravitational})}
    \label{fig:R_dist}
\end{figure}
\paragraph{Matched filtering statistics} 
\label{subsec:matched_filtering_stat}
After detecting a gravitational wave at a specific SNR, denoted as $\rho_m$, it is essential to assess the statistical significance of this detection, which relies on the statistical properties of the noise. When a signal is buried within Gaussian noise, the presence of any triggers due to this noise can be eliminated by setting a high threshold for $\rho_m$. However, real-world scenarios often involve non-Gaussian noise, which possesses heavy tails, even at high $\rho_m$ values, making it challenging to distinguish genuine signals from spurious non-Gaussian features. This challenge emphasizes the need for multi-detector networks to perform coincidence tests, helping mitigate issues arising from non-Gaussian noise. Recalling the SNR $\rho_m$, which represents the ratio of the expected value of $\hat{d}$ when a signal is present to the root mean square (RMS) value of $\hat{d}$ when the signal is absent, we can express $\hat{d}$ from Equation~\eqref{eq:define_d}. Introducing $\rho = \hat{d}/N,$ where $N$ is the RMS value of $\hat{d}$ when the signal is absent, we find that $\rho_m = \langle \rho \rangle.$ From Equation~\eqref{eq:define_d}, it's evident that $\rho$ is a random variable following a standard normal distribution when the signal is absent,
\begin{equation}
    P(\rho \mid h = 0)d\rho = \frac{1}{\sqrt{2\pi}}\exp^{-\rho^2/2}d\rho
    \label{eq:rho_dist_wo_signal}
\end{equation}
whereas in the presence of a GW signal with SNR equal to $\bar{\rho}$, $\rho = \bar{\rho} + \hat{n}/N$, where $\hat{n} = \int dt n(t)K(t)$ which implies
\begin{equation}
    P(\rho\mid \bar{\rho})d\rho = \frac{1}{\sqrt{2\pi}} \exp(\rho - \bar{\rho})^2/2 d\rho
    \label{eq:rho_dist_w_signal}
\end{equation}
Defining $R$\footnote{This is equivalent to SNR in energy so that $R$ is always positive.} $\equiv \rho^2$, we can write
\begin{equation}
    P(R\mid \bar{R}) dR = \frac{1}{\sqrt{2\pi R}} \exp[-(R + \bar{R})/2] \cosh[\sqrt{R\bar{R}}] dR.
    \label{eq:P_R_dist}
\end{equation}
As illustrated in Figure~\ref{fig:R_dist}, it becomes evident that we can differentiate a genuine GW signal from fluctuations caused by Gaussian noise by setting a threshold on the parameter $R$ at a designated value denoted as $R_t$. This threshold should be chosen sufficiently far into the tail of the distribution, ensuring the retention of a substantial portion of the true signals.  At the same time, there also exists a false alarm probability, 
\begin{equation}
    \begin{split}
        p_{\text{FA}} &= \int_{R_t}^{\infty} dR P(R\mid \bar{R}=0)\\
        &= 2\int_{\rho_t}^{\infty}d\rho \exp^{-\rho^2/2}
    \end{split}
    \label{eq:false_alarm_prob}
\end{equation}
and a false dismissal probability, 
\begin{equation}
    p_{\text{FD}} = \int_{0}^{R_t} dR P(R\mid \bar{R})
    \label{eq:false_dismiss_prob}
\end{equation}
The threshold $R_t$ can be fixed by setting a maximum false alarm rate we are willing to risk. We can generalize Eq.~\eqref{eq:P_R_dist}, extending it for cases when we have multiple detectors and their corresponding SNRs, each following a Gaussian distribution such that $\rho^2 = \rho_1^2 + \rho_2^2 + ... + \rho_n^2$, one can show that
\begin{equation}
    P_n(R\mid \bar{R}) dR = \frac{1}{\sqrt{2\pi R}} \left(\frac{R}{\bar{R}}\right)^{(n-2)/4}\exp[-(R + \bar{R})/2] I_{n/2 - 1}[\sqrt{R\bar{R}}] dR.
    \label{eq:P_R_dist_with_multi_dets}
\end{equation}
where $I_{(\cdot)}$ modified bessel function of first kind. It is a non-central $\chi^2$ distribution with $n$-degrees of freedom. In this section, we shall discuss how we can detect a coalescence of compact binaries.
\subsubsection{Compact binary coalscences}
The merger of compact binary systems like BBH, BNS, and neutron star-black hole binaries (NSBH) is particularly favorable for broad-band GW detectors~\cite{adv_ligo_2015, AdVirgo} because these signals spend a substantial number of cycles within the detector's sensitive frequency band. Of course, it crucially depends on the component masses of the binary. Matched filtering, when used effectively, can prove invaluable for extracting these signals, even when they are deeply embedded in a noisy background. To achieve this, it's crucial to have a precise understanding of the waveform associated with specific binary parameters. Using Eq.~\eqref{eq:hp_hc_inspiraling_with_tau} and Eq.~\eqref{eq:ht_as_lin_comb}, we can write $h(t)$ as 
\begin{equation}
    h(t) = A\left(\frac{\pi f_{\text{gw}}(t)}{c}\right)^{2/3} \cos[\Phi(f_{\text{gw}}(t)) + \phi]
    \label{eq:total_signal}
\end{equation}
where $\phi$ is some constant phase which is related to $\Phi$ at colascence and $A= \sqrt{A_{+}^2 + A_{\times}^2}$. There are some extrinsic parameters, such as time shifts and phase, which can be eliminated. 
\begin{figure}[!hbt]
    \centering
    \includegraphics[width=0.75\linewidth]{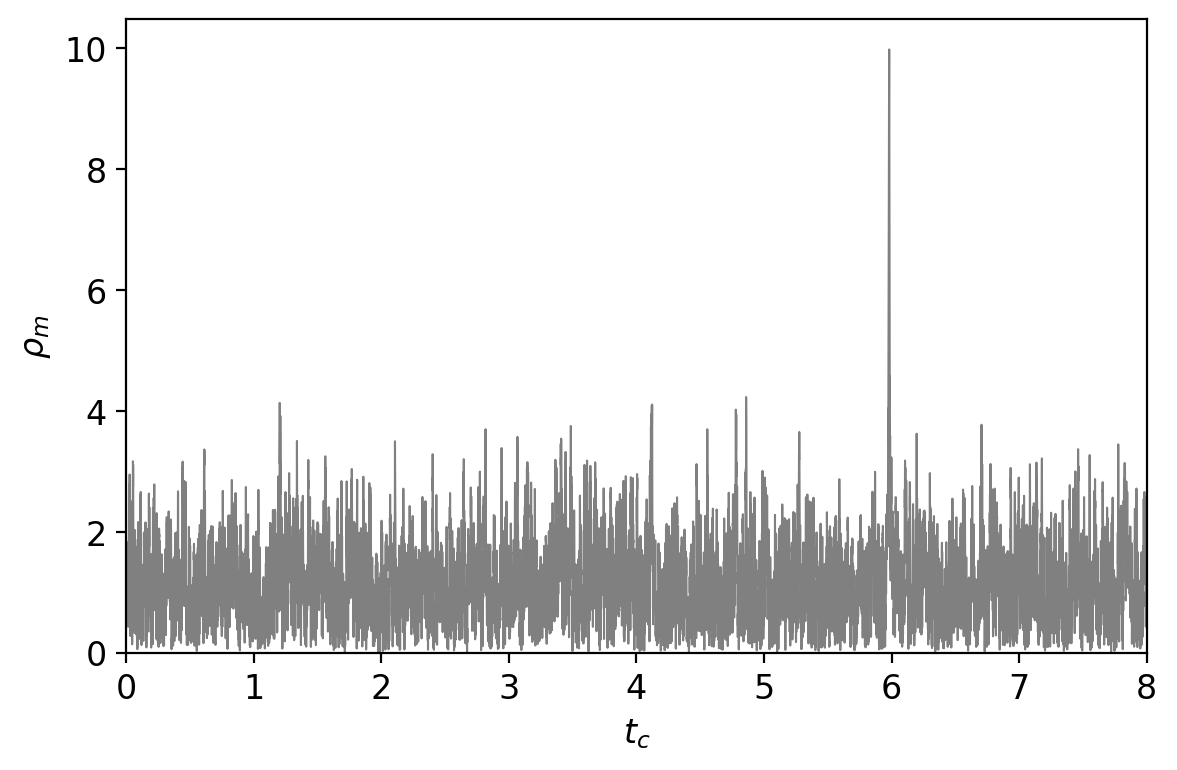}
    \caption{The matched-filter SNR for a BBH signal with component masses $10\: M_{\odot}$ each, at a distance of $700$ Mpc injected into the noise of the LIGO detector. The time of coalescence $(t_c = 6)$ is characterized by the peak of the SNR (made using PyCBC~\cite{usman2016pycbc, biwer2019pycbc}).}
    \label{fig:mf_snr_tc}
\end{figure}
To see this, observe that the frequency domain GW signal at some time of coalescence $t_{c}$ can be calculated by simply using a time-translation, $\tilde{h}(f;\vec \theta)\exp(i2\pi f t_{c})$, where $\tilde{h}(f;\vec \theta)$ is the Fourier transform of the GW signal $h(t; \vec \theta, t_{c} =0 )$. Using the definition of inner product, we find
\begin{equation}
    \langle h(\vec \theta, t^{*})\mid d\rangle = 4\text{Re}\int_{0}^{\infty} df\frac{\tilde{h}^{*}(f;\vec \theta)\tilde{d}(f)}{S_n(f)}\exp(i2\pi f t_{0})
\end{equation}
So, in a single Fast Fourier Transform (FFT), we can evaluate $\langle h(\vec \theta, t^{*})\mid d\rangle$ at all possible time shifts, implying we can immediately locate the $t_c$ corresponding to highest matched-filter SNR as evident from the Fig.~\ref{fig:mf_snr_tc}. Apart from the time of coalescence, we can also analytically maximize $\ln \mathcal{L}$ over $A$ (see Sec.~\ref{subsec:max_like_estm}) and $\phi$. Additionally, one can show that the SNR average over the sky angles and inclination is
\begin{equation}
    \langle \rho\rangle = \frac{2}{5} \left(\frac{5}{6}\right)^{1/2} \frac{1}{\pi^{2/3}} \frac{c}{r}\left(\frac{G\mathcal{M}}{c^3}\right)^{5/6} \times \sqrt{\left[\int_{0}^{f_{\text{max}}} df \frac{f^{-7/3}}{S_n(f)}\right]}.
    \label{eq:angle_inc_averaged_snr}
\end{equation}
This formula can be inverted to calculate the maximum observable distance, r, at which a binary coalescence can be detected with a specific angle-inclination-averaged SNR, denoted as $\langle \rho \rangle$. This relationship between distance and SNR implies that a greater number of cycles, $\mathcal{N}_c$, that a GW signal spends within the detector's sensitivity band results in a higher accumulated SNR. Additionally, it can be demonstrated that the relative error in measuring the chirp mass $\mathcal{M}$ is inversely proportional to the number of in-band cycles, indicating that events with a higher number of cycles will have lower chirp mass errors compared to those with fewer in-band cycles. In the following section, we will discuss the GW interferometers and provide a brief overview of current and future GW detectors.
\subsection{GW Detectors: Interferometers}
The concept of employing interferometers for gravitational wave (GW) detection was initially introduced by M. Gertsenshtein and V.I. Pustovoit in 1962, and it was subsequently advanced by R. Weiss, R. Drever, and other researchers. In this section, we will discuss the underlying principles of interferometer-based GW detectors and provide a concise survey of various GW detectors.
\subsubsection{Michelson inteferometer}
\begin{figure*}[!htp]
    \centering
    \includegraphics[width=0.8\linewidth]{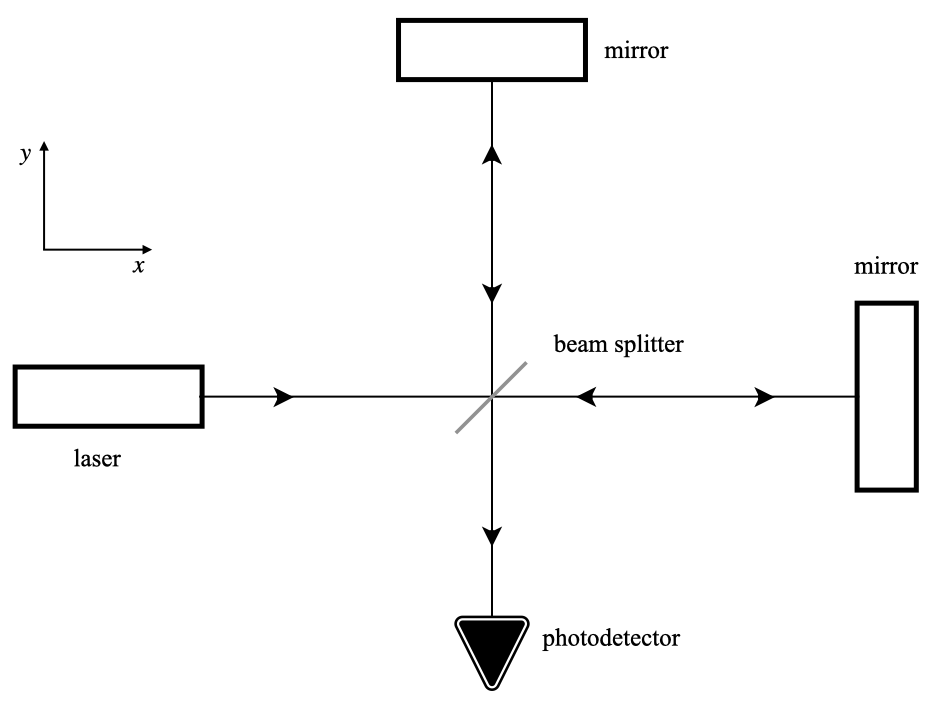}
    \caption{A simple setup of a Michelson interferometer (reproduced from~\cite{maggiore2008gravitational}).}
    \label{fig:michel_ifo}
\end{figure*}

The Michelson interferometer~\cite{1881AmJS...22..120M}, employed by Michelson and Morley in their renowned $1887$ experiment~\cite{1887AmJS...34..333M} to investigate the presence of ether, is a specific type of interferometer. This instrument, as depicted in Fig.~\ref{fig:michel_ifo}, was devised to detect changes in the time taken by light to travel through its arms. It is comprised of a monochromatic light source (typically a laser) that emits light directed through a beam-splitter, dividing it into two beams of equal amplitude. These two beams travel along arms that are oriented perpendicular to each other. At the end of each arm, a mirror with total reflectivity redirects the light beams back towards the beam splitter, where they are recombined. A portion of this recombined beam travels to a photodetector, which quantifies its intensity, while the remaining part goes back to the light source.
Let's denote the length of the two arms by $L_x$ and $L_y$, respectively. There exist an Electic field of the input laser light whose spatial component is given by
\begin{equation}
    E_0e^{-\iota \omega_L + i \vec k_L \cdot \vec x}
\end{equation}
where $\omega_L$ is the laser's frequency,  $k = \omega_L/c$ represents wave number and $\lambda_L = 2\pi/k_L$ defines the wavelength of the laser light. Consider a photon arriving at the beam splitter at some initial time $t_0$. The part of the Electric field that travels through $x$-arm and bounces off the mirror at the end of the arm and returns back to the beam splitter at time $t = t_0 + 2L_x/c$, while the beam which traveled along the $y$-arm arrived back at the beam splitter at $t' = t_0 + 2L_y/c$. Defining $\vec x = 0$ at the beam splitter, the two electric fields which recombine at the beam splitter are given by
\begin{subequations}
    \begin{align}
        E_1 &= -\frac{1}{2}E_0e^{-i\omega_L t + 2ik_LL_x}\\
        E_2 &= \frac{1}{2}E_0e^{-i\omega_L t + 2ik_LL_y}
    \end{align}
\end{subequations}
So that the recombined electric field is 
\begin{equation}
    E_{\text{out}} = E_1 + E_2 = -iE_0e^{-i\omega_L t + ik_L (L_x + L_y)} \sin [k_L(L_x - L_y)]
\end{equation}
and the corresponding power measured by the photodetector is given by
\begin{equation}
    |E_{\text{out}}|^2 = E_0^2\sin^2[k_L(L_y - L_x)].
\end{equation}
The above equation indicates that any change in the arm's length causes a corresponding variation of the power detected at the photodetector. Now, we turn to discuss the interaction of the detector with GWs within the TT gauge. As discussed in earlier sections, positions are determined by freely falling objects in the TT gauge. Consequently, the coordinates of these freely falling objects remain constant even in the presence of GWs~\footnote{It is important to note that while the mirrors are not in true free fall and are instead held in place by suspensions to counteract the Earth's gravity, these forces are essentially static in comparison to the time-scales of the GWs under study. Nonetheless, the mirrors can be considered to be in free fall in the horizontal plane.}. 
This implies that the GWs do not affect the mirror and beam splitter coordinates. For the sake of simplicity, we will assume only a ``$+$'' polarization of the GW originating from the $z$-direction. In the plane perpendicular to the direction of propagation, we can express the plus polarization as $h_+(t) = h_0\cos(\omega_{\text{gw}}t)$. For a photon following a null geodesic path, we have the following expression:
\begin{equation}
    dx = \pm cdt \left[1 - \frac{1}{2} h_{+}(t)\right],
\end{equation}
where the plus sign denotes the time it takes for light to travel from the beam splitter to the mirror, while the negative sign represents the time for the return journey. If we consider a photon departing from the beam splitter at time $t_0$ and arriving at the mirror at time $t_1$, and then returning to the beam splitter at time $t_2$, we can determine $t_2$ in terms of $t_0$ and $h_{+}(t)$ using the equation above as:
\begin{equation}
    \begin{split}
        \Delta t \equiv t_2 - t_0 = &= \frac{2L_x}{c} + \frac{1}{2}\int_{t_0}^{t_2} dt' h_{+}(t')\\
        &= \frac{2L_x}{c} + \frac{L_x}{c}h(t_0 + L_x/c) \frac{\sin(\omega_{\text{gw}}L_x/c)}{\omega_{\text{gw}}L_x/c}.
    \end{split}
\end{equation}
Similarly, for the $y$-arm, the above expression becomes,
\begin{equation}
    \Delta t \equiv t_2 - t_0 = \frac{2L_y}{c} + \frac{L_y}{c}h(t_0 + L_y/c) \frac{\sin(\omega_{\text{gw}}L_y/c)}{\omega_{\text{gw}}L_y/c}.
\end{equation}
Inverting the above two equations in the first order in $h_0$ to get the corresponding value of $t_0$ for a given value of $t \equiv t_2$, we find
\begin{subequations}
    \begin{align}
       t^{(x)}_0 &= t - \frac{2L_x}{c} - \frac{L_x}{c}h(t - L_x/c)\sin c(\omega_{\text{gw}}L_x/c),\\
       t^{(y)}_0 &= t - \frac{2L_y}{c} - \frac{L_y}{c}h(t - L_y/c)\sin c(\omega_{\text{gw}}L_y/c).
    \end{align}
\end{subequations}
The corresponding electric fields are
\begin{subequations}
    \begin{align}
        E^{(x)}(t) = -\frac{1}{2}E_0e^{-i\omega_L (t - 2L/c) + i\phi_0 + i\Delta \phi_x(t)},\\
        E^{(y)}(t) = -\frac{1}{2}E_0e^{-i\omega_L (t - 2L/c) - i\phi_0 + i\Delta \phi_y(t)}
    \end{align}
\end{subequations}
where $L = (L_x + L_y)/2$, $\phi_0 = k_L(L_x - L_y)$, and $\Delta \phi_x$ given by
\begin{equation}
    \begin{split}
        \Delta \phi_x(t) &= h_0 k_L L \sin c(\omega_{\text{gw}}L/c) \cos(\omega_{\text{gw}} (t - L/c)),\\
        &\equiv |\Delta \phi_x|\cos(\omega_{\text{gw}}t + \alpha),
    \end{split}
    \label{eq:phase_change_x}
\end{equation}
with $\alpha = -\omega_{\text{gw}}L/c$. 

The total phase difference caused by GWs in the interferometer can be shown as $\Delta \phi_{\text{Mich}} \equiv \Delta \phi_x - \Delta \phi_y = 2\Delta \phi_x$ where $\Delta \phi_y = -\Delta \phi_x$ and corresponding total electric field is given by
\begin{equation}
    E_{\text{tot}}(t) = -iE_0e^{-i\omega_L(t - 2L/c)}\sin [\phi_0 + \Delta \phi_x(t)].
    \label{eq:total_efield}
\end{equation}
where $\phi_0$ is a parameter chosen by the experimenter. In the limit $\omega_{\text{gw}}L/c \ll 1$ which is equivalent to $\lambda_{\text{gw}} \gg L$, Eq.~\eqref{eq:phase_change_x} becomes $\Delta \phi_x(t) \simeq h(t - L/c)k_LL$ which implies that in this limit, GWs effect on the phase shift is equivalent to a variation in $L_x - L_y$ and can be written as 
\begin{equation}
    \frac{\Delta (L_x - L_y)}{L} \simeq h(t - L/c)
\end{equation}
From the Eq.~\eqref{eq:total_efield}, we find the total power observed at the photodetector is
\begin{equation}
    \begin{split}
        P &= P_0\sin^2[\phi_0 + \Delta \phi_x(t)]\\
        &= \frac{P_0}{2}(1 - \cos[2\phi_0 + \Delta \phi_{\text{Mich}}(t)]).
    \end{split}
\end{equation}
From the above equation, it is evident that the $\Delta \phi_{\text{Mich}}$ is clearly dependent on the $L$ apart from $h_0$. So the optimal length of the arms is $\omega_{\text{gw}}L/c = \pi/2$ or in terms of $f_{\text{gw}}$ given by
\begin{equation}
    L \simeq 750 \text{km} \left(\frac{100 \text{Hz}}{f_{\text{gw}}}\right).
    \label{eq:optimal_length}
\end{equation}
In the case of this arm length $L$, the time shift caused by GW consistently has the same sign throughout the entire round trip within the arm. This leads to an accumulation of the effect. However, in the case of longer arms, the amplitude of GW causes the sign to reverse during the round trip, which results in the cancellation of the phase shift that the light has accumulated. To incorporate this optimal path length into an interferometer of practical size, a Fabry-Perot cavity~\cite{1901ApJ....13..265F, 1899ApJ.....9...87P} is used, which we discuss in the next section.
\subsubsection{Fabry-Perot cavity}
From the Equation ~\eqref{eq:optimal_length}, it becomes evident that in order to detect GWs with a frequency on the order of $100$ Hz, it would necessitate arm lengths (denoted as L) extending several hundred kilometers. However, achieving such extensive arm lengths for Earth-based interferometers is both practically and financially infeasible. The solution lies in attaining an optimal path length of $750$ km using an interferometer with arms that are only a few kilometers long. For example, the LIGO interferometer features arms with a length of 4 km, while Virgo's arms are 3 km long. This is made possible through the utilization of a Fabry-Perot (FP) cavity. Let's briefly explore how the FP cavity enables us to reach the required optimal path length.
\begin{figure*}[!htp]
    \centering
    \includegraphics[width=\linewidth]{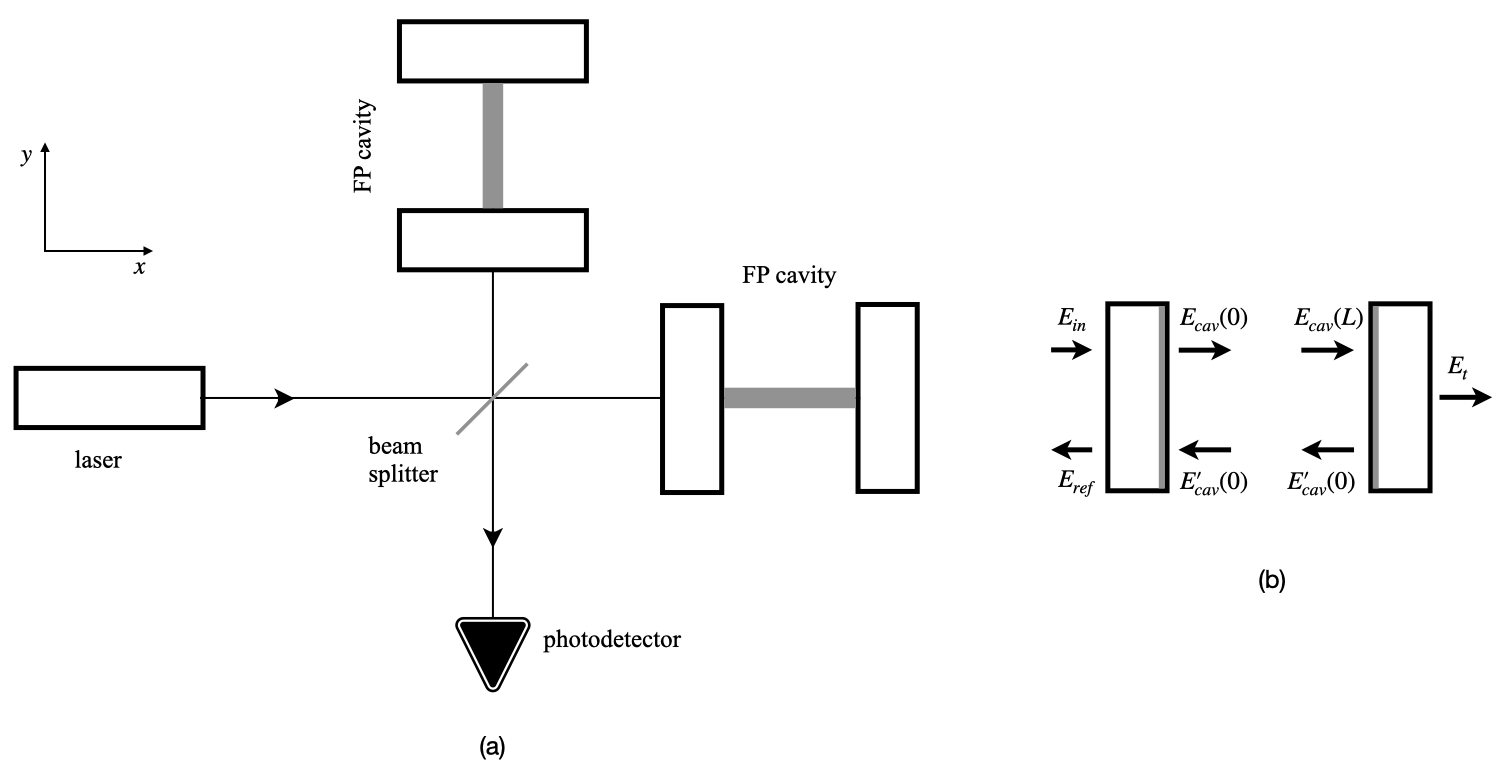}
    \caption{\textit{Left panel}: A simple setup of a Michelson interferometer with Febry-Perot cavities. \textit{Right panel}: A schematic of Febry-Perot cavity (reproduced from~\cite{maggiore2008gravitational})}
    \label{fig:michel_ifo_fp}
\end{figure*}

An FP cavity is constructed with two parallel mirrors, both featuring high-reflectivity coatings on their interior surfaces, as depicted in the second panel of Figure ~\ref{fig:michel_ifo_fp}. The incoming electric field $E_{in}$ enters from the left side of the cavity, with a portion being reflected by the coating while the remainder is transmitted. The transmitted portion $E_{cav}(0)$ traverses through the cavity, where it experiences partial reflection and partial transmission. The reflected portion $E_{cav}(L)$ journeys back to the initial mirror, once again undergoing partial reflection and transmission and so on. The total fields for reflection and transmission inside the cavity are composed of multiple beams resulting from numerous reflections. Using reflection and transmission coefficients denoted as $r_1$ and $t_1$ for the first mirror and $r_2$ and $t_2$ for the second mirror, we can express $E_{cav}(0)$ in terms of $E_{in}$ and $E'_{cav}(0)$ as follows:
\begin{subequations}
    \begin{align}
        E_{cav}(0) &= t_1E_{in} + r'_1 E'_{cav}(0),\\
        E_{ref}(0) &= r_1E_{in} + t'_1 E'_{cav}(0).
    \end{align}
\end{subequations}
where $E_{in} = E_0e^{-i \omega_L t}$. We can write similar equations at the second mirror as
\begin{subequations}
    \begin{align}
        E_{t} &= t_2E_{cav}(L),\\
        E'_{cav}(L) &= -r'_2E_{cav}(L).
    \end{align}
\end{subequations}
In the above equations, we can set $r'_{1(2)} = -r_{1(2)}$ and $t'_{1(2)} = t_{1(2)}$~\footnote{This can be proved by assuming no gap between the two mirrors. For a perfectly reflecting mirror, $r'_1 = -1$ when the reflection occurs from a less dense to a more dense medium while $r = 1$ when it happens from the denser side to the less dense side.}. The solution within the cavity consists of plane waves, which means that the cavity fields at positions $x=L$ and $x=0$ at the same time are related by the following equation:
\begin{subequations}
    \begin{align}
        E_{cav}(L) &= e^{ik_L L}E_{cav}(0),\\
        E'_{cav}(L) &= e^{-ik_L L}E'_{cav}(0)
    \end{align}
\end{subequations}
Solving the above equations for $E_{ref}$, $E_{t}$, and $E_{cav}(0)$, we find
\begin{subequations}
    \begin{align}
        E_{ref} &= E_0e^{-i \omega_L t_0} \frac{r_1 - r_2(1 - p_2)e^{2ik_LL}}{1 - r_1r_2e^{2ik_L L}}\label{eq:e_ref},\\
        E_{t} &= E_0e^{-i\omega_L t_0} \frac{t_1 t_2 e^{ik_L L}}{1 - r_1r_2e^{2ik_L L}}\label{eq:e_t},\\
        E_{cav}(0) &= E_0e^{-i\omega_Lt_0}\frac{t_1}{1 - r_1r_2e^{2ik_L L}}
    \end{align}
\end{subequations}
where $p_1$ represents losses in the first mirror and is related to $r_1$ and $t_1$ by $r_1^2 + t_1^2 = 1 - p_1$. From the above equations, we can find the power of the transmitted field $P_t$ as
\begin{equation}
    P_t = |E_t|^2 = E_0 \frac{t_1^2 t_2^2}{1 + (r_1r_2)^2 - 2r_1r_2\cos(2k_LL)}
\end{equation}
The separation between the peaks in the function above is referred to as the free spectral range of the cavity and is quantified as $\Delta \omega_L = \pi c/L$. Meanwhile, the full width of the peaks at half maximum (FWHM) is expressed as $\delta \omega_L = c(1 - r_1r_2)/L \sqrt{r_1r_2}$. We can define the finesse of the cavity, denoted as $\mathcal{F}$, as the ratio of the free spectral range to the FWHM,
\begin{equation}
    \mathcal{F} = \frac{\pi\sqrt{r_1r_2}}{1 - r_1r_2}
    \label{eq:finesse}
\end{equation}

We can calculate another important quantity called storage time, denoted as $\tau_s$, in the context of the FP. It is defined as the average duration a photon remains inside the cavity. It can be expressed in terms of the finesse as follows:
\begin{equation}
    \tau_s \simeq \frac{L}{c}\frac{\mathcal{F}}{\pi}
\end{equation}
In fact, the intensity of light that comes out from the first mirror after undergoing multiple round trips is dependent on $ \tau_s$ and decreases with time. From Eq.~\eqref{eq:e_ref}, we can write down $E_{ref}$ at resonances as
\begin{equation}
    E_{ref} = E_0e^{-i\omega_L t_0}\frac{r_1 - r_2(1 - p_1)}{1 - r_1 r_2}
\end{equation}
where resonances are defined to be the conditions when $2k_L L = 2\pi n$ where $n = 0, \pm 1, \pm 2,....$. This condition implies that the beams bouncing back and forth inside the cavity undergo constructive interference, resulting in a significantly amplified field within the cavity. Note that there is no reflected light if $r_1 = r_2(1 - p_1)$. The light immediately reflected back undergo destructive interference with the light that gets reflected after going through either one or multiple round trips within the cavity, causing all the incident light to leak out from the second mirror. This scenario is referred to as critical cavity coupling, and it's highly undesirable in the context of a GW interferometer because we are interested in the reflected light. Now, from Eq.~\eqref{eq:finesse}, we can write $r_1r_2 = 1 - \pi/\mathcal{F} + \mathcal{O}(\pi^2/\mathcal{F}^2)$ and introduce coupling rate $\sigma$ defined as $\sigma = p\mathcal{F}/\pi$ where $p = 1 - (1 - p_1)r_2^2$ which encapsulates the losses from both mirrors. Using these expressions, we can write 
\begin{equation}
    \frac{r_1 - r_2 (1 - p_1)}{1 - r_1 r_2} = \frac{\sigma - 1}{r_2}
\end{equation}
\begin{figure*}[!htp]
    \centering
    \includegraphics[width=0.75\linewidth, height=0.45\linewidth]{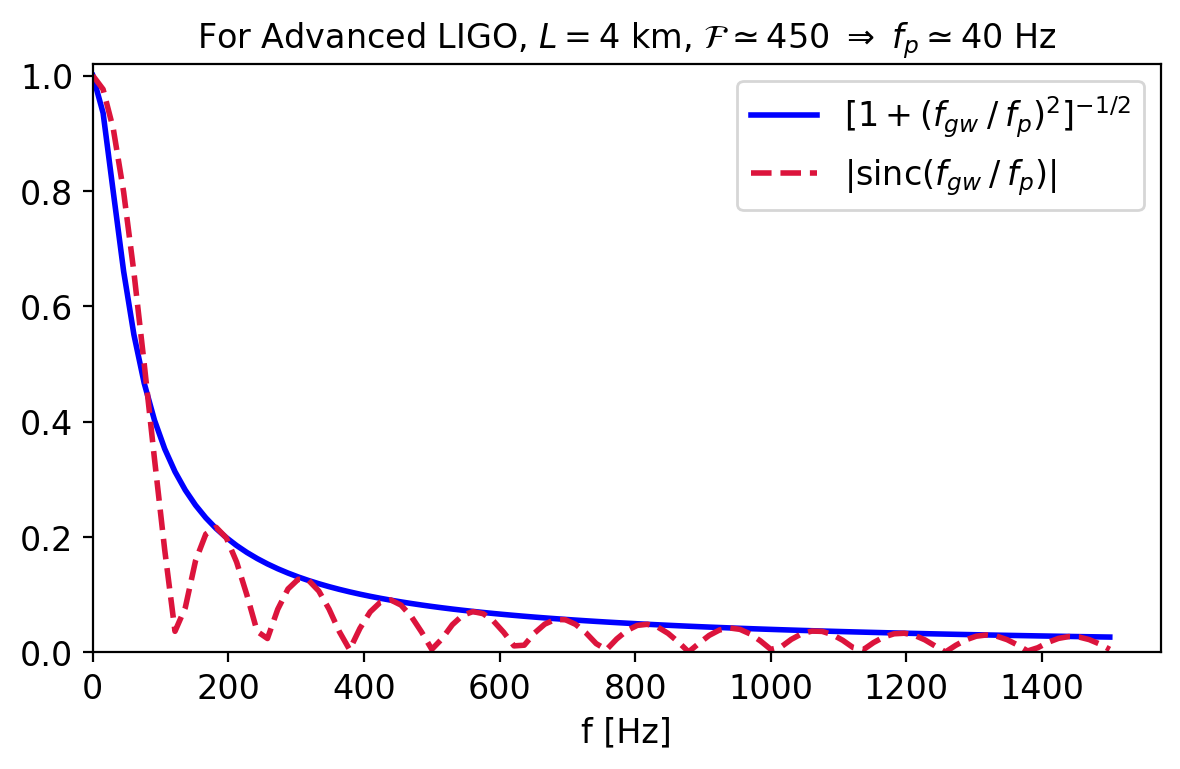}
    \caption{Plot of the factors affecting the phase of the FP cavity and Michelson interferometer. It is clear that the sinc function oscillates rapidly beyond a certain frequency, canceling the phase contributions in contrast to the other factor.}
    \label{fig:fp_vs_mi_sense}
\end{figure*}
From the above equation, we can see that $\sigma = 1$ corresponds to critical coupling. For $0 < \sigma < 1$, the cavity is overcoupled, equivalent to having a higher sensitivity to changes of $2k_L L$, which is crucial for the phase of the reflected field. For the arms of Advanced LIGO~\cite{Advanced_LIGO_2015} and Advanced Virgo~\cite{AdVirgo}, $p \sim 2 \times 10^{-5}$ with $\mathcal{F}$ to be $\simeq 450$, giving $\sigma \sim 10^{-3}$, which indicates that LIGO and Virgo's cavities are well overcoupled. Without going into the technical details of the effect of GWs on FP cavities, we write down the phase shift in an FP interferometer due to GWs as 
\begin{equation}
    |\Delta \phi_{\text{FP}}| \simeq h_0\frac{4\mathcal{F}}{\pi}k_L L \frac{1}{\sqrt{1 + (f_{\text{gw}}/f_p)^2}}.
    \label{eq:FP_phase}
\end{equation}
where $f_p$ is called pole frequency and defined as $f_p \equiv 1/4\pi \tau_s \simeq c/4\mathcal{F}L$ and the above expression is valid if $\omega_{\text{gw}}L/c \ll 1$ or equivalently,
\begin{equation}
    \begin{split}
        f_{\text{gw}} &\ll \frac{c}{2\pi L},\\
        &\simeq 12 \text{kHz} \left(\frac{4 \text{km}}{L}\right).
    \end{split}
\end{equation}
In the Fig.~\ref{fig:fp_vs_mi_sense}, $[1 + (f_{gw}\:/\: f_p)^2]^{-1/2}$ (see Eq.~\eqref{eq:FP_phase}) is compared to $|\text{sinc}(f/f_p)|$. The equivalent length $L_{\text{Mich}}$ of the Michelson interferometer needed to match sensitivity of a FP cavity with length $L$ and Finesse $\mathcal{F}$ is given by $L_{\text{Mich}} = 2\mathcal{F}L/\pi$. To give a comparison, for an ordinary Michelson interferometer to achieve similar sensitivity as Advanced LIGO~\cite{Advanced_LIGO_2015} (an FP cavity-based interferometer) with $L = 4$ km, $\mathcal{F} \simeq 450$, we would need a whopping $L_{\text{Mich}} \simeq 1100$ km long arms-lengths. 

In summary, the use of an FP cavity significantly enhances detector's sensitivity (by a factor of $(2/\pi)\mathcal{F}$) to a phase shift . This enhancement allows us to achieve a response to GWs equivalent to that of a Michelson interferometer with arms spanning on the order of hundreds of kilometers, which is optimal for gravitational waves with frequencies $f_{\text{gw}} = \mathcal{O}(10^2)$ Hz. We can achieve this by replacing the long arms with FP cavities of just a few kilometers in length and finesse of $\mathcal{F} = \mathcal{O}(10^2)$.
In the next section, we will provide a brief overview of the various noise sources in the LIGO-Virgo interferometers.
\subsubsection{Noise sources}
\label{subsec:noise_sources}
As discussed in the previous section, we measure the phase shift, which encodes the signatures of the GWs. We expect to have a GW amplitude of $h_0 \sim \mathcal{O}(10^{-21})$, the corresponding displacement $\Delta L = h_0 L/2$ of the mirrors for inteferometer with $L = 4$ km comes out to be $\mathcal{O}(10^{-18})$ m. It amounts to a Michelson phase shift of $\Delta \phi_{\text{Mich}} \sim \mathcal{O}(10^{-11})$ rad but for a FP cavity with $\mathcal{F} \simeq 450$, we get the the FP phase shift to be $\phi_{\text{FP}} \sim \mathcal{O}(10^{-8})$ rad. Unfortunately, the measurement of such minute phase shifts is influenced by various noise sources, which impose limitations on the interferometer's sensitivity. This strain sensitivity is quantified by ${S_n}^{1/2}(f)$ (measured in units of $\text{Hz}^{-1/2}$). Some of the prominent noise sources are as follows:
\begin{itemize}
    \item \textbf{Shot noise} This noise source arises because laser light is composed of discrete quanta known as photons. Measurment of the power of a light source is equivalent to counting the number of photons that arrive within a specific time frame. Considering an FP cavity with a finesse $\mathcal{F}$ and length $L$, for an optimal orientated periodic GW source radiating at frequency $f$ with only plus polarization, the strain sensitivity due to short noise is given by
    \begin{equation}
        S_n(f)^{1/2}\bigg|_{\text{shot}} = \frac{1}{8\mathcal{F}L}\left(\frac{4\pi \hbar \lambda_L c}{\eta P_{\text{bs}}}\right)^{1/2}\sqrt{1 + (f/f_p)^2},
        \label{eq:strain_sense_shot_noise}
    \end{equation}
    where $P_{\text{bs}}$ represents the beam splitter's power after recycling, which is equivalent to  $CP_0$ with $C$ being the power recycling factor for the laser, $P_0$ is the input power from the laser, and $\eta$ is the factor limiting the power of extraction of electrons at the photodiode. 
    
    \item \textbf{Radiation pressure} As indicated by Equation~\eqref{eq:strain_sense_shot_noise}, shot noise reduction can be achieved by either increasing  $P_0$ or increasing the recycling factor $C$. However, the photons striking the mirror and reflecting back also exert radiation pressure on the mirror, causing fluctuations in response to the number of photons reaching the mirror, ultimately resulting in mirror vibrations. This pressure grows as $\sqrt{P_{\text{bs}}}$ while shot noise deceased as $1/\sqrt{P_{\text{bs}}}$. This suggests that beyond a certain increase in $P_{\text{bs}}$, radiation pressure starts dominating over shot noise. The radiation pressure's strain sensitivity is given by:
    \begin{equation}
        S_n(f)^{1/2}\bigg|_{\text{rad}} = \frac{16\sqrt{2}\mathcal{F}}{ML(2\pi f)^2} \sqrt{\frac{\hbar P_{\text{bs}}}{2\pi \lambda_L c}} \frac{1}{\sqrt{1 + (f/f_p)^2}}.
        \label{eq:strain_sense_rad_press}
    \end{equation}
    where $M$ is the mass of the mirror.
    \begin{figure*}[!htp]
    \centering
    \includegraphics[width=0.75\linewidth, height=0.50\linewidth]{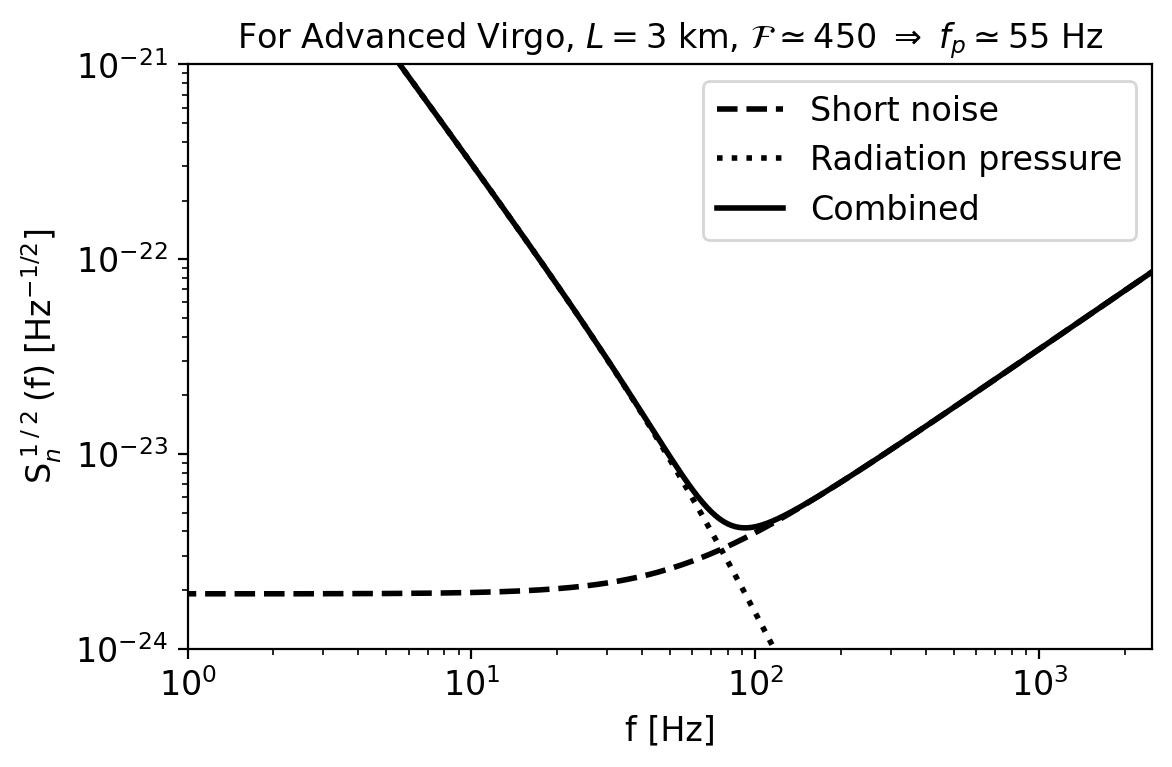}
    \caption{The strain sensitivity $S_n^{1/2}(f)$ representing shot noise, radiation pressure, and the combined noise.}
    \label{fig:noise_sense}
    \end{figure*}
    \item \textbf{The standard quantum limit} We can write the combined noise encompassing both shot noise and radiation pressure as the following: (see Fig.~\ref{fig:noise_sense}):
    \begin{equation}
        S_n(f)|_{\text{opt}} = S_n(f)|_{\text{shot}} + S_n(f)|_{\text{rad}}.
    \end{equation}
    As we discussed earlier, shot noise and radiation pressure are competing against each other. The position of an object is measured using photons, which in turn is disturbed by radiation pressure exerted by themselves. Let's rewrite the above equation as the following:
    \begin{equation}
        S_n(f)|_{\text{opt}} = \frac{1}{L\pi f_0}\sqrt{\frac{\hbar}{M}} \left[\left(1 + \frac{f^2}{f_p^2}\right) + \frac{f_0^4}{f^4}\frac{1}{1 + (f/f_p)^2}\right]^{1/2}
    \end{equation}
    where $f_0 = \frac{8\mathcal{F}}{2\pi} \sqrt{P_{\text{obs}}/\pi \lambda_L cM}$. The above $S_n(f)|_{\text{opt}}$ can be minimized with respect to $f_0$ and optimal value of $f_0$ is one for which $S_n(f)|_{\text{shot}} = S_n(f)|_{\text{rad}}$ which is given by
    \begin{equation}
        1 + \frac{f^2}{f_p^2} = \frac{f_0^2}{f_p^2}.
    \end{equation}
    Pluggin' it in the above equation to get the optimal value of $S_n(f)|_{\text{opt}}$, also known as the standard quantum limit (SQL),
    \begin{equation}
        S_n(f)^{1/2}|_{\text{SQL}}(f) = \frac{1}{2\pi fL}\sqrt{\frac{8\hbar}{M}}.
    \end{equation}
    It represents the lower bound of the family of functions $S_n^{1/2}(f;f_0)|_{\text{opt}}$ parameterized by $f_0$
    \begin{figure*}[!htp]
    \centering
    \includegraphics[width=0.75\linewidth, height=0.50\linewidth]{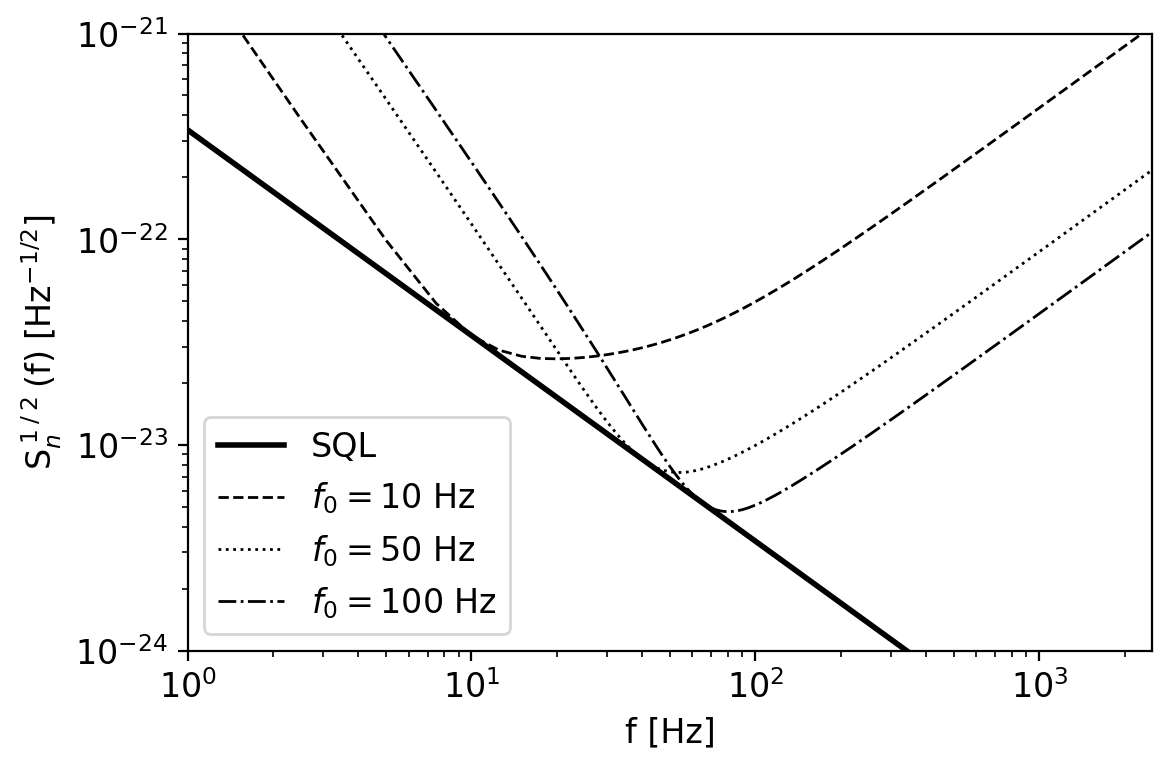}
    \caption{$S_n(f)_{\text{opt}}$ for different values of $f_0$ along with SQL.}
    \label{fig:noise_sense_sql}
    \end{figure*}
    \item \textbf{Seismic, Newtonian} 
    In addition to the aforementioned noise sources, it's important to consider those mirror movements that can be caused by factors unrelated to gravitational waves. These sources of noise fall into the category of displacement noise, and they give rise to a corresponding strain sensitivity denoted as $x(f)$, which can be described by the following equation:
    \begin{equation}
        x(f) \simeq A \left(\frac{1 \text{Hz}}{f^{\nu}}\right) \text{m} \text{Hz}^{1/2},
    \end{equation}
    where index $\nu \simeq 2$ and $A \sim \mathcal{O}(10^{-7})$. Dividing the above equation by $L$ gives the noise strain sensitivity, which is at least $\sim O(10^{10})$ larger than the aimed values. This clearly indicates the need to significantly attenuate seismic noise, often accomplished through a series of pendulum systems. However, for ground-based detectors, seismic noise can only be effectively reduced for GW frequencies greater than approximately $10$ Hz. Consequently, these detectors are unable to detect GWs below around $10$ Hz. Newtonian noise, on the other hand, originates from time-varying Newtonian gravitational forces caused by moving objects and is induced by micro-seismic noise that couples to the interferometer's mirrors. Unfortunately, we cannot entirely eliminate Newtonian noise, as gravitational forces cannot be shielded. While it is not the primary source of noise in current GW interferometers, it may become a limiting factor in future detectors as we strive to detect even lower frequencies. In addition to these noise sources, there are other factors that can impact the strain sensitivity of GW interferometers. For more in-depth information, interested readers can refer to the following references~\cite{maggiore2008gravitational, Abbott_2020_noise}. 
    \end{itemize}
\subsubsection{GW detectors: Initial to advanced and beyond}
\begin{figure*}[!hbt]
\centering
\includegraphics[width=\linewidth, height=0.70\linewidth]{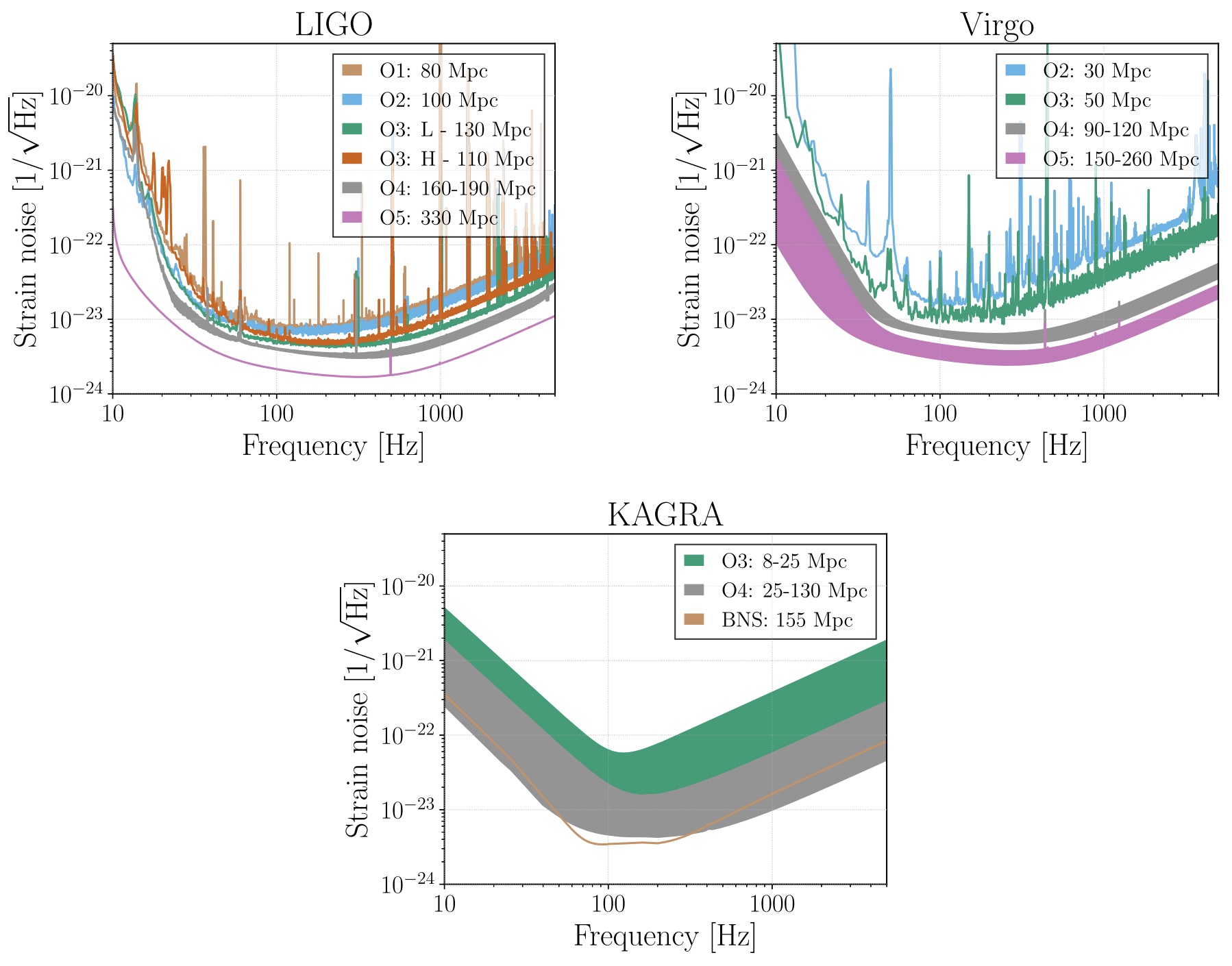}
\caption{aLIGO, AdV, and KAGRA target sensitivities are shown as a function of frequency. The quoted range is for $1.4 M_{\odot} + 1.4 M_{\odot}$ BNS merger~\cite{abbott2020prospects}.}
\label{fig:noise_sense_observ_scene_gb_gw}
\end{figure*}

\begin{figure*}[!htp]
\centering
\begin{subfigure}{0.49\linewidth}
    \centering
    \includegraphics[width=\linewidth, height=0.75\linewidth]{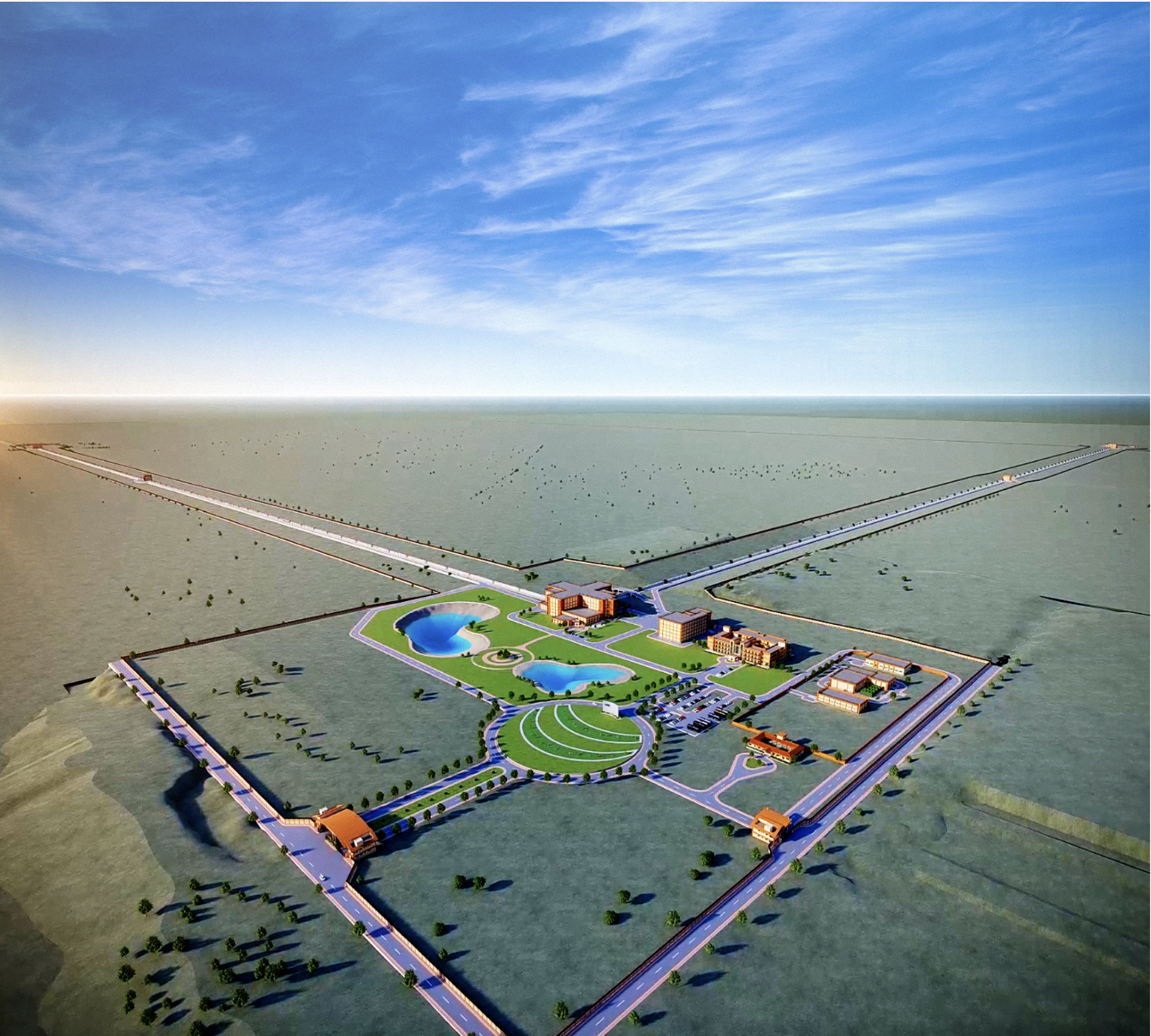}
    \caption{An artistic impression of LIGO-Aundha (source $\sim$ \href{https://www.ligo.caltech.edu/news/ligo20230417}{link})}
    \label{fig:LIGO_india}
\end{subfigure}\hfill
\begin{subfigure}{0.49\linewidth}
    \includegraphics[width=\linewidth]{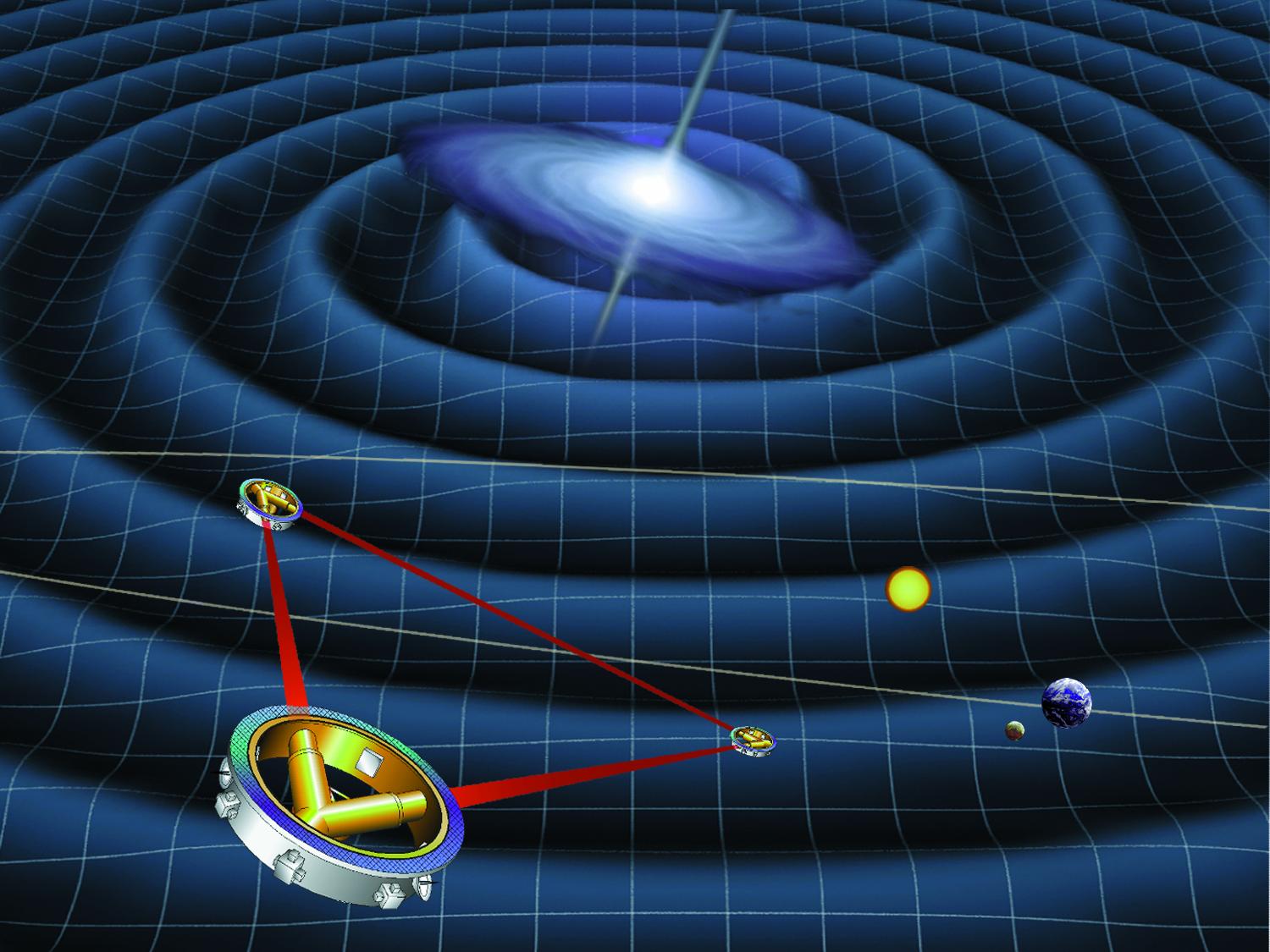}
    \caption{An artistic impression of LISA (source $\sim$ ESA)}
    \label{fig:LISA}
\end{subfigure}
\end{figure*}

\begin{figure*}[!htp]
    \centering
    \includegraphics[width=0.75\linewidth, height=0.45\linewidth]{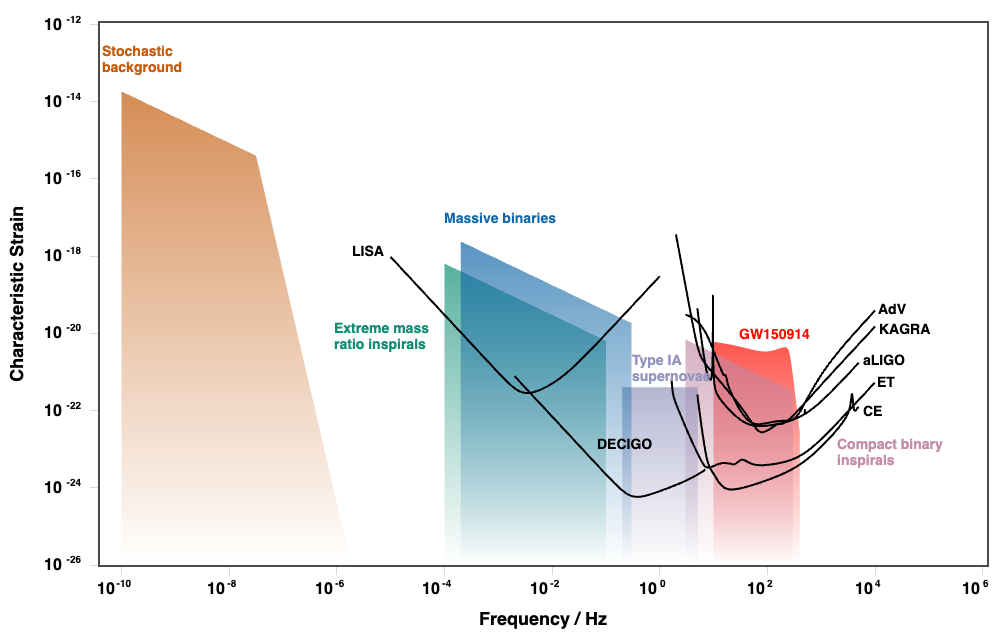}
    \caption{Noise strain sensitivity in various GW detectors, encompassing both ground-based and space-based observatories, is depicted. It is evident that LISA, DECIGO, and other ground-based detectors collectively span a wide spectrum of GW frequencies. This comprehensive coverage is particularly valuable for conducting multiband GW analyses~\cite{abbott2020prospects} (made using \href{http://gwplotter.com/}{gwplotter} as explained in~\cite{Moore_2014}).}
    \label{fig:strain_sensitiv}
\end{figure*}

\paragraph{Ground based detectors}
As previously explained, the fundamental structure of ground-based gravitational wave detectors consists of a Fabry-Perot (FP) cavity-based interferometer. There are two LIGO detectors: one situated in Livingston, Louisiana, and the other in Hanford, Washington. Each of these detectors has $4$ km long arms and is separated by approximately $10$ ms in terms of light travel time.
\begin{figure*}[!htp]
\centering
\begin{subfigure}{0.49\linewidth}
    \centering
    \includegraphics[width=0.9\linewidth]{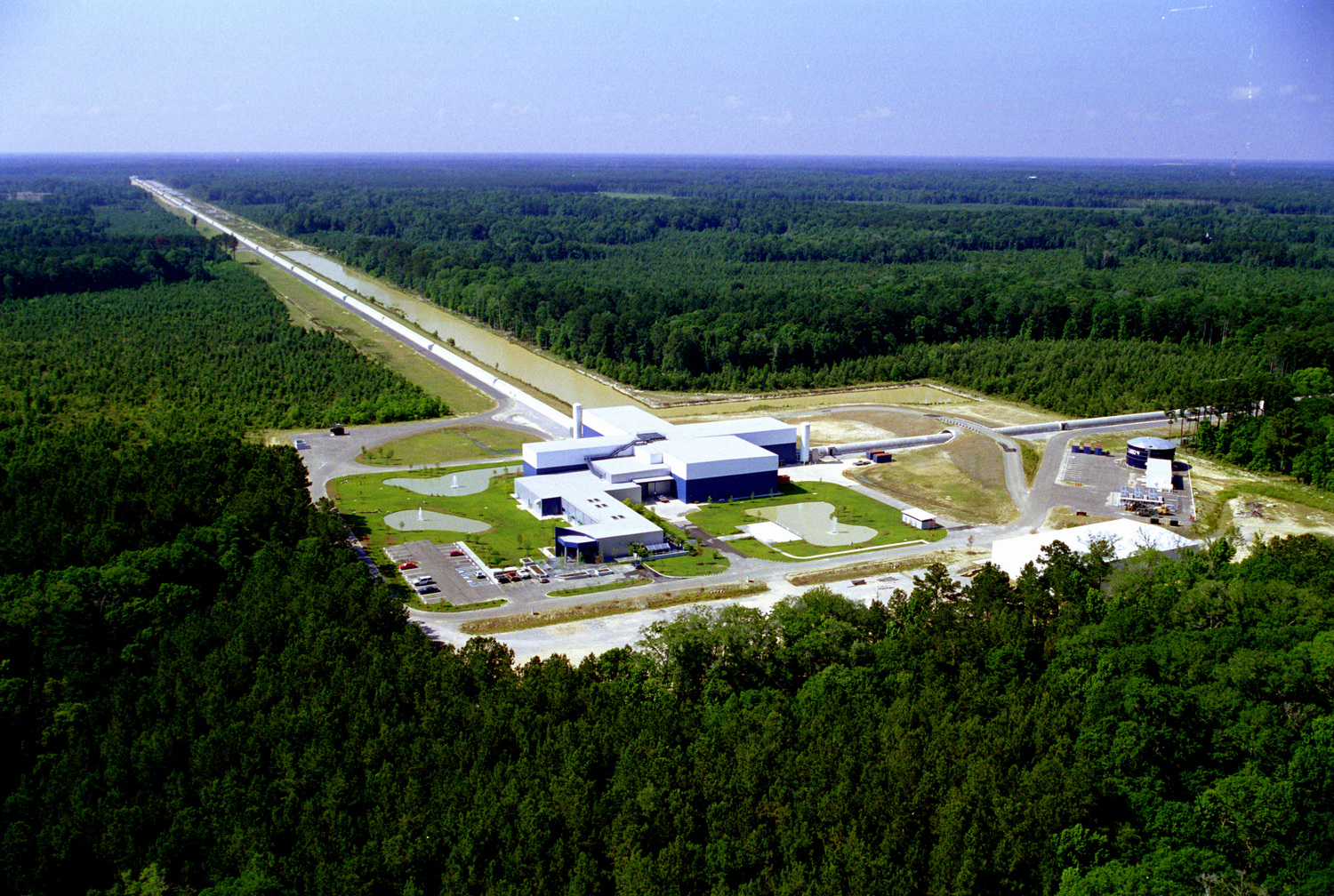}
    \caption{LIGO Livingston}
\end{subfigure}\hfill
\begin{subfigure}{0.49\linewidth}
    \includegraphics[width=0.9\linewidth]{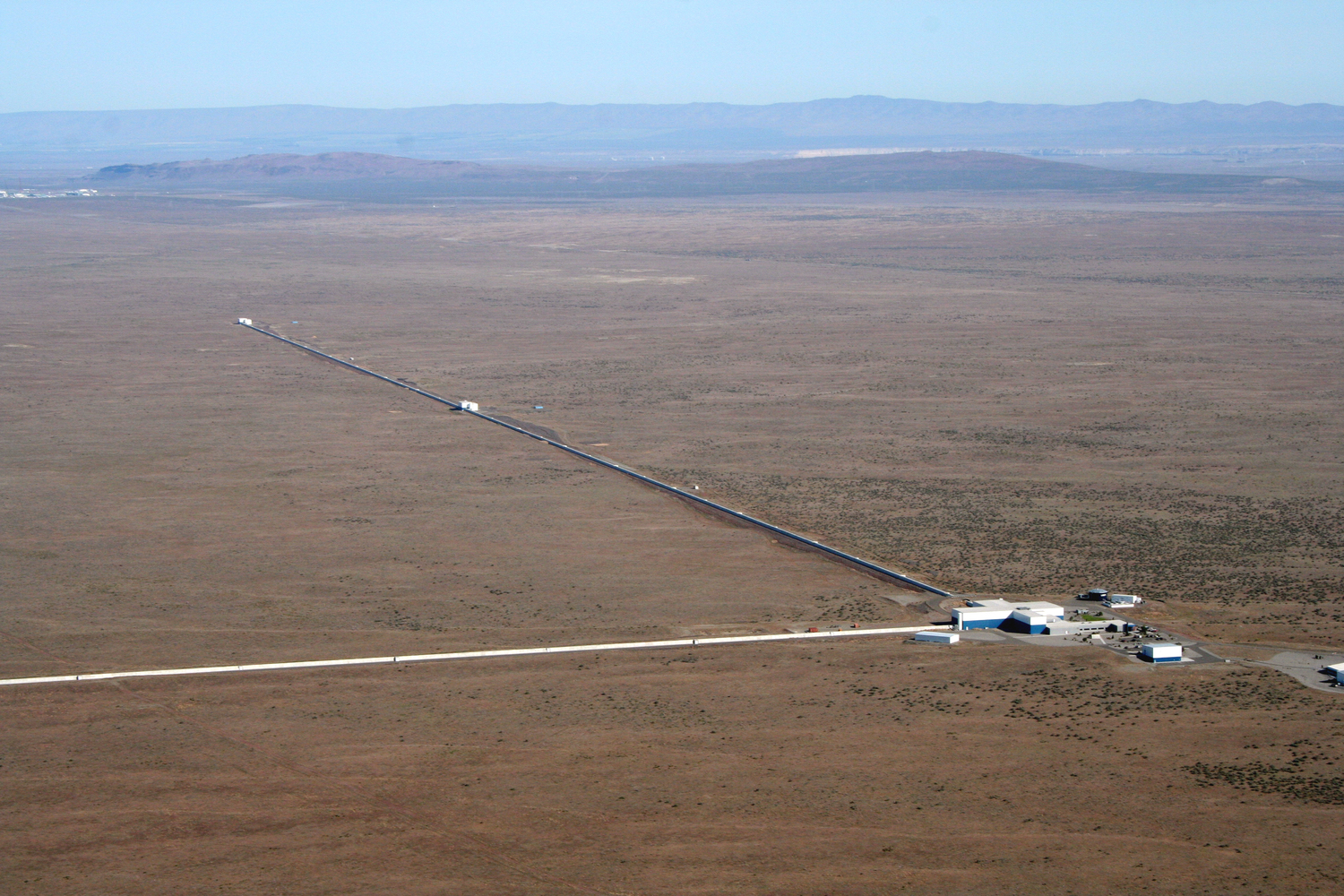}
    \caption{LIGO Hanford}
\end{subfigure}
\caption{LIGO detectors~\cite{adv_ligo_2015}} 
\label{fig:ligo_detectors}
\end{figure*}
\begin{figure*}[!htp]
\centering
\includegraphics[width=0.60\linewidth]{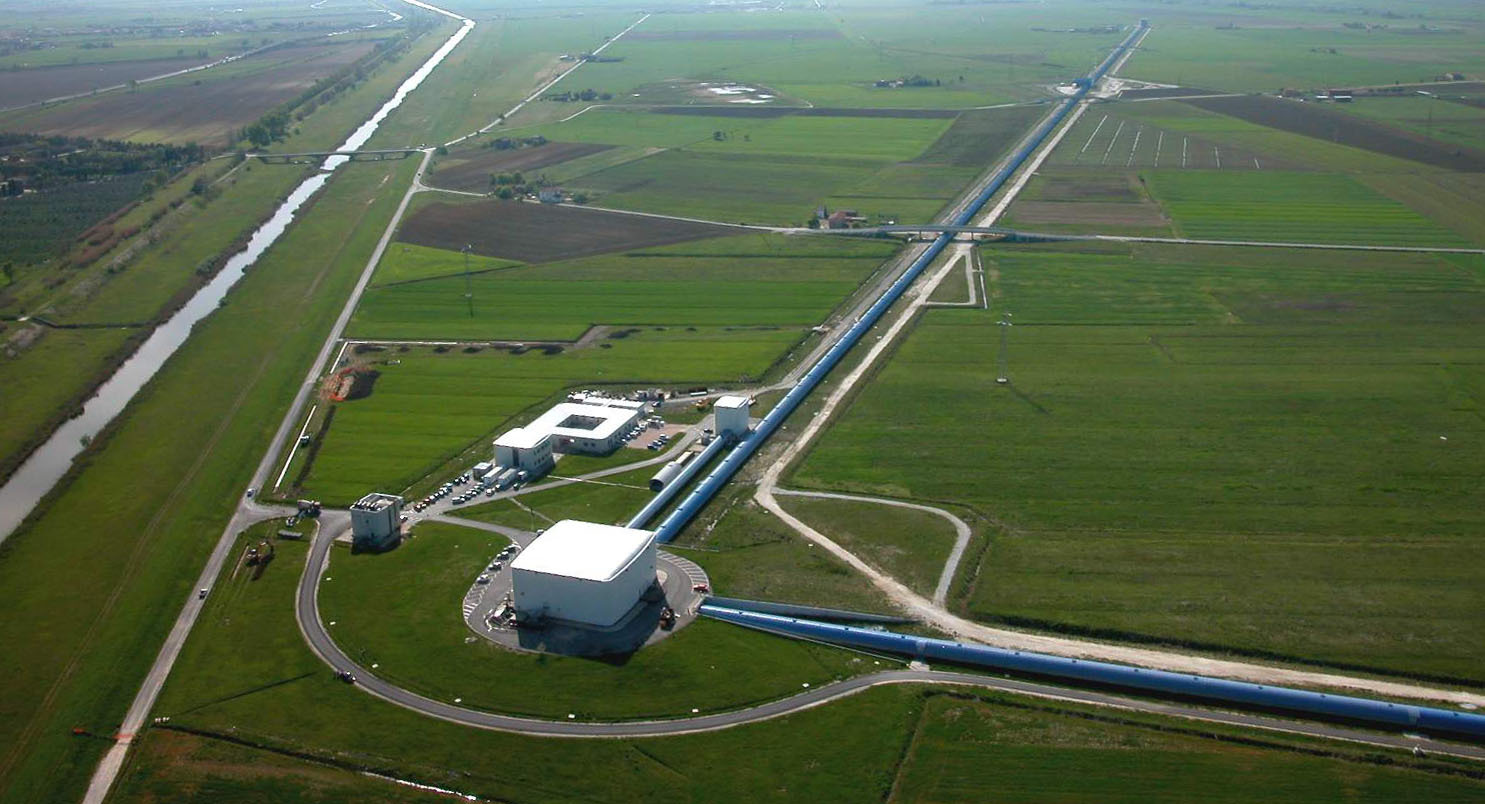}
\caption{Virgo detector~\cite{AdVirgo}}
\label{fig:virgo_detector}
\end{figure*}
The researchers and scientists who are collaborating on this project are members of the LIGO scientific collaboration (LSC). The aerial views of the two LIGO detectors~\cite{adv_ligo_2015} are shown in Fig.~\ref{fig:ligo_detectors}. The Virgo interferometer~\cite{AdVirgo}, located near Pisa, Italy, is hosted by the European Gravitational Observatory (EGO), which is a collaboration between France and Italy. It has an arm length of $3$ km, which is shorter than the arm lengths of the two LIGO interferometers. An aerial view of the same is shown in Fig.~\ref{fig:virgo_detector}. Situated near the city of Hida in Gifu Prefecture, Japan, the Kamioka Gravitational Wave Detector (KAGRA)~\cite{Somiya_2012, PhysRevD.88.043007, 10.1093/ptep/ptaa125} is another interferometer-based GW detector. It has a $3$ km long arm. Figure~\ref{fig:noise_sense_observ_scene_gb_gw} illustrates the target strain sensitivities of the LIGO, Virgo, and KAGRA detectors for various observing runs. As of now, O4 is ongoing, while O5 represents forthcoming observing runs in which all the detectors aim to reach their respective theoretical design sensitivities.
Another gravitational wave (GW) detector following the LIGO design~\cite{LIGO_India, LIGO_India_location, UNNIKRISHNAN_2013, abbott2020prospects, unnikrishnan2023ligoindia} is scheduled for construction in Aundha, Maharashtra, India, as illustrated in Figure~\ref{fig:LIGO_india}. This facility will feature $4$ km long arms and is anticipated to become part of the worldwide network of GW detectors by the year $2030$. In addition to the ground-based GW detectors mentioned earlier, commonly referred to as the second generation (``2G") detectors, there are also plans for third-generation (``3G") detectors like the Einstein Telescope~\cite{2010CQGra..27h4007P, 2011CQGra..28i4013H} and Cosmic Explorer~\cite{Abbott_2017_cosmic_expl, reitze2019cosmic}.
\paragraph{Space borne detectors}
As previously discussed, ground-based gravitational wave detectors cannot access frequencies below 10 Hz due to the presence of seismic and Newtonian noise. To explore the lower frequency range, especially in the millihertz (mHz) region where supermassive black holes could be potential GW sources, space-based observatories are essential due to the absence of seismic noise. Several planned space-based detectors, including the Laser Interferometer Space Antenna (LISA)~\cite{babak2021lisa} and DECi-hertz Interferometer Gravitational-wave Observatory (DECIGO)~\cite{2011CQGra..28i4011K, kawamura2020current, Sato_2017}, aim to accomplish this. LISA is a collaborative project between the European Space Agency (ESA) and the National Aeronautics and Space Administration (NASA). It consists of three spacecraft positioned in an equilateral triangle configuration with sides measuring $5$ million km, and the center of this triangle is located approximately $50$ million km behind Earth. LISA has been optimized for detecting GW frequencies in the range of $0.1$ mHz to $0.1$ Hz. An artist's representation of LISA is provided in Figure~\ref{fig:LISA}. DECIGO is also a proposed space-based GW detector, a Japanese initiative designed to be sensitive to GW frequencies between $0.1$ Hz and $10$ Hz. The characteristic strain sensitivity of all these detectors can be seen in Fig.~\ref{fig:strain_sensitiv}, with DECIGO effectively bridging the gap between the sensitive frequency bands of LIGO and LISA. 

In the next section, we shall discuss various strategies to sample probability distributions, which are the backbone of the Bayesian source reconstruction of GW sources. 
%

\section{Parameter estimation}

\subsection{Inverse problems}

One of the objectives in the scientific investigation of a particular problem is to establish a connection between the physical parameters defining a model, denoted as $\mathcal{X}$, and the observations in the form of data, denoted as $d$, obtained through experimental procedures. Assuming an adequate understanding of the underlying physics, it is possible to define a function $\mathcal{G}$ in such a way that
\begin{equation}
    \mathcal{G}\mathcal{X} = d
    \label{eq:inverse_prob_1}
\end{equation}

Typically, real observations are invariably affected by noise, stemming from various factors like unaccounted influences on instrument readings, numerical approximations, and more. It is reasonable to consider the data as a combination of pure, noise-free observations denoted as $d_{true}$, accompanied by a noise component $\sigma$
\begin{equation}
\begin{split}
 d 
    &= \mathcal{G}(\mathcal{X}_{true}) + \varsigma \\
    &= d_{true} + \varsigma
\end{split}
\label{eq:inverse_prob_2}
\end{equation}

The forward problem involves finding $d$ given $\mathcal{X}$. However, our focus lies in estimating $\mathcal{X}$ given $d$, a task commonly referred to as the inverse problem. In numerous scenarios, we can express model parameters as an $n$-component vector denoted by $\vec{\mathcal{X}}$. Conversely, if the data is presented as a finite number of data points, it can be represented as an $m$-element vector denoted by $\vec d$. Such scenarios are termed discrete inverse problems or parameter estimation problems. Solving inverse problems is challenging as there could be multiple models that adequately fit the data. Nevertheless, once a solution is obtained, it must be evaluated for its physical plausibility, its ability to predict data, and its alignment with other constraints. We will now discuss a very well-known inverse problem, linear regression. 
\paragraph{Linear regression}
Linear regression is the problem of finding a model, linear in its parameters, that approximately fits a set of data. Given a set of $m$ observations $\vec d$ and $n$ model parameters represented as $n$-component vector $\vec{\mathcal{X}}$ which we want to estimate, we can express this discrete linear inverse problem as
\begin{equation}
    \mathcal{G}\vec{\mathcal{X}} = \vec d
    \label{eq:linear_reg_1}
\end{equation}

For simplicity, we assume that $\mathcal{G}$ has full column rank~\footnote{A $m$ by $n$ matrix $\textbf{\textit{M}}$ has a full column rank if the rank of $\textbf{\textit{M}}$ is equal to n, where the rank of the matrix $\textbf{\textit{M}}$ is defined as the dimension of column space of \textbf{\textit{M}}.}. In many cases, the dimension of the column space of $\mathcal{G}$ is smaller than $\vec{\mathcal{X}}$, and a noisy data $\vec d$ will generally lie outside of the column space of $\mathcal{G}$. In these cases, there is no solution $\vec{\mathcal{X}}$ that satisfies Eq.~\eqref{eq:linear_reg_1} exactly. An approximate solution $\vec{\mathcal{X}}$ can be found such that it minimizes some measure of misfit between the $\vec d$ and $\mathcal{G}\vec{\mathcal{X}}$. In this context, we define residual vector $\vec r$
\begin{equation}
    \vec r = \mathcal{G}\vec{\mathcal{X}} - \vec d 
    \label{eq:resd_vector}
\end{equation}

One of the measures of misfit is the $2$-norm of the residual vector, which is defined as follows:
\begin{equation}
    ||\vec r||_2 = \sqrt{{\vec r}^{\:T} {\vec r}},
    \label{eq:2-norm}
\end{equation}
and the model that minimizes this $2$-norm is called the least-squares solution. The least squares solution has the advantage that it is geometrically intuitive to understand, and it is statistically the most likely solution if the data errors are normally distributed. It can be obtained by projecting $\vec d$ onto column space of $\mathcal{G}$.
\begin{equation}
    \mathcal{G} {\vec x}_{ls} = \textbf{\textit{p}}_{\textbf{\textit{A}}}\vec d
    \label{eq:least_sqr_eq_1}
\end{equation}
where $\textbf{\textit{p}}_{\textbf{\textit{A}}}\vec d$ denotes the projection of $\vec d$ onto column space of $\mathcal{G}$ and subscript \textit{ls} denotes least squares. Then $\mathcal{G} {\vec x}_{ls} - \vec d$, becomes perpendicular to the column space of $\mathcal{G}$ that inturn implies that each of the columns of $\mathcal{G}$ will be orthogonal to $\mathcal{G} {\vec x}_{ls} - \vec d$ which can be succintly expressed as
\begin{equation}
    \mathcal{G}^T(\mathcal{G} {\vec x}_{ls} - \vec d) = 0
    \label{eq:least_sqr_eq_2}
\end{equation}
Rearranging the above equation leads to a system of equations known as normal equations:
\begin{equation}
    \mathcal{G}^T\mathcal{G} {\vec x}_{ls} = \mathcal{G}^T\vec d
    \label{eq:normal_equations}
\end{equation}
From the above equation, we find the least-squares solution to be 
\begin{equation}
    \vec{\mathcal{X}}_{L2} = (\mathcal{G}^T\mathcal{G})^{-1} \mathcal{G}^T\vec d
    \label{eq:l2_soln}
\end{equation}
It can be shown that if $\mathcal{G}$ has a full column rank, then $(\mathcal{G}^T\mathcal{G})^{-1}$ exists.

\paragraph{L2-norm: A maximum likelihood solution}
Considering that the data is derived from measurements that inherently include random errors, it becomes essential to seek a statistically sound solution. One of these approaches is maximum likelihood estimation, which raises the question: given the observations, their statistical attributes, and a mathematical model for the forward problem, what model would be most likely to produce these observations?

Let's assign a probability density ${p_i(d_i\mid \vec{\mathcal{X})}}$ to the individual observations $d_i$, given a model $\vec{\mathcal{X}}$. Assuming the observations are independent, we can assign a joint probability density to all observations as follows:

\begin{equation}
    p(\vec d\mid \vec{\mathcal{X}}) = \prod_{i=1}^{m} p_i(d_i\mid \vec{\mathcal{X}})
    \label{eq:joint_prob_density_inv_prob}
\end{equation}

The term ${p(\vec d\mid \vec{\mathcal{X}})}$ in the equation above is also known as the likelihood function and can be denoted as $\mathcal{L}(\vec d\mid \vec{\mathcal{X}})\equiv p(\vec d\mid \vec{\mathcal{X}})$. It's also apparent that for models with an extremely low likelihood of producing the observed data, the likelihood value would approach zero. Conversely, models that are relatively more likely to generate the observed data would yield a higher likelihood value. This is what the maximum likelihood principle also proposes. It is interesting that for such discrete linear inverse problems, given the data errors are independent and normally distributed, the maximum likelihood solution is indeed the least-square solution. To see this, let's assume that the data errors are independent and are distributed normally with zero mean and standard deviation $\sigma_i$ for each observation $d_i$. The joint probability density or likelihood is given by
\begin{equation}
    \mathcal{L}(\vec d\mid \vec{\mathcal{X}}) = \frac{1}{(2\pi)^{m/2} \prod_{i=1}^m \sigma_i}\prod_{i=1}^{m} e^{-\frac{1}{2} (d_i - (\mathcal{G}\vec{\mathcal{X}})^2_i)/\sigma_i^2}
    \label{eq:likelihood_exp_for_lstsq_soln_1}
\end{equation}

The maximization of the above equation can be considered as a minimization of the following equation:

\begin{equation}
    \text{min}\sum_{i=1}^m \frac{(d_i - (\mathcal{G}\vec{\mathcal{X}})_i)^2}{\sigma_i^2}
    \label{eq:likelihood_exp_for_lstsq_soln_2}
\end{equation}

Defining $W = \text{diag}(1/\sigma_1, 1/\sigma_2, ...., 1/\sigma_m)$, we can write down the weighted system of equations as the following:
\begin{equation}
    \mathcal{G}_w\vec{\mathcal{X}} = \vec{d}_w
    \label{eq:weighted_system_of_eqns}
\end{equation}
where $\mathcal{G}_w = W\mathcal{G}$ and $\vec{d}_w = W\mathcal{G}$

The least-squares solution corresponding to the Eq.~\eqref{eq:weighted_system_of_eqns} leads to 

\begin{equation}
    \vec{\mathcal{X}}_{\:l2} = (\mathcal{G}^T_w\mathcal{G}_w)^{-1}\mathcal{G}_w^T\:\vec{d}_w
    \label{eq:least_sq_maxm_likelihood}
\end{equation}

Now, calculating the resulting residual vector $\vec{d}_w - \mathcal{G}_w\vec{\mathcal{X}}_{\:l2}$ and evaluating the corresponding square of the $2$-norm leads to 

\begin{equation}
    ||\vec{d}_w - \mathcal{G}_w\vec{\mathcal{X}}_{\:l2}||^2_2 = \sum_{i=1}^m \frac{(d_i - (\mathcal{G}\vec{\mathcal{X}}_{\:l2})_i)^2}{\sigma_i^2}
\end{equation}

Therefore, the least-squares solution to $\mathcal{G}_w\vec{\mathcal{X}} = \vec{d}_w$ is indeed the maximum likelihood solution. Another important aspect of the least-squares solution is that it is unbiased. To see that let's calculate the expectation value of $\vec{\mathcal{X}}_{\:l2}$ solution:
\begin{equation}
    E[\vec{\mathcal{X}}_{\:l2}] = (\mathcal{G}^T_w\mathcal{G}_w)^{-1}\mathcal{G}^T_w\:E[\vec{d}_w]
    \label{eq:expectation_ml2}
\end{equation}

Since $E[\boldsymbol{\varsigma}] = 0$, it implies $E[\vec{d}_w] = \vec{d}_{\:true, w}$ and $\mathcal{G}_w\vec{\mathcal{X}}_{\:true} = \vec{d}_{\:true}$, it leads to the following:
\begin{equation}
\begin{split}
 \mathcal{G}^T_w\mathcal{G}_w\vec{\mathcal{X}}_{\:true} 
    &= \mathcal{G}^T_w\vec{d}_{\:true, w} \\
    &= \vec{\mathcal{X}}_{\:true}
\end{split}
\label{eq:unbiased_2_norm_soln}
\end{equation}

In the following sections, we will discuss the Bayesian analysis, which is an alternative approach to solving inverse problems.

\subsection{Bayesian analysis}
Bayesian analysis deals with conclusions about a parameter $\theta$ given some observed data $y$. In order to make some statements about $\theta$ given $y$, we need to first define a \textit{joint probability density} for $\theta$ and $y$. It can be written as a product of two densities: a prior density $p(\theta)$ and the data distribution $p(y\mid \theta)$. 
\begin{equation}
\begin{split}
 p(\theta, y) 
    &= p(\theta)\: p(y\mid \theta) \\
    &= p(y)\: p(\theta \mid y)
\end{split}
\label{eq:conditional_prob}
\end{equation}
where the second line of the above equation follows from the properties of conditional probability.
Rearranging Eq.~\eqref{eq:conditional_prob} yields the following equation:
\begin{equation}
    p(\theta\mid y) = \frac{p(y\mid \theta)\: p(\theta)}{p(y)}
    \label{eq:bayes_theorem}
\end{equation}
which is known as \textit{Bayes' theorem} and ${p(y) = \sum_{\theta}p(\theta)\: p(y\mid \theta)}$, which is independence of $\theta$ and for a fixed $y$ can be considered a constant. It is also known as the \textit{model evidence}. 

Before delving into Bayesian analysis, it's essential to briefly discuss the two fundamental distinctions between classical approaches, as outlined in the previous section, and the Bayesian approach. Firstly, in the classical approach, we seek to estimate a specific but unknown model denoted as $\textit{\textbf{m}}_{true}$, whereas in the Bayesian approach, the model itself is not deterministic but rather a random variable. The solution is presented in the form of a probability distribution, allowing us to address various inquiries about the model. Secondly, the Bayesian approach inherently incorporates prior information about the solution, such as strict constraints or intuition derived from experience. This information can be mathematically expressed through the prior density/distribution for the model.

In principle, the posterior density ${p(\theta\mid y)}$ can be calculated using Bayes' theorem, assuming a given model that generates the data $y$. We can generate a grid on the parameter space $\theta$, covering the important regions of the posterior density and calculating the right-hand side of the Eq.~\eqref{eq:bayes_theorem} would give us an estimate of posterior density $p(\theta)$ at those values of $\theta$. For a higher dimensional parameter space, generating a dense multidimensional grid over the parameter space and evaluation of ${p(\theta\mid y)}$ would become prohibitively expensive. For instance, a model with $15$ parameters would require a multidimensional grid over $\theta$ with $N^{15}$ points in the parameter space, where $N$ is the number of points in each dimension. Even for a modest value of $N=10$, the number of points would be a whopping $10^{15}$. The evaluation of $p(\theta \mid y)$ at each of those $10^{15}$ points would take $\sim 32000$ years, assuming each evaluation takes $\sim 1$ ms. Of course, $10$ points in each dimension might not be dense enough to pick up the important features of the posterior density. This simple example highlights the importance of alternative strategies for estimating the posterior density ${p(\theta \mid y)}$. There exist many sampling methods to draw representative samples from the posterior density. We will now discuss some of these methods in the following sections. 

\subsection{Sampling}
\subsubsection{Direct approximation}
Given the target density $p(\theta)$, evaluated at a set of evenly spaced values $\theta_1, \theta_2,....,\theta_N$, that cover a broad range of the parameter space for $\theta$, one can draw samples from $p(\theta)$ using the inverse-CDF method, where CDF stands for the cumulative distribution function (CDF). The CDF of a one-dimensional distribution, $p(\theta)$, is defined by
\begin{equation}
    F(\theta) = \begin{cases}
    \sum_{\theta^{*} \leq \theta}, p(\theta^{*})& \text{if $p$ is discrete}.\\
    \int_{-\infty}^{\theta} p(\theta^{*})\, d\theta^{*}, & \text{if $p$ is continuous}.
    \end{cases}
\end{equation}
The inverse CDF can be used to obtain random samples from the distribution $p(\theta)$ as follows:
\begin{enumerate}
    \item First, a random value, $U$ is drawn from $\mathcal{U}(0,1)$ which denotes a uniform distribution on $[0, 1]$. 
    \item Set $\theta = F^{-1}(U)$. The $F$ may not necessarily be a one-to-one function, but $F^{-1}(U)$ is unique with probability 1. The value $\theta$ will be a random draw from $p(\theta)$ as long as $F^{-1}(U)$ is simple to calculate (see Theorem $2.1.10$, p.$54$.~\cite{casella2002statistical}).
\end{enumerate}

Let's take an example of exponential distribution, ${p(\theta)= \lambda \: \exp(-\lambda\: \theta)}$. We want to draw some samples from this distribution, which is evaluated at some evenly spaced $\theta$ values. We first estimate an empirical CDF using these $\theta$ values and then interpolate it using a cubic spline to get a smooth CDF function. Once we have the CDF interpolant, we can invert it, as mentioned earlier, to generate random samples from $p(\theta)$.
\begin{figure*}[!hbt]
    \centering
    \includegraphics[width=0.85\textwidth, clip=True]{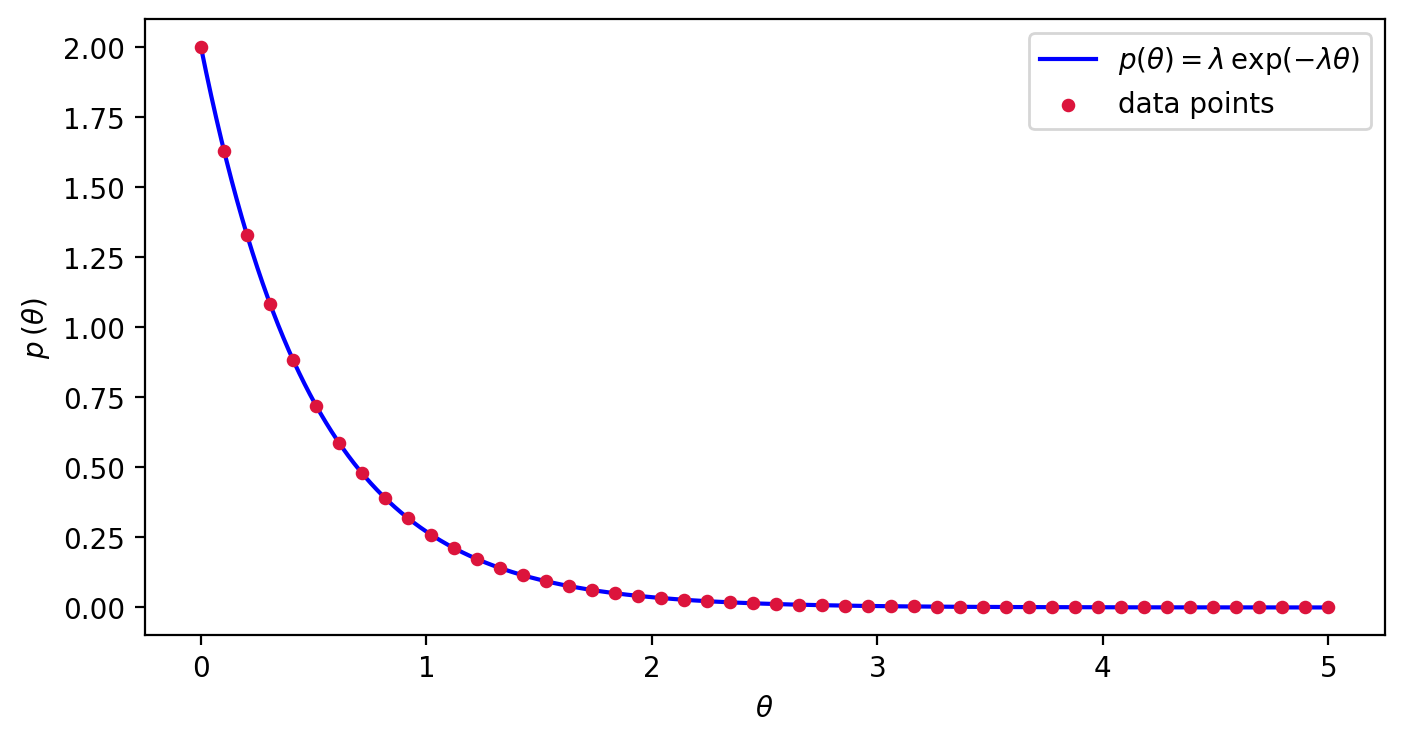}
    \caption{The red dots represent the values of $\theta$ where target density $p(\theta)$ is given. The task is to use these $\theta$ values and their corresponding densities $p(\theta)$ to draw random samples from the $p(\theta)$ itself.}
    \label{fig:direct_approx_expo_dist}
\end{figure*}
\begin{figure*}[!hbt]
    \centering
    \begin{subfigure}{0.49\linewidth}
        \includegraphics[width=\linewidth]{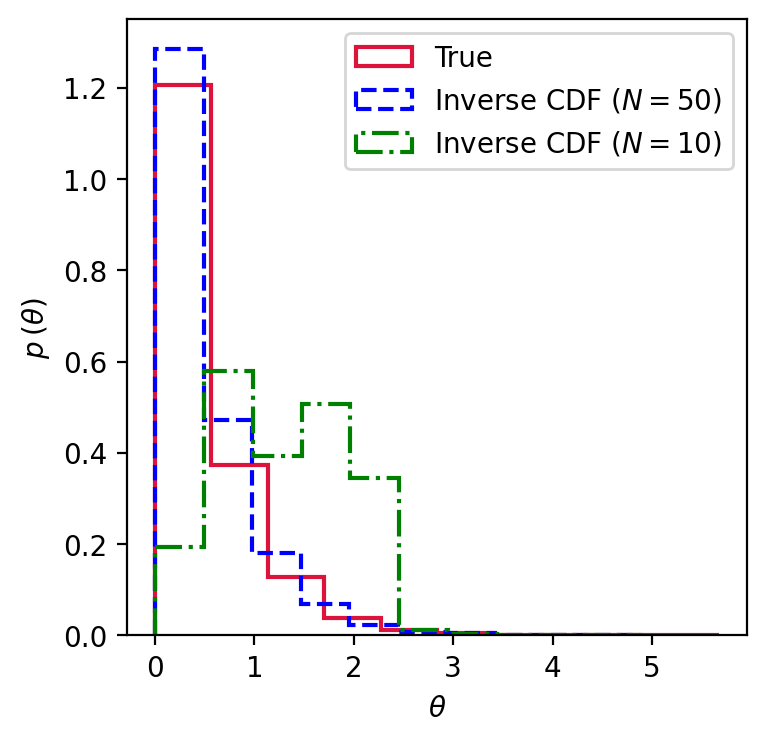}
    \end{subfigure}\hfill
    \begin{subfigure}{0.49\linewidth}
        \includegraphics[width=\linewidth]{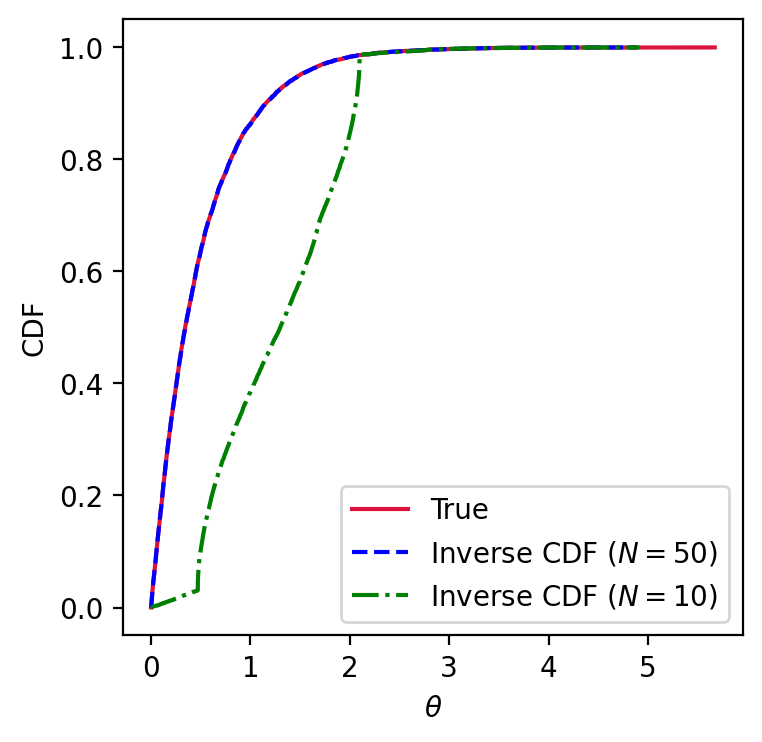}
    \end{subfigure}
    \caption{In the left panel, the red histogram shows samples directly drawn from the true distribution, while blue and green histograms of the samples are obtained using the inverse CDF method for $N=50$ and $N=10$ data points, respectively. On the right panel, the corresponding CDFs are plotted.} 
    \label{fig:direct_approx_pdfs_cdfs}
\end{figure*}

As evident from Fig.~\ref{fig:direct_approx_pdfs_cdfs}, $N=10$ data points are not enough for the inverse CDF method to accurately sample from the target density $p(\theta)$, while with $N=50$ data points, the samples obtained from the inverse CDF method agree well with the true distribution. This illustration shows the limitation of the direct approximation method. If the points $\theta_i$ are spaced closely enough and mostly cover high-density areas, this method works well. Clearly, in high-dimensional multivariate problems, the number of points needed would become so large that it would become prohibitively expensive to compute  $p(\theta)$ at every point of the multidimensional grid. 

\subsubsection{Rejection sampling}
Rejection sampling~\cite{b51783ba-efdf-3457-8f30-343b049f284c} is another method to sample from a distribution when it might not be possible to draw samples from it directly. Suppose we want to draw a single random draw from a distribution $p(\theta)$. To perform rejection sampling, a positive function $q(\theta)$ for all $\theta$, is required for which $p(\theta) > 0$ that has the following properties:

\begin{itemize}
     \item A sample can be drawn from a probability distribution proportional to $q$. 
    \item There must exist some known constant $M$ for which the \textit{importance ratio} ${p(\theta)/q(\theta) \leq M}$ for all $\theta$. 
\end{itemize}
This algorithm proceeds in the following steps:
\begin{enumerate}
    \item A random sample $\theta$ is drawn from the distribution proportional to $q(\theta)$.
    \item $\theta$ is accepted as a draw from $p(\theta)$ with probability $p(\theta)/M\,q(\theta)$.
\end{enumerate}
\begin{figure*}[!hbt]
    \centering
    \includegraphics[width=0.85\textwidth, clip=True]{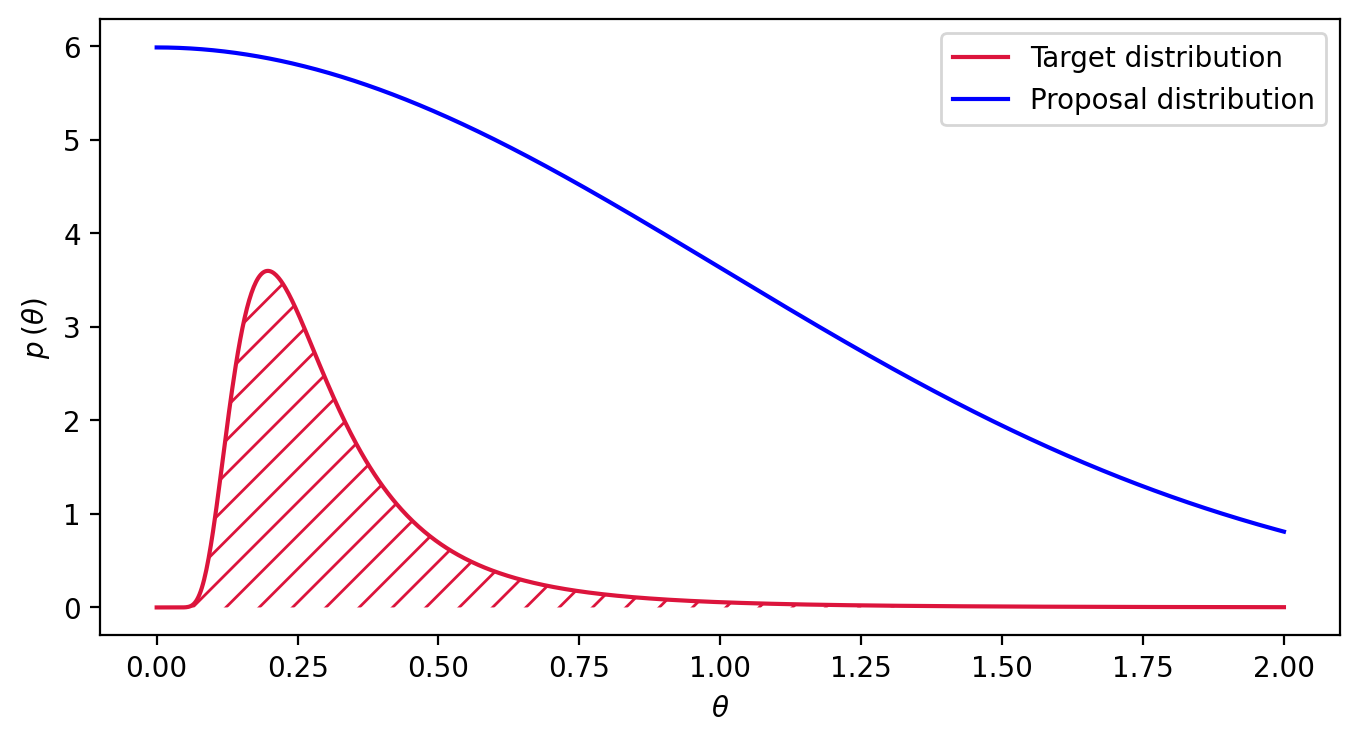}
    \caption{The red curve represents the target distribution $p(\theta)$, which is a inverse-Gamma distribution. The blue curve is the proposal distribution $q(\theta)$, which in this case is a Gaussian distribution. The probability that a sample $\theta$ is accepted is the ratio of the height of the lower curve to the height of the higher curve at the value $\theta$.}
    \label{fig:rejection_samp_expo_dist}
\end{figure*}
\begin{figure*}[!hbt]
    \centering
    \begin{subfigure}{0.49\linewidth}
        \includegraphics[width=\linewidth]{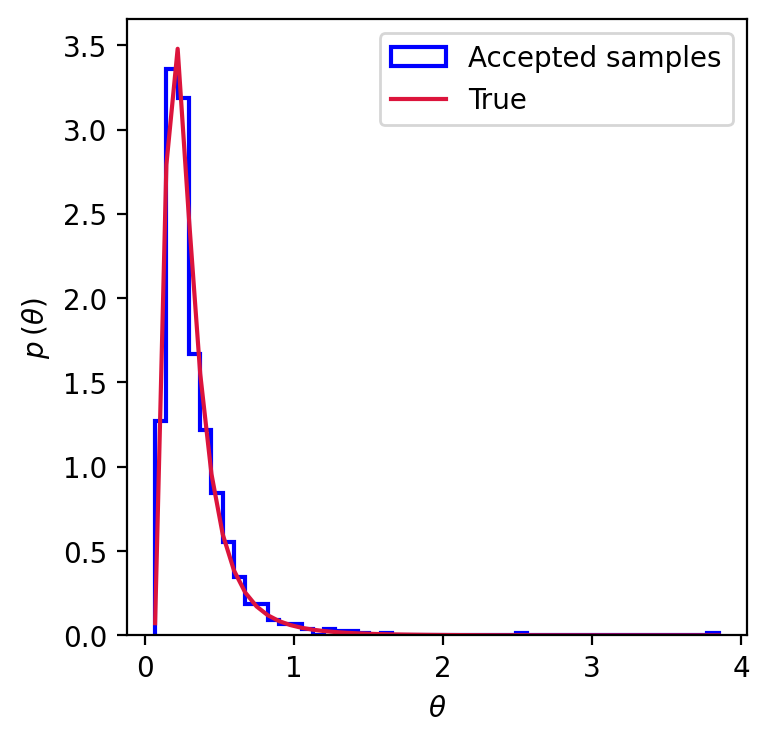}
    \end{subfigure}\hfill
    \begin{subfigure}{0.49\linewidth}
        \includegraphics[width=\linewidth]{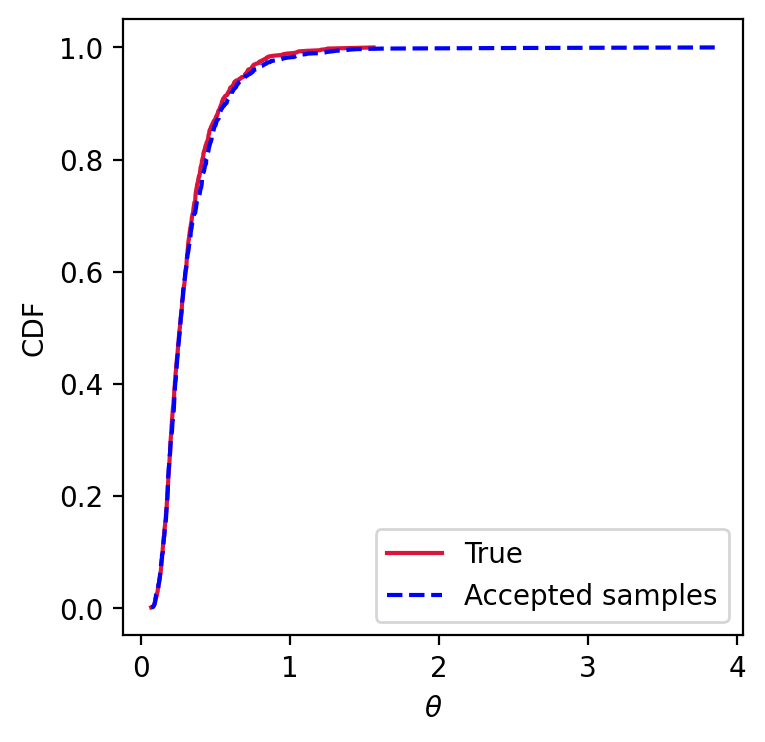}
    \end{subfigure}
    \caption{In the left panel, the red curve is the true distribution $p(\theta)$, while the blue histogram shows the accepted samples in the rejection sampling algorithm. On the right panel, their corresponding CDFs are shown.} 
    \label{fig:rejection_samp_pdfs_cdfs}
\end{figure*}
Fig.~\ref{fig:rejection_samp_expo_dist} shows the target distribution (an inverse-Gamma distribution) given by
\begin{equation}
    p(\theta) = \frac{x^{-a-1}}{\Gamma(a)}\: \exp\left[-\frac{1}{x}\right]
\end{equation}
whereas the proposal distribution $q(\theta)$ is a gaussian distribution $\mathcal{N}(\mu, \sigma^2)$ with a given mean $\mu$ and standard deviation $\sigma$.

\vspace{1mm}
An effective option for the proposal distribution, $q(\theta)$, might involve a distribution that is approximately proportional to $p(\theta)$. Naturally, if we have $q$ proportional to $p$ with an appropriate $M$, each draw during step $2$ would be accepted with probability $1$. This algorithm possesses a self-monitoring feature, as only a few samples are accepted when the method is not operating efficiently.

\subsubsection{Importance sampling}
It is a method to estimate the expectation values using a random sample drawn from an approximate distribution to the target distribution. Let's consider a case where we want to compute $E(f(\theta))$ but cannot generate random samples from the target distribution, $p(\theta)$. 

If $q(\theta)$ is the distribution from which we can draw random samples, then we can estimate $\mathbb{E}(f(\theta))$ as follows:
\begin{equation}
\begin{split}
\mathbb{E}(f(\theta)) 
    &= \frac{\int f(\theta)\: p(\theta)\: d\theta}{\int p(\theta)\: d\theta} \\
    &= \frac{\int \left[f(\theta)\: p(\theta)\: / \: q(\theta)\right]\: q(\theta)\: d\theta}{\int \left[ p(\theta)\: /\: q(\theta)\right]\: q(\theta)\: d\theta}.
\end{split}
\label{eq:importance_sampl}
\end{equation}
which can be estimated using $N$ number of draws $\theta_1,...,\theta_N$ from $q(\theta)$ by the expression,
\begin{equation}
    \frac{\frac{1}{N}\: \Sigma_{n=1}^{N}\: f(\theta_n)\: w(\theta_n)}{\frac{1}{N}\: \Sigma_{n=1}^{N}\:w(\theta_n)}
\label{eq:importance_expectation}
\end{equation}
where ${w(\theta_n) = p(\theta_n)/g(\theta_n)}$ are called \textit{importance ratios} or \textit{importance weights}. Note that $p(\theta)$ may not necessarily be normalized. 

\begin{figure*}[hbtp]
\begin{subfigure}{0.49\linewidth}
    \includegraphics[width=\textwidth]
    {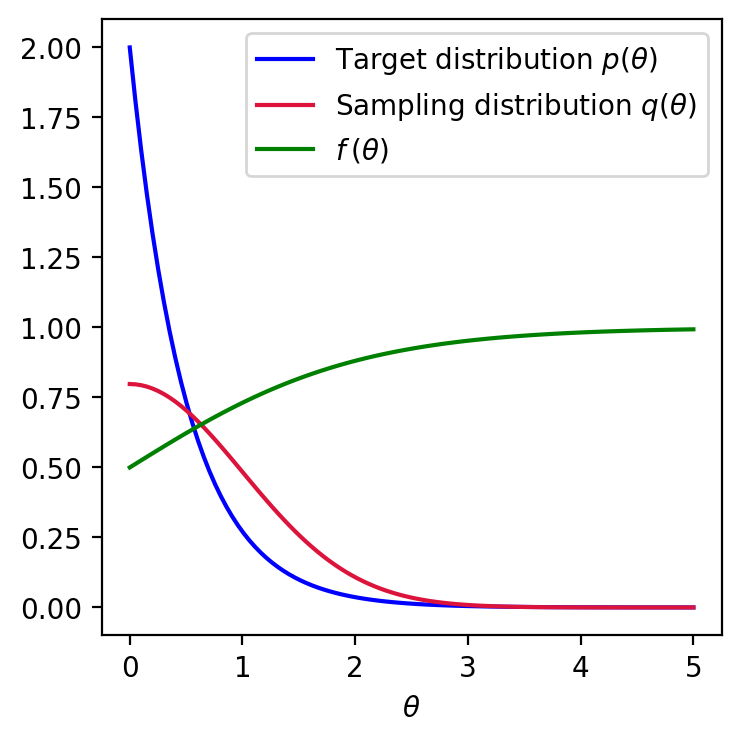}
\end{subfigure}\hfill
\begin{subfigure}{0.49\linewidth}
    \includegraphics[width=\textwidth, height=\textwidth]{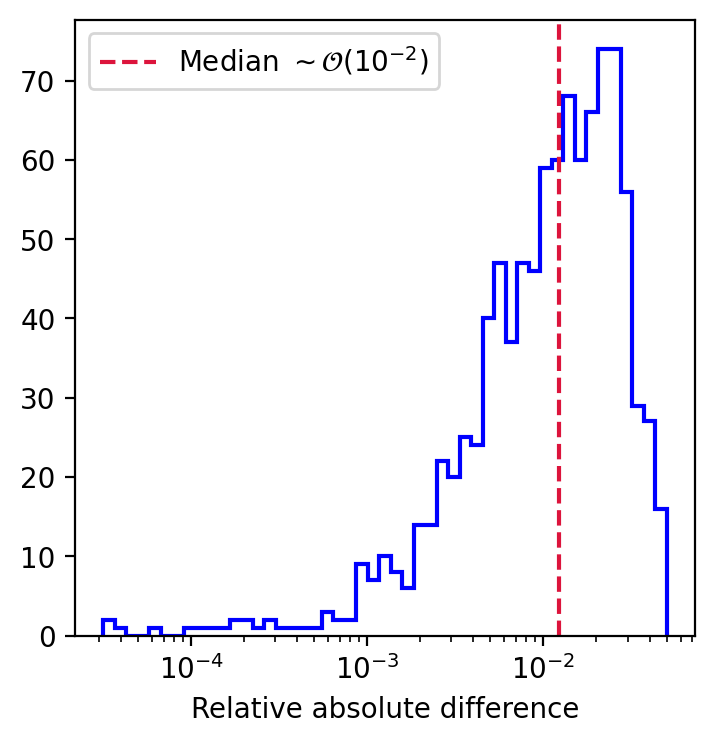}
\end{subfigure}\hfill
\caption{The left panel shows the target (blue) and sampling distribution (red), and the function $f(\theta)$ (green) whose expectation value is to be estimated. The right panel shows a histogram of the relative absolute difference between $\mathbb{E}(f(\theta))_{p(\theta)}$ and $\mathbb{E}(f(\theta))_{q(\theta)}$. Evidently, the relative difference is mostly $\lesssim 5 \%$.}
\label{fig:importance_samp_dist_rel_diff}
\end{figure*}

For an illustration of this method (see the left panel in Fig.~\ref{fig:importance_samp_dist_rel_diff}), suppose we want to estimate  $\mathbb{E}(f(\theta))$, where $f(\theta) = \frac{1}{1 \:+\: \exp(-\theta)}$ and our target distribution is an exponential distribution. Let's choose the sampling distribution $q(\theta)$ to be a truncated normal distribution (so that all the samples are positive) with $\mu = 0$ and $\sigma = 1$.

From Fig.~\ref{fig:importance_samp_dist_rel_diff}, we can conclude the estimation of the integral in Eq.~\eqref{eq:importance_sampl} can be estimated fairly accurately for a suitable choice of the sampling distribution $q(\theta)$. Moreover, a fairly precise estimate of the integral can be obtained if we choose $q(\theta)$ such that $\frac{fp}{q}$ is nearly a constant. One of the limitations of importance sampling~\cite{910e8e7d-2e70-317c-adf6-e21cd50755a1} is that it may not be useful when the importance ratios vary substantially, particularly when they are small with a high probability but with a low probability are huge. It usually happens if $fp$ has wide tails compared to $q$ as a function of $\theta$.
\subsubsection{Markov chain (MC) simulation}
A Markov chain~\cite{feller1991introduction} is defined by a sequence of random elements $\theta_1, \theta_2, ...$ if the conditional distribution of $\theta_{n+1}$ given $\theta_1, \theta_2,..,\theta_n$ depends only on $\theta_n$. It has a \textit{stationary transition distribution} if the conditional distribution of $\theta_{n+1}$ given $\theta_n$ does not depend on $n$. A Markov chain simulation is typically used when it is not possible or not computationally efficient to directly sample $\theta$ from the target density $p(\theta\mid y)$. Instead, samples are drawn \textit{iteratively} from a distribution that becomes closer to $p(\theta \mid y)$ as simulation proceeds. The key idea is to construct a Markov chain whose stationary distribution is target distribution $p(\theta)$, and the simulation runs for long enough to reach a point where the stationary distribution is very close to the target density. In other words, in each iteration, we are sampling from a distribution that is becoming close to the target distribution. An MC simulation starts from some point $\theta_0$ and then in each iteration, a new point $\theta_i$ is proposed from a \textit{transition/proposal distribution} ${T_i(\theta_i\mid \theta_{i-1})}$, where $i$ denotes the iteration number. The proposal distribution is constructed in such a way that the MC coverges to the target distribution $p(\theta)$. In the following sections, we will discuss some basic MC simulation methods such as Gibbs sampling, Metropolis, and Metropolis-Hastings algorithms.

\paragraph{Gibbs sampling}
Gibbs sampling~\cite{casella1992explaining} is a kind of MC algorithm that is found to be useful in several multidimensional problems. It is also called alternating conditional sampling because it involves drawing samples for one of the model parameters from a distribution conditioned on the rest of the model parameters. In other words, the parameter space is divided into $d$ components or subvectors, ${\theta = (\theta_1, \theta_2,......,\theta_d)}$ and in every iteration $t$ the sample for each parameter $\theta_j$ is drawn from a distribution ${p(\theta_j\mid \theta^{t-1}_{-j}, y})$, where $\theta^{t-1}_{-j}$ represents all the components of $\theta$, except for $\theta_j$, at their current values:
\begin{equation}
    \theta^{t-1}_{-j} = (\theta^t_1,...,\theta^t_{j-1}, \theta^{t-1}_{j+1},....,\theta^{t-1}_{d}).
\end{equation}
Thus, each subvector $\theta_j$ is updated conditional on the latest values of the other components of $\theta$, which are iteration $t$ values for the components already updated and the iteration $t-1$ values for the others. Let's try to understand the Gibbs sampling algorithm with the help of a simple example. Consider a single observation $\vec y = (y_1, y_2)$ drawn from a bivariate normal distribution with an unspecified mean $\vec \theta = (\theta_1, \theta_2)$ and covariance variance matrix defined as 
$\Sigma=$ $\left(
\begin{smallmatrix}
 1 & \rho \\
 \rho & 1 \\
\end{smallmatrix}
\right)$
. Assuming a uniform prior distribution for the parameter $\theta$, the resulting posterior distribution becomes 
\begin{equation}
    p(\vec \theta \mid \vec y) \sim \mathcal{N}(\vec y, \Sigma)
\end{equation}

Using the properties of multivariate normal distribution, we can write the conditional posterior distributions for each parameter as follows:
\begin{equation}
\begin{aligned}
    p(\theta_1\mid \theta_2, y) &\sim \mathcal{N}({y_1 + \rho (\theta_2 - y_2), (1 - \rho^2)}) \\
    p(\theta_2\mid \theta_1, y) &\sim \mathcal{N}({y_2 + \rho (\theta_1 - y_1), (1 - \rho^2)}) 
\end{aligned}
\end{equation}
Gibbs sampling proceeds in the following steps:
\begin{enumerate}
    \item A random sample $\theta_1$ is drawn from $p(\theta_1\mid \theta^{\text{init}}_2, y)$ where $\theta^{\text{init}}_2$ is the some initial value of $\theta_2$.
    \item Now a sample $\theta_2$ is drawn from $p(\theta_2\mid \theta_1, y)$ where $\theta_1$ is the value drawn in the first step.
    \item Then put $\theta_2 = \theta^{\text{init}}_2$, and then all the steps repeat till the iterations end. 
\end{enumerate}

For this example, we choose $\rho = -0.6$ and generate four independent sequences which start from $(\pm 5.5, \pm 5.5)$.
\begin{figure*}[!hbt]
    \centering
    \begin{subfigure}{0.333\linewidth}
        \includegraphics[width=\linewidth]{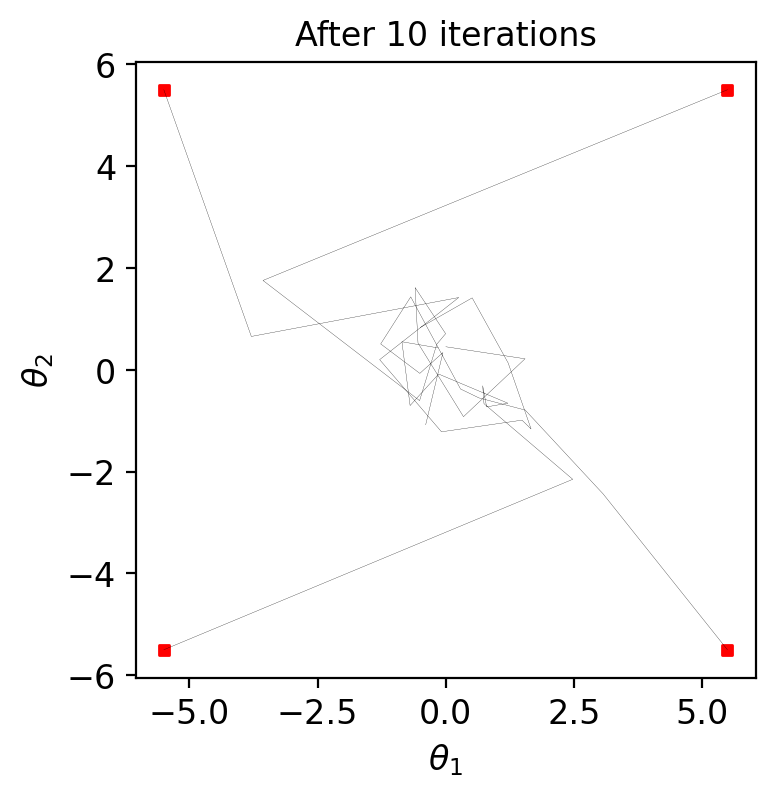}
    \end{subfigure}\hfill
    \begin{subfigure}{0.333\linewidth}
        \includegraphics[width=\linewidth]{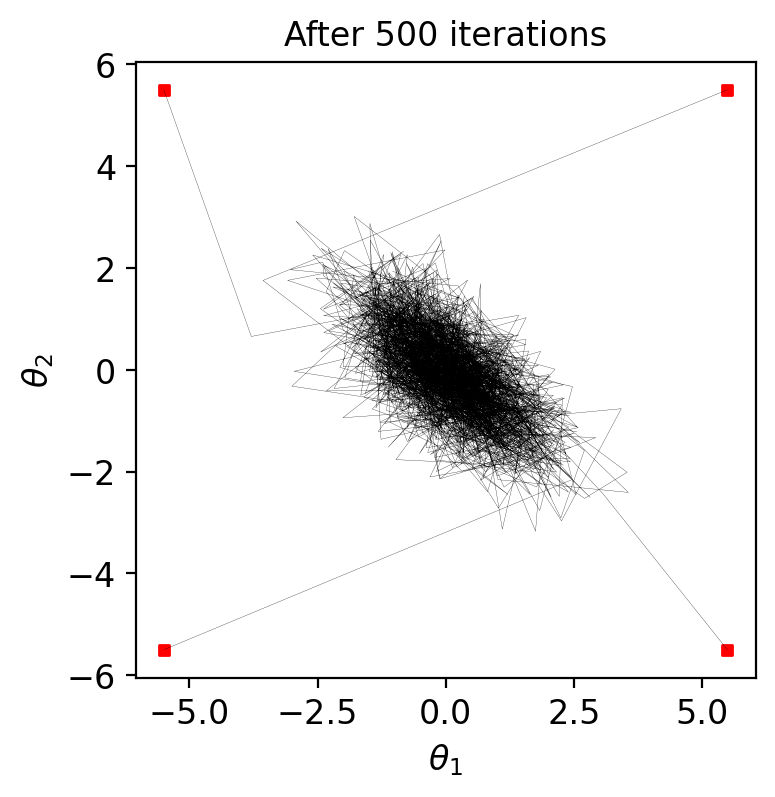}
    \end{subfigure}\hfill
    \begin{subfigure}{0.333\linewidth}
        \includegraphics[width=\linewidth]{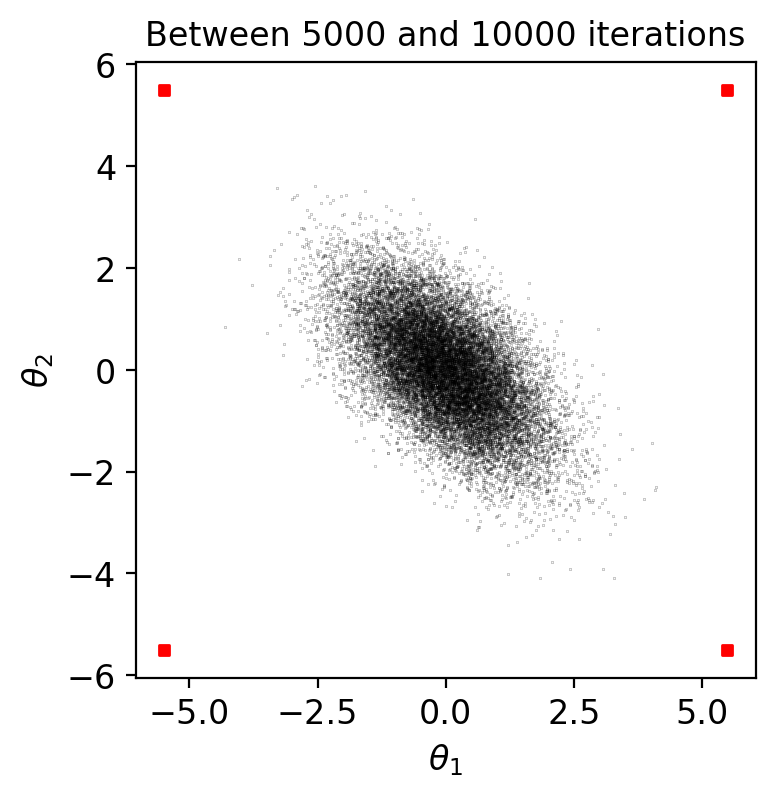}
    \end{subfigure}
    \caption{Gibbs sampling for four independent sequences starting from overdispersed locations in the parameter space, represented by solid red squares,  for a bivariate normal distribution with correlation ${\rho = -0.6}$. The first panel shows the component-wise updating of Gibbs iterations. The second panel shows the distribution of $\theta_1$ and $\theta_2$ after $500$ iterations, which have reached close to convergence. The last panel displays the latter half of the sequences, which have now converged, and the obtained samples accurately represent the target distribution (example reproduced from~\cite{gelman2013bayesian}).} 
    \label{fig:gibbs_sampling_distro}
\end{figure*}
\paragraph{Metropolis algorithm and Metropolis-Hastings algorithm}
Metropolis algorithms~\cite{1953JChPh..21.1087M, 2003AIPC..690...22R} belong to a family of MC simulations used for generating samples from Bayesian posterior distributions. We will now discuss a brief overview of the basic Metropolis algorithm, followed by its extension to the Metropolis-Hastings algorithm.

\paragraph{The Metropolis algorithm}
The Metropolis algorithm uses a random walk with acceptance/rejection criteria to converge to the target distribution. It proceeds as described in the following steps:

\begin{enumerate}
    \item Choose a starting point $\theta_0$. It can also be drawn from a starting distribution $p_0(\theta)$ such that ${p(\theta_0\mid y) > 0}$.
    \item For $t = 1, 2, ..,$ where $t$ indicates the iteration number.
    \begin{itemize}
        \item Draw a sample $\theta^*$ from a proposal distribution at iteration $t$, ${J_t(\theta^*\mid \theta^{t-1})}$. The proposal distribution must be \textit{symmetric}, satisfying the condition ${J_t(\theta_a\mid \theta_b) = J_t(\theta_b\mid \theta_a)}$ for all $\theta_a, \:\theta_b$, and iterations.
        \item Estimate the ratio of densities,
        \begin{equation}
            r = \frac{p(\theta^*\mid y)}{p(\theta^{t-1}\mid y)}.
            \label{eq:metropolis_ratio}
        \end{equation}
        \item Set
        \begin{equation}
        \theta^t = \begin{cases}
        \theta^*, & \text{with probability min($r, 1$)}.\\
        \theta^{t-1}, & \text{otherwise}.
        \end{cases}
        \end{equation}
        \\
        The above step can be performed by generating a uniform random number.
    \end{itemize}
\end{enumerate}

To demonstrate the Metropolis algorithm, we again use a simple example of bivariate normal distribution where the target distribution is a bivariate normal distribution, ${p(\theta\mid y)= \mathcal{N}(\theta\mid 0, \Sigma)}$, where $\Sigma=$ $\left(
\begin{smallmatrix}
 1 & \rho \\
 \rho & 1 \\
\end{smallmatrix}
\right)$ with $\rho = 0.9$. The proposal distribution is also chosen to be a bivariate normal distribution centered on the current iteration with a jumping scale equal to $1/4$ the size: $J_t(\theta^*\mid \theta^{t-1}) \sim \mathcal{N}(\theta^*\mid \theta^{t-1}, 0.25^2\: \Sigma)$, which is also symmetric.
\\
\begin{figure*}[!hbt]
    \centering
    \begin{subfigure}{0.333\linewidth}
        \includegraphics[width=\linewidth]{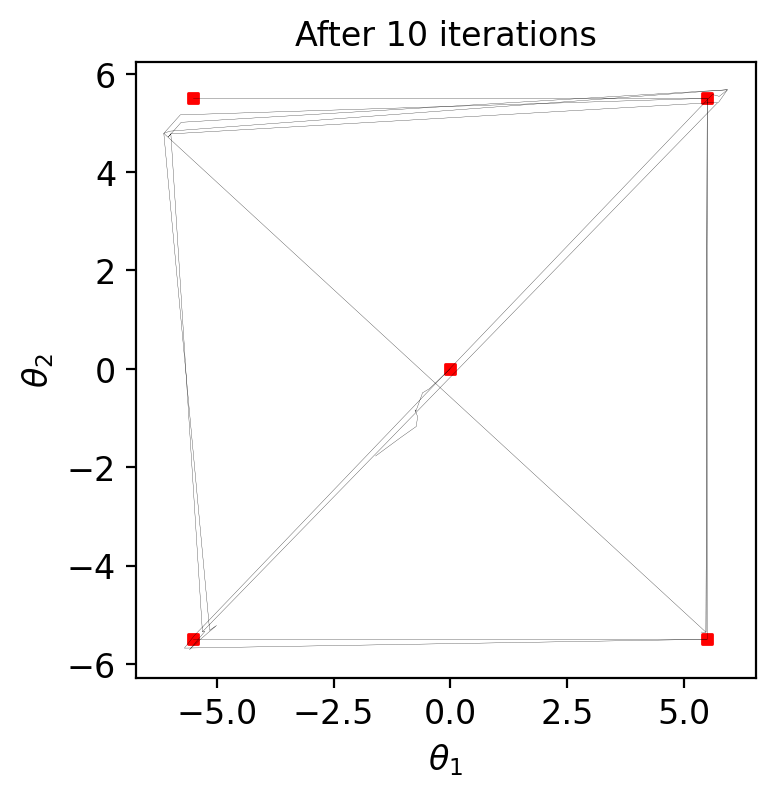}
    \end{subfigure}\hfill
    \begin{subfigure}{0.333\linewidth}
        \includegraphics[width=\linewidth]{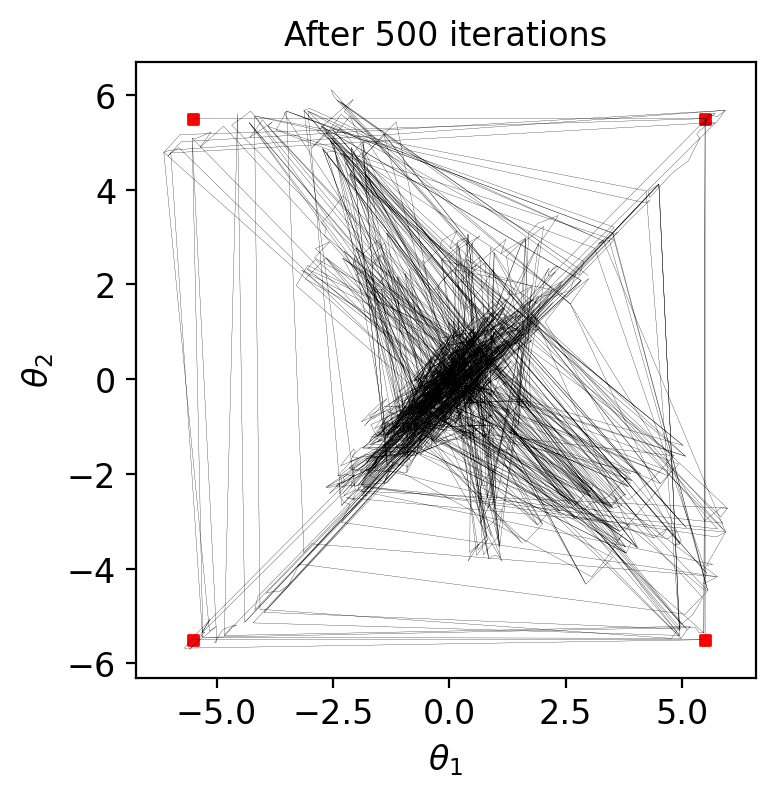}
    \end{subfigure}\hfill
    \begin{subfigure}{0.333\linewidth}
        \includegraphics[width=\linewidth]{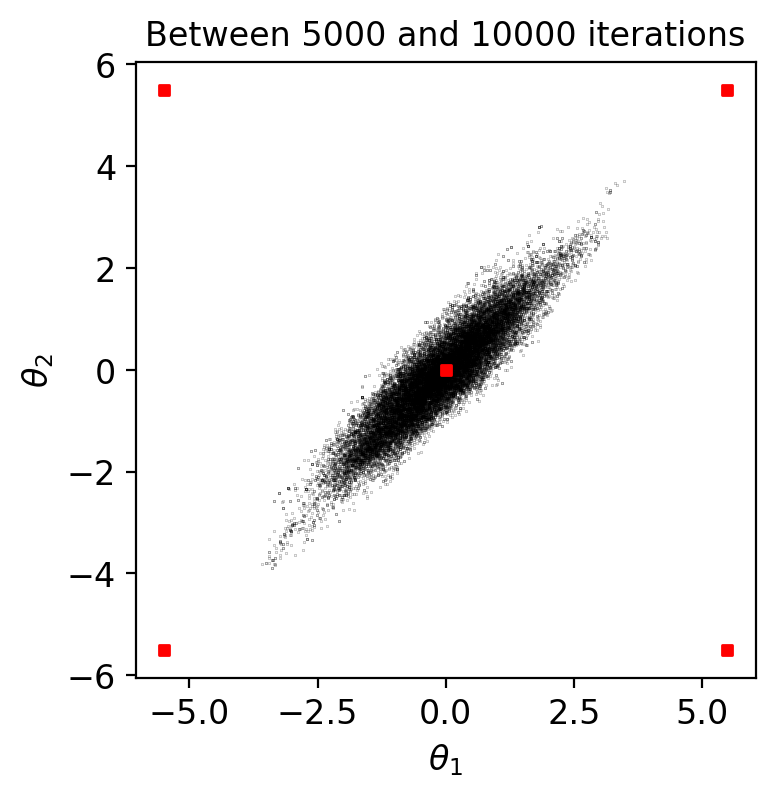}
    \end{subfigure}
    \caption{Metropolis algorithm for five independent sequences starting from overdispersed locations in the parameter space, represented by solid red squares,  for a bivariate normal distribution with correlation ${\rho = 0.9}$. The first panel shows the sampled distribution from the first $10$ iterations, which is far from convergence. The second panel shows the distribution after $500$ iterations and looks close to convergence. In the last panel, samples from the latter half of the sequences are shown, which have now converged and represent the highly correlated draws from the target distribution (example reproduced from~\cite{gelman2013bayesian}).} 
    \label{fig:mc_sampling_distro}
\end{figure*}

\paragraph{The Metropolis-Hastings algorithm}
The Metropolis algorithm, as described in the last section, can be generalized to the Metropolis-Hastings algorithm by relaxing the requirement of symmetric proposal distributions. In other words, $J_t(\theta_a\mid \theta_b)$ may not necessarily be equal to $J_t(\theta_b\mid \theta_a)$. Asymmetric proposal distributions can be useful in speeding up the random walk. Additionally, the $r$ in Eq.~\eqref{eq:metropolis_ratio} is replaced by a ratio of ratios:
\begin{equation}
    r = \frac{p(\theta^*\mid y)/J_t(\theta^*\mid \theta^{t-1})}{p(\theta^{t-1}\mid y)/J_t(\theta^{t-1}\mid \theta^{*})}.
    \label{eq:metropolis_hastings_ratio}
\end{equation}

Let's see how the Metropolis algorithm, indeed, converges to the target distribution. To begin, the generated sequences should constitute a Markov Chain (MC) with a unique stationary distribution. This is fulfilled when the MC meets the criteria of irreducibility, aperiodicity, and non-transience. With the exception of some straightforward instances, these latter conditions are satisfied for most random walks on appropriate distributions. Furthermore, the condition of irreducibility necessitates that the proposal distribution $J_t$ can transition to all states with a non-zero probability. The Metropolis algorithm guarantees that the stationary distribution associated with MC is the target distribution. In order to prove it, let's consider that the algorithm starts at an iteration $t-1$ where a sample is drawn $\theta^{t-1}$ from the target distribution $p(\theta\mid y)$. Suppose two points $\theta_a$ and $\theta_b$, are drawn from $p(\theta \mid y)$ and labelled in such a way that $p(\theta_a\mid y) \leq p(\theta_b\mid y)$. The corresponding transition probability from $\theta_a$ to $\theta_b$ is given by
\begin{equation}
    p(\theta^{t-1}=\theta_a, \theta^t = \theta_b) = p(\theta_a\mid y)\: J_t(\theta_b\mid \theta_a), 
    \label{mc_proof_1}
\end{equation}
where the probability of acceptance is $1$ by construction. Now, the transition probability from $\theta_b$ to $\theta_a$,
\begin{equation}
\begin{split}
 p(\theta^{t}=\theta_a, \theta^{t-1} = \theta_b)
    &= p(\theta_b\mid y)\: J_t(\theta_a \mid \theta_b) \left(\frac{p(\theta_a\mid y)}{p(\theta_b\mid y)}\right) \\
    &= p(\theta_a \mid y)\: J_t(\theta_a \mid \theta_b),
\end{split}
\label{eq:mc_proof_2}
\end{equation}
which is the same as Eq.~\eqref{mc_proof_1} since we assumed a symmetric proposal distribution. Following similar arguments, we can prove the validity of the Metropolis-Hastings algorithm. 
\subsubsection{Nested sampling}
Markov Chain Monte Carlo (MCMC) techniques~\cite{berg2004introduction, gilks1995markov, 10.5555/1051451}, while effective in generating samples for a diverse range of problems, can encounter challenges when dealing with posteriors characterized by widely separated modes. Moreover, due to its generation of samples in proportion to the posterior, calculating the model evidence (refer to Eq.~\eqref{eq:bayes_theorem}) using these samples becomes challenging. The Nested sampling~\cite{skilling2006nested} is a novel approach to estimate both the posterior and evidence. This method addresses the limitations of MCMC by partitioning the posterior into numerous nested slices, generating samples from each slice, and then combining these samples with appropriate weights to reconstruct the target (posterior) distribution. Let us understand how Nested sampling is able to estimate evidence and the corresponding posterior as a byproduct of evidence estimation.

In contrast to MCMC methods, which primarily target direct sampling from the posterior, Nested sampling is designed for estimating the evidence, which is defined as
\begin{equation}
\begin{split}
 \mathcal{Z}
    &\equiv \int_{\Omega_{\vec \theta}}\: p(\vec \theta)\ d\vec \theta \\
    &= \int_{\Omega_{\vec \theta}}\: \mathcal{L}(\vec \theta)\: \pi(\vec \theta) \:d\vec \theta,
\end{split}
\label{eq:intro_nested_evidence}
\end{equation}
and refactor the integral above to encompass the prior volume denoted as $X$ within the enclosed parameter space.
\begin{equation}
 \mathcal{Z} = \int_{\Omega_{\vec \theta}}\: \mathcal{L}(\vec \theta)\: \pi(\vec \theta) \:d\vec \theta = \int_{0}^{1}\: \mathcal{L}(X) \:dX
\label{eq:intro_nested_evidence_refactor}
\end{equation}
where $\mathcal{L}(X)$ corresponds to an equal-likelihood contour which defines the edge of the volume $X$, whereas the prior volume is defined as the fraction of the prior with the likelihood $\mathcal{L}(\vec \theta) \geq \lambda$ and can be written as
\begin{equation}
    X(\lambda) \equiv \int_{\Omega_{\vec \theta}\: : \: \mathcal{L}(\vec \theta) \geq \lambda}\: \pi(\vec \theta)\ d\vec \theta
    \label{eq:restricted_prior_volume}
\end{equation} 
A normalized prior implies that $X(\lambda = 0) = 1$ and $X(\lambda = \infty) = 0$, defining the bounds of integration for Eq.~\eqref{eq:intro_nested_evidence_refactor}. 

\paragraph{Generating samples}
Nested sampling, instead of sampling directly from the posterior, samples from the prior $\pi(\vec \theta)$ subjected to a hard likelihood constraint $\lambda$, which might be difficult for an arbitrary prior $\pi(\theta)$ because of the drastic variations of the density across the parameter space. Nonetheless, the process can be simplified when the prior follows a standard uniform distribution (i.e., flat between $0$ and $1$) across all dimensions, resulting in a constant density within the interior of $\lambda$. This can be achieved employing a suitable ``prior transform'' function $\mathcal{T}$ that maps a set of parameters $\vec \Theta$, characterized by a uniform prior distribution over the $D$-dimensional unit cube, to the desired parameter $\vec \theta$. It simplifies our task from directly sampling from the posterior $p(\vec \theta)$ to a much easier task of repeatedly sampling uniformly within the transformed, constrained prior.

\begin{equation}
        \pi^{'}_{\lambda}(\vec \Theta) \equiv \begin{cases}
        1/X(\lambda) & \mathcal{L}(\vec \theta = \mathcal{T}(\Theta))\geq \lambda\\
        0 & \text{otherwise}
        \end{cases}
    \label{eq:constrained_prior}
\end{equation}

Since the exact value of $X(\lambda)$ at a given likelihood level $\lambda = \mathcal{L}(\vec \theta)$ is unknown, we can construct an estimator $\hat{X}$ with a known statistical distribution. It is evident from Eq.~\eqref{eq:constrained_prior}, $X(\mathcal{L})$ defines a CDF over $\mathcal{L}$. Therefore, we can define the corresponding probability distribution function (PDF) for $\mathcal{L}$ as 

\begin{equation}
    P(\mathcal{L}) \equiv \frac{dX(\mathcal{L})}{d\mathcal{L}} = \frac{d}{d\mathcal{L}}\int_{\Omega_{\vec \theta}}\pi_{\mathcal{L}}(\vec \theta)\: d\vec \theta
\end{equation}

Assuming it is possible to sample $\mathcal{L}$ from its PDF $P(\mathcal{L})$, we can use the Probability Integral Transform (PIT)\footnote{The PIT refers to a theorem stating that if a random variable $X$ follows a continuous distribution with a CDF given denoted by $F_X$, then a random variable defined as $Y = F_X(X)$ has a standard uniform distribution (see Theorem $2.1.10$, p.$54$.~\cite{casella2002statistical}).} to sample $X(\mathcal{L})$ from a standard uniform distribution $\mathcal{U}(0, 1)$:

\begin{equation}
    \mathcal{L}^{'} \sim P(\mathcal{L})  \Rightarrow X(\mathcal{L}^{'}) \sim \mathcal{U}(0,1)
\end{equation}
where $\mathcal{U}(0,1)$ is the standard Uniform distribution. It can also be extended for cases when the sampling is performed with a given threshold. 

\begin{equation}
    \mathcal{L}^{'} \sim P(\mathcal{L}\mid \mathcal{L} > \lambda) \Rightarrow \frac{X(\mathcal{L}^{'})}{X(\lambda)} \sim \mathcal{U}(0,1)
\end{equation}
which implies we directly sample from $\pi_{\lambda}(\vec \theta)$ (bypassing $\lambda$ and $P(\mathcal{L})$) to satisfy PIT:

\begin{equation}
    \vec \theta^{\:'} \sim \pi_{\lambda}(\vec \theta) \Rightarrow \frac{X(\mathcal{L}(\vec \theta^{\:'})}{X(\lambda)} \sim \mathcal{U}(0,1)
\end{equation}

Now, for a given prior volume $X_{i-1}$ corresponding to a given likelihood level $\lambda_{i-1} = \mathcal{L}(\vec \theta_{i-1})$ after $i-1$ iterations of this procedure, the current prior volume $X_i$ will be

\begin{equation}
    \hat{X}_i = U_i\: \hat{X}_{i-1} = \prod_{j=1}^{i}\: U_j
\end{equation}
where $U_1, U_2,..., U_i \stackrel{\text{iid}}{\sim} \mathcal{U}(0,1)$ are independent and identically distributed (iid) random variables drawn from the $\mathcal{U}(0,1)$. Now, in order to decide whether to terminate the iterations, we need to estimate $\hat{X}_i$. The expectation value of $\hat{X}_i$ could be a good choice:

\begin{equation}
    \mathbb{E}[\hat{X}_i] = \mathbb{E}\left[\prod_{j=1}^{i}\: U_j\right] = \prod_{j=1}^{i}\mathbb{E}[U_j] = \left(\frac{1}{2}\right)^i
\end{equation}

Similarly, we can also evaluate the expectation value of $\ln \hat{X}_i$ (also known as the geometric mean),

\begin{equation}
    \ln \mathbb{E}[\hat{X}_i] = \sum_{j=1}^{i}\: \mathbb{E}[\ln U_j] \sim - \sum_{j=1}^{i} \mathbb[E_j] = -i
\end{equation}
where we have used the fact that 

$U\sim \mathcal{U}(0, 1) \Rightarrow -\ln U \sim \text{Expo}$, where Expo is the standard exponential distribution and 

\begin{equation}
    E_1, E_2, ...., E_i \stackrel{\text{iid}}{\sim} \text{Expo}
\end{equation}

So far, we have considered sampling using only one point (also known as a live point), which evolves at each iteration in accordance with the constrained prior. We can either perform multiple nested sampling runs with a single live point or simply perform a single nested sampling run with many live points. In fact, as shown in~\cite{speagle2020dynesty, sergey_koposov_2023_7600689, higson2019dynamic}, sampling with $K$ live points is equivalent to combining $K$ sets of independent samples derived using one live point each. We will now calculate the estimate $\hat{X}_i$ of the prior volume in case of $K$ live points. At any given iteration $i$, the current set of prior volumes $\{X_i^{[1]}, X_i^{[2]}, ..., X_i^{[K]}\}$ corresponding to the $K$ live points are uniformly distributed within the prior volume from the previous iteration $X_{i-1}$ so that
\begin{equation}
    X_i^{[j]} = U^{[j]}X_{i-1}
\end{equation}
where $U^{[1]}, U^{[2]},..., U^{[K]} \:\stackrel{\text{iid}}{\sim} \:\mathcal{U}(0,1)$. Now we replace the live point with the lowest likelihood $\mathcal{L}_i^{\text{min}}$ (also called ``dead point'') corresponding to the largest prior volume for which make an ordered list of prior volumes 

\begin{equation}
    X_i^{(j)} = U^{(j)}X_{i-1}
\end{equation}
where $(j)$ now indicates the position in the ordered list (from smallest to largest) and 

\begin{equation}
    U^{(j)} \sim \beta(j, K + 1 - j)
\end{equation}
where $\beta(., .)$ is a beta distribution. Therefore, we can write an estimate of the prior volume based on $K$ live points at iteration $i$ as
\begin{equation}
    \hat{X}_i = \prod_{j=1}^iU_{j}^{(K)}
\end{equation}
where $U_i^{(K)}, U_i^{(K)},..., U_i^{(K)}$ are the iid draws of the $K^{\text{th}}$ standard uniform order statistic with marginal distribution $\beta(K,1)$. Its arithmetic and geometric mean are given as follows:
\begin{equation}
\begin{split}
 \mathbb{E}(\hat{X}_i)&= \left(\frac{K}{K + 1}\right)^i \\
 \mathbb{E}(\ln \hat{X}_i)&= -\frac{i}{K}
\end{split}
\label{eq:expectation_prior_vol_multi_live_points}
\end{equation}
\begin{figure*}[!hbt]
    \centering
    \includegraphics[width=0.85\linewidth]{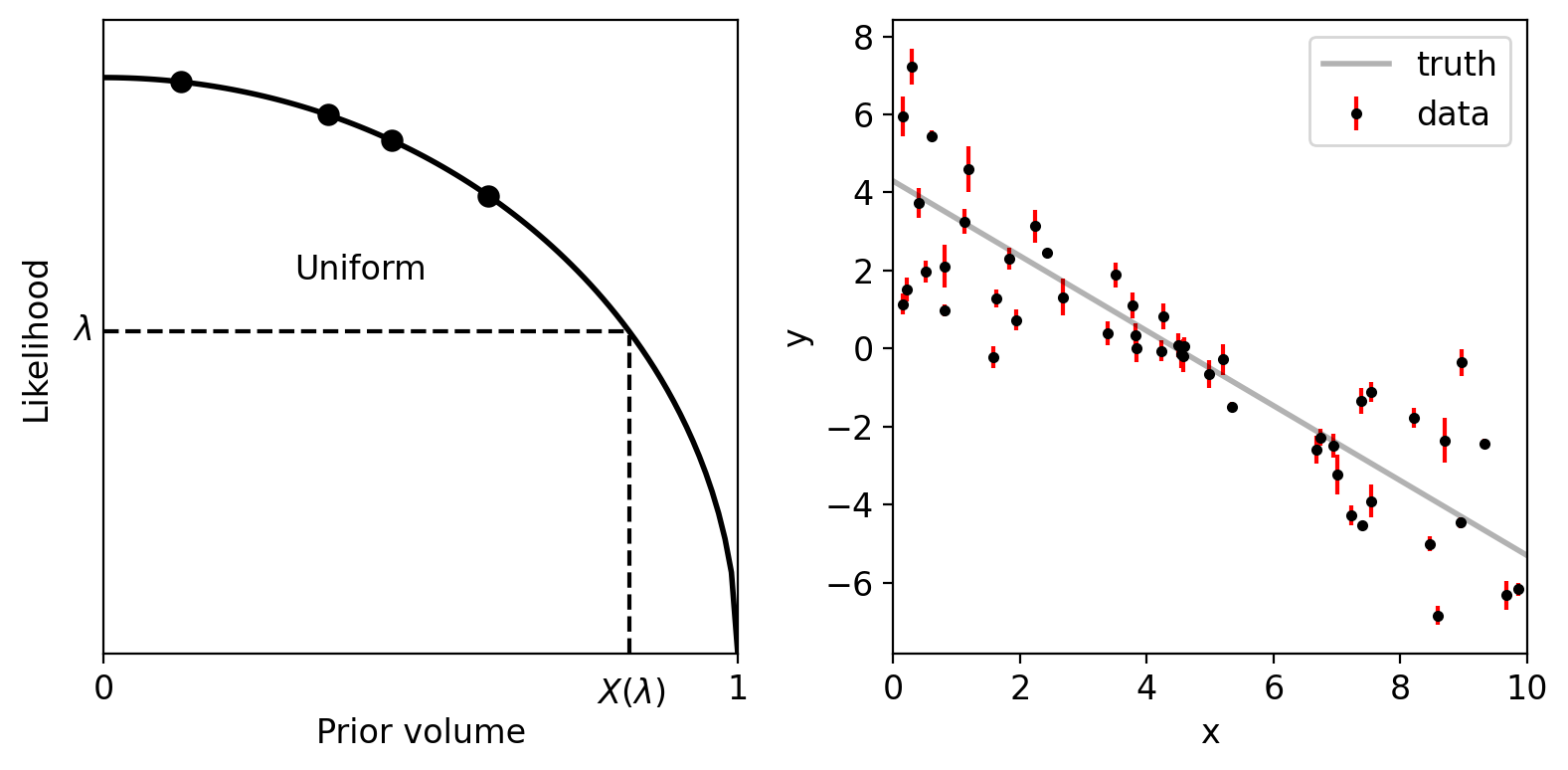}
    \caption{The figure in the left panel shows how $4$ live points are sampled uniformly in $X < X(\lambda)$ (or in $\mathcal{L} < \mathcal{L}^*$). In the right panel, the linear model generating the observations is shown. The data also contains noise following a normal distribution. The gray line represents the true model, whereas the black dots (with error bars) denote the data. We want to estimate the model parameters using nested sampling. (left figure reproduced from~\cite{sivia2006data})}
    \label{fig: nested_sampling_example_data}
\end{figure*}
From the above equations, we can compute an estimate of the change in the prior volume at a given iteration $i$ as 
\begin{equation}
    \mathbb{E}[\Delta \ln \hat{X}_i] = \mathbb{E}[\ln \hat{X}_i - \ln \hat{X}_{i-1}] = -\frac{1}{K}
\end{equation}
This is also known as \textit{exponential shrinkage} of the prior volume. 
\paragraph{Termination Condition}

Since nested sampling aims to estimate the evidence, a natural stopping criterion is when the current set of live points covers almost the entire posterior distribution. In other words, we can stop sampling at a given iteration if
\begin{equation}
    \Delta \ln \hat{\mathcal{Z}}_i \equiv \ln (\hat{\mathcal{Z}}_i + \Delta \hat{\mathcal{Z}}_i) - \ln (\hat{\mathcal{Z}}_i) < \epsilon
    \label{eq:nested_sampling_stopping_crit}
\end{equation}
where $\Delta \hat{\mathcal{Z}_i}$ is the estimated remaining evidence we have yet to integrate over, and $\epsilon$ determines the tolerance. In principle, we can construct a strict upper bound on the remaining evidence $\Delta \hat{\mathcal{Z}_i} \lesssim \mathcal{L}^{\text{max}}X_i$, where $\mathcal{L}^{\text{max}}$ is the maximum-likelihood value across the sample space and $X_i$ is the prior volume at the current iteration. Since both $\mathcal{L}^{\text{max}}$ and $X_i$ are not known exactly, we can work with their corresponding best estimates available. $\mathcal{L}^{\text{max}}$ across the sample space can be replaced by maximum likelihood among the live points at iteration $i$ and $\hat{X}_i$ is the estimated prior volume.

\paragraph{Evidence estimation}
The evidence integral can be approximated using a set of $N$ dead points via
\begin{equation}
    \mathcal{Z} = \int_{0}^{1}\mathcal{L}(X)dX \approx \hat{\mathcal{Z}} = \sum_{i=1}^{N}\mathcal{L}_i(X_{i-1} - X_i) = \sum_{i=1}^N \hat{w}_i
\end{equation}
where $\hat{w}_i$ is estimated weight of each dead point. 

\paragraph{Posterior estimation}
As a byproduct of evidence estimation, we can estimate the posteriors from the same set of $N$ dead points by assigning each sample its associated \textit{importance} weight as follows:
\begin{equation}
    p(\vec \theta_i) = p(X_i) \equiv p_i \approx \hat{p}_i = \frac{\hat{w}_i}{\sum_{i=1}^N\hat{w}_i} = \frac{\hat{w}_i}{\hat{\mathcal{Z}}}
\end{equation}

We use a simple example of linear regression to estimate the model parameters using nested sampling. The data-generating model is described by $y = mx + b$, where $m$ and $c$ are the parameters to be estimated. The noise in the data is modeled by a Gaussian distribution with variance underestimated by a fraction amount $f$, which we will also estimate along with other model parameters. The corresponding likelihood function is given by

\begin{equation}
    \ln \mathcal{L}(y\mid x, \sigma, m, b, f) = -\frac{1}{2} \sum_n\left[\frac{(y_n- mx_n - b)^2}{s_n} + \ln (2\pi_n^2)\right]
\end{equation}
where $s_n^2 = \sigma_n^2 + f^2(mx_n + b)^2$. We choose uniform priors over $m, b$, and $\ln f$. It can be represented in the following form:

\begin{equation}
        p(m) = \begin{cases}
        1/(m_{max} - m_{max}), & \text{if} \:  m_{min} < m < m_{max}\\
        0, & \text{otherwise}.
        \end{cases}
\end{equation}

Likewise, we can write down the uniform prior distribution over other parameters. These types of priors are often referred to as 'uninformative' because they reflect a lack of prior knowledge before performing the PE. Nevertheless, it is important to ensure that these priors encompass the significant regions of the posterior distribution. In scientific applications like GW data analysis, these priors are frequently inspired by astrophysical considerations. For example, all known neutron stars fall within the mass range of $1\: M_{\odot}$ to $2.35 \: M_{\odot}$. Consequently, when analyzing binary neutron star systems, our priors must encompass this range.
\begin{figure*}[!hbt]
    \centering
    \includegraphics[width=\linewidth]{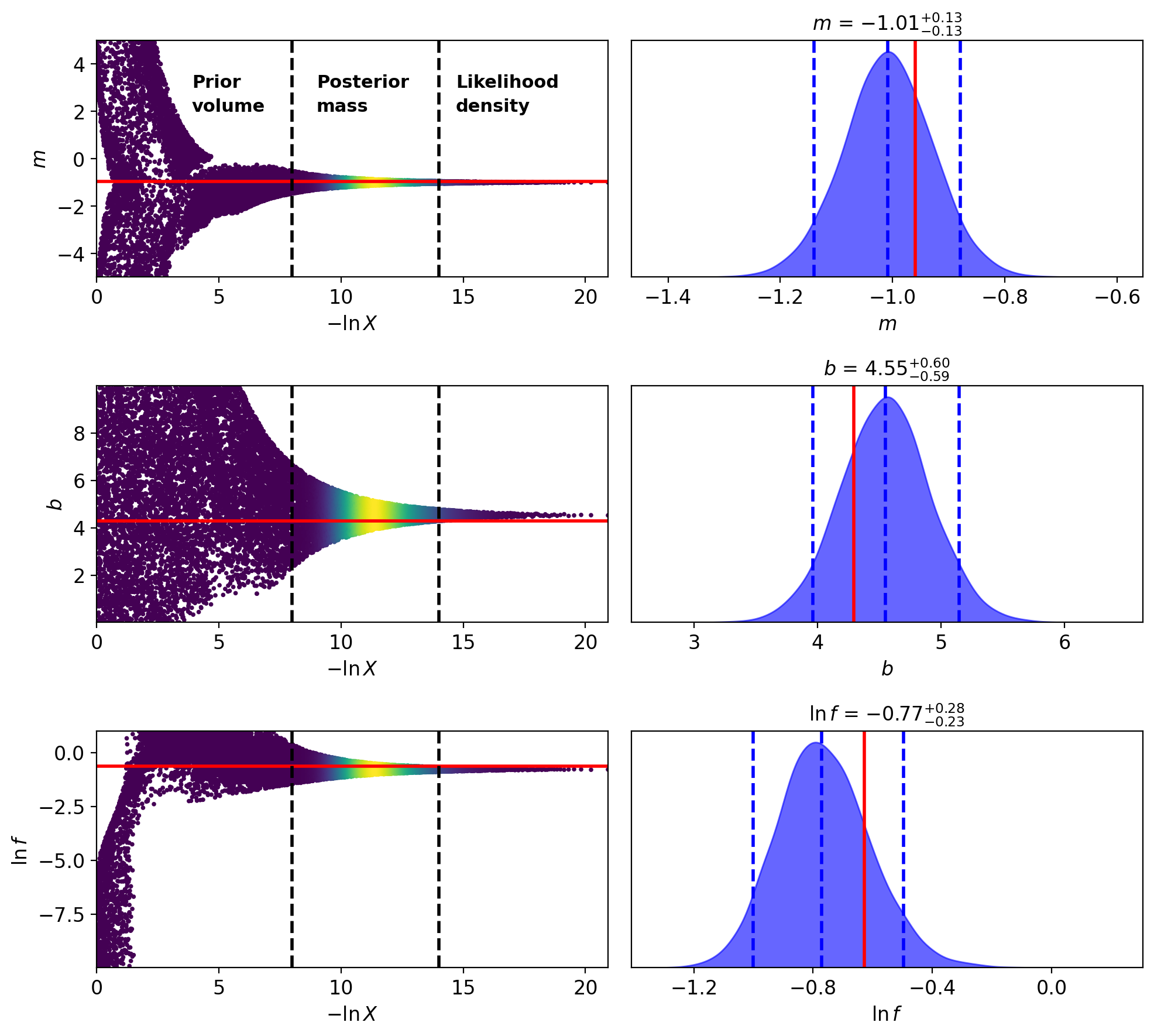}
    \caption{PE results using nested sampling are shown. The first column in the figure represents trace plots of the ``dead'' points positions as a function of $\ln X$, colored by the importance weight. The red solid line depicts the true value of the parameters. The PDF of the importance weight can be roughly categorized into distinct regions based on the influence of prior volume, posterior mass, and likelihood density. The posterior mass holds the highest relevance in posterior estimation, while evidence estimation also relies on the prior volume. The posterior distributions over $m$, $b$, and $\ln f$ are presented in the second column of the figure, with each posterior's title indicating the median values and their respective $90\%$ CIs. The true values are illustrated using a solid blue line.}
    \label{fig:nested_sampling_results}
\end{figure*}
We used \texttt{dynesty}~\cite{speagle2020dynesty, sergey_koposov_2023_7600689}, a Python implementation of a nested sampling algorithm for running the PE. Fig.~\ref{fig:nested_sampling_results} shows the posterior distributions over $m, b$ and $\ln f$. The true value (as shown by the solid red line) lies within the $90\%$ credible intervals (CI) of the posterior distributions.

\subsubsection{Effect of priors and likelihoods on posterior}
\begin{figure*}[!hbt]
    \centering
    \includegraphics[width=0.95\linewidth]{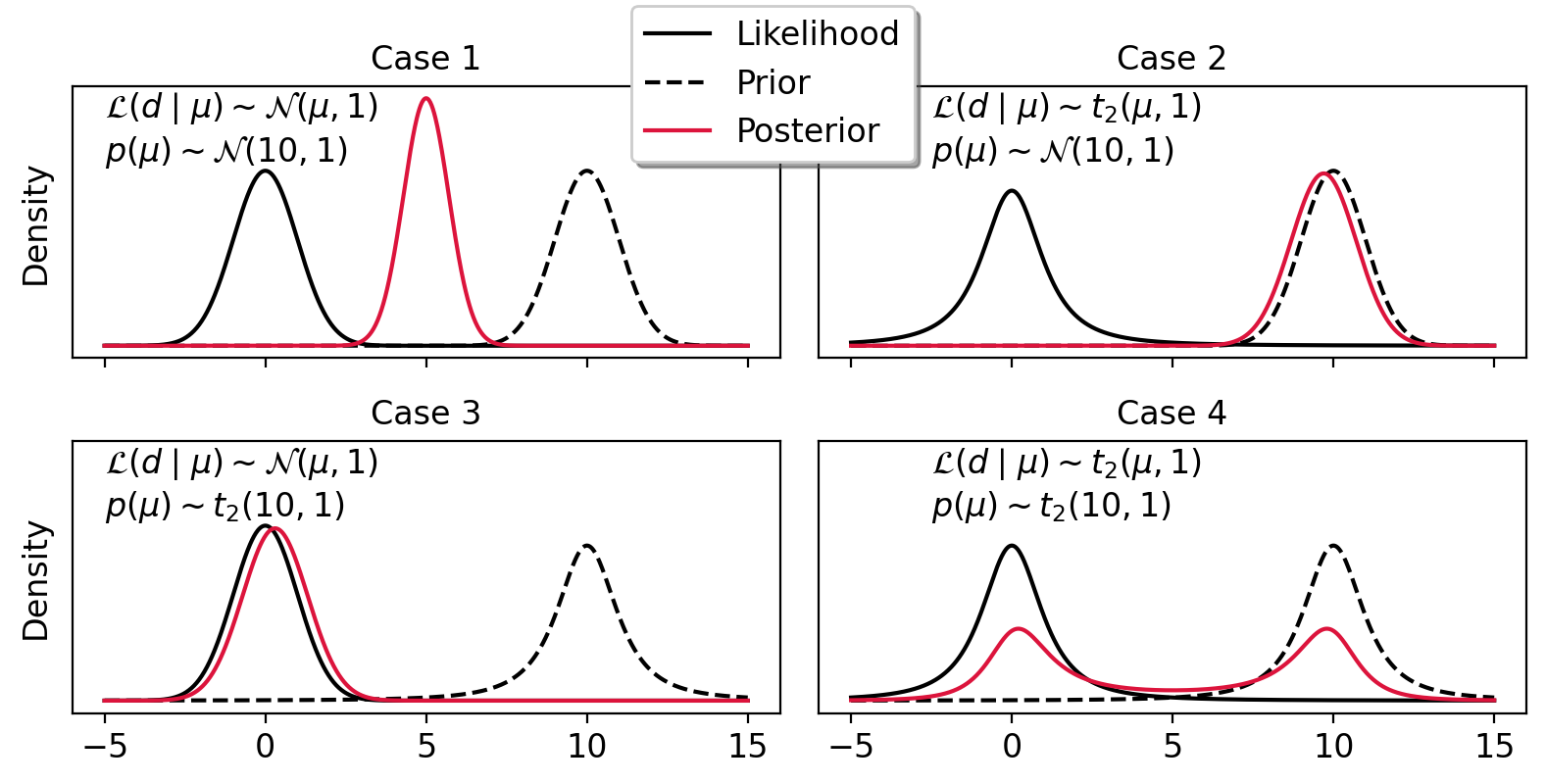}
    \caption{The effect of priors on the shape of the posterior distributions is shown.}
    \label{fig:effect_of_priors}
\end{figure*}

An essential aspect concerning the priors and likelihood is the potential conflict arising from their concentration in different regions of the parameter space. In such scenarios, the shape of the distribution tails significantly impacts the resulting posterior distribution. To demonstrate it, let's consider four cases (models) to fit one observation at $d=0$ (taken from problem $13\text{M}6$ on page $432$~\cite{mcelreath2020statistical}). As evident from Fig.~\ref{fig:effect_of_priors}, the priors have a dramatic effect on the shape of the posterior. The Gaussian distribution is characterized by narrow tails, while the Student-t distribution exhibits heavier tails. When either the prior or the likelihood has heavier tails, the resulting posterior is skewed towards the side with heavier tails (as seen in case 2 and case 3). Conversely, when both the prior and the likelihood have narrow tails (case 1), the posterior represents a compromise between the two. In the case where both distributions have heavier tails (case 4), each distribution views the other mode as more plausible, resulting in a compromise in the posterior, albeit not the optimal one.

In the next chapter, we discuss a fast PE method based on a combination of numerical linear algebra and meshfree approximation to reconstruct the compact binary sources.
%
\section{Appendix}
\label{appendix:definitions_GR_chap1}
The Christoffel symbol is given by
\begin{equation}
    \Gamma^{\rho}_{\mu\nu} = \frac{1}{2}g^{\partial \sigma}(\rho_{\mu}g_{\sigma \nu} + \partial_{\nu}g_{\sigma \mu} - \partial_{\sigma}g_{\mu \nu})
    \label{eq:reimann_tens}
\end{equation}
The Riemann tensor is defined as
\begin{equation}
    R^{\mu}_{\,\, \nu\rho\sigma} = \partial_{\rho}\Lambda^{\mu}_{\nu\sigma} - \partial_{\sigma}\Lambda^{\mu}_{\nu\rho} + \Lambda^{\mu}_{\alpha\rho}\Lambda^{\alpha}_{\nu\sigma} - \Lambda^{\mu}_{\alpha\sigma}\Lambda^{\alpha}_{\nu\rho}
    \label{eq:reim_tensor}
\end{equation}
We can obtain Ricci tensor by contracting with metric $g^{\mu\nu}$ as the following:
\begin{equation}
    R_{\mu\nu} \coloneqq R^{\lambda}_{\,\mu \lambda \nu} = g^{\lambda \sigma}R_{\sigma\mu\lambda\nu}
    \label{eq:ricc_tens}
\end{equation}
and Ricci scaler can be obtained by further contracting $R_{\mu\nu}$ as the following:
\begin{equation}
    R = g^{\mu\nu}R_{\mu\nu}
    \label{eq:ricci_scal}
\end{equation}
To obtain the energy-momentum tensor $T^{\mu\nu}$, we can vary the matter action $S_{M}$ under a metric transformation $g_{\mu\nu}\rightarrow g_{\mu\nu} + \delta g_{\mu\nu}$, in accordance to
\begin{equation}
    \delta S_{M} = \frac{1}{2c} \int d^4x \sqrt{-g}T^{\mu\nu} \delta g_{\mu\nu}
    \label{eq:energy_moment_tens}
\end{equation}
The covariant derivative of a vector field $A^{\mu}(x)$ along the curve $x^{\mu}(\lambda)$ is defined as 
\begin{equation}
    \frac{DA^{\mu}}{D\tau} = \frac{dA^{\mu}}{d\tau} + \Gamma^{\mu}_{\nu\rho} A^{\nu}\frac{dx^{\rho}}{d\tau}
    \label{eq:cov_der_def}
\end{equation}
The Green's function is defined as 
\begin{equation}
    \Box_{x}G(x-x^{'}) = \delta^4(x-x^{'})
    \label{eq:green_fn_def}
\end{equation}
The various momenta of stress-tensor $T^{ij}$ are
\begin{equation}
    \begin{split}
    S^{ij}(t) &= \int d^3 x T^{ij}(t,\boldsymbol{x}),\\
    S^{ij,k}(t) &= \int d^3 x T^{ij}(t,\boldsymbol{x})\,x^{k},\\
    S^{ij,kl}(t) &= \int d^3 x T^{ij}(t,\boldsymbol{x})\,x^{k}\,x^{l}\\
    \label{eq:momentas_of_stres_energy}
    \end{split}
\end{equation}
Useful Identities:
\begin{equation}
    \begin{aligned}
        \int \frac{d\Omega}{4\pi}\, n_i\, n_j &= \frac{1}{3}\delta_{ij}\\
        \int \frac{d\Omega}{4\pi}\, n_i\, n_j\,n_k\,n_l &= \frac{1}{15}(\delta_{ij}\delta_{kl} + \delta_{ik}\delta_{jl} + \delta_{il}\delta_{jk})
    \end{aligned}
    \label{eq:direction_ids}
\end{equation}

\chapter{Fast and faithful interpolation of numerical relativity surrogate waveforms using meshfree approximation}
\label{chap:chapter_2}



\paragraph{\textbf{Abstract}}
Several theoretical waveform models have been developed over the years to capture the gravitational wave emission from the dynamical evolution of compact binary systems of neutron stars and black holes. 
As ground-based detectors improve their sensitivity at low frequencies, the real-time computation of these waveforms can become computationally expensive, exacerbating the steep cost of rapidly reconstructing source parameters using Bayesian methods. 
This paper describes an efficient numerical algorithm for generating high-fidelity interpolated compact binary waveforms at an arbitrary point in the signal manifold by leveraging computational linear algebra techniques such as singular value decomposition and meshfree approximation. The results are presented for the time-domain \texttt{NRHybSur3dq8} inspiral-merger-ringdown (IMR) waveform model that is fine tuned to numerical relativity simulations and parameterized by the two component-masses and two aligned spins. For  demonstration, we target a specific region of the intrinsic parameter space inspired by the previously inferred parameters of the \texttt{GW200311\_115853} event -- a binary black hole system whose merger was recorded by the network of advanced-LIGO and Virgo detectors during the third observation run. We show that the meshfree interpolated waveforms can be evaluated in $\sim 2.3$ ms, which is about $\times 38$ faster than its brute-force (frequency-domain tapered) implementation in the \textsc{PyCBC} software package at a median accuracy of  $\sim \mathcal{O}(10^{-5})$. The algorithm is computationally efficient and scales favourably with an increasing number of dimensions of the parameter space. This technique may find use in rapid parameter estimation and source reconstruction studies.


\section{Introduction}
\label{sec:intro_chap1}
This chapter is based on the publication \textit{Fast and faithful interpolation of numerical relativity surrogate waveforms using meshfree approximation}, \href{https://arxiv.org/abs/2403.19162}{arXiv:2403.19162 (2024)}.\newline

Compact binary systems such as binary black holes or binary neutron stars are one of the most important sources for ground-based gravitational wave (GW) detectors. The GW signal emitted by such sources can be theoretically modeled in the secular inspiral or the post-merger ring-down phase by solving the Einstein's field equations. While analytical solutions are not available in the highly non-linear, so-called `merger'-domain of the evolution of these sources, the inspiral and ringdown solutions are often calibrated to a set of numerical relativity waveforms to construct complete semi-analytical waveforms covering the inspiral, merger and ringdown (IMR) phases of the evolution of such compact binary sources.
Several such waveform models have been developed both in the frequency and time-domain, such as the family of frequency-domain IMR waveforms~\cite{IMRPhenomD, IMRPhenomXAS, IMRPhenomPv2, IMRPhenomXHM, IMRPhenomXPHM}, and the EOB-family~\cite{Boh_2017, Cotesta_2020} of waveforms, that are routinely used for GW data analysis. 
More recently, numerical relativity (NR) based surrogate models have also been developed such as the \texttt{NRSur7dq2}~\cite{NRSur7dq2}, \texttt{NRSur7dq4}~\cite{NRSur7dq4}, and the \texttt{NRHybSur3dq8}~\cite{NRHybSur3dq8} models, which are among the most accurate waveform models available but are computationally expensive to generate in comparison to their frequency domain counterparts such as the \texttt{IMRPhenomXAS} and $\texttt{SEOBNRv4}\_\texttt{ROM}$ models. For reference, \texttt{NRHybSur3dq8} takes $\sim 75$~ms~\footnote{The timing is for the implementation in the \textsc{PyCBC} software package}. to generate a waveform with component masses ${m_{1, 2} = (50, 20) \, M_\odot}$ and aligned spin parameters ${\chi_{1z, 2z} = (0.05, 0.05)}$ starting at a seismic cutoff frequency of $15$~Hz. 

Our research is inspired by a suggestion by Verma et al.~\cite{NRHybSur3dq8}, where the authors table the idea of speeding up the evaluation of the \texttt{NRHybSur3dq8} time-domain surrogate waveforms by creating a faster frequency-domain variant. 
In the past, fast frequency-domain surrogates using the technique of model order reduction have been applied to the SEOBNRv4~\cite{Boh_2017} waveform model leading to a significant reduction in computational complexity of evaluating these waveforms. Such an approach could potentially accelerate parameter estimation (PE) for compact binary sources using these accurate waveform models.

There have been several attempts in the past to construct surrogate models of other waveform models in the past. 
Cannon et al.~\cite{cannon2012interpolating} constructed the approximate non-spinning IMR waveforms~\cite{Ajith_2011} parameterized by only the two-component masses - by first projecting these waveforms over a  set of singular value of decomposition (SVD) basis vectors, followed by a grid-based two-dimensional Chebyshev interpolation of the SVD coefficients.
Chua et al.~\cite{PhysRevLett.122.211101} used artificial neural networks (ANNs) to generate a four-dimensional reduced order model (ROM) of the $2.5$ post-Newtonian (PN) frequency domain \texttt{TaylorF2} waveform model~\cite{PhysRevD.79.104023} parameterized by the two-component masses and the two aligned spins. In their approach, the waveforms are represented as the weighted sums over the reduced basis vectors. The ANNs are then trained to accurately map the GW source parameters into the basis coefficients. 
Other studies related to building NR surrogate models utilized ANNs~\cite{Khan_2021}, Gaussian Process Regression (GPR)~\cite{Williams_2020}, and deep learning~\cite{Lee_2021} architectures.

In this work, we aim to significantly reduce the computational cost of generating frequency-domain \texttt{NRHybSur3dq8} waveforms by using a combination of the SVD-decomposition and meshfree approximation. We focus on illustrating our approach by concentrating on a region of the four-dimensional (two component masses and two aligned spins) parameter space around the inferred source parameters of the  \texttt{GW200311\_115853}~\cite{PhysRevX.13.011048} event detected during the third observing run of the LVK collaboration~\cite{Abbott_2020_Prospects}. We construct meshfree interpolants for both the amplitude and phase of these waveforms separately. These interpolants are then used to rapidly evaluate the amplitude and phase of the waveform at any arbitrary query points in the chosen parameter space. These interpolated waveforms can be evaluated in $\sim 2.3$ ms in comparison to $89$ ms ($\sim 38 \times $ faster) with the standard (frequency-domain tapered) implementation of time-domain \texttt{NRHybSur3dq8} in \textsc{PyCBC}~\cite{usman2016pycbc} with a median error of  ${\sim \mathcal{O}(10^{-5})}$. 
Further, we also performed a Bayesian PE study of a simulated \texttt{GW200311\_115853} like event coherently injected in Gaussian noise to mimic data from the network of LIGO and Virgo detectors. The PE runs were carried out using the meshfree interpolated waveforms and took ${\sim 16.4 \, \text{minutes}}$  to complete on a $64$ CPU-cores setup and the posterior distributions over the source parameters were found to be broadly consistent with their injected values.  

The rest of the paper is organized as follows: Section~\ref{sec:wave_model} introduces the \texttt{NRHybSur3dq8} waveform model and the preprocessing required before we build meshfree interpolants. Section~\ref{sec:interp_generation}
explains an iterative strategy to find a suitable set of basis vectors using SVD to span the space of amplitude and phase and construct the meshfree interpolants of the resulting SVD coefficients. In Section~\ref{sec:choice_of_rbf_kernel}, we describe various RBF kernels that can be used to generate meshfree interpolants. Section~\ref{sec:results} demonstrates the results of a PE performed on a simulated BBH event using meshfree interpolants.  Finally, we summarize the results in Section~\ref{sec:conclusion} and discuss the limitations of the current implementation and suggest some ideas for overcoming this limitation in follow-up studies.

\section{NR Waveform model}
\label{sec:wave_model}
\texttt{NRHybSur3dq8} is a time-domain, aligned spin surrogate model for ``hybridized" non-precessing NR waveform, which is valid for stellar compact binaries with total masses as low as $2.5\, M_{\odot}$. In this context, the term ``hybridized" refers to a combined waveform incorporating both a post-Newtonian (PN) and effective one-body (EOB) waveform during the early times, attached smoothly to numerical relativity (NR) waveform at late times. The training of this model involves hybridized waveforms derived from $104$ numerical relativity (NR) waveforms, supporting mass ratios $q \leq 8$, and spin values $\chi_{1z}$; $\chi_{2z} \leq 0.8$. Here, $\chi_{2z}$$(\chi_{2z})$ represents the spin of the heavier (lighter) black hole in alignment with the direction of orbital angular momentum. The waveform model also supports spin-weighted spherical harmonic modes with $l \leq 4$ and $(5, 5)$ mode but excludes the $(4, 0)$ or $(4, 1)$ modes. In this work, we specifically focus on the dominant, commonly referred to as the "quadrupole" mode $(2, 2)$ present in this waveform model. 
\begin{figure*}[!hbt]
\centering
\includegraphics[width=\textwidth]{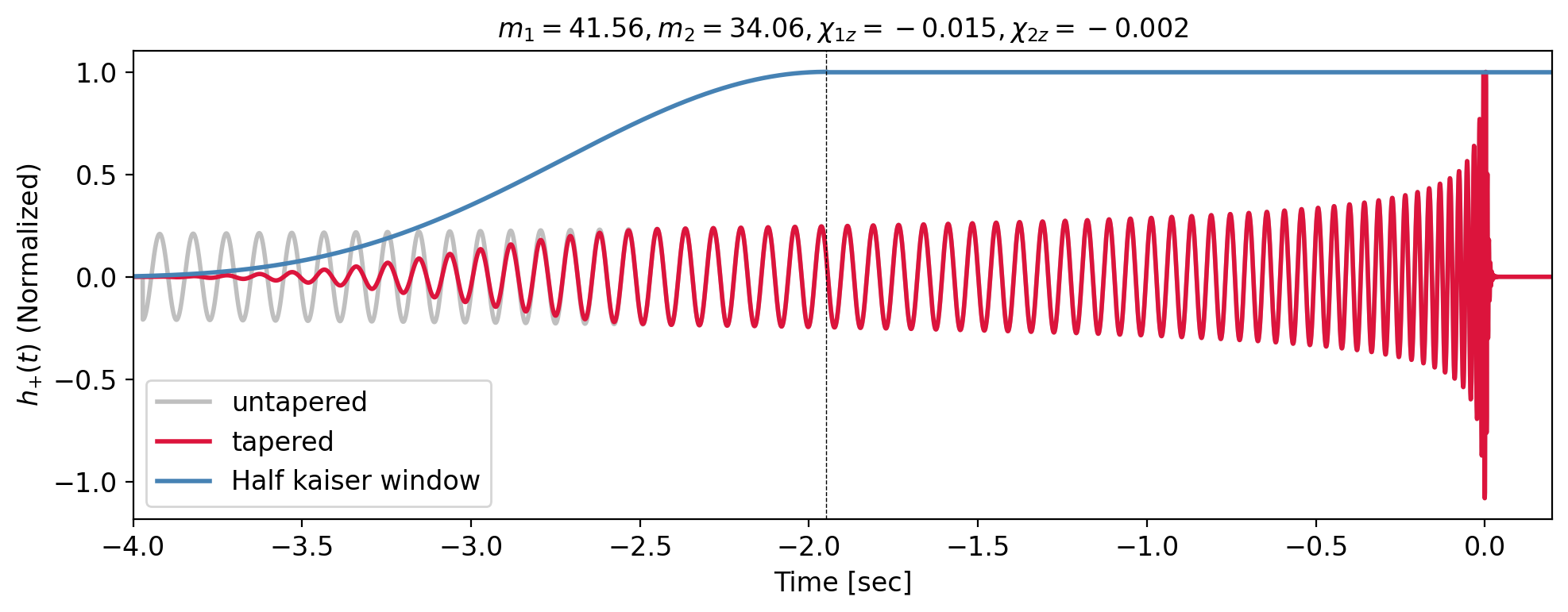}
\caption{The \texttt{NRHybSur3dq8} is generated at a frequency ($f_{start} = 10$ Hz) lower than the seismic cutoff frequency ($f_{low} = 15$ Hz), which leads to a longer duration waveform. It is generated for a simulated BBH event whose intrinsic parameters are shown in the title. A time-domain tapering using a half-Kaiser window is performed on the longer waveform to smoothly attenuate the waveform amplitude to zero, that helps in alleviating the Gibbs phenomenon arising due to a jump discontinuity while converting the time-domain waveform to a frequency-domain waveform. The black dashed vertical line represents the epoch at which the amplitude of the tapered and un-tapered waveform becomes equal (or equivalently, the value of the half-Kaiser window becomes $1$). The sampling frequency is set at $2048$ Hz.}
\label{fig:wf_tap_vs_untap_td}
\end{figure*} 
\begin{figure*}[!hbt]
\begin{subfigure}{0.49\linewidth}
    \centering
    \includegraphics[width=\linewidth]{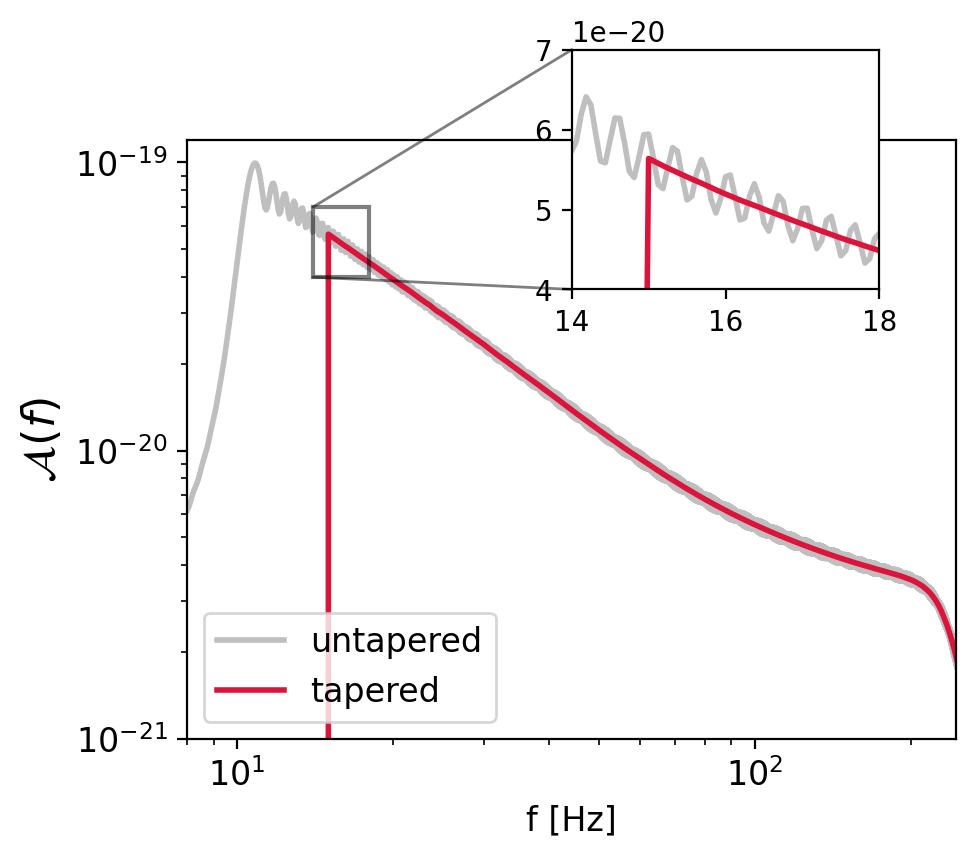}
    \caption{}
    \label{fig:wf_tap_vs_untap_fd}
\end{subfigure}\hfill
\begin{subfigure}{0.49\linewidth}
    \centering
    \includegraphics[width=\linewidth]{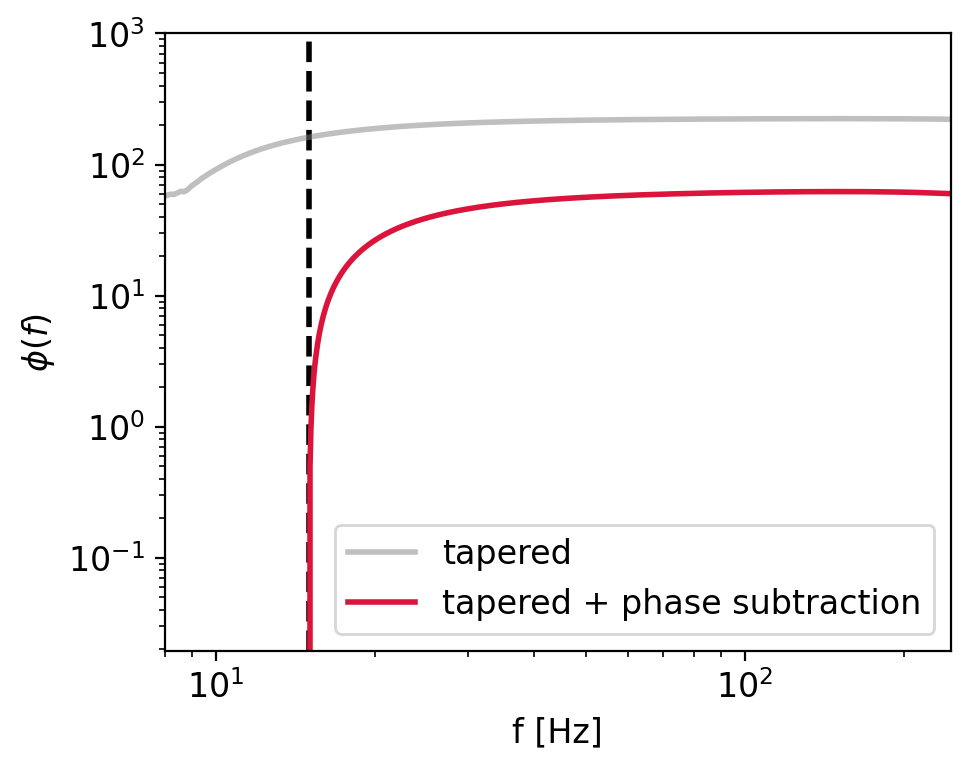}
    \caption{}
    \label{fig:wf_tap_vs_untap_fd_phase}
\end{subfigure}
\caption{\textit{left}: The amplitude of the untapered waveform (\textit{solid red}) is plotted with that of the tapered (\textit{solid grey}) waveform. Note the ringing artifacts appearing due to the Gibbs phenomenon in the untapered waveform. It is almost negligible in the amplitude of the tapered waveform. \textit{right}: After converting the tapered waveform into the frequency domain, there is some accumulated phase below $f_{low} = 15$ Hz, which we subtract from the total phase so that the phase of the frequency domain waveform is zero at $f_{low} = 15$ Hz. The black-dashed vertical line represents $f_{low} = 15$ Hz.}
\end{figure*}

The GW polarizations, denoted as ``plus" $(h_{+})$ and ``cross" $(h_{\times})$, can be concisely represented as a unified complex time series, denoted as ${h = h_{+} - i \,h_{\times}}$. This complex time series can be expressed as a linear combination of spin-weighted harmonic modes $h_{lm}$~\cite{Newman:1966ub, Goldberg:1966uu}. Consequently, the GW along any direction $(\iota, \phi_0)$ in the binary source's frame can be expressed as follows:
\begin{equation}
    h(t, \iota, \phi_0) = \sum_{l=2}^{\infty}\sum_{m=-l}^{l}\, h_{lm}(t)^{-2}Y_{lm}(\iota, \phi_0)
    \label{eq:wave_spheric_modes}
\end{equation}
Here, $^{-2}Y_{lm}$ denotes the $\text{spin} = -2$ weighted spherical harmonics, $\iota$ represents the inclination angle between the orbital angular momentum of the binary and the line of sight to the detector, and $\phi_0$ corresponds to the initial binary phase. The corresponding frequency domain waveform for the dominant mode can be written as the following:
\begin{equation}
    \tilde{h}_{+/\times} = \mathcal{A}_{+/\times}(f)\exp[j\,\psi_{+/\times}(f)]
    \label{eq:frequency_domain_waveform}
\end{equation}
where $\mathcal{A}_{+/\times}(f)$ and $\psi_{+/\times}(f)$ are amplitude and phase as a function of frequency, respectively. Since \texttt{NRHybSur3dq8} is an aligned-spin waveform model~\cite{PhysRevD.49.1707, PhysRevD.59.124016, Faye_2012, PhysRevLett.74.3515, khan2016frequency, husa2016frequency}, the relation ${\tilde{h}_{\times} \propto -i\tilde{h}_{+}}$ holds and therefore we only consider $\tilde{h}_{+}$ and the cross-polarization ($\tilde{h}_{\times}(f)$) can be calculated from the above relation. From now onwards, we will drop the subscript `+/$\times$' from $\mathcal{A}_{+/\times}$ and $\psi_{+/\times}$ and simply denote the amplitude and phase by $\mathcal{A}$ and $\psi$ which correspond to plus polarization.

As a pre-processing step, we first calculate the duration of the waveform starting at the seismic cutoff frequency ($f_{low}$). Subsequently, the starting frequency ($f_{start}$) of the waveform generation is decreased until we reach twice the duration of the original waveform. Then the time-domain \texttt{NRHybSur3dq8} waveform is generated starting at $f_{start}$ followed by a time-domain tapering to smoothly decrease the amplitude to zero (see Fig.~\ref{fig:wf_tap_vs_untap_td}). The length of the tapering window is taken as a fraction ($0.8$ in this case) of the difference between the new and original duration of the waveform. The longer duration of the waveform makes sure that we do not lose any portion of the waveform in the frequency band to be used in the PE while tapering. We use a Kaiser window~\cite{kuo1966system} (as implemented in \textsc{}{PyCBC}) for time-domain tapering, which can be expressed as
\begin{equation}
    w(n) = I_0\left(\beta \, \sqrt{1 - \frac{4n^2}{(M-1)^2}}\right)/I_0(\beta)
    \label{eq:kaiser_window}
\end{equation}
with $n \leq |\frac{M-1}{2}|$ and where $I_0$ is the modified zeroth-order Bessel function, $M$ is the number of points in the output window, and $\beta$ is the shape parameter, which determines the trade-off between main-lobe width and side lobe level in the Fourier response of the window. As the beta becomes large, the main lobe width increases while the side lobe level decreases. For this analysis, we use a half-Kaiser window for tapering (see Fig.~\ref{fig:wf_tap_vs_untap_td}). The tapering helps in reducing the Gibbs phenomenon while converting the tapered waveform in the frequency domain (see Fig.~\ref{fig:wf_tap_vs_untap_fd}). Additionally, we also subtract the phase accumulated up to $f_{low}$ from the total phase of the frequency domain waveform (tapered) to make sure the phase at $f_{low}$ is zero, which significantly helps in the interpolation of the SVD coefficients corresponding to the phase (see Fig.~\ref{fig:wf_tap_vs_untap_fd_phase}). 

In the next section, we lay down an iterative strategy to find suitable basis vectors to span the space of amplitude and phase using SVD. Subsequently, we fit the resulting SVD coefficients for amplitude and phase separately using a linear combination of RBFs and monomials. Since the amplitude and phase are smoothly varying functions of frequency (see Fig.~\ref{fig:wf_tap_vs_untap_fd} and Fig.~\ref{fig:wf_tap_vs_untap_fd_phase}), the corresponding SVD coefficients are expected to exhibit smooth variation over the intrinsic parameter space as well and, therefore, suitable for interpolation. Combining the interpolated coefficients at the arbitrary query points within the interpolating region with the corresponding basis vectors gives the interpolated amplitude and phase and, hence, the interpolated waveform.
\section{Generation of meshfree interpolants}
\label{sec:interp_generation}
To construct the interpolants for evaluating the meshfree waveforms, the initial step involves choosing a patch of intrinsic parameter space for which interpolants are to be built. The dimensionality of the patch depends on the dimensionality of the intrinsic parameter space over which interpolation is to be performed. Rather than opting for a single patch with large boundaries, a more effective approach (as suggested in~\cite{P_rrer_2014}) is to create multiple overlapping patches with smaller boundaries to comprehensively encompass the desired parameter space. For each patch, meshfree interpolants are independently constructed, enhancing the accuracy of individual interpolants. In this analysis, our selection of patches is inspired by a simulated BBH event with \texttt{GW200311\_115853}-like parameters. We use \texttt{NRHybSur3dq8}, an aligned spin numerical relativity (NR) surrogate waveform model, which also includes support for subdominant modes as well. As a proof of principle study, this analysis focuses only on the leading-order mode; however, the procedure presented here can be extended to incorporate higher-order modes. The procedure to generate interpolants comprises the following steps:
\subsection{Selection of patch(es)}
\begin{figure*}[!hbt]
\begin{subfigure}{0.49\linewidth}
    \centering
    \includegraphics[width=\linewidth]{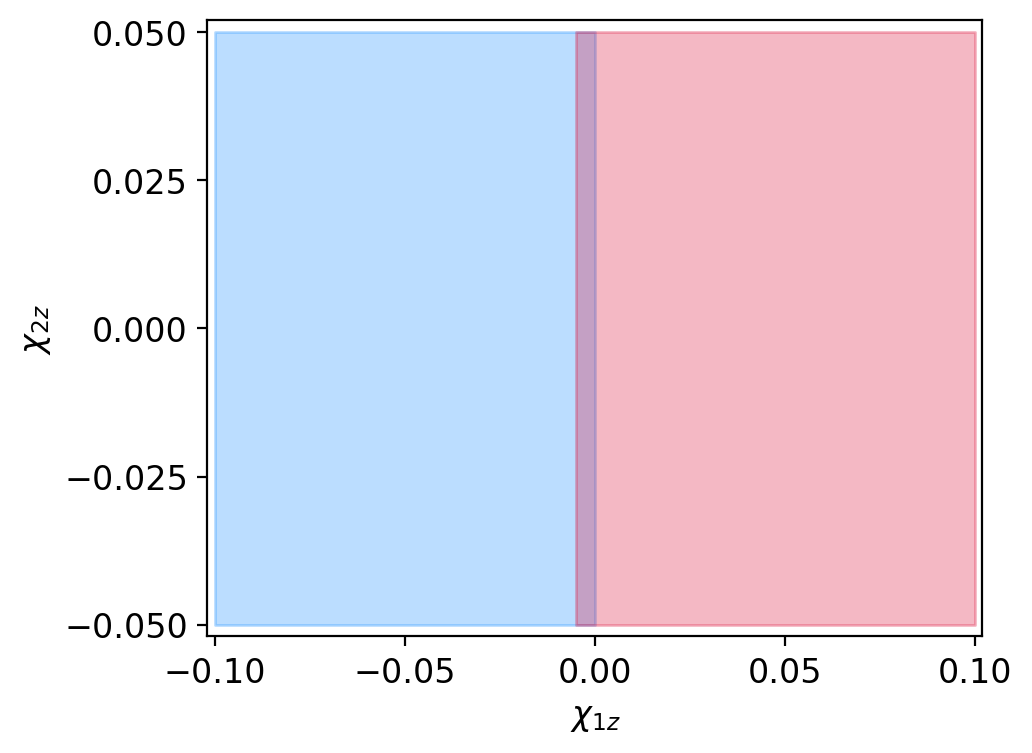}
    \caption{}
    \label{fig:overlap_patch}
\end{subfigure}\hfill
\begin{subfigure}{0.49\linewidth}
    \centering
    \includegraphics[width=\linewidth, height=0.75\linewidth]{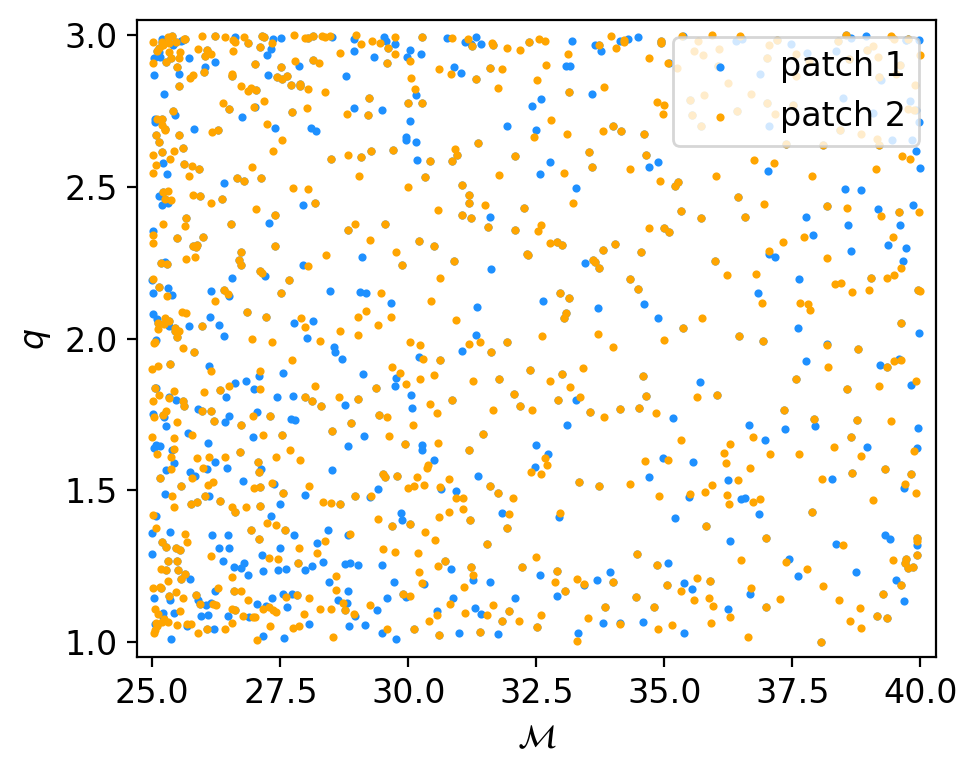}
    \caption{}
    \label{fig:final_nodes_both_patches}
\end{subfigure}
\caption{\textit{left}: We choose two overlapping patches over $\chi_{1z}$ to cover the desired range in $\chi_{1z}$. A similar procedure can be performed to extend the range of the other parameters (say ${\mathcal{M}, \, q}$ etc.). \textit{right}: Final nodes after iterative SVD process terminates for both patches. The nodes found using the ``Iterative SVD'' procedure tend to cluster near the lower mass boundaries.}
\end{figure*}

This task involves identifying the patch (or patches) within the parameter space for the construction of waveform interpolants\footnote{A similar approach was taken by Morisaki et al. in their work~\cite{morisaki2023rapid} for constructing ROM bases corresponding to different patches of parameter space.}. While it could be any segment of the pertinent parameter space in general, for this study, we opt for a patch that encompasses the injected values of parameters associated with the simulated BBH event. We define two four-dimensional hyper-rectangular patches with the following ranges: ${\mathcal{M} \in [25, \, 40]}$, ${q\in [1, \, 3]}$, ${\chi_{1z} \in [-0.1, \, 0.1]}$, and ${\chi_{2z} \in [-0.05, \, 0.05]}$. Both patches share common ranges for $\mathcal{M}$, $q$, and $\chi_{2z}$. However, for the first patch, ${\chi_{1z} \in [-0.1, \, 0]}$, and for the second patch, ${\chi_{1z} \in [-0.005, \, 0.1]}$. Note that these two patches overlap in the $\chi_{1z}$ dimension (see Fig.~\ref{fig:overlap_patch}).

\subsection{Placing RBF nodes using `Iterative SVD' method}
\label{subsec:itr_svd}
This step starts by randomly spraying a minimum\footnote{The initial number of nodes to start the algorithm can be decided by the degree of the monomial (see Eq.~\eqref{eq:amp_phi_rbfcoeff}) and dimensionality of the intrinsic parameter space.} number of nodes ($N$) across the four-dimensional sample space and generating frequency domain waveforms (as specified in Sec.~\ref{sec:wave_model}) at these specified nodes. Subsequently, $\mathcal{A}$ and $\psi$ are extracted from $\tilde{h}_{+}(f)$ for each of the $N$ waveforms. The obtained amplitudes ($\mathcal{A}$) and phases ($\psi$) are then stacked in a row-wise fashion to construct two matrices, $\mathbf{A}$ and $\mathbf{\Psi}$. These matrices are subjected to singular value decomposition (SVD), which, in turn, provides the relevant basis for spanning the space defined by these amplitudes and phases.
\begin{equation}
    \mathcal{A} = \sum_{\mu = 1}^{N} C^{\mathcal{A}}_{\mu} \, \vec u^{\,\mathcal{A}}_{\mu} \quad\mathrm{and}\quad \psi = \sum_{\mu = 1}^{N} C^{\psi}_{\mu} \, \vec u^{\,\psi}_{\mu}
    \label{eq:svd_amp_phase}
\end{equation}
\begin{figure}[!hbt]
    \centering
    \includegraphics[width=0.55\linewidth]{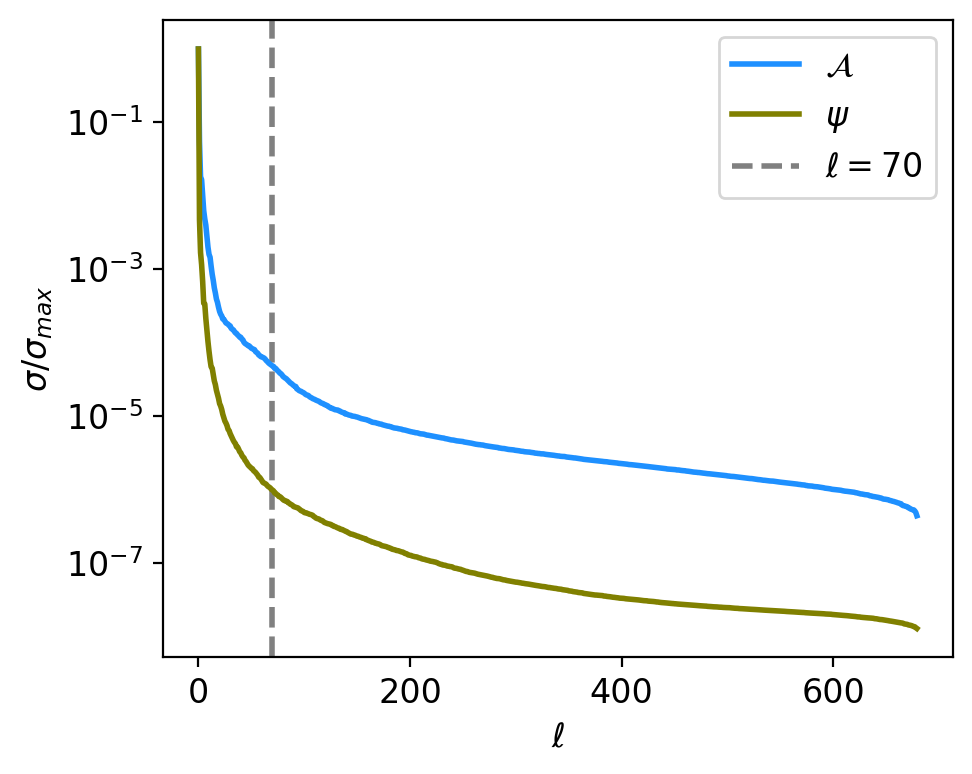}
    \caption{Singular values spectrum corresponding to amplitude $\mathcal{A}$ and phase $\varphi$. Note that singular values drop to $\sim \mathcal{O}(10^{-6})$ within $\sim 70$ basis vectors.}
    \label{fig:sing_vals}
\end{figure}
Since by construction, the basis vectors $(\vec u_{\mu}^{\mathcal{A}}, \vec u_{\mu}^{\,\psi})$ are in the decreasing order of importance, it turns out that we need to retain only top few basis vectors (say upto $\ell$) to accurately reconstruct amplitude and phase as illustrated by the singular values spectrum in Fig.~\ref{fig:sing_vals}. Consequently, we limit the construction of interpolants to the SVD coefficients $C^{\mathcal{A}}_{\mu}$ and $C^{\psi}_{\mu}$ corresponding to top-$\ell$ basis vectors. Later on, we can choose to retain an even lesser number of basis vectors while reconstructing amplitude and phase at arbitrary query points. These coefficients, denoted as $C^{\mathcal{A}}_{\mu}$ and $C^{\psi}_{\mu}$, exhibit smooth variations as functions of $\vec \lambda$, where ${\vec \lambda \equiv \{\mathcal{M}, q, \chi_{1z}, \chi_{2z}\}}$ represents the interpolating parameter space. These coefficients can be written as a linear combination of RBFs ($\phi$) and monomials ($p$) of specified degree as follows~\cite{doi:10.1142/6437}:
\begin{subequations}
    \begin{align}
    C^{\mathcal{A}\, (q)}_{\mu} &= \sum_{n=1}^N\, a^{\mathcal{A}}_{n}\, \phi(\|\vec \lambda^q - \vec \lambda^{n}\|_2) + \sum_{j = 1}^{M}\, b^{\mathcal{A}}_{j}\, p_j(\vec \lambda^q) \\
    C^{\psi\, (q)}_{\mu} &= \sum_{n=1}^N\, a^{\psi}_{n}\, \phi(\|\vec \lambda^q - \vec \lambda^{n}\|_2) + \sum_{j = 1}^{M}\, b^{\psi}_{j}\, p_j(\vec \lambda^q) 
    \label{eq:amp_phi_rbfcoeff}
    \end{align}
\end{subequations}
The addition of these monomial terms enhances the accuracy of RBF interpolation, as shown in~\cite{2016JCoPh}. Since these coefficients are known at the $N$ nodes, we can invert the above equations to calculate the coefficients $a^{\mathcal{A}}$, $a^{\psi}$, $b^{\mathcal{A}}$, and $b^{\psi}$ corresponding to each basis vector. Using these coefficients, we can quickly evaluate the $C^{\mathcal{A}\, (q)}_{\mu}$ and $C^{\psi\, (q)}_{\mu}$ (for $\mu^{th}$ basis) at an arbitrary query point within the interpolating parameter space and then combine it with corresponding basis vectors ($\vec u^{\,\mathcal{A}}$ and $\vec u^{\,\psi}$) to get back the interpolated amplitude $\mathcal{A}$ and phase $\psi$. Then, we combine both amplitude and phase in accordance with Eq.~\eqref{eq:frequency_domain_waveform} to get the frequency domain waveform. 
    
Once we have the interpolants ready, a set of $10^3$ query points is generated within the interpolating patch, and corresponding true frequency domain waveforms and meshfree interpolated waveforms are generated at these query points. Subsequently, the mismatches between the true and interpolated waveforms are calculated. A specified number of query points ($10$ for this work) exhibiting the worst mismatch are added back into the original set of initial nodes. The process then reverts to Subsection~\ref{subsec:itr_svd} and continues until either a limit of maximum nodes is reached or the maximum mismatch at the current iteration falls below a predefined threshold (e.g., ${\leq 10^{-4}}$) on maximum mismatch~\footnote{Note that the mismatches are calculated between meshfree waveforms and waveforms first generated in time-domain at $f_{start} = 10$ Hz, then tapered and converted into frequency domain followed by phase subtraction as in Section~\ref{sec:wave_model}.}. We define the match between the two waveforms $h_1$ and $h_2$ as
\begin{equation}
    \mathcal{O}(h_1, h_2) = \frac{\langle h_1 \mid h_2\rangle}{\sqrt{\langle h_1 \mid h_1\rangle\langle h_2 \mid h_2\rangle}}
    \label{eq:match}
\end{equation}
where,
\begin{equation}
    \langle h_1 \mid h_2\rangle = \int_{f_{\text{low}}}^{f_{\text{high}}} \frac{\tilde{h}_{1}(f)\tilde{h}^{\ast}_{2}(f)}{S_n(f)} \, df   
\end{equation}
which defines a noise-weighted inner product with PSD, $S_n$ and $*$ denotes the complex conjugation. The match is also maximized over a time and phase shift between two waveforms. Finally, the mismatch is defined as $1 - \mathcal{O}(h_1, h_2)$.

We call this process ``Iterative SVD'' because it progressively refines the meshfree waveform approximation by reinserting the query points with the highest mismatch back into the initial set of nodes, followed by evaluation of the true frequency domain waveforms at the appended set of nodes. Currently, with each iteration of the "iterative SVD" node placement algorithm, as additional RBF nodes are introduced, the data matrices $\mathbf{A}$ and $\mathbf{\Psi}$ are expanded by incorporating more rows (one for each newly added node) followed by a Singular Value Decomposition (SVD) conducted on these expanded data matrices. This approach is practical for handling the decomposition of the small-sized data matrices. However, one can update the basis vectors and singular values at each iteration with more sophisticated algorithms ~\cite{BRAND200620, Stange2008OnTE} that update the left and right subspace and singular values from the existing SVD decomposition of a dense matrix as new rows are added to it. For updating the subspaces and singular values, Brand et al. \cite{BRAND200620} developed an algorithm for computing a thin SVD of a data matrix of size (${m \times n}$) in a single pass with linear time complexity $\mathcal{O}(mnr)$, where $r \leq \sqrt{min (m, n)}$. The authors proposed fast and memory-efficient sequential algorithms for tracking the singular values and subspaces, initialized with a general identity included in the existing decomposing of the data matrix. Adding a new column would modify the identity representation and provide \mbox{rank one} updates via the modified Gram–Schmidt algorithm. 

Note that if the required number of RBF nodes is small, changing the strategy for computation of the subspace and coefficients using the ``updated SVD'' approach would not significantly reduce the overall computational cost. On the other hand, for a large number of nodes, the incorporation of such sophisticated linear algebra algorithms in our proposed framework would be beneficial. We will explore this possibility in detail in future work.

\begin{figure}[!hbt]
    \centering
    \includegraphics[width=0.55\linewidth]{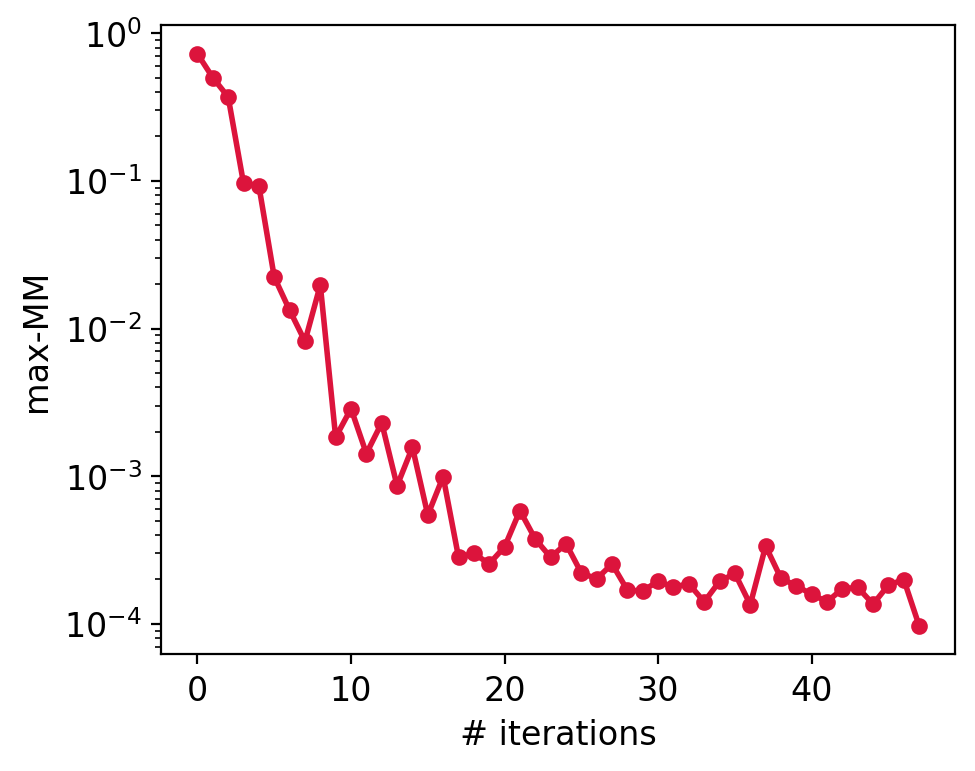}
    \caption{Mismatch between the worst approximated meshfree waveform and true waveform at each iteration. As evident from the figure, with each iteration, the meshfree approximation becomes better.}
    \label{fig:max_mismatch_at_eachitrn}
\end{figure}

Next, the SVD is performed for the new amplitude and phase matrices. This is a similar approach that was used in previous studies for building reduced-order models (ROMs) of a number of waveform approximants~\cite{canizares2013gravitational, canizares2015accelerated, Morisaki_2020, morisaki2023rapid, effroqs23}. In each iteration, SVD is employed to identify improved basis vectors for $\mathcal{A}$ and $\psi$. As shown in Fig.~\ref{fig:max_mismatch_at_eachitrn}, the maximum mismatch calculated between the worst meshfree waveform and the corresponding true waveform decreases as we add the query points corresponding to the worst meshfree waveforms at each iteration into the set of nodes used to generate the interpolants. The calculation of mismatches assumes a {\texttt{aLIGOZeroDetHighPower}}~\cite{aLIGO_ZDHP} Power Spectral Density (PSD) as implemented in \textsc{PyCBC}. We repeat the same ``Iterative SVD'' procedure for the other interpolating patches (if needed). For the analysis in this paper, we choose two patches (see Section~\ref{sec:results}). As evident from Fig.~\ref{fig:final_nodes_both_patches}, RBF nodes (for both patches) found by the ``Iterative SVD'' tend to cluster more towards the low-mass boundaries implying that the points near the lower mass boundaries contribute the most in terms of the suitable basis vectors for each patch. Similar patterns are also seen in the work focusing on building a surrogate of PN waveforms~\cite{NRHybSur3dq8} to find the desired training dataset of the parameters and building ROM using ANNs~\cite{chua2019reduced}, using a greedy strategy. 
Similar patterns are also seen in the work focusing on building a surrogate of PN waveforms~\cite{NRHybSur3dq8} to find the desired training dataset of the parameters and building ROM using ANNs~\cite{chua2019reduced}, using a greedy strategy. 

Note that initially, the maximum mismatch at each iteration is decreasing rapidly. After $\sim 25$ iterations, the rate of decrease in the mismatch reduces despite adding $10$ worst points at each iteration. It implies that it is close to having the minimum number of nodes required to satisfy the maximum mismatch threshold criteria mentioned earlier. After $\sim 48$ iterations, the maximum mismatch falls below $10^{-4}$, and algorithm stops. However, to confirm whether it has indeed crossed this criterion (since there are fluctuations), we can test the accuracy of interpolants again by spraying a different set of random query points within the interpolating parameter space and confirm that the maximum mismatch is still less than $10^{-4}$. Otherwise, we can again start the algorithm until this threshold criterion is met by at least two sets of random query points.

\section{Choice of RBF kernel}
\label{sec:choice_of_rbf_kernel}
\begin{figure}[!hbt]
    \centering
    \includegraphics[width=0.55\linewidth]{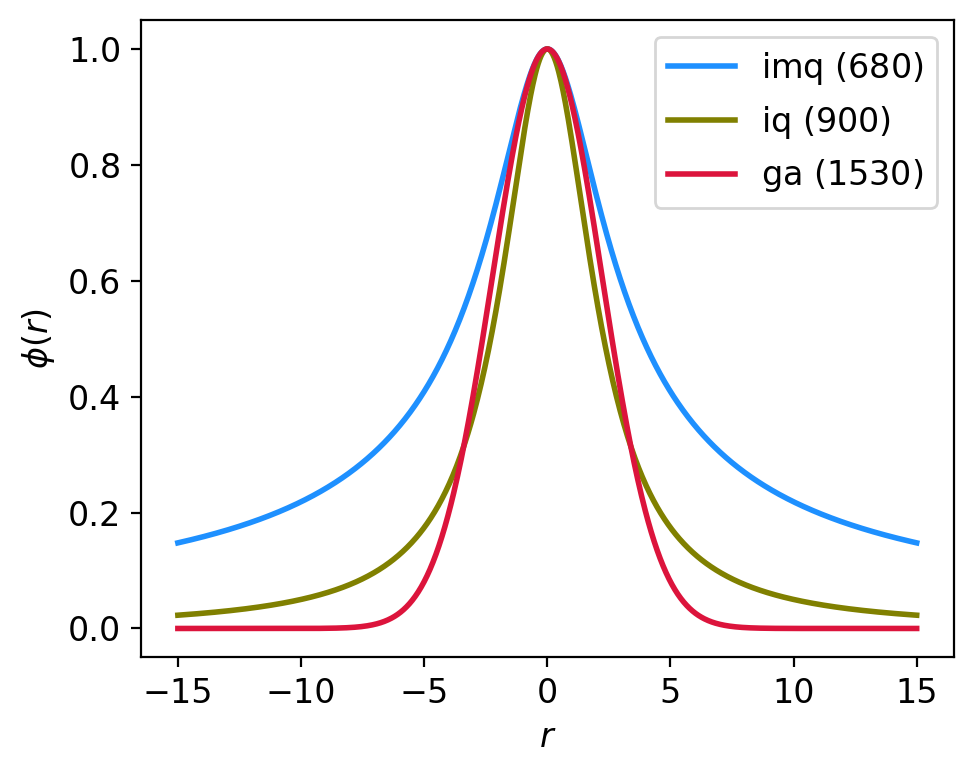}
    \caption{Profiles of different RBF kernels. In the legend, numbers in the bracket are the final number of nodes required (for patch $1$) to fall below the maximum mismatch threshold of $10^{-4}$ in the iterative SVD process. For the `imq' kernel, we required the least number of nodes to achieve the threshold.}
    \label{fig:diff_rbf_kernels}
\end{figure}

There are different RBF kernels that can be used as a basis (see Eq.~\eqref{eq:amp_phi_rbfcoeff}) for meshfree approximation. In this work, we tried the following three different RBF kernels (see table~\ref{tab:rbf_kernels}): (i) Inverse multiquadric (`imq'), (ii) Inverse Quadric (`iq'), and (iii) Gaussian (`ga'). As mentioned earlier in sec.~\ref{sec:interp_generation}, to terminate the iterative SVD, we set a threshold of either a maximum number of nodes (3000 here) or a maximum mismatch of $10^{-4}$ or less at a given iteration. 
\begin{table}[!hbt]
    \centering
        \begin{tabular}{l c c c}
            \hline \hline
            RBF kernel type & $\phi(r)$ & \multicolumn{2}{c}{$\epsilon$} \\ \cmidrule[0.8pt](rr{0.95em}){3-4}
            & & Patch 1 & Patch 2 \\ \hline
            inverse multiquadric (`imq') & $1/\sqrt{(1 + (\epsilon\,r)^2)}$ &  0.4016 & 0.4010 \\
            Gaussian (`ga')              &  $\exp[-(\epsilon\, r)^2]$ & 0.3172 & 0.2909 \\
            inverse quadric (`iq')       &  $1/(1 + (\epsilon\,r)^2)$& 0.4464 & 0.4378 \\
            \hline \hline
        \end{tabular}
    \caption{Different RBF kernels}
    \label{tab:rbf_kernels}
\end{table}
The interpolant generation is stopped whenever any of the above-stated conditions is satisfied.  In the case of `ga'(`iq') kernels, the final number of nodes for patch $1$ and patch $2$ was $1530(900)$ and $1620(590)$, respectively, when the maximum mismatch at an iteration fell below the threshold of $10^{-4}$. However, in the case of `imq' RBF kernel, the final number of nodes for patch one was $680$, while for the second patch, it was $770$ for the same threshold. A natural question arises: what is so special about the `imq'? A possible explanation is the spread of these RBF kernels. As shown in Fig.~\ref{fig:diff_rbf_kernels}, the `ga', `imq', and `iq' have similar profiles near the center ($r=0$). However, as we go away from the center, the `ga' and `iq' fall very rapidly in comparison to the `imq' kernel. It implies that the `imq' kernel has a significant overlap with the `imq' kernel centered at the other nearby nodes (centers) in comparison to the other two RBF kernels, and therefore, it explains the lesser number of nodes required to fall below the maximum mismatch threshold at a given iteration for `imq' in comparison to a relatively higher number of nodes for the other RBF kernels.
For the next section, we choose `imq' as the RBF kernel for generating meshfree interpolants. 
\section{Results}
\label{sec:results}
As previously mentioned, there are two overlapping patches in $\chi_{1z}$, and we independently construct interpolants for each of the patches. In the first patch where ${\chi_{1z} \in [-0.1, \,0]}$, we initiate with ${N = \binom{\nu + d}{\nu}}$ nodes, with $\nu$ representing the degree of the monomial terms in Eq.~\eqref{eq:amp_phi_rbfcoeff} and $d$ denoting the dimensionality of the parameter space (here, $\nu = 6$ and $d = 4$ for the four-dimensional parameter space interpolation). The `imq' RBF kernel is selected, where $\epsilon$ is the shape factor influencing the RBFs' spread at the nodes, determined through (LOOCV) procedure~\cite{hastie2009elements} (see table~\ref{tab:rbf_kernels} for values of $\epsilon$). In the second patch covering ${\chi_{1z} \in [-0.005, \,0.1]}$, the RBF parameters chosen for the first patch are retained, with the exception of a different $\epsilon$ value for this patch. The seismic cutoff frequency ($f_{low}$) is set at $15$ Hz while $f_{start}$ is equal to $10$ Hz.
Once the interpolants for both patches are ready (see Sec.~\ref{sec:interp_generation}), we can use them to generate the interpolated waveforms at the arbitrary query points within the interpolating sample space. Note that the generation of interpolants is a one-time offline process, and it can be completed well in advance before the parameter estimation of the GW event is initiated. While multiple patches are necessary to adequately cover the desired parameter space ranges, the generation of interpolants is highly parallelizable and can benefit from a multicore setup to expedite this stage. 
\begin{figure}[!hbt]
    \centering
    \includegraphics[width=0.55\linewidth]{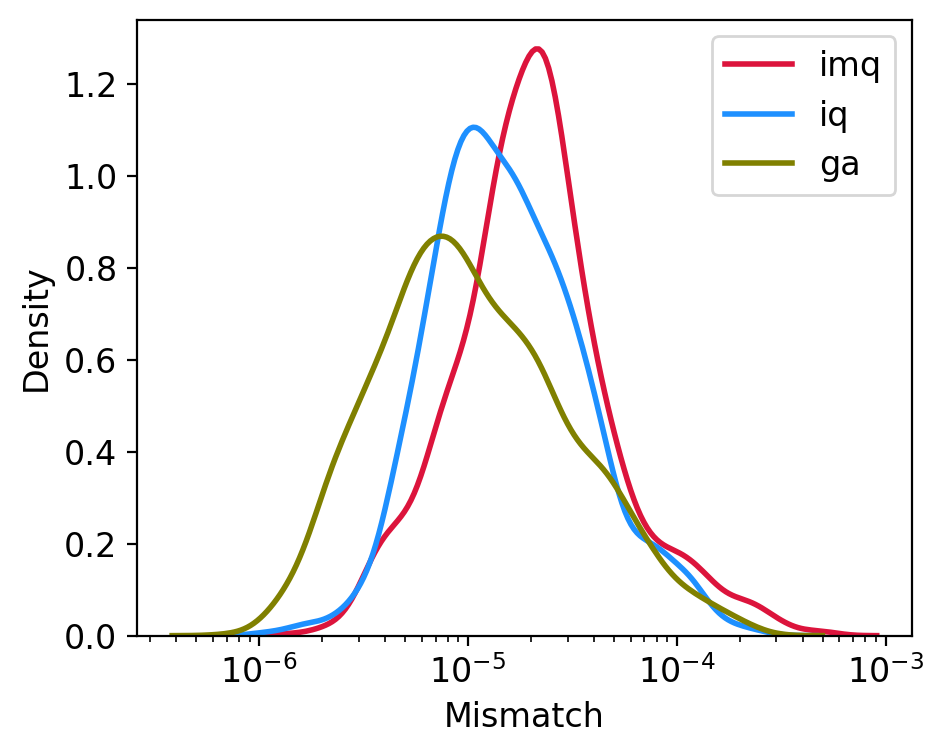}
    \caption{The mismatches calculated between interpolated and true waveforms at the posterior samples are shown. For comparison, we also show the mismatches for meshfree waveforms using `ga' and `iq' RBF kernels. The median mismatched (for all kernels) is $\mathcal{O}(10^{-5})$. Note that the accuracy achieved with `ga' is a little better than with the other kernels due to the high number of nodes that were used to construct `ga' kernel-based meshfree interpolants. However, we do not find any significant difference in the corresponding posterior distributions.}
    \label{fig:mismatches_at_post_samps}
\end{figure}
In the specific case of this analysis, the generation of interpolants for each patch was accomplished in ${\sim 20}$ minutes using a CPU setup with $32$ cores each. We use the publically available RBF Python package~\cite{RBF_github} to generate RBF interpolants.

To assess the accuracy of the interpolated waveforms, we generated $20$ sets of $10^3$ query points each, randomly distributed in the intrinsic parameter space within the interpolating region (see table~\ref{tab:priordistr}). We then generated both the true and interpolated waveforms at these query points and computed the mismatches between them, assuming a flat Power Spectral Density (PSD). The median mismatch is found to be $\mathcal{O}(10^{-5})$, which shows the high level of accuracy of the meshfree waveforms in approximating the true waveforms across the interpolating sample space
\begin{table}[!hbt]
\centering
\begin{tabular}{lcl}
\hline \hline
 Parameters     &   Range    &   Prior distribution \\
\hline
 $\mathcal{M}$  &   $[25, 40]$ &   $\propto \mathcal{M}$ \\
 $q$            &   $[1, 3]$             & $ \propto \left [ (1 + q)/q^3 \right ]^{2/5}$     \\
 $\chi_{1z}$   &   $[-0.1, 0.1]$   & Uniform  \\
 $\chi_{2z}$   &   $[-0.05, 0.05]$   & Uniform  \\
 $V_{com}$     & $[5e3, 1e11]$                & Uniform\\
 $t_c$          & $t_{\text{trig}} \pm 0.12$& Uniform\\
 $\alpha$       & $[0, 2\pi]$               & Uniform\\
 $\delta$       & $\pm \pi/2$       & $\sin^{-1} \left [ {\text{Uniform}}[-1,1]\right ]$\\
 $\iota$        & $[0, \pi]$                & Uniform in $\cos \iota$\\
 $\psi$         & $[0, 2\pi]$               & Uniform angle\\
\hline \hline
\end{tabular}
\caption{Prior parameter space over the ten-dimensional parameter space $\vec \Lambda$.}
\label{tab:priordistr}
\end{table}

In terms of the computational speed-ups, the meshfree waveforms can be evaluated in $\sim 2.3$ ms in comparison to $\sim 89$ ms\footnote{frequency domain tapered version of \texttt{NRHybSur3dq8} time-domain model} implying a speed-up factor of $\sim 38$. Note that these timings are for the Python implementation of these waveforms. A ``C implementation'' of the true waveforms can be evaluated in $\sim 10$ ms as shown by ~\cite{NRHybSur3dq8} (about $ 4.3 \times$ slower than the meshfree waveform's Python implementation). Since these numbers are also dependent on the hardware configurations of the machines on which these speed tests are performed, a fair comparison of meshfree speed-ups with their C implementation is not possible until we make a C version of meshfree waveforms.

Finally, to test the accuracy of the meshfree model in the context of parameter estimation, we use $\texttt{SEOBNRv4}\_\texttt{ROM}$~\cite{Boh_2017} waveform model to inject $16$ seconds long simulated BBH event into the Gaussian noise with PSDs taken from {\texttt{aLIGOZeroDetHighPower}}~\cite{aLIGO_ZDHP} for both LIGO-Livingston and LIGO-Hanford detectors, and \texttt{AdvVirgo}~\cite{AdVirgo} for Virgo detector with network matched-filter signal-to-noise ratio (SNR) $\sim 18$. The seismic cutoff frequency ($f_{low}$) is set at $15$ Hz, and the high cutoff frequency is equal to the ringdown frequency of the lowest component masses within the chosen range of chirp masses and mass ratios. A sampling frequency of $2048$ Hz is considered. We employ dynesty~\cite{speagle2020dynesty, higson2019dynamic}, sampler, a Python implementation of the Nested sampling algorithm, to sample the posterior distribution. The prior distributions and the boundaries of the parameters (to be estimated) are shown in Table~\ref{tab:priordistr}. We choose the following sampler settings for dynesty: ${\texttt{nLive} = 500}$, ${\texttt{nWalks} = 500}$, ${\texttt{dlogz} = 0.1}$, {\texttt{sample} = ``rwalk''}.  Here, the parameter \texttt{nlive} represents the number of live points, which determines the resolution of the sampled posterior distribution. 
\begin{figure}[!hbt]
    \centering
    \includegraphics[width=0.55\linewidth]{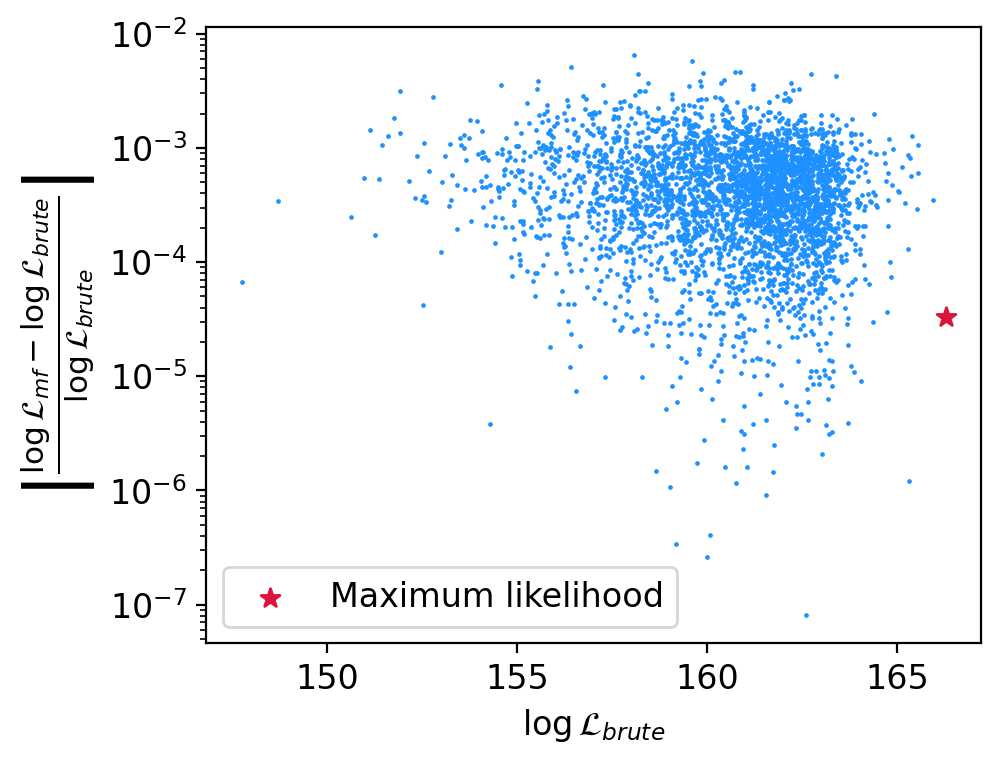}
    \caption{Absolute relative errors in the likelihood as a function of true $\log \mathcal{L}$. The absolute relative error is defined as ${|(\log \mathcal{L}_{\text{mf}} - \log \mathcal{L}_{\text{brute}})/\log \mathcal{L}_{\text{brute}}|}$. The $\log \mathcal{L}_{\text{mf}}$ is calculated using the interpolated waveforms, while $\log \mathcal{L}_{\text{brute}}$ is calculated using the true waveforms. These likelihoods were calculated at the posterior samples obtained from the PE of the simulated BBH event. The red star represents the maximum likelihood posterior sample. Note that the majority of relative errors (especially near the peak) are well within $1\%$, demonstrating the accuracy of the interpolants.}
    \label{fig:like_errors}
\end{figure}

\begin{figure*}[!hbt]
    \centering
    \includegraphics[width=1\textwidth]{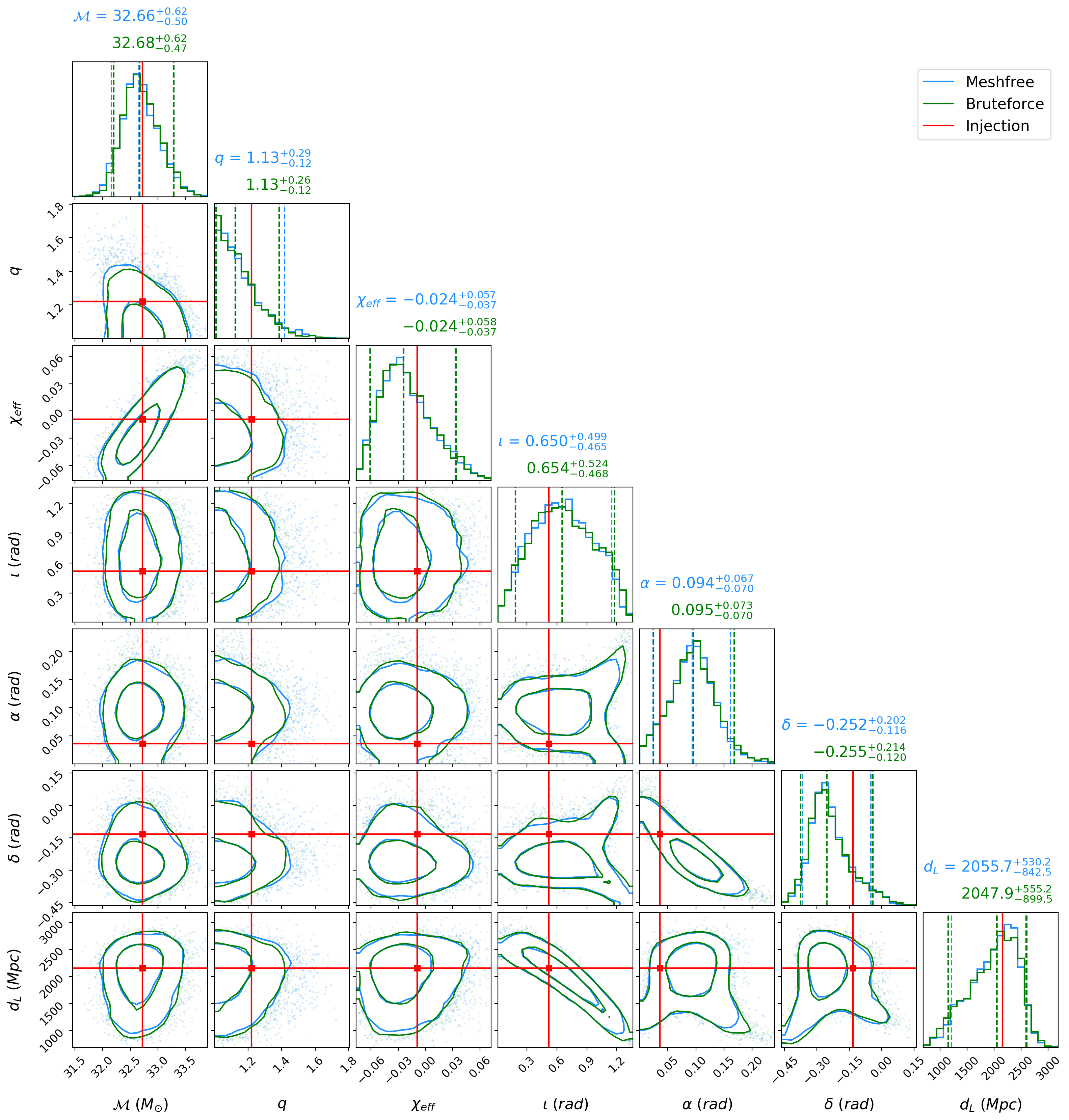}
    \caption{Corner plot of the posterior distributions(with $50\%$ and $90\%$ CI and contours) obtained by the meshfree interpolated waveforms for a simulated BBH event. The vertical red line represents the injected values. It took \mbox{$\sim 16.4$ min} to complete on $64$ CPU cores. In comparison, the PE run with the direct evaluation of the frequency-domain version of the tapered time-domain \texttt{NRHybSur3dq8} waveform model took \mbox{$\sim 5$ h and $50$ min} on the same number of CPU cores and the resulting posterior distributions are broadly consistent with those obtained using meshfree interpolated waveforms.} 
    \label{fig:corner_plot_meshfree_vs_bruteforce}
\end{figure*}

Smaller values of ${\texttt{nLive}}$ might give rise to a poorly sampled distribution (hence evidence) with much faster convergence. Instead, taking a larger value can give us a finely sampled distribution at the expense of lower convergence speeds. \texttt{walks} specifies the minimum number of steps required before a new live point is proposed, which replaces the live point with the lowest likelihood in the nested sampling. \texttt{sample} indicates the chosen approach for generating samples, and \texttt{dlogz} characterizes the remaining prior volume's contribution to the total evidence. In this PE study, the sampling terminates once \texttt{dlogz} reaches a threshold of $0.1$. For a more comprehensive understanding of dynesty's nested sampling algorithm and its practical implementation, one can refer to the following references ~\cite{speagle2020dynesty, sergey_koposov_2023_7600689}.  
As evident from Fig.~\ref{fig:corner_plot_meshfree_vs_bruteforce}, posterior distributions of various binary parameters of simulated BBH event contain the injected values well within the  $90\%$ CI. This PE analysis took $\sim 16.4$ minutes on a $64$ CPU cores setup. Fig.~\ref{fig:mismatches_at_post_samps} also shows the probability distribution function (PDF) of the mismatches of the interpolated waveforms with the true waveforms generated at the posterior samples obtained from the PE and median mismatch is $\sim \mathcal{O}(10^{-5})$, demonstrating the good accuracy of the meshfree waveforms.
To quantify the effect of these approximate waveforms on the likelihood calculation, we also evaluated the relative errors in the likelihood, as shown in Fig.~\ref{fig:like_errors} and found a median absolute relative error of $\sim \mathcal{O}(10^{-3})$ further showing the effectiveness of the meshfree waveforms.
All the tests were performed on AMD EPYC 7542 CPU@2.90GHz processors.
\section{Conclusion and Future Outlook}
\label{sec:conclusion}
In this work, we have constructed a fast, interpolated frequency domain model that approximates the time-domain \texttt{NRHybSur3dq8} waveform model with high fidelity. We show that such interpolated waveforms can be evaluated in ${\sim 2.3}$ ms ($\times 38$) times faster than the standard implementation of this waveform model in \textsc{PyCBC}. We constructed the waveform interpolants by first selecting a patch of the parameter space. Then, we spray randomly scattered nodes in the sample space and evaluate $\mathcal{A}$ and $\psi$ from the frequency domain waveform corresponding to \texttt{NRHybSur3dq8} at those nodes. Subsequently, we find a suitable basis spanning the space of both amplitude and phase by performing the SVD of the amplitude and phase matrices resulting from stacking the amplitudes and phases at the nodes.  The resulting SVD coefficients corresponding to amplitude and phase are fit with a linear combination of RBFs and monomials (``interpolants''), followed by the mismatched calculation of the interpolated waveforms with the true waveforms at arbitrary query points in the selected patch of parameter space. The worst $10$ query points in terms of mismatch are added back into the set of initial nodes, and the whole process (called ``Iterative SVD'') repeats till either the maximum mismatch at the given iteration falls below a set threshold, or the number of nodes crosses a fixed number of maximum nodes. In each iteration, the accuracy of the interpolants increases, which in turn enhances the meshfree approximation of the waveforms. Note that the choice of $10$ worst points, as mentioned earlier, is completely arbitrary and chosen to accelerate the interpolants generation process. In order to test the validity of the meshfree interpolated waveforms, we also performed a parameter estimation of a simulated BBH event using the meshfree interpolated waveforms and found posterior distributions are consistent with the injected values. It is important to highlight that our method does not require importance sampling as used in deep learning methods such as \textsc{DINGO}~\cite{DINGO} and \textsc{NESSAI}~\cite{NESSAI}.

Currently, a limitation we face involves the multiplication of SVD coefficients with the basis vectors. As the duration of the waveform increases, the computational expense of this multiplication rises, consequently impacting the evaluation cost of the meshfree interpolated waveform. In this context, we can also use the Empirical interpolation method (EIM)~\cite{BARRAULT2004667, Yvon, doi:10.1137/090766498, canizares2015accelerated} to reduce the number of frequency points included in the basis vectors. It is yet to see whether this kind of strategy could be helpful within the meshfree framework in reducing the dimensionality of the basis vectors, which is especially needed for longer-duration waveforms. Further, in this work, we used an iterative SVD scheme that updates the basis vectors with a new set of points by applying independent SVD of the updated amplitude and phase matrix. In the follow-up work, we will replace our iterative SVD scheme with advanced updated SVD algorithms ~\cite{BRAND200620, Stange2008OnTE} and will compare the computation cost of performing independent SVD in each iteration and updating the initial basis and singular values from the first iteration using updated SVD algorithms. Another notable aspect is that the distance employed in Eq.\eqref{eq:amp_phi_rbfcoeff} is Euclidean, even though the parameter space we operate in is inherently non-Euclidean.  Nevertheless, we choose nodes in chirp-time coordinates ($\theta_0$ and $\theta_3$) instead of component masses or chirp mass and mass ratio as the metric varies slowly in chirp-time coordinates.  Additionally, it's worth noting that this work solely focuses on the aligned spin waveform model \texttt{NRHybSur3dq8}. Subsequent extensions of this work will explore broadening the meshfree framework to incorporate NR surrogate models encompassing subdominant modes, considering spins, and addressing the aforementioned limitations. We also performed similar exercises for other aligned spin waveform models, such as \texttt{IMRPhenomXAS} and $\texttt{SEOBNRv4}\_\texttt{ROM}$. Since these waveform models are already fast to evaluate, we don't get any significant speed-up in their corresponding meshfree waveform evaluation. However, the speed-up can be further enhanced by employing strategies such as adaptive frequency resolution~\cite{Morisaki:2021ngj} and EIM procedure, as mentioned earlier. In future follow-ups of this work, we would also consider extending this meshfree framework to include NR surrogate models that include subdominant modes, precessing spins, and possibly eccentric models.

\section*{Acknowledgements}
We thank Srashti Goyal for carefully going through the manuscript and giving helpful comments. We thank Prayush Kumar for his help in the initial stages of this work. We also thank Abhishek Sharma and Sachin Shukla for their useful suggestions and comments. L.~P. is supported by the Research Scholarship Program of Tata Consultancy Services (TCS). A.~R is supported by the research program of the Netherlands Organisation for Scientific Research (NWO).  A.~S. gratefully acknowledges the generous grant provided by the Department of Science and Technology, India, through the DST-ICPS (Interdisciplinary Cyber Physical Systems) cluster project funding. We thank the HPC support staff at IIT Gandhinagar for their help and cooperation. The authors are grateful for the computational resources provided by the LIGO Laboratory and supported by the National Science Foundation Grants No. PHY-0757058 and No. PHY-0823459. This material is based upon work supported by NSF's LIGO Laboratory, which is a major facility fully funded by the National Science Foundation. 

This research has made use of data or software obtained from the Gravitational Wave Open Science Center~\cite{gwosc_web}, a service of the LIGO Scientific Collaboration, the Virgo Collaboration, and KAGRA. This material is based upon work supported by NSF's LIGO Laboratory, which is a major facility fully funded by the National Science Foundation, as well as the Science and Technology Facilities Council (STFC) of the United Kingdom, the Max-Planck-Society (MPS), and the State of Niedersachsen/Germany for support of the construction of Advanced LIGO and construction and operation of the GEO600 detector. Additional support for Advanced LIGO was provided by the Australian Research Council. Virgo is funded through the European Gravitational Observatory (EGO), the French Centre National de Recherche Scientifique (CNRS), the Italian Istituto Nazionale di Fisica Nucleare (INFN), and the Dutch Nikhef, with contributions by institutions from Belgium, Germany, Greece, Hungary, Ireland, Japan, Monaco, Poland, Portugal, Spain. KAGRA is supported by the Ministry of Education, Culture, Sports, Science and Technology (MEXT), Japan Society for the Promotion of Science (JSPS) in Japan; National Research Foundation (NRF) and Ministry of Science and ICT (MSIT) in Korea; Academia Sinica (AS) and National Science and Technology Council (NSTC) in Taiwan.

Code availability: Codes used in this analysis are publicly available in a Github repository~\cite{github_meshfree_waveform}.
\chapter{Fast likelihood evaluation using meshfree approximations for reconstructing compact binary sources}
\label{chap:chapter_3}

\paragraph{\textbf{Abstract}}
Several rapid parameter estimation methods have recently been advanced to deal with the computational challenges of the problem of Bayesian inference of the properties of compact binary sources detected in the upcoming science runs of the terrestrial network of gravitational wave detectors. Some of these methods are well-optimized to reconstruct gravitational wave signals in nearly real-time, necessary for multi-messenger astronomy. In this context, this work presents a new, computationally efficient algorithm for fast evaluation of the likelihood function using a combination of numerical linear algebra and mesh-free interpolation methods. The proposed method can rapidly evaluate the likelihood function at any arbitrary point of the sample space at a negligible loss of accuracy and is an alternative to the grid-based parameter estimation schemes. We obtain posterior samples over model parameters for a canonical binary neutron star system by interfacing our fast likelihood evaluation method with the nested sampling algorithm. The marginalized posterior distributions obtained from these samples are statistically identical to those obtained by brute force calculations. We find that such Bayesian posteriors can be determined within a few minutes of detecting such transient compact binary sources, thereby improving the chances of their prompt follow-up observations with telescopes at different wavelengths. It may be possible to apply the blueprint of the meshfree technique presented in this study to Bayesian inference problems in other domains.


\section{Introduction}
\label{sec:intro_chap2}
This chapter is based on the publication \textit{Fast likelihood evaluation using meshfree approximations for reconstructing compact binary sources}, \href{https://journals.aps.org/prd/abstract/10.1103/PhysRevD.108.064055}{Phys. Rev. D \textbf{108}, 064055 (2023)}.\newline

The detection of gravitational waves (GW) from the GW170817~\cite{abbott2017gravitational} binary neutron star (BNS) system, followed by the prompt multi-wavelength (gamma-rays to radio) observation of its electromagnetic~(EM) counterpart, has led to several fundamental discoveries; and is hailed as a significant breakthrough in astronomy. These discoveries include the validation of long-held hypotheses that BNS mergers are ideal sites for r-process nucleosynthesis and produce short gamma-ray bursts, the first GW-based constraints on the equation of state of nuclear matter in such stars, and the measurement of Hubble constant independent of the cosmic distance ladder. 

The inevitable improvement of the detectors' sensitivity in future observation runs is likely to have a two-fold impact on the prospects of multi-messenger observations: 
{\emph{firstly}}, the increased bandwidth of the detectors (especially improved sensitivity at low frequencies) will result in a tremendous increase in the computational cost of Bayesian inference of source parameters, including sky localization essential for prompt observation of EM counterparts. Although the {\texttt{BAYESTAR}}~\cite{singer2016rapid} algorithm could be used to produce rapid sky maps, it has been recently shown~\cite{Finstad_2020} that coherent parameter estimation (PE) can localize the sources better by an average reduction of $14 \, \text{deg}^{2}$ in the uncertainty, underlining the importance of developing fast PE algorithms.
{\emph{Second}}, the reach of the terrestrial network of GW detectors will extend out to several Gpc to the effect that one would have far too many detections of BNS/NSBH signals to contend with whilst generating prompt sky-location maps~\cite{abbott2020prospects}; so much so that 
one may have to prioritize the GW sources for EM follow-up based on prospects of new science from a rapid estimation of their mass and spin components as shown by Margalit $\&$ Metzger~\cite{Margalit_2019}, thereby helping EM observatories to use resources optimally. 
Several fast PE algorithms have been developed recently, such as the coherent multi-detector extension of the relative binning/heterodyne method by Finstaad and Brown (2020)~\cite{Finstad_2020}, which produces the posterior within twenty minutes for BNS systems with 32 CPU cores. Well-trained machine learning PE methods~\cite{Dax2021, gabbard2022bayesian, Dax2023} can significantly reduce the runtimes and produce the posteriors in nearly real-time.
In the past, algorithms for accelerated parameter estimation have mainly focussed on speeding up the overlap integral. These include reduced-order models (ROMs)~\cite{canizares2015accelerated, Qi_2021, Soichiro_2020}, machine-learning aided ROMs~\cite{chua2020learning}, Gaussian process regression based interpolation~\cite{https://doi.org/10.48550/arxiv.1805.10457} and relative binning~\cite{cornish2021heterodyned, Venumadhav2018, Finstad_2020} algorithms. Our approach takes inspiration from the grid-based likelihood interpolation method~\cite{smith2014rapidly} based on orthonormal Chebyshev polynomials. The grid-based techniques have a drawback in that the number of interpolation nodes grows exponentially with the dimensionality of the parameter space.

In this work, we propose a new and alternative approach to grid-based likelihood interpolation method ~\cite{smith2014rapidly}, a computationally efficient method for evaluating the likelihood function (a key ingredient in Bayesian inference) using meshfree interpolation methods with dimension reduction techniques. We directly interpolate the likelihood function over the parameter space, bypassing the generation of templates and brute-force computation of the overlap integral altogether.  
Our scheme can quickly approximate the log-likelihood function with high accuracy and produce statistically indistinguishable posteriors over source parameters. 
Further, both the {\texttt{GstLAL}} search framework~\cite{GstLAL_2010} and the meshfree method use the idea of dimension reduction using SVD~\cite{golub2013matrix}, it may be prudent to incorporate this method with the low-latency {\texttt{GstLAL}} search pipeline for rapid, automated follow-ups of the detected events.

\section{Bayesian inference}
\label{sec:bayesianinference_chap2}

Given data ${\boldsymbol{d} = \boldsymbol{h}(\vec \Lambda_{\text{true}}) + \boldsymbol{n}}$ recorded at a detector containing an astrophysical GW signal $\boldsymbol{h}(\vec \Lambda_{\text{true}})$ embedded in additive Gaussian noise $\boldsymbol{n}$, one is interested in solving the inverse problem to estimate the source parameters. Bayesian inference is a stochastic inversion method where the posterior probability density $p(\vec \Lambda \mid \boldsymbol{d})$ over the source parameters is related to the likelihood function ${\mathcal{L}(\boldsymbol{d} \mid \vec \Lambda)}$ of observing the data through the Bayes' theorem: 
\begin{equation}
p(\vec \Lambda \mid \boldsymbol{d}) = \frac{\mathcal{L}(\boldsymbol{d} \mid \vec \Lambda) \, p(\vec \Lambda)}{p(\boldsymbol{d})} \, ,
\label{Eq:bayestheorem}
\end{equation}
where $p(\vec \Lambda)$ is the prior distribution over the model parameters ${\vec \Lambda \equiv \{ \vec \lambda^{\text{ext}}, \vec \lambda \}}$. 
In our notation, $\vec \lambda$ denotes the intrinsic parameters such as component masses and spins. The set of extrinsic parameters is denoted by $\vec \lambda^{\text{ext}}$.
We are particularly interested in estimating the extrinsic parameter $t_{c}$ denoting the fiducial time of coalescence of the two masses. $t_{c}$  will be mentioned explicitly wherever required, as it is treated in a special way in our analysis.

The forward generative frequency-domain restricted waveform model for non-precessing compact binaries can be expressed as ${\boldsymbol{h} (\vec \Lambda) = {\mathcal{A}}\, h_+(f_{k}; \vec\lambda)}$, where the complex amplitude ${\mathcal{A}}$ depends only on the extrinsic parameters and $h_+(f_{k}; \vec\lambda)$ is the `+' polarization of the signal that depends only on the intrinsic parameters~\cite{Foreman_Mackey_2013}. Here ${\{f_{k}\}_{k = 0}^{N_{s}/2}}$ defines positive Fourier frequencies, and $N_s$\footnote{${N_{s} = \text{signal duration} \times \text{sampling frequency}}$} is the number of sample points.
A GW signal, observed by an interferometric detector, can be considered as a linear combination of the two polarizations weighted by the antenna pattern function. The $h_{+}(f_{k}; \vec\lambda)$ polarization is related to the $h_{\times}(f_{k}; \vec\lambda)$ polarization for non-precessing GW signal as: ${h_{+}(f_{k}; \vec\lambda) \propto i h_{\times}(f_{k}; \vec\lambda)}$~\cite{findchirp_2012}. This allows us to write the detector response in terms of any one of the polarizations alone (we have chosen the $h_{+}(f_{k}; \vec\lambda)$ polarization). 
Using this model, the posterior ${p(\vec \Lambda \mid \boldsymbol{d})}$ can be directly evaluated at every point in $\vec \Lambda$ using Eq.~(\ref{Eq:bayestheorem}). However, in view of the high-dimensionality of $\vec \Lambda$, it is more efficient to sample the posterior using stochastic sampling algorithms such as Nested-Sampling~\cite{skilling2006nested}, or Markov Chain Monte Carlo (MCMC)~\cite{Foreman_Mackey_2013}.
From Eq.~(\ref{Eq:bayestheorem}), it is evident that for a quick estimation of the posterior distribution, it is imperative to rapidly evaluate the likelihood function.

We work with the phase-marginalized log-likelihood function~\cite{thrane_2019}:
\begin{equation}
\begin{split}
\ln \mathcal{L} (\vec \Lambda, t_{c})	= \ln I_{0} \left [ |{\mathcal{A}}| \, z(\vec\lambda, tc) \right ] 
								- \frac{1}{2} ||{\boldsymbol{h}(\vec \Lambda)}||_{2}^2 
\end{split}
\label{Eq:log-likelihood}
\end{equation}
where $I_0(\cdot )$ is the 0-th order modified Bessel function of the first kind, 
and $z(\vec\lambda, t_{c})$ is the frequency-domain overlap-integral: 
\begin{equation}
z(\vec\lambda, t_c) = 4 \, \Delta f \, \left | \sum_{k = 0}^{N_{s}/2} {
\frac{{d}^{*}(f_{k}) \, {h}_+(f_{k}, \vec\lambda)}{S_{h}(f_{k})} \, e^{-2 \pi i f_{k} t_{c}}}  \right |,  
\label{Eq:overlap}
\end{equation}
inversely by $S_{h}(f_{k})$, the detector's one-sided noise power spectral density (PSD).
The data and template vectors are sampled at discrete frequencies $\{f_{k}\}_{k = 0}^{N_{s}/2}$. 

The complexity of evaluating the overlap integral scales directly with the number of data samples, which in turn, scales with the seismic cut-off frequency (approximately) as ${N_s \sim f_{\text{low}}^{-8/3}}$.  
As we progress from the O4 observational run (${f_{\text{low}}= 20}$ Hz) to O5 at design sensitivity (${f_{\text{low}}= 10}$ Hz), evaluating ${p(\vec \Lambda \mid \boldsymbol{d})}$ is likely to take  at least $\times 6.3$ longer. In addition, additional costs will be incurred in constructing longer templates at the proposal points. Therefore, the likelihood calculation can be expensive. However, our method is immune to this issue as our scheme directly approximates the likelihoods at different sample points.
In this work, we have used non-orthonormal radial basis functions (RBF) (Gaussian kernels) centered at interpolation nodes that can be randomly scattered over the volume of the intrinsic parameter space. In this manner, we have effective control of their number in higher dimensional parameter spaces. For completeness, we will begin by exploring the grid-based likelihood interpolation scheme initially introduced by Smith et al.~\cite{cannon2012interpolating, smith2014rapidly}, and subsequently, we will delve into the mesh-free interpolation scheme introduced in this thesis.
%

\section{Likelihood interpolation}
The computational cost of Bayesian inference comprises two parts: the first is incurred in waveform generation (see Appendix~\ref{appendix:scaling_of_speed_up}), followed by likelihood evaluation at a point proposed by the sampler. Typically, a sampler proposes a large number (${\sim 10^6 - 10^7}$) of points to adequately capture the posterior distribution - which makes this part computationally expensive. The other part of the total computational cost can be attributed to the overheads of the sampling method itself. Since the latter cost depends on the efficiency of the sampling algorithm used (and its software implementation) and is significantly less in comparison to the overall cost of PE, we shall ignore it in our discussions. Using likelihood interpolation, we simply bypass the generation of template waveforms and directly interpolate the likelihood over the intrinsic parameter space, thereby reducing the computational cost of likelihood calculation drastically. Both grid-based and meshfree interpolation schemes consist of two stages. In the first stage (start-up stage), we generate interpolating functions for the likelihood while in the second stage (online stage), we rapidly calculate the likelihood, which can be interfaced with a stochastic sampler to estimate the posterior distribution over the sampling parameters. We assume that the parameter estimation is ``seeded'' by the most significant trigger $\vec \Lambda^\ast$ from an upstream detection pipeline~\cite {messick2017analysis, usman2016pycbc}. For an injected signal, this can simply be the injection parameters. The sampling algorithm draws new proposals from a sample space, which is taken to be a moderate-sized ``hyper-rectangle'' in $\vec \Lambda$, centered around the most significant search trigger. 

From Eq.~\eqref{Eq:log-likelihood} and~\eqref{Eq:overlap}, it is evident that we need to interpolate two pieces: $z(\vec \lambda, t_{c})$ and ${\sigma^2(\vec \lambda) \equiv ||h_+(\vec \lambda)||_2^2}$; and combine them with the amplitude ${\mathcal{A}}$ to calculate the log-likelihood ratio at a given `query' point $\vec \lambda^q$ and a particularly given value of $t_c$. We will first discuss likelihood interpolation using a grid-based interpolation scheme, which makes use of Cheyshev polynomials to interpolate the likelihood. Then, we will discuss the meshfree interpolation scheme, followed by the application of these two methods on simulated BBH and BNS systems. 

\label{sec:likelihood_interpolation}
\subsection{Grid-based likelihood interpolation}
\label{sec:grid_based_e_likelihood_interpolation_chap2}
\subsubsection{Start-up stage}
\begin{figure*}[!htp]
\begin{subfigure}{0.5\linewidth}
    \includegraphics[width=\linewidth]{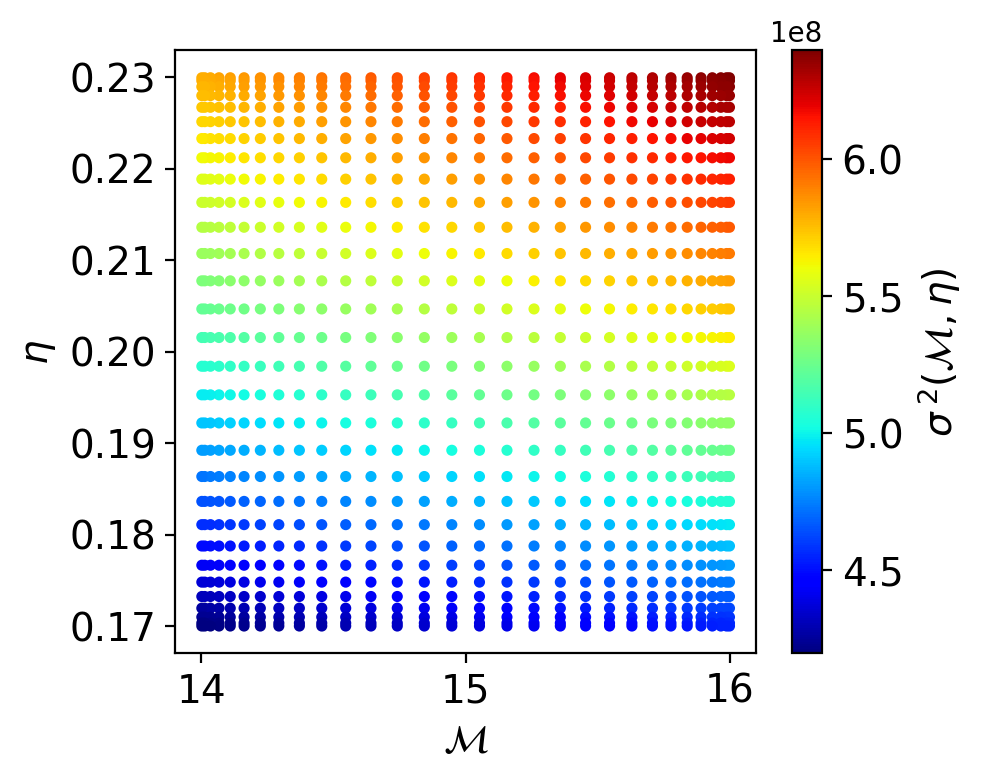}
    \caption{Template norm square surface}
    \label{fig:sigmasq_surf_heavy_cheby}
\end{subfigure}\hfill
\begin{subfigure}{0.5\linewidth}
    \includegraphics[width=\linewidth]{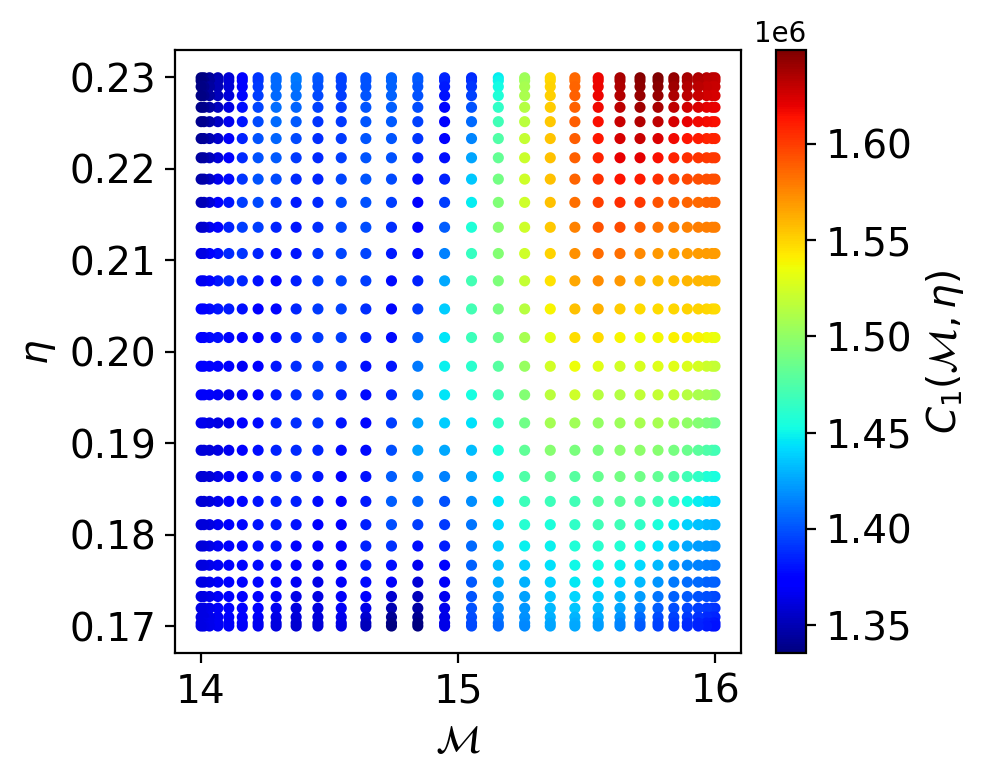}
    \caption{SVD coefficient surface}
    \label{fig:C_surf_heavy_cheby}
\end{subfigure}
\caption{} 
\label{}
\end{figure*}

As previously mentioned, during the start-up stage, we construct likelihood interpolants approximating the true likelihood. In the case of a grid-based interpolation scheme, the initial step involves establishing a grid of nodes within the sample space (intrinsic parameter space). Since we are using Chebyshev polynomials (see Appendix ~\ref{appendix:cheby_poly}) for interpolation, a prudent choice for nodes would be "Chebyshev nodes" known for mitigating Runge's phenomenon. The physical nodes are selected from the sample space and are scaled so that each dimension spans the interval $[-1, 1]$, positioned at the $J_{\text{max}}$-th order Chebyshev nodes. The locations of these nodes for the $i$-th dimension are expressed as follows:
\begin{equation}
    \lambda_i = \cos\left(\pi \frac{i + 1/2}{J_{\text{max}} + 1}\right),
    \label{eq:cheby_nodes_locs}
\end{equation}
where $i$ ranges from $0$ to $J_{\text{max}}$. It can be straightforwardly extended to higher dimensions with each component of $\vec \lambda$ following Eq.~\eqref{eq:cheby_nodes_locs}. As evident from the Fig.~\ref{fig:sigmasq_surf_heavy_cheby}, the template norm ${\sigma^2(\vec \lambda)}$ is a smoothly varying scalar field over $\vec \lambda$. Here $\vec \lambda$ consists of a tuple of values $\{\lambda_i\}$ where $i=1,2...$ representing each dimensions.

We can interpolate $\sigma^2$ at a query point $\vec \lambda^q$ by first evaluating the values explicitly at the interpolation nodes $\vec \lambda^\alpha$, and then expressing ${\sigma^2(\vec \lambda^q)}$ at an arbitrary point $\vec \lambda^q$ as a linear combination of Chebyshev polynomials evaluated at these nodes. 
\begin{equation}
    \sigma^2(\vec \lambda^q) = \sum_{L=0}^{L_{\text{max}}}\sum_{M=0}^{M_{\text{max}}} P_{LM}T_{L}(\lambda^q_i)T_{M}(\lambda^q_j)
    \label{eq:cheby_sigmasq_at_query}
\end{equation}
where $P_{LM}$ are the projections of the $\sigma^2$ onto the Chebyshev polynomials and are given by
\begin{equation}
    P_{LM} = \sum_{l=0}^{L_{\text{max}}}\sum_{m=0}^{M_{\text{max}}} T_{L}(\lambda^{\alpha}_l)T_{M}(\lambda^{\alpha}_m)\sigma^2(\lambda^{\alpha}_l, \lambda^{\alpha}_m).
\end{equation}
The unknown coefficients $P_{LM}$ of this linear combination can be uniquely found by enforcing the interpolation criteria as $\sigma^2$ are known at $N$ interpolating nodes where $N = \prod_{i=1}^{d} N_i$ with $N_i$ representing nodes in $i$-th dimension and $d$ is the total number of dimensions. 

On the other hand, as the overlap integral has to be evaluated at an arbitrary point $(\vec \lambda^q, t_{c})$, it will turn out to be more convenient to interpolate it as a vector. In this case, a set of overlap-integral vectors $\vec{z}_\alpha$ are first constructed at the interpolation nodes, sampled on a uniform grid over $t_c$. The vectors $\vec{z}_\alpha$ have elements $z_\alpha[k] \equiv z(\vec \lambda^\alpha, k \, \Delta t)$,  where $\Delta t$ is the sampling interval, $k$ is an integer $\in {\text{int}} ([t_c^\ast \pm \tau]/\Delta t)$, $t_c^\ast$ represents the `reference' coalescence time as triggered by the search pipeline and $2\tau$ is the dimension of the sample-space (hyper-rectangle in $\vec \Lambda$) along the $t_{c}$ direction.
Since the bulk of the support for the posterior distribution comes from near the peak of these time series, we choose samples that are centered around the triggered value. From Eq.~\eqref{Eq:overlap}, it is clear that $\vec{z}_\alpha$'s can be constructed efficiently using FFT correlations. Once the set of vectors $\{ \vec{z}_\alpha \}$ is available, they can be projected over a suitable set of basis vectors $\{ \vec u_\mu \}$ with linear coefficients $\{C_\mu(\vec \lambda^\alpha)\}$ as the following:
\begin{equation}
	\vec z_\alpha  =
				\sum_{\mu = 1}^{N} C^{\alpha}_\mu  \, \vec {u}_{\mu} \, ,
\label{eq:SVD}
\end{equation}
\begin{figure}
    \centering
    \includegraphics[width=0.60\linewidth]{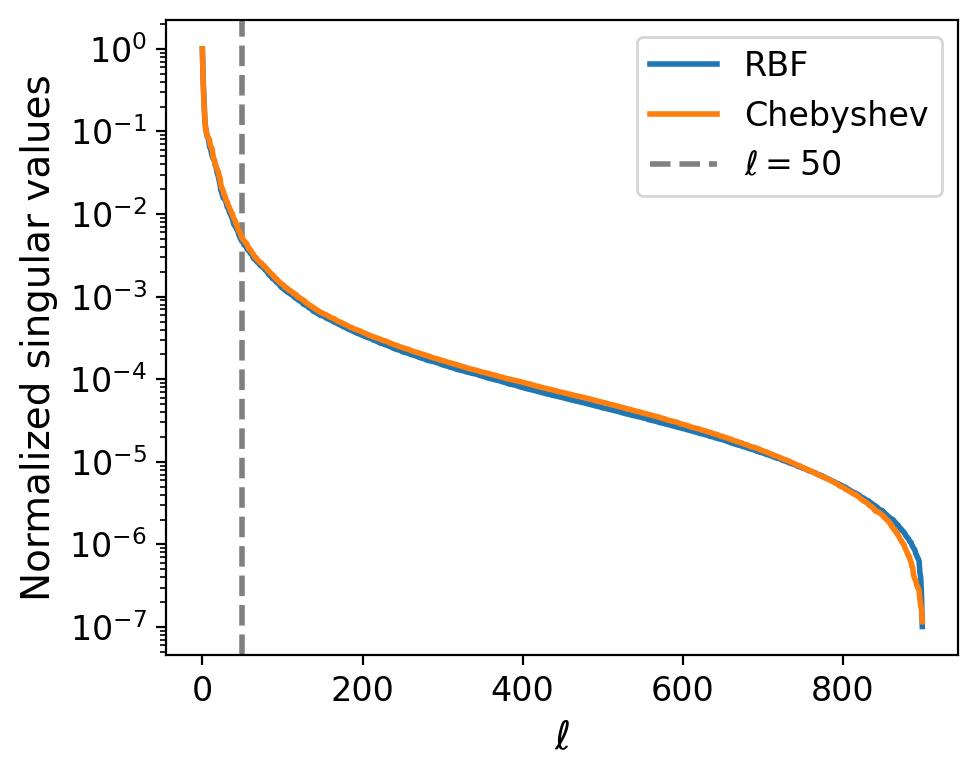}
    \caption{Normalized singular value profiles for both Chebyshev and Meshfree interpolation schemes. In both cases, $N=900$ nodes are chosen where Chebyshev nodes lie on a grid, and meshfree nodes are uniformly scattered across the sample space. The boundaries for the interpolating parameters are also the same for both cases.}
    \label{fig:singular_vals_cheby_vs_rbf}
\end{figure}
where $C^{\alpha}_{\mu}$ is the value of the SVD coefficient at $\vec \lambda^{\alpha}$ for the $\mu$-th basis vector $\vec u_{\mu}$. These basis vectors $\vec u_{\mu}$ can be obtained by taking a singular value decomposition (SVD) of the matrix obtained after row-wise stacking of $\vec z_{\alpha}$. These SVD coefficients appear in the above equation in the decreasing order of their relative importance as determined from the spectrum of singular values. Since there is a strong correlation between $\vec z_{\alpha}$, the overlap vectors lie in the span of the top-$\ell$ basis vectors ($\ell \ll N$), which implies that a vector $\vec z_{q}$ at a query point in the sample space can also be spanned by the same set of basis vectors (See Fig.~\ref{}). As shown in Fig.~\ref{fig:C_surf_heavy_cheby}, the SVD coefficients $C^{\alpha}_{\mu}$ are smoothly varying scalar fields over $\vec \lambda$ (sampled at the interpolation nodes), we can use Chebyshev polynomial to interpolate SVD coefficient corresponding to each basis vector $\vec u_{\mu}$ as the following:
\begin{equation}
    C^{q}_{\mu} = \sum_{L=0}^{L_{\text{max}}}\sum_{M=0}^{M_{\text{max}}} P_{LM\mu}T_{L}(\lambda^q_i)T_{M}(\lambda^q_j)
    \label{eq:cheby_svdcoeff_at_query}
\end{equation}
where coefficients $P_{LM\mu}$ are given by
\begin{equation}
    P_{LM\mu} = \sum_{l=0}^{L_{\text{max}}}\sum_{m=0}^{M_{\text{max}}} T_{L}(\lambda^{\alpha}_l)T_{M}(\lambda^{\alpha}_m)C_{\mu}(\lambda^{\alpha}_l, \lambda^{\alpha}_m).
\end{equation}
Now, these coefficients can be found by again imposing the interpolation criteria as they are known at $N$ interpolating nodes. Once we have the interpolants for both $\sigma^2(\vec \lambda^q)$ and $C^{q}_{\mu}$, we can evaluate them at the query points proposed by a sampler in the online stage.
\subsubsection{Online-stage}
\label{subsec:cheby_online_stage_chap2}
In the online stage, a stochastic sampler proposes the query points $\vec \lambda^q$ at which the likelihood is evaluated. In a standard (bruteforce) calculation of the likelihood, first, the template waveform is generated at the query point $\vec \lambda^q$ and then overlap integral, and template norm square are calculated. In the interpolation scheme, we have the interpolating functions which we are required to be evaluated at the query point $\vec \lambda^q$, which are significantly cheaper to evaluate in comparison to bruteforce calculation. Then we combine the $C^q_{\mu}$ with the basis vectors $\vec \mu$ in accordance with Eq.~\eqref{eq:SVD} noting that only the first few basis vectors are required to reconstruct $\vec z_{\alpha}$ within a small error. Note that the interpolating time-series at $\vec \lambda^q$ is uniformly sampled over the interval $[t^{*}_c \pm \tau]$. As such, it is possible that the query $t_c$ does not coincide with the discrete-time samples of $\vec{z}_q$. In such a case, we use a one-dimensional cubic-spline interpolation to evaluate ${z}(\vec \lambda^q, t_c)$ using a few `nearby' grid samples of $\vec z_q$ as input. This implies that $\vec z_q$ has to be reconstituted only at a few ($\sim 10$) consecutive sample points, which considerably accelerates the matrix-vector multiplication in Eq.~\eqref{eq:SVD}.

The interpolant for the square of the template norm can be directly evaluated to get the interpolated value ${\sigma^2(\vec \lambda^q)}$. Similarly, the $\ell$ interpolants for SVD coefficients (Eq.~\eqref{Eq:rbfcoeffs_chap3}) are evaluated to get a set of interpolated coefficients, which are then combined with the corresponding top-$\ell$ basis vectors $\vec{u}_\mu$ (see Eq.~\eqref{eq:SVD}) to obtain the interpolated overlap-integral $\vec{z}_q$. Combining the interpolated values $z(\vec \lambda^q, t_c)$ and ${\sigma^2(\vec \lambda^q)}$ with the extrinsic-parameter dependent complex amplitude ${\mathcal{A}}$ (see Eq.~\eqref{Eq:log-likelihood}), we finally obtain the log-likelihood ratio $\ln \mathcal{L}$ at the arbitrary point $(\vec \Lambda^q, t_{c})$. Now, let us discuss the meshfree interpolation scheme.

\subsection{Meshfree likelihood interpolation}
\label{sec:meshfree_likelihood_interpolation_chap2}
\begin{figure*}[!htp]
\begin{subfigure}{0.5\linewidth}
    \includegraphics[width=\linewidth]{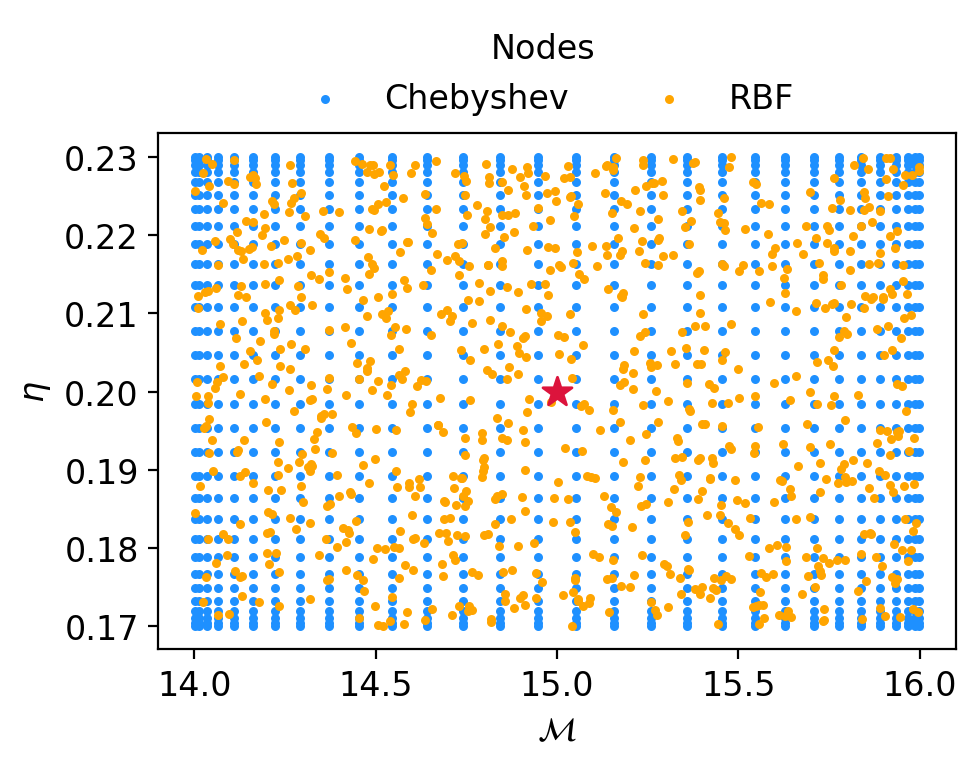}
    \caption{Nodes in $\mathcal{M}$ and $\eta$}
    \label{fig:cheby_vs_rbf_nodes_chirpeta}
\end{subfigure}\hfill
\begin{subfigure}{0.5\linewidth}
    \includegraphics[width=\linewidth]{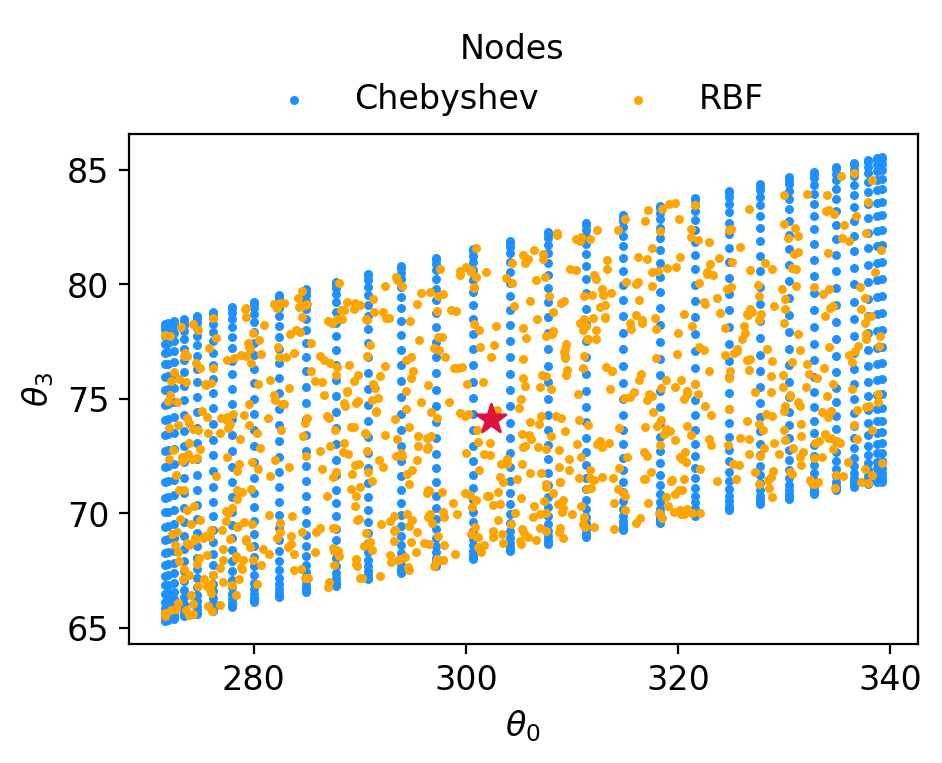}
    \caption{Nodes in $\theta_0$ and $\theta_3$}
    \label{fig:cheby_vs_rbf_nodes_t0t3}
\end{subfigure}
\caption{Chebyshev and meshfree nodes in the sample parameter space. The red star indicates the BBH injection. For Chebyshev interpolation, we use nodes in $\mathcal{M}$ and $\eta$ coordinates while for meshfree interpolation, we use $\theta_0$ and $\theta_3$ coordinates as our nodes since the metric varies slowly in $\theta_0$ and $\theta_3$ coordinates in comparison to $\mathcal{M}$ and $\eta$.} 
\label{fig:cheby_vs_rbf_nodes}
\end{figure*}

Similar to the grid-based interpolation scheme, the meshfree scheme can also be divided into two stages: (i) a preparatory, start-up stage where we explicitly determine the RBF interpolating functions (interpolants) from the pre-computed likelihood values at the interpolation nodes and (ii) an online stage where these interpolants are evaluated on the fly to `predict' the likelihood values at arbitrary query points in the sample space. Unlike the grid-based scheme (see Fig.~\ref{fig:cheby_vs_rbf_nodes}), the nodes are not laid on a specific grid. Instead, for the meshfree scheme, the nodes are randomly distributed around the most prominent trigger identified by the detection pipeline. The subsequent sections will elaborate on the two stages integral to this approach.

%
%
\subsubsection{Start-up stage}
\label{subsec:startup_chap2}

In this stage, the nodes are randomly sprayed over the sample space, and the meshfree interpolants are constructed. Similar to the grid-based scheme, we find a suitable set of basis vectors that span the space of $N$ input overlap vectors $\{\vec{z}_\alpha\}$ by stacking these vectors row-wise and performing the SVD of the resultant matrix as shown in the Eq.~\eqref{eq:SVD}. Again, a strong correlation between the $\vec{z}_\alpha$'s implies that the overlap vectors lie in the span of the top-$\ell$ basis vectors ($\ell \ll n$). A vector $\vec{z}_q$ at a query point in the sample space can also be spanned by the same set of basis vectors. Note that for a fixed index $\mu$, the coefficients $C^{\alpha}_\mu$ represent a surface whose values are known only at the input nodes. Along with Eq.~\eqref{eq:SVD}, this implies that the interpolation of the inner-product vector at an arbitrary query point essentially boils down to interpolating the value of the coefficients $C^q_{\mu} \equiv C_\mu(\vec \lambda^q)$. The meshfree interpolants for the SVD coefficient corresponding to the $\mu^{\text{th}}$ basis vector can be expressed~\cite{doi:10.1142/6437} as a linear combination of RBF kernels centered on the scattered, distinct nodes $\vec \lambda^\alpha$, augmented by monomials ranging up-to a specific order:
\begin{equation}
    C^{q}_{\mu} = \sum_{\alpha = 1}^{n} r_{\alpha} \, \phi(||\vec \lambda^q - \vec \lambda^\alpha||_2) \, 
        									+ \,\sum_{j = 1}^{M} b_{j} \, f_j(\vec \lambda^q) \, ,
\label{Eq:rbfcoeffs_chap3}
\end{equation} 
where $\phi$ is the Gaussian kernel (see the Appendix~\ref{appendix:diff_rbf_kernel} for various RBF kernels and their corresponding effect on the likelihood reconstruction accuracy) centered on $\vec \lambda^{\alpha}  \in \mathbb{R}^d$, and $\{f_j\}$'s are monomials that span the space of polynomials of some preset target degree $\nu$ in $d$ variables. In this analysis, we choose $\phi = \exp(-\epsilon\, ||\vec \lambda^q - \vec \lambda^\alpha||_2)$ where $\epsilon$ is the shape parameter which determines the extent of the Gaussian RBF.  Also, ${\boldsymbol{r} = [r_{1}, r_{2}, \ldots, r_{n}]^T}$ and ${\boldsymbol{b} = [b_{1}, b_{2}, \ldots, b_{M}]^T}$ are the set of $(n+M)$ coefficients that need to be uniquely determined to determine the interpolant. Since $C^{q}_{\mu}$ are known at the interpolation nodes, it allows us to enforce $n$ interpolation conditions. $M$ additional conditions $\sum_{k=1}^n r_k \,f_j(\vec \lambda^k) = 0, \, j=1, \ldots, M$ are added to ensure a unique solution. Together, these lead to a system of equations:
\begin{equation}
	\begin{bmatrix}
		\boldsymbol{K} & \boldsymbol{F} \\
		\boldsymbol{F}^{T} & \boldsymbol{O} 
	\end{bmatrix} \
		\begin{bmatrix}
		\boldsymbol{r} \\ 
		\boldsymbol{b}
	\end{bmatrix} \ 
	=  
	\begin{bmatrix}
		C^{\alpha}_{\mu} \\ 
		\boldsymbol{0} 
	\end{bmatrix}
\label{Eq:RBF-SLE}
\end{equation}
where the matrices $\boldsymbol{K}$ and $\boldsymbol{F}$ have components $K_{ij} = \phi(||\vec \lambda^{i}-\vec \lambda^{j}||_2)$ and $F_{ij} = f_j(\vec \lambda^{i})$ respectively; $\boldsymbol{O}_{M\times M}$ is a zero-matrix and $\boldsymbol{0}_{M\times 1}$ is a zero-vector. Eq.~\eqref{Eq:RBF-SLE} can be solved uniquely for the unknown coefficients $\boldsymbol{r}$ and $\boldsymbol{b}$, thus completely determining the meshfree interpolant in Eq.~\eqref{Eq:rbfcoeffs_chap3}. The solution for $\boldsymbol{r}$ and $\boldsymbol{b}$ can be shown to be unique if $\boldsymbol{F}$ has full column rank. A minimum set of ${n = \binom{\nu + d}{\nu}}$ interpolation nodes are required to be uniformly distributed over the $d$-dimensional intrinsic parameter space to determine the coefficients uniquely. Euclidean distances between pairs of points seem to work well, possibly due to the small volume of the sample space in a typical PE analysis. Parameter-space metric-based distances could also be used. A similar procedure is followed to create a separate meshfree interpolant for ${\sigma^2(\vec \lambda^q)}$.
Note that by construction, the meshfree interpolation eliminates both the factors that contribute to the high computational cost, namely, (a) generating the forward signal model and (b) explicitly calculating the overlap integral at every query point proposed by the sampling algorithm. 

\subsubsection{Online stage}
\label{subsec:online_chap2}
In this stage, the $(\ell+1)$ interpolants (prepared in the offline stage earlier) are evaluated on the fly at arbitrary query points $(\vec \lambda^q, t_c)$ proposed by the sampling algorithm. 
The interpolant for the square of the template norm can be directly evaluated to get the interpolated value ${\sigma^2(\vec \lambda^q)}$. Similarly, the $\ell$ interpolants for SVD coefficients (Eq.~\eqref{Eq:rbfcoeffs_chap3}) are evaluated to get a set of interpolated coefficients, which are then combined with the corresponding top-$\ell$ basis vectors $\vec{u}_\mu$ (see Eq.~\eqref{eq:SVD}) to obtain the interpolated overlap-integral $\vec{z}_q$. As mentioned earlier in the Sec.~\ref{subsec:cheby_online_stage_chap2}, the $\vec{z}_q$ is known at the discrete values of times in the interval $[t_c^\ast \pm \tau]$ and a query $t_c$ may not coincide with those discrete time samples. So we again generate a Cubic spline interpolant of $\vec z_{q}$ centered around query $t_c$, implying that we only need to reconstruct $\vec z_q$ at a few sample times ($\sim 10$) greatly accelerating the matrix multiplication process in Eq.~\eqref{eq:SVD}. Finally, we combine the interpolated values of $z(\vec \lambda^q, t_c)$ and ${\sigma^2(\vec \lambda^q)}$ with extrinsic-parameter dependent complex amplitude ${\mathcal{A}}$ (see Eq.~\eqref{Eq:log-likelihood}) and obtain the log-likelihood ratio $\ln \mathcal{L}$ at the arbitrary point $(\vec \Lambda^q, t_{c})$.
\section{Numerical experiments}
\label{sec:numerical_experiments_chap2}
\subsection{2D likelihood interpolation}
\begin{figure*}[!htp]
\begin{subfigure}{0.5\linewidth}
    \includegraphics[width=\linewidth]{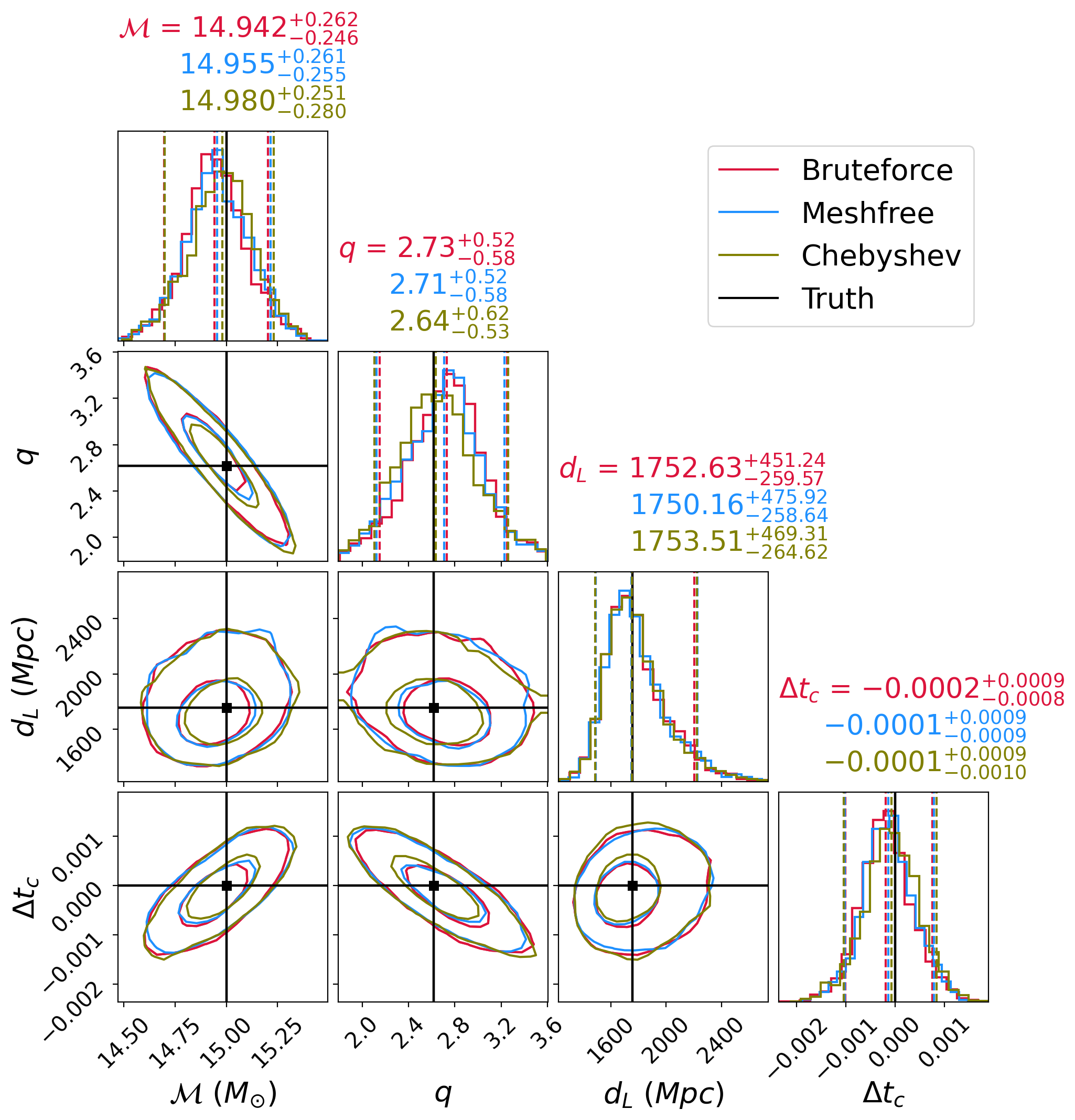}
    \caption{Marginalized posterior distributions for BBH system}
    \label{fig:pe_cheby_vs_rbf_brute_heavy}
\end{subfigure}\hfill
\begin{subfigure}{0.5\linewidth}
    \includegraphics[width=\linewidth]{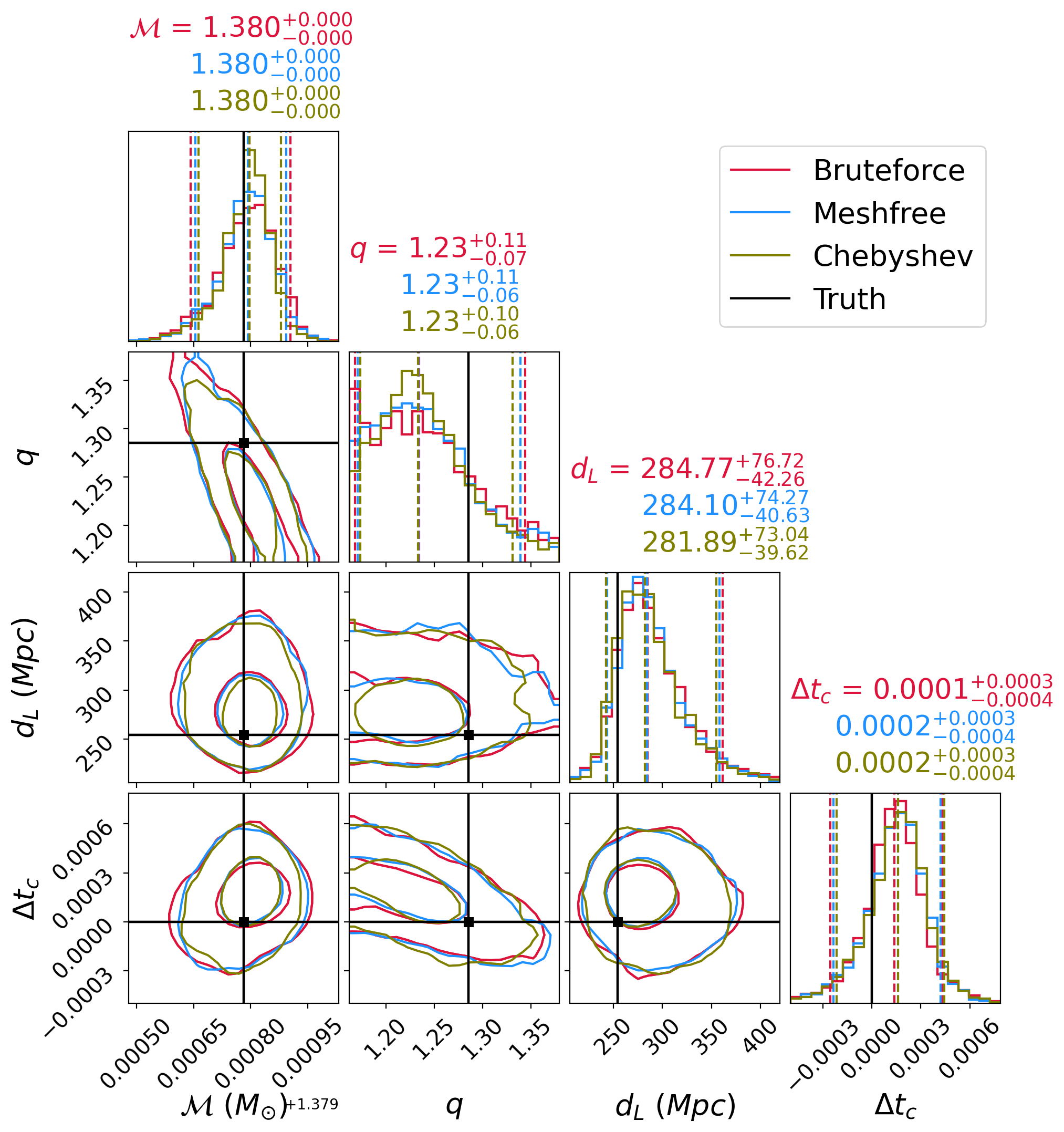}
    \caption{Marginalized posterior distributions for BNS system}
    \label{fig:pe_cheby_vs_rbf_brute_light}
\end{subfigure}
\caption{Posterior distributions as obtained by performing PE using bruteforce, Chebyshev, and meshfree likelihood interpolation schemes. Note that the chosen priors over the intrinsic parameter space are narrow which is currently a limitation of our meshfree method. We are currently working on this limitation by employing different stragies to widen the prior widths and will be presented in the future versions of our method.} 
\label{fig:posteriors_cheby_vs_rbf_brute}
\end{figure*}
To demonstrate the accuracy and efficiency of both meshfree and grid-based interpolation methods in reconstructing source parameters, we generated simulated data for two non-spinning compact binary systems: (i) a Binary Black Hole (BBH) system and (ii) a Binary Neutron Star (BNS) system. The details of the injections and the simulated data are provided in the table below. Both signals were injected into the \texttt{aLIGOZeroDetHighPower} noise, simulating a configuration of aLIGO design sensitivity for a single detector, namely 'L1,' such that the matched-filter (SNR) is approximately ${\sim 10}$.
\begin{table}[hbt]
\centering
\begin{tabular}{c|c|c|c|c|c|rc}
\hline\hline
& $ \mathcal{M}/M_{\odot}$ & $\eta$ & $d_L$ (Mpc) &  $t_c$ & duration (secs) & SNR \\ \hline
BBH   & $15$ & $0.20$ & $1758$ & $1126258006$ & 8 & 10 \\ \hline
BNS   & $1.379$ & $0.246$ & $254$ & $1126258250$ & 256 & 10 \\
\hline\hline
\end{tabular}
\caption{Injection parameters of the simulated GW events. For both cases, we chose $\alpha = \delta = \iota = \pi/4$, and $\phi_c = \psi = 0$. The seismic cutoff frequency was chosen to be $20$ Hz, whereas the higher cutoff frequencies were $1024$ Hz and $1500$ Hz for BBH and BNS systems, respectively.}
\label{tab:injection_params}
\end{table}
We performed Bayesian inference on these simulated data using both (a) {\emph{bruteforce}} likelihood calculation used in PyCBC inference and (b) and by using both grid-based interpolation and the proposed meshfree likelihood interpolation scheme outlined in earlier sections. We used publicly available software~\cite{RBF_github} for radial basis functions and the {\texttt{Dynesty}}~\cite{speagle2020dynesty, sergey_koposov_2023_7600689} nested-sampling package for carrying out the Bayesian inference analysis. We varied two intrinsic parameters (component masses) and two extrinsic parameters (luminosity distance and coalescence time), keeping other parameters fixed. For the grid-based scheme, we opted for a Chebyshev grid of dimensions $30 \times 30$ in $\mathcal{M}$ and $\eta$. In contrast, for the meshfree method, we distributed uniformly random nodes across the same boundaries as in the Chebyshev case, as illustrated in Fig.~\ref{fig:cheby_vs_rbf_nodes}. In both approaches, we utilized a total of $N=900$ interpolating nodes. In the meshfree case, we selected $\phi = \exp(-\epsilon \, r^2)$ as the RBF kernel and employed $\theta_0$ and $\theta_3$ as interpolating nodes, derived straightforwardly from the corresponding nodes in $\mathcal{M}$ and $\eta$ coordinates. In both scenarios, the top $\ell = 50$ basis vectors were employed. For the BBH case, we set $\nu = 1$ and $0$ for $\sigma^2$ and $C^{q}_{\mu}$ interpolants, respectively. For the BNS system, $\nu = 2$ and $0$ were chosen for $\sigma^2$ and $C^{q}_{\mu}$ interpolants, respectively. Another pivotal parameter influencing the accuracy of the interpolating functions in the meshfree method is $\epsilon$, determining the spread of the Gaussian at each node. In the BBH case, we employed $\epsilon = 0.75$ and $0.5$ for $\sigma^2$ and $C^{q}_{\mu}$ interpolants, respectively. For the BNS system, $\epsilon = 0.5$ and $0.6$ were selected for $\sigma^2$ and $C^{q}_{\mu}$, respectively. We found an absolute error of $\sim 10^{-2}$ across the sample space in approximating the log-likelihood function for both schemes. As depicted in Fig.~\ref{fig:posteriors_cheby_vs_rbf_brute}, the posterior distributions of source parameters using both Chebyshev and meshfree methods align well with the bruteforce method.
\begin{figure}
    \centering
    \includegraphics[width=0.55\linewidth]{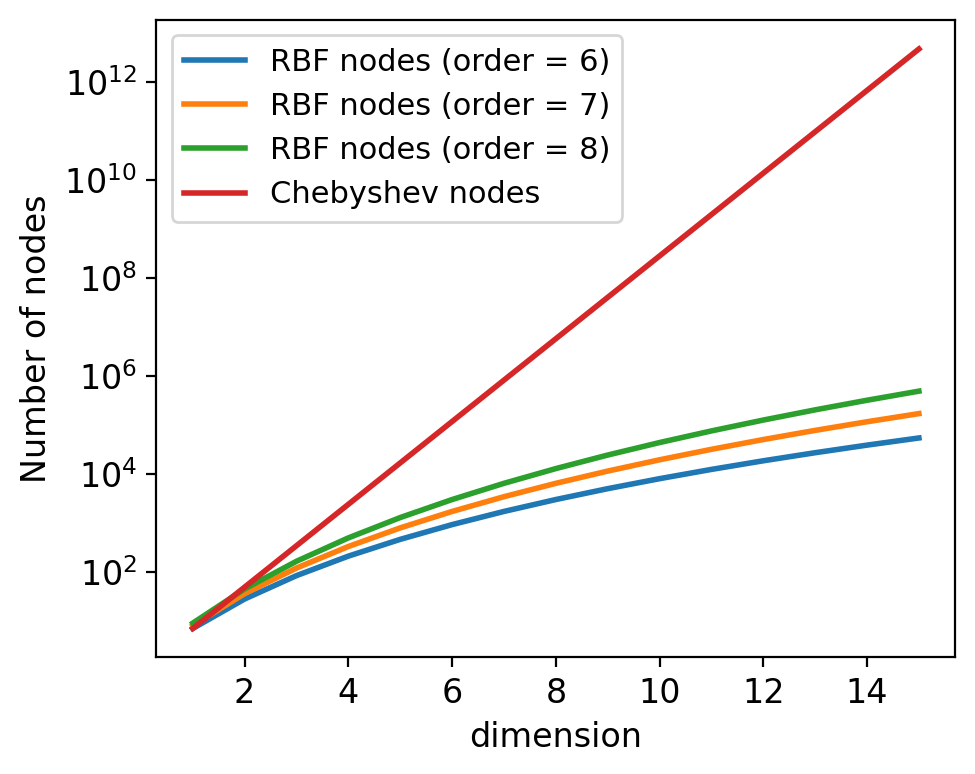}
    \caption{Scaling of nodes with the number of dimensions. The number of Chebyshev nodes increases exponentially in stark contrast to those of meshfree nodes.}
    \label{fig:nodes_scaling_cheby_rbf}
\end{figure}
It's worth noting that for higher dimensions, the grid-based scheme requires an exponentially increasing number of Chebyshev nodes compared to meshfree nodes, as illustrated in Fig.~\ref{fig:nodes_scaling_cheby_rbf}. Deploying a grid-based scheme for likelihood interpolation becomes impractical due to the escalating computational expense associated with generating interpolants and evaluating likelihoods. In contrast, the meshfree method, by its design, is independent of any grid and thus suitable for higher-dimensional problems. In the subsequent subsection, we illustrate an example involving a simulated BNS system and estimate its source parameters using our proposed meshfree method, where the dimensionality of the interpolating parameters is four. 
\subsection{4D likelihood interpolation}
To demonstrate the accuracy and speed of the meshfree method in reconstructing the source parameters, we prepared synthetic $360$ seconds long data $\boldsymbol{d}$, sampled at $4096$ Hz. For this, a simulated GW signal $h$ from a canonical BNS system with component masses $m_{1, 2} = 1.4 \, M_{\odot}$ and mass-weighted effective dimensionless spin $\chi_{\text{eff}} = 0.05$ was injected (using the {\texttt{IMRPhenomD}}~\cite{khan2016frequency} signal model) in colored Gaussian noise using the noise power spectral density model~\cite{aLIGO_ZDHP} of {\texttt{aLIGO}} detectors. 
The distance to the source was adjusted for a moderate matched-filtering SNR of $10$.
The seismic cut-off frequency was chosen to be $20$ Hz to mimic data from the upcoming O4 science run.

\begin{table}[hbt]
\centering
\def\arraystretch{1.47}
\begin{tabular}{c|c|c|c|rc|}
\hline\hline
& $ \mathcal{M}/M_{\odot}$ & $\eta$ & $\chi\strut_{\text{eff}}$ &  $\text{SNR}$\\ \hline
Injection   & $1.2187$ & $0.25$ & $0.05$ & $10.00$  \\ \hline
Standard    & $1.2187\;\strut^{1.2188}_{1.2185}$ & $0.249\;\strut^{0.250}_{0.246}$ & $0.050\;\strut^{0.0507}_{0.0493}$ & $9.67$  \\ \hline
Meshfree    & $1.2187\;\strut^{1.2185}_{1.2185}$ & $0.249\;\strut^{0.250}_{0.246}$ & $0.050\;\strut^{0.0507}_{0.0493}$ & $9.67$  \\
\hline\hline
\end{tabular}
\caption{Reconstruction of a canonical BNS event.}
\label{tab:Inj_details}
\end{table}
\begin{figure*}
    \centering
    \includegraphics[width=1\linewidth]{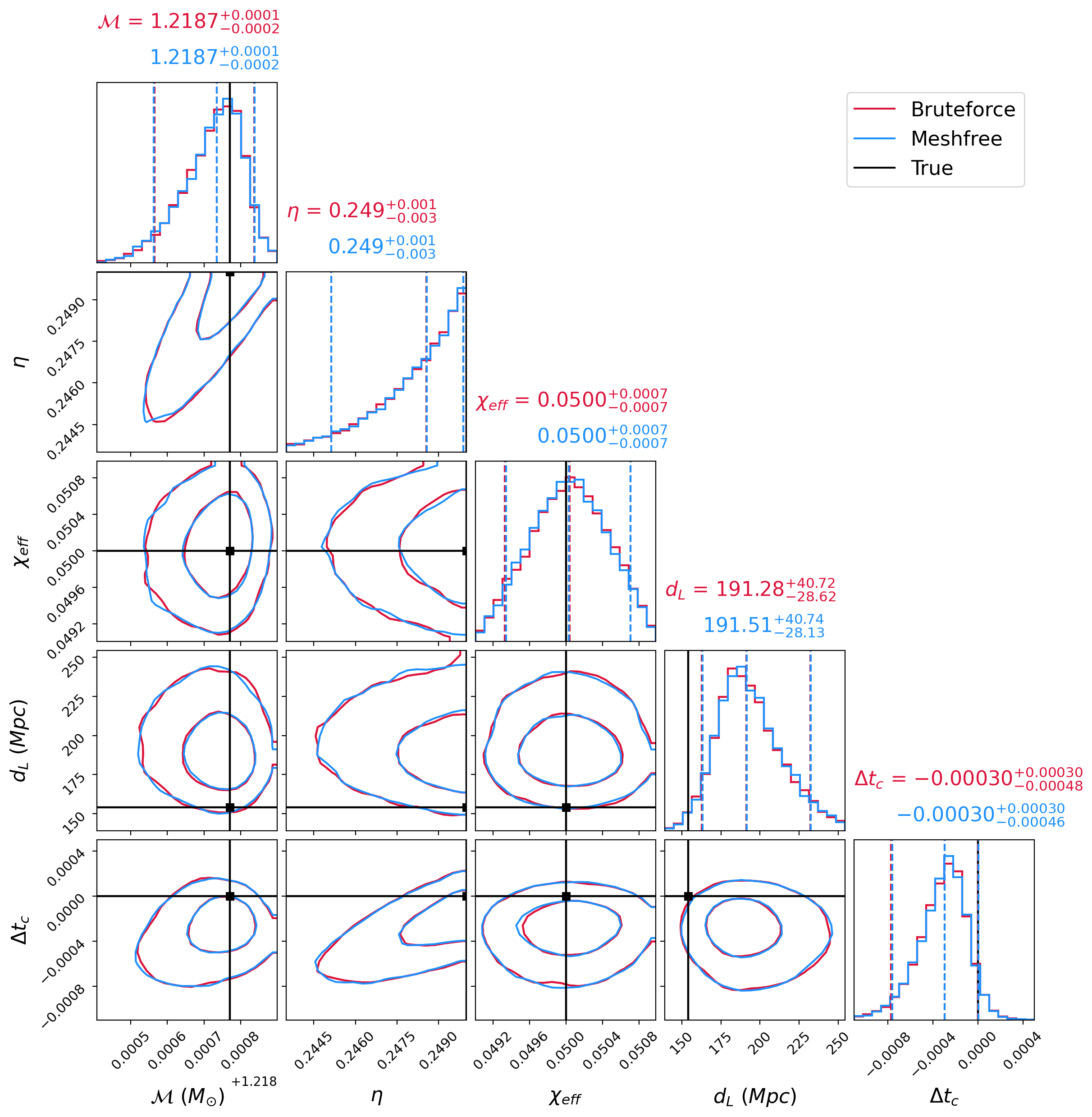}
    \caption{Marginalized PDF for chirp mass $\mathcal{M}$, symmetric mass ratio $\eta$, and effective spin $\chi_{eff}$, $d_L$, and $\Delta \text{t}_{c}$ parameters of a simulated BNS event at a seismic cutoff of $20$ Hz. The injection parameters are shown as black lines. The $50\%$ and $90\%$ contours for meshfree method (blue) and Burteforce (red) are also shown. The plot-overlaid marginalized PDF obtained from the proposed meshfree method (blue) and Bruteforce (red) is virtually indistinguishable.}
    \label{fig:4d_like_interp_corner_bruteforce_vs_rbf}
\end{figure*}
\begin{figure*}
    \centering
    \includegraphics[width=1\linewidth]{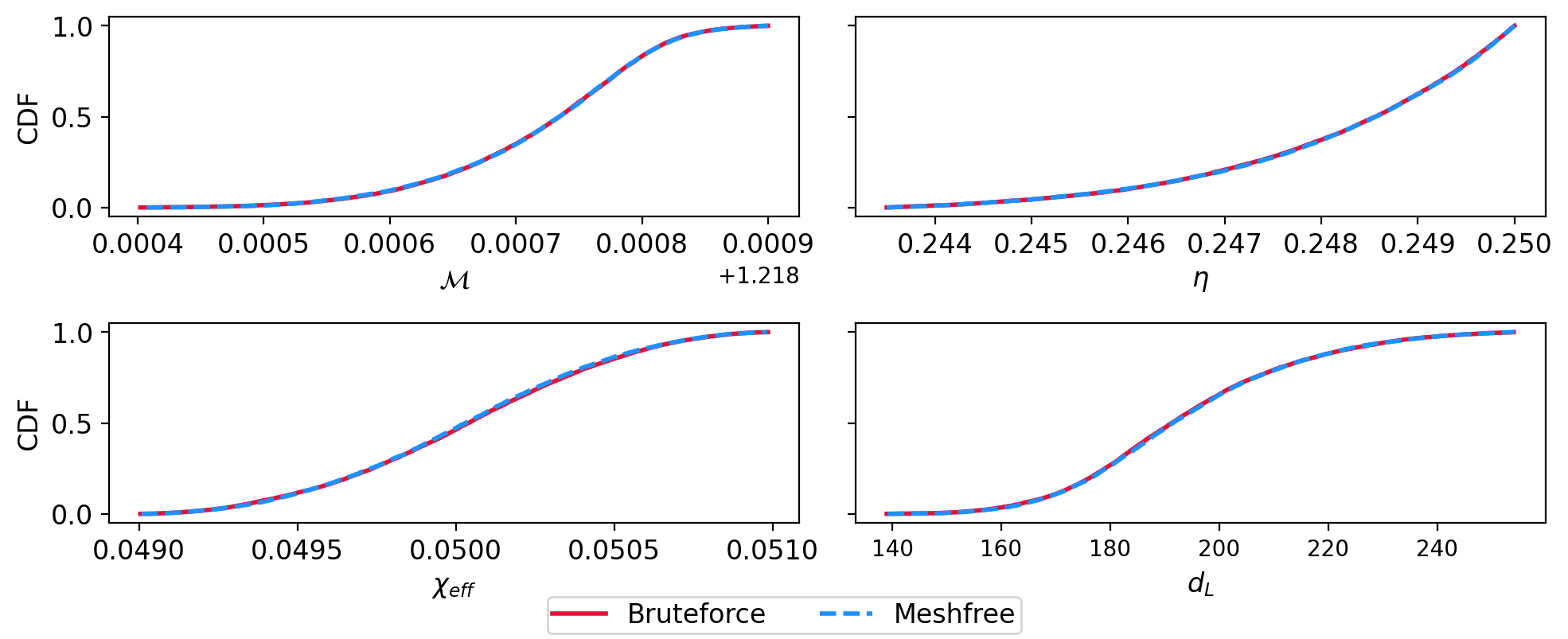}
    \caption{The cumulative distribution functions (CDFs) of $\mathcal{M}$, $\eta$, and $\chi_{eff}$, $d_L$ using both meshfree (dashed blue) and bruteforce (red) method are shown and are in good agreement with each other.}
    \label{fig:4d_like_interp_corner_bruteforce_vs_rbf_cdfs}
\end{figure*}
We performed Bayesian inference on this simulated data using both (a) {\emph{direct}} likelihood calculation used in PyCBC inference and (b) and by using the proposed meshfree likelihood interpolation scheme outlined in earlier sections. We used publicly available software~\cite{RBF_github} for radial basis functions and the {\texttt{Dynesty}}~\cite{speagle2020dynesty, sergey_koposov_2023_7600689} nested-sampling package for carrying out the Bayesian inference analysis.
We varied four intrinsic parameters (component masses and aligned-spin magnitudes) and two extrinsic parameters (luminosity distance and coalescence time), keeping other parameters fixed.

For this exercise, we used $n = 800$ random input nodes over $\vec \lambda$. The top $\ell = 120$ basis vectors and a polynomial order $\nu = 6$ with a corresponding nominal median relative error $\sim 10^{-5}$ across the sample space in approximating the log-likelihood function.
The accuracy trade-offs of likelihood reconstruction by varying basis size is shown in the Appendix~\ref{appendix:accuracy_trade_off}.

The meshfree parameter reconstruction was completed in $5.3$~min in comparison to $31.7$~h taken by the direct calculation. The likelihood function was evaluated $676$ times faster using the meshfree method. Some of the estimated parameters have been compared in Table~\ref{tab:Inj_details}, which show identical values obtained by both methods. The marginalized PDF over five parameters $\mathcal{M}$, $\eta$, $\chi_{eff}$, $d_L$, and $\Delta t_{c}$ are shown in the corner plot Fig.~\ref{fig:4d_like_interp_corner_bruteforce_vs_rbf}. The Fig.~\ref{fig:4d_like_interp_corner_bruteforce_vs_rbf_cdfs} shows cumulative density (CDF) profiles of these distributions obtained from both Bruteforce and the meshfree method plotted together. While both the PDF and CDF profiles look virtually indistinguishable, we also calculate statistical measures of similarity~\cite{stats_book} using the PDFs obtained from the two methods for further validation: both the Kolmogorov-Smirnov statistic ($0.0130$) and the Bhattacharyya distance ($0.0006$) between the chirp-mass PDF profiles support the fact that the two distributions are nearly identical. We get similar results for posterior distributions of other parameters.

This numerical example shows that the meshfree method can generate a statistically indistinguishable replica of the posterior distributions in a GW Bayesian inference problem at a small fraction of the total computational cost.

\section{Speed-up analysis}
\label{sec:speed-up_chap2}

We calculated the ratio of (average) time taken to compute the log-likelihood by these techniques at a fixed accuracy of reconstruction. Several simulated data sets were used in this study, generated by injecting signals having different parameters in colored Gaussian noise using the {\texttt{aLIGO}} noise model at a fixed matched-filtering SNR of $10$.  Seismic cutoff frequencies at $20$ Hz ($10$ Hz) were considered to mimic data from upcoming O4 (O5) science runs. 

\begin{figure}[!hbt]
    \centering
    \includegraphics[width=0.55\linewidth]{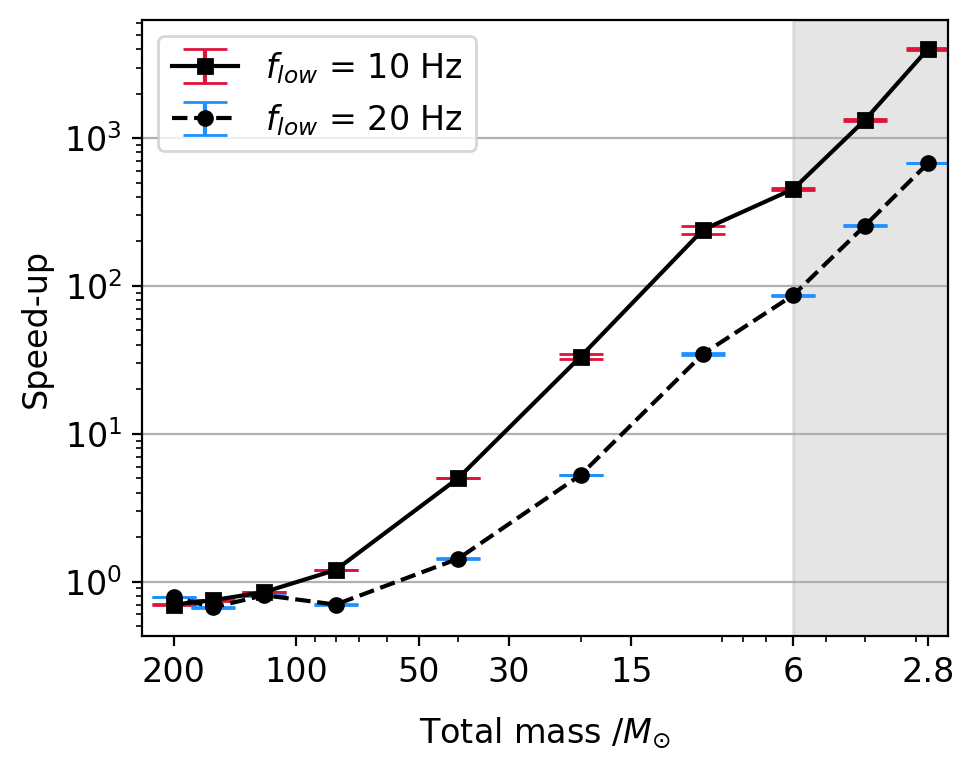}
    \caption{Speed-up factors (averaged over several thousand evaluations) in calculating the log-likelihood function with the meshfree method for different equal-mass ($q=1$) systems with (dimensionless) spin magnitude of $0.05 \, (0.2)$ for BNS (BBH) systems. The shaded region denotes BNS systems. The combination of $(n, \ell, \nu)$ parameters were tuned for a median error of $\sim 10^{-5}$.}
    \label{fig:speed_up_plot}
\end{figure}
\begin{table}[!hbt]
\centering
\def\arraystretch{1.075}
\begin{tabular}{p{0.25\columnwidth}|c|c|c|rc}
\hline\hline
seismic cut-off 
	& \multicolumn{1}{c|}{$M /M_{\odot}$} 
	& \multicolumn{1}{c|}{$t_\text{mf}$ /ms} 
	& \multicolumn{1}{c|}{$t_\text{pycbc}$ /ms} 
	& \multicolumn{1}{c}{speed-up} \\ 
\hline
\multirow{3}{*}{$f_{\text{low}} = 10\,\text{Hz}$}
	& 2.8 & 0.90 & 3604.76 & 4005.2 \\
	& 4.0 & 1.06 & 1405.84 & 1326.2 \\
	& 20.0 & 0.88 & 29.27 & 33.2 \\ 
\hline
\multirow{3}{*}{$f_{\text{low}} = 20\,\text{Hz}$}
	& 2.8 & 0.67 & 452.93 & 676.0 \\
	& 4.0 & 0.81 & 207.35 & 256.0 \\ 
	& 20.0 & 0.63 & 3.32 & 5.3 \\
\hline\hline
\end{tabular}
\caption{\label{tab:speed_up}
The first and second columns define the seismic cut-off frequency and total mass of the injected GW signal, respectively. The third and fourth columns show the median time (in ms) taken for a single evaluation of the log-likelihood function using the standard PyCBC method and our approach at a nominal relative error of $\mathcal{O}(10^{-5})$. The last column provides the relative speed-up of our method in comparison with the standard likelihood calculation. Median time (in ms) taken for a single evaluation of the log-likelihood function using the standard PyCBC method and the meshfree approach at a nominal relative error of $\sim 10^{-5}$. 
}
\end{table}

There is an obvious trade-off between accuracy and speed-up of the meshfree method, which are determined by the choice of $(n, \ell, \nu)$ parameters. Larger values can lead to more accurate likelihood estimates, albeit at a higher computational cost and vice-versa. 
We used a heuristic combination to guarantee median (relative) errors $\lesssim 9 \times 10^{-5}$ in the estimated log-likelihood values across the entire sample space.  The log-likelihood function was evaluated and timed for a large number of random points uniformly distributed over the $(\vec \lambda, t_{c})$ space. 

Table~\ref{tab:speed_up} summarizes the speed-ups corresponding to the two different seismic cutoff frequencies for three compact binary systems with equal component masses. It is further elucidated in Fig.~\ref{fig:speed_up_plot}, where the speed-up comparison is drawn between a larger number of equal-mass binary systems covering a wider range of parameters. From Table~\ref{tab:speed_up} and Fig.~\ref{fig:speed_up_plot}, it is clear that the likelihood computation can be sped up $\sim 4000$ times faster for canonical BNS systems at a nominal error of $\sim 10^{-5}$ using the meshfree method against standard likelihood implementation in PyCBC inference. However, this would not reflect the true benefit of our proposed method, as PyCBC inference is not optimized to calculate the fast posterior distribution for LIGO. Our method computes the posterior within a few minutes for the BNS system. Thus, our proposed scheme has the potential to perform rapid reconstruction of source parameters in upcoming observation runs for {\texttt{aLIGO}} detectors similar to the other existing optimized methods. The low-mass systems with a large number of in-band cycles would benefit most from the meshfree method. In contrast to the standard method, the time taken by the meshfree method is relatively unaffected by the chirp time of the signals. All the tests were performed on a single-core AMD EPYC 7542 CPU@2.90GHz CPU. \\

\section{Conclusion and Outlook}
\label{sec:conclusion_chap2}

We have presented an alternative approach to the grid-based method~\cite{smith2014rapidly}, which is computationally efficient for accurately evaluating the log-likelihood function. 
Our meshfree method can be easily integrated into well-known sampling algorithms (e.g., MCMC, Nested Sampling) to substantially accelerate the Bayesian inference of source parameters of coalescing compact binary sources, triggering prompt observation of their EM counterparts in the future. 
Using synthetically generated data of a GW signal from a merging BNS system, we have demonstrated that the posterior distributions are statistically identical to those obtained using the standard PyCBC inference ~\cite{biwer2019pycbc}. 
For BNS systems, the likelihood function can be calculated $\sim 4000$ times faster at any point proposed by the sampling algorithm than the direct likelihood calculation implemented in PyCBC inference.
At this point, we want to remind the readers that we made the comparison of our scheme against PyCBC inference only to verify the robustness of our scheme. PyCBC inference is not used for the fast PE run by LVK. Therefore, for a fair speed-up comparison, we must compare our methods against those schemes ~\cite{Finstad_2020, Dax2021, gabbard2022bayesian, canizares2015accelerated, Qi_2021, Soichiro_2020, https://doi.org/10.48550/arxiv.1805.10457, cornish2021heterodyned, Venumadhav2018} computing the real-time posterior distribution. However, the comparison with the optimized schemes is beyond the scope of the current work. In the follow-up works, we will compare our method against those schemes in detail.
Numerical experiments with a coherent, multi-detector implementation of the meshfree algorithm suggest that we can solve the coherent BNS PE problem (over a ten-dimensional parameter space) in $\sim 164$ seconds using $64$ CPU cores.
Further optimizations and runs over the full parameter space are underway. These details are available in a follow-up paper~\cite{pathak2023prompt}. 
It may be prudent to incorporate this method with the low-latency {\texttt{GstLAL}}~\cite{messick2017analysis} search pipeline for rapid, automated follow-ups of the detected events since both use the idea of dimension reduction using SVD.
In the current implementation of this algorithm, we need to create meshfree RBF interpolants from scratch for each new event triggered by the search pipelines. However, this task is embarrassingly parallel and can benefit from multiple CPU cores to expedite the preparatory stage. Future refinements could involve using the Fisher matrix~\cite{vallisneri2008use, pankow2015novel} as a guide to identifying the sample space volume and using sophisticated interpolation node distribution algorithms~\cite{Manca_2010}. Finally, the techniques of dimension reduction and meshfree approximation could be applied to situations where the likelihood function varies smoothly over the sample space. It is thus possible to adapt this idea to Bayesian inference in fields as diverse as cosmology, biochemical kinetic processes, and systems biology. 

In the next chapter, we discuss the multidetector extension of this method to enable sky-localization of sources along with other angles as well.

\section*{Acknowledgements}
We thank M.~K.~Haris for various suggestions and for his help at the conceptual stages of this work. We thank M.~Vallisneri and T.~Dent for carefully reading the manuscript and for offering several comments and suggestions to improve the presentation and content of the paper. We also thank other LSC colleagues, S.~Mitra, A.~Ganguly, and S.~Kapadia, for their useful comments. L.~P. is supported by the Research Scholarship Program of Tata Consultancy Services (TCS). A.~R is supported by the research program of the Netherlands Organisation for Scientific Research (NWO). A.~S. gratefully acknowledges the generous grant provided by the Department of Science and Technology, India, through the DST-ICPS cluster project funding. We would like to thank the HPC support staff at IIT Gandhinagar for their help and cooperation. 
%
%
\section{Appendix}
\subsection{Accuracy trade-off: Effect of varying basis size}
\label{appendix:accuracy_trade_off}
We present the results of a simulation to demonstrate the accuracy of the mesh-free method on the basis of size. We inject a GW signal from a compact binary system with component masses ($1.4 M_{\odot}$, $1.4 M_{\odot}$) in simulated noise generated using the advanced LIGO design sensitivity curve with a lower seismic cut-off frequency of $10$ Hz. On the one hand, we compute the log-likelihood-ratio (LLR) using the standard functions available in the PyCBC package (considered to be the ‘ground truth’ for this simulation) and compare it with the approximate value using the mesh-free interpolation method as outlined in our manuscript. 
\begin{figure*}[!htp]
\begin{subfigure}{0.5\linewidth}
    \includegraphics[width=\linewidth]{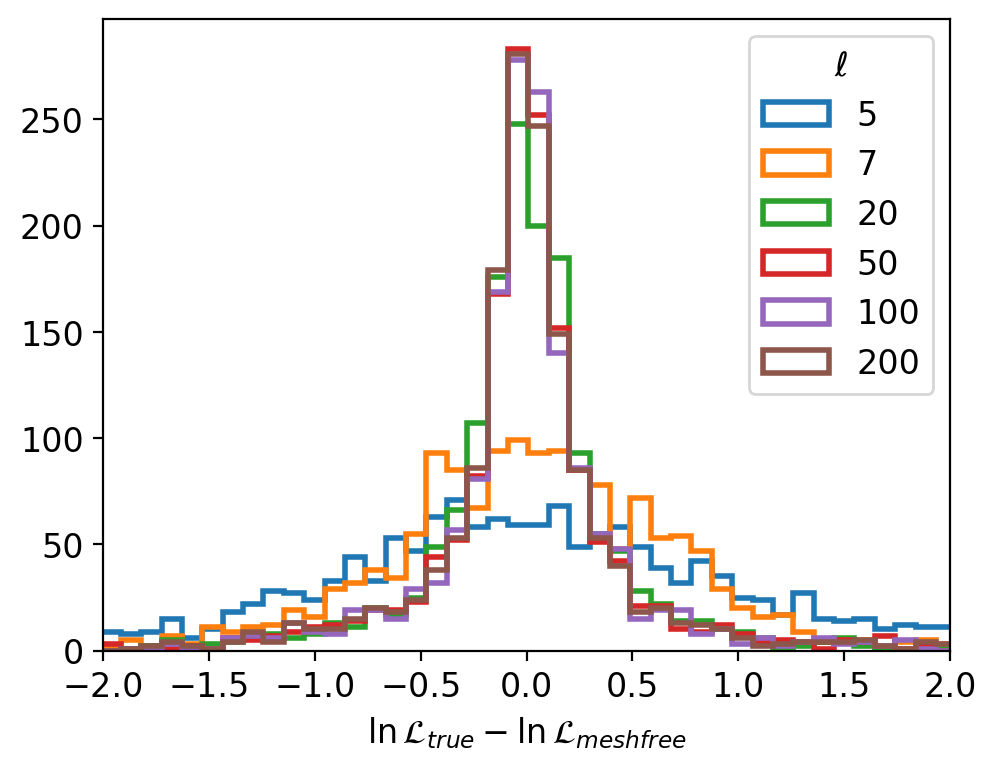}
    \caption{PDFs over log-likelihood errors}
    \label{fig:pdf_log_like_err}
\end{subfigure}\hfill
\begin{subfigure}{0.5\linewidth}
    \includegraphics[width=\linewidth]{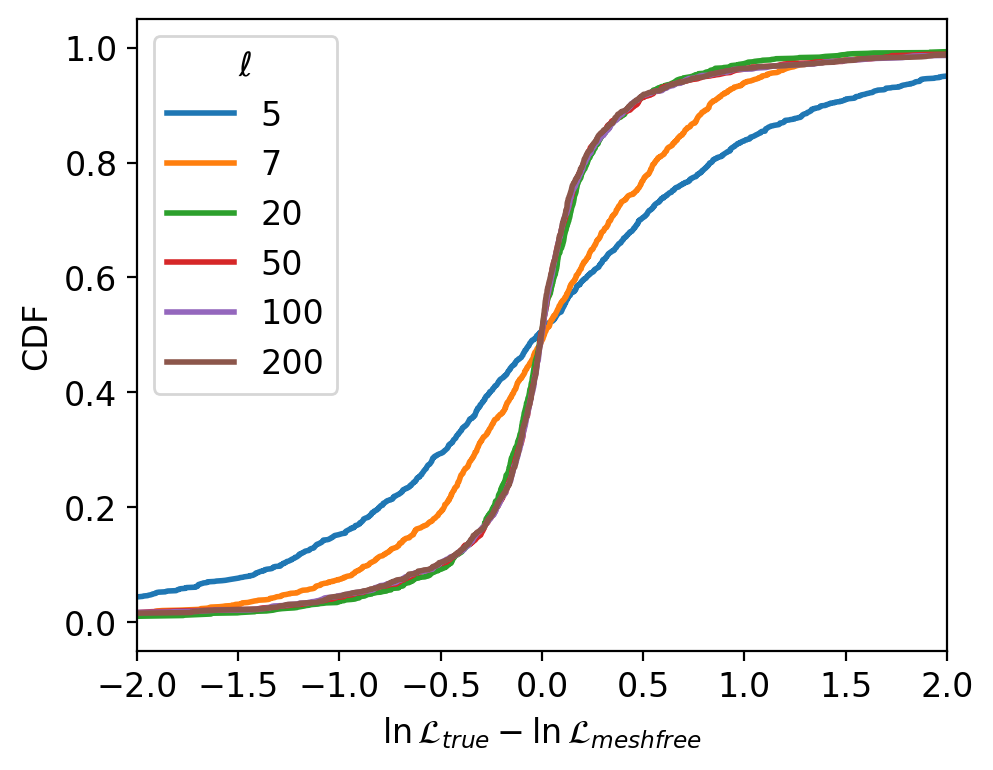}
    \caption{CDFs of the corresponding PDFs over log-likelihood errors}
    \label{fig:cdf_log_like_err}
\end{subfigure}
\caption{Chebyshev and meshfree nodes in the sample parameter space. The red star indicates the BBH injection. For Chebyshev interpolation, we use nodes in $\mathcal{M}$ and $\eta$ coordinates while for meshfree interpolation, we use $\theta_0$ and $\theta_3$ coordinates as our nodes since the metric varies slowly in $\theta_0$ and $\theta_3$ coordinates in comparison to $\mathcal{M}$ and $\eta$.} 
\label{fig:log_like_err_dist}
\end{figure*}
The meshfree method uses $10^{3}$ randomly chosen initial nodes over the sample space. We vary the number of the basis vectors for reconstructing the likelihood and estimate the difference between interpolated likelihood and the ‘ground-truth’ likelihood value. 
For these simulations, we observe that the top $20$ basis vectors are sufficient to approximate the likelihood with sufficient accuracy. However, we demonstrated the absolute error between the true and approximated values for $5$, $7$, $20$, $50$, $100$, and $200$ basis vectors, respectively. Fig.~\ref{fig:pdf_log_like_err} shows the probability distribution function (PDF) of the difference between true and estimated likelihood, and Fig.~\ref{fig:cdf_log_like_err} shows the corresponding cumulative distribution function (CDF). It is clear from these figures that as we increase the number of basis vectors, the error distribution becomes more concentrated around zero as compared to the relatively flattened distribution for smaller basis sizes.

\subsection{Scaling of the speed-up factor with waveform duration}
\label{appendix:scaling_of_speed_up}
The meshfree likelihood evaluation has no dependence on the length of the waveform. On the other hand, in the standard method, a significant time is spent on the waveform generation, followed by the evaluation of the likelihood integral. The latter depends on the length of the waveform. As shown in Table II, the speed-up ratios decrease with higher masses (shorter waveforms). 
\begin{figure}[!hbt]
    \centering
    \includegraphics[width=0.55\linewidth]{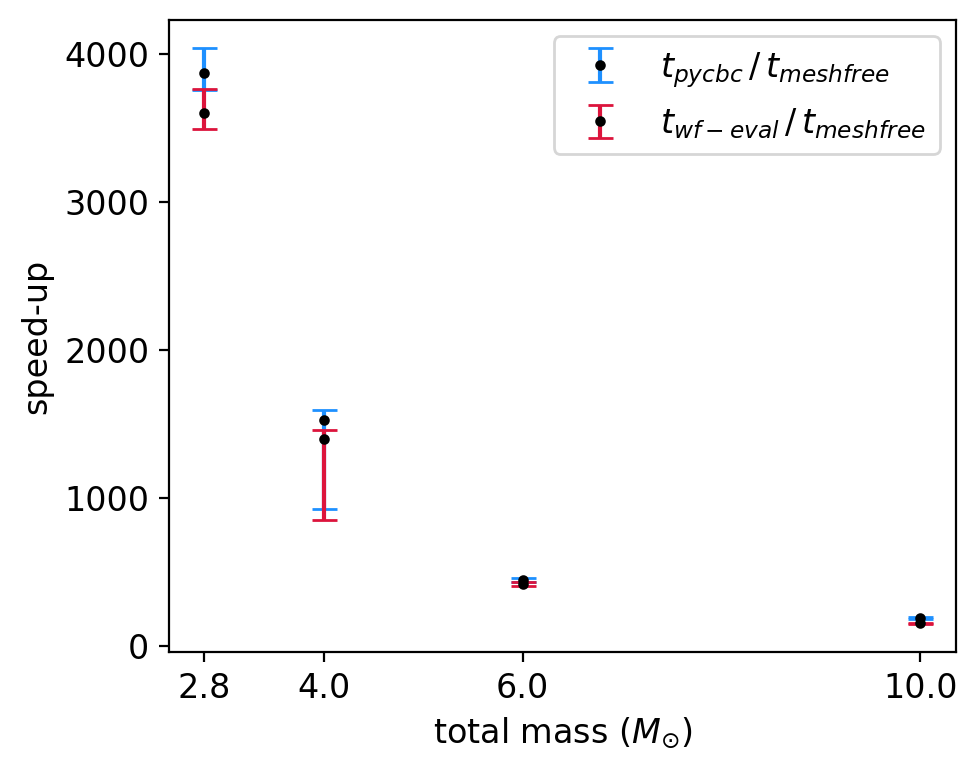}
    \caption{The relative speed-up between the likelihood evaluation using PyCBC and the proposed meshfree method is shown (blue). Also, The ratio of overall time for generating waveforms (total mass: $2.8 M_{\odot} - 10 M_{\odot}$) using PyCBC and the likelihood evaluation via meshfree scheme has been compared (red). The dominant cost in the traditional likelihood calculation arises from the generation of long-duration waveforms.}
    \label{fig:speed-up-waveform}
\end{figure}
To verify this, we calculated the likelihood evaluation time using both bruteforce ($\text{t}_{\text{pycbc}}$) and meshfree method ($\text{t}_{\text{rbf}}$) and also evaluated the waveform generation time (${\text{t}_{\text{wf}}}_{\text{eval}}$) for the four compact binary sources at $10^{4}$ points each. As shown in Fig.~\ref{fig:speed-up-waveform}, the waveform generation part is the dominant cost in traditional calculations. For a BNS system, it takes about $\sim 7$ times longer to generate the waveform as compared to the time for evaluating the likelihood integral. As we go towards heavy CBC systems, the speed-up ratio decreases as expected.

\subsection{RBF kernels and meshfree interpolation}
\label{appendix:diff_rbf_kernel}
In our analysis we chose $\phi = \exp(-\epsilon\, ||\lambda^q - \lambda^{\alpha}||_2)$. For the sake of completeness, we also performed the 2D meshfree likelihood interpolation using three different RBFs apart from the Gaussian RBF kernel (see table~\ref{tab:rbf_kernels_chapter_3}).
\begin{table}[!hbt]
    \centering
    \begin{tabular}{cccc}
        \hline\hline
        symbol & $\phi$ & SVD coeff & $\sigma^2$\\
        \hline
         `ga'& $\exp[-(\epsilon\, r)^2]$ &  0.5& 0.75\\
         `iq'&  $[1 + (\epsilon\, r)^2]^{-1}$& 0.25& 0.5 \\
         `imq'&  $[1 + (\epsilon\, r)^2]^{-1/2}$& 0.35& 0.60 \\
         `mq'&  $-[1 + (\epsilon\, r)^2]^{1/2}$& 0.25& 0.5 \\
         \hline
    \end{tabular}
    \caption{RBF kernels}
    \label{tab:rbf_kernels_chapter_3}
\end{table}
As shown in the table above, the RBF kernels $\phi$ also depend on the shape parameter ($\phi$), which affects the interpolation accuracy. The shape parameter can be optimized using the ``Leave One Out Cross-Validation'' (LOOCV) procedure~\cite{hastie2009elements}. In this procedure, we estimate the optimized value of the epsilon (given a range) by constructing the interpolant at each value of $\epsilon$ using all the nodes but one at which we evaluate the interpolant and calculate the error at that point, and it is repeated for all the nodes. Then, the rms value of the error is calculated using the errors at all the nodes. So, at each value of $\epsilon$, we have a root-mean-square (RMS) error, and the value of the $\epsilon$ giving the minimum RMS error is the optimized value. However, the values shown in the table~\ref{tab:rbf_kernels_chapter_3} are chosen heuristically. In chapter~\ref{chap:chapter_2}, we find the shape parameter using the LOOCV method as implemented in the RBF python package~\cite{RBF_github}.

\begin{figure*}[!hbt]
    \centering
    \includegraphics[width=0.55\linewidth]{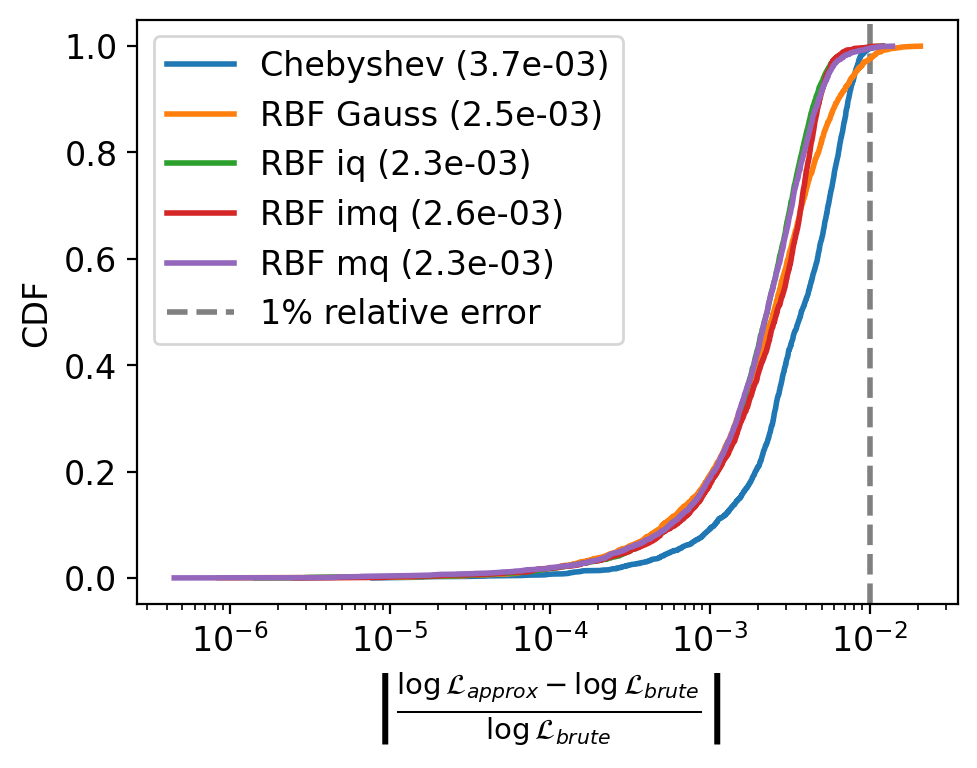}
    \caption{Relative error in likelihood reconstruction for Chebyshev interpolation, and meshfree interpolation with different RBF kernels. The median relative errors are shown in the legend, and all are $\sim \mathcal{O}(10^{-3})$ for all the kernels.}
    \label{fig:relative_error_likelihood_2d_meshfree}
\end{figure*}
\subsection{Chebyshev polynomials}
\label{appendix:cheby_poly}
Chebyshev polynomials for a single dimension are given as
\begin{equation}
T_L(x) = \frac{(x - \sqrt{x^2 - 1})^L + (x + \sqrt{x^2 - 1})^L}{2w}
\end{equation}
where $w = \sqrt{(1 + \delta_{L0})(L_{\text{max}} + 1)/2}$ is a normalization factor for the chebyshev polynomials and $\delta_{L0}$ is the Kroenecker delta. These polynomials satisfy the discrete orthogonality condition,
\begin{equation}
    \sum_{n=0}^{L_{\text{max}}}T_{K}(x_n)T_{M}(x_n) = \delta_{KM}
\end{equation}
\begin{figure}[!hbt]
    \centering
    \includegraphics[width=0.55\linewidth]{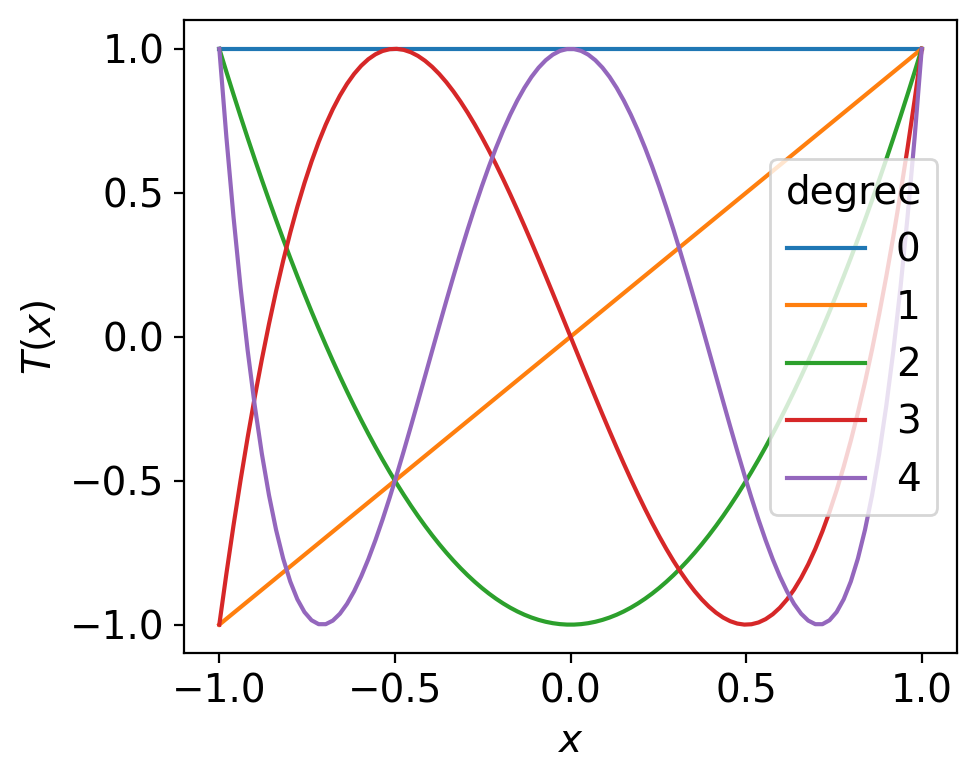}
    \caption{Chebyshev polynomials with different degrees}
    \label{fig:cheby_poly_plot}
\end{figure}
\chapter{Prompt sky localization of compact binary sources using meshfree approximation}
\label{chap:chapter_4}

\paragraph{\textbf{Abstract}}
The number of gravitational wave signals from the merger of compact binary systems detected in the network of advanced LIGO and Virgo detectors is expected to increase considerably in the upcoming science runs. Once a confident detection is made, it is crucial to reconstruct the source's properties rapidly, particularly the sky position and chirp mass, to follow up on these transient sources with telescopes operating at different electromagnetic bands for multi-messenger astronomy. In this context, we present a rapid parameter estimation (PE) method aided by mesh-free approximations to accurately reconstruct properties of compact binary sources from data gathered by a network of gravitational wave detectors. This approach builds upon our previous algorithm [L. Pathak \textit{et al.}, Fast likelihood evaluation using meshfree approximations for reconstructing compact binary sources, \href{https://journals.aps.org/prd/abstract/10.1103/PhysRevD.108.064055}{Phys. Rev. D \textbf{108}, 064055 (2023)}] to expedite the evaluation of the likelihood function and extend it to enable coherent network PE in a ten-dimensional parameter space, including sky position and polarization angle. Additionally, we propose an optimized interpolation node placement strategy during the start-up stage to enhance the accuracy of the marginalized posterior distributions.
With this updated method, we can estimate the properties of binary neutron star (BNS) sources in approximately 2.4~(2.7) minutes for the \texttt{TaylorF2}~(\texttt{IMRPhenomD}) signal model by utilizing 64 CPU cores on a shared memory architecture. 
Furthermore, our approach can be integrated into existing parameter estimation pipelines, providing a valuable tool for the broader scientific community. We also highlight some areas for improvements to this algorithm in the future, which includes overcoming the limitations due to narrow prior bounds.

\section{Introduction}
\label{sec:intro_chap4}
This chapter is based on the publication \textit{Prompt sky localization of compact binary coalescences using meshfree approximation}, \href{https://journals.aps.org/prd/abstract/10.1103/PhysRevD.109.024053}{Phys. Rev. D \textbf{109}, 024053 (2024)}.\newline

Gravitational wave signals emitted by compact binary systems made of neutron stars and black holes are expected to take up to a few minutes to sweep through the sensitive band of the advanced LIGO and Virgo detectors, leading up to the epoch of their eventual cataclysmic merger. The merger of two neutron stars/Neutron star-black holes (NSBH)~\cite{Eichler:1989ve, 1986ApJ_308L_47G, kilonovae_metzger, Li_1998, PhysRevD.107.124007}) can be followed by the emission of electromagnetic (EM) radiation at different wavelengths that fade over different time scales ranging from a few minutes to several months, as seen in the fortuitous detection of the binary neutron star system (BNS) GW170817~\cite{abbott2017gw170817} by the network of advanced LIGO-Virgo detectors~\cite{Harry:2010zz, Abbott_2020, Acernese_2014, 2015, PhysRevD.93.112004} in the second science run. This discovery resulted in the first multi-wavelength observation of post-merger electromagnetic (EM) emissions from the merger of two neutron stars. The successful localization and discovery, in this case, were primarily due to the source's fortunate proximity: GW170817 was relatively close (about 40 Mpc), falling well within the sky- and orientation-averaged ranges of the two LIGO detectors, resulting in the loudest signal ever detected by a GW network. The discovery of the EM counterpart of GW170817 improved understanding of the physics of short gamma-ray bursts (SGRBs)\cite{lamb2018grb}, confirmation of the formation of heavy elements after the merger and other areas of new physics, and provided invaluable opportunities to explore the mysteries of the universe. Therefore, BNS systems are the prime targets for multi-messenger astronomy.

In upcoming observing runs, the LIGO-Virgo detectors~\cite{Harry:2010zz, Abbott_2020, Acernese_2014, 2015, PhysRevD.93.112004} are expected to be sensitive above $10$ Hz, enabling a more extensive exploration of cosmic volume compared to their current versions. Consequently, the detection rate is expected to increase to the point where signals from compact binary mergers can be observed daily. It is project that there could be $10^{+52}_{-10}$/$1^{+91}_{-1}$ BNS/NSBH events detected in the O4 run~\cite{abbott2020prospects}. To efficiently observe EM counterparts in the future, it will be crucial to prioritize events based on their chirp masses, as suggested by Margalit et al.~\cite{Margalit_2019}. Furthermore, future GW events could provide opportunities to study various physical properties of binary systems, such as those with eccentricity~\cite{Lenon_2020}, spin precession~\cite{PhysRevD.49.6274, Hannam_2022}, and more. As such, prompt localization and source reconstruction would be vital for studying electromagnetic (EM) counterparts of binary neutron star systems (BNS) or neutron star-black hole (NSBH) binary systems in the future. Moreover, a fast PE technique can facilitate the execution of extensive inference investigations~\cite{wolfe2023small}, like population inference, within a reasonable time frame, which could be impractical when employing conventional brute-force PE methods.

The standard technique for PE uses Bayesian inference, which involves calculating the likelihood of observing the data under the hypothesis that an astrophysical signal (with some model parameters) is embedded in additive Gaussian noise with zero mean and unit variance. This likelihood calculation involves two computationally expensive parts: first, generating the model (template) waveforms~\cite{IMRPhenomXPHM, NRSur7dq4, ramosbuades2023seobnrv5phm} at sample points proposed by the sampling algorithm and, thereafter, determining the overlap between the data and the waveform. Generating these templates is the most computationally expensive part of the likelihood calculation, especially for low-mass systems like binary neutron stars (BNS). This computational burden is amplified by the improved sensitivity of detectors, resulting in a larger number of waveform cycles within the detectors' sensitive band for BNS systems. Additionally, incorporating physical effects into the waveform can further increase the computational cost of generating the waveform and sampling from the expanded parameter space. 
Wherever possible, factorizing~\cite{islam2022factorized} the log-likelihood function into pieces that exclusively depend on the extrinsic and intrinsic parameters, respectively, can lead to a significant reduction in the number of waveform generation and thereby speed-up the PE analysis. For example, such factorizations can be made for non-precessing waveforms, as considered in this work. 

In the current operational framework of the LIGO-Virgo-KAGRA (LVK) collaboration, the primary tool for rapid sky localization is {\texttt{BAYESTAR}}~\cite{singer2016rapid} - a Bayesian, non-Markov Chain Monte Carlo (MCMC) sky localization algorithm. This method exhibits remarkable speed, furnishing posterior probability density distributions across sky coordinates within a few tens of seconds following the detection of a gravitational wave signal from the merger of a compact binary source.
However, through a comprehensive Bayesian PE analysis encompassing both intrinsic and extrinsic model parameters, Finstad et al.~\cite{Finstad_2020} have established that the accuracy of sky localization can be significantly augmented, achieving an enhancement of approximately ${14 \, \text{deg}^2}$ over those obtained from the {\texttt{BAYESTAR}} algorithm alone.
Beyond the evident advantages of sky-location precision, a full PE analysis also furnishes a highly accurate estimation of the chirp mass for compact binary systems. This additional information plays a pivotal role in making judicious decisions regarding electromagnetic (EM) follow-up observations - which would be crucial in future observing runs of the network of advanced detectors, further underscoring the merits of developing fast-PE algorithms. Deep-learning-based sky-localization tools (CBC-SkyNet~\cite{chatterjee2022premerger}) have been recently developed that can obtain ``pre-merger'' sky localization areas that are comparable in accuracy to {\texttt{BAYESTAR}}.

Numerous rapid PE algorithms have surfaced in the gravitational-wave literature, which revolves around two overarching concepts:
\begin{enumerate}
    \item[i.] Likelihood-Based Approaches: The first category involves approaches that rapidly evaluate the likelihood function. This set of methods encompasses various methods such as Reduced Order Models (ROMs)\cite{canizares2013gravitational, canizares2015accelerated, smith2016fast, Morisaki_2020, morisaki2023rapid}, Heterodyning (or relative binning)~\cite{Venumadhav2018, cornish2021heterodyned, islam2022factorized}, the simple-pe~\cite{fairhurst2023fast} algorithm, and methods based on Gaussian process regression (GPR)~\cite{rasmussen2006gaussian} like RIFT~\cite{https://doi.org/10.48550/arxiv.1805.10457}. Other techniques, such as adaptive frequency resolution-based likelihood evaluation~\cite{Morisaki:2021ngj} and mass-spin reparametrization-based rapid PE~\cite{Lee:2022jpn}, have also been proposed.
    Improved algorithms allowing for more efficient reduced-order quadrature bases have been recently proposed~\cite{effroqs23}.
    
    \item[ii.] Likelihood ``Free'' Approaches: The second category consists of methods that bypass direct likelihood evaluation and instead learn the posterior distributions using advanced machine-learning (ML) techniques such as Deep Learning, Normalizing Flows, and Variational Inference~\cite{Chua_2020, Green_2020, green2020complete, gabbard2022bayesian}.
\end{enumerate}
Hybrid techniques that combine likelihood heterodyning with tools to enhance the convergence of gradient-based MCMC samplers have also been proposed~\cite{wong2023fast}. Other techniques involving score-based diffusion models~\cite{score_based_likelihood} to learn an empirical noise distribution directly from the detector data have also been proposed. In this context, our mesh-free approach aligns with the first category of rapid PE methods since it is designed to swiftly assess the likelihood. Note that the RIFT~\cite{https://doi.org/10.48550/arxiv.1805.10457} method also interpolates the likelihood directly using GPR. In comparison, the meshfree method first represents the likelihood function on an orthonormal basis and then uses radial basis functions to interpolate the coefficients. So, inspite of the apparent similarity, the context in which the “interpolation” technique is applied is completely different in the two methods. The use of GPR (a supervised training method) to ``learn'' the likelihood function at different points of the parameter space is in the spirit of ML methods in contrast with the classical approach followed by the meshfree method.

In our previous work~\cite{PhysRevD.108.064055}, we introduced the meshfree method as a means of rapidly inferring parameters for compact binary coalescence (CBC) sources for a single detector by considering only six parameters: the two-component masses, two aligned spins, luminosity distance, and coalescence time.
In this study, (a) we extend the method to include additional parameters such as sky location, inclination, and polarization, thereby enabling a coherent multi-detector PE analysis. (b) Furthermore, we have modified the Radial Basis Function (RBF) meshfree interpolation node placement scheme, which directly impacts the accuracy of the PE results. In contrast to our previous work, which utilized uniformly distributed nodes over the sample space, we now employ a combination of multivariate Gaussian and uniform distributions for efficient node placement. 
(c) Our results demonstrate that when run on $32\, (64)$ cores, the meshfree method can produce accurate marginalized posteriors for the GW170817 event and locate it in the sky within $\sim 3.4\, (2.4)$ minutes of detecting the event. The resulting posteriors obtained from the meshfree method are statistically indistinguishable from those obtained using the bruteforce method implemented in \pycbc~\cite{alex_nitz_2022_6912865}. Please note that the brute force method is not optimized and is used only as a yardstick for accuracy.

The rest of the paper is structured into the following sections: Section \ref{sec:param_est_chap4} introduces the PE basics, where we briefly discuss the Bayesian approach for inferring GW parameters from the data and define the coherent network likelihood. Section \ref{sec:meshfree_chap4} explains the start-up and online stages and discusses the likelihood evaluation procedure using meshfree interpolation. The next section \ref{subsec:gw17_chap4} presents the results of a detailed analysis of the meshfree method for the GW170817 BNS event. 
In Section \ref{subsec:simulated Data_chap4}, we examine the effectiveness of our method by testing it on simulated events covering a wide range of signal-to-noise ratios (SNR). Finally, we summarize the results in Section \ref{sec:concl_outl_chap4} and discuss the limitations and future follow-ups of the current implementation. Finally, we summarize the results in Section~\ref{sec:concl_outl_chap4} and discuss the limitations of the current implementation (such as restricting the sampler over narrow prior bounds). We suggest some ideas for overcoming this limitation in follow-up studies.

\section{Parameter estimation}
\label{sec:param_est_chap4}
\subsection{Bayesian inference}
\label{subsec:bayes_infe_chap4}
Given a stretch of GW strain data $\boldsymbol{d}$ from detectors containing a GW signal $\boldsymbol{h}(\vec \Lambda)$, embedded in additive Gaussian noise $\boldsymbol{n}$, we want to estimate the posterior distribution ${p(\vec \Lambda \mid \boldsymbol{d})}$ over the source parameters $\vec \Lambda$. The posterior distribution, in turn, is related to the likelihood function ${\mathcal{L}(\boldsymbol{d}\mid \vec \Lambda)}$ via the well-known Bayes' theorem:
\begin{equation} 
p(\vec \Lambda \mid \boldsymbol{d}) = \frac{\mathcal{L}(\boldsymbol{d} \mid \vec \Lambda) \ p(\vec \Lambda)}{p(\boldsymbol{d})}
\label{eq:Bayes}
\end{equation}
where, 
$p(\vec \Lambda)$ is the prior distribution over model parameters ${\param \equiv \{ \vec \lambda, \vec \theta, t_{c} \}}$. Here $\vec \lambda$ is a set of intrinsic parameters such as the component masses $m_{1,2}$ and dimensionless aligned spins $\chi_{1z,2z}$ while $\vec \theta$ represents the set of extrinsic parameters such as the source's sky location i.e. right ascension ($\alpha$) and declination ($\delta$), the inclination ($\iota$) of the orbital plane of the binary with respect to the line of sight, the polarization angle ($\psi$), the luminosity distance ($d_{L}$) of the source from the Earth and the geocentric epoch of coalescence $t_c$. 

In principle, given a numerical prescription for calculating the likelihood function ${\mathcal{L}(\boldsymbol{d}\mid \vec \Lambda)}$ under a waveform model and assume prior distributions over the model parameters, we can evaluate the left-hand side of Eq.~\eqref{eq:Bayes} at any given point in the sample space upto an overall normalization factor. However, with a large number of parameters, $\vec \Lambda$ (typically $\sim 15$ parameters), evaluating ${p(\vec \Lambda \mid \boldsymbol{d})}$ on a fine grid over the sample space becomes increasingly tedious and eventually intractable with finite computational resources. Therefore, a more intelligent strategy is employed where one uses stochastic sampling techniques to estimate the posterior distribution ${p(\vec \Lambda \mid \boldsymbol{d})}$. There are a number of schemes to sample the posterior distribution, such as Markov Chain Monte Carlo (MCMC)~\cite{Foreman_Mackey_2013} and its variants, Nested Sampling~\cite{skilling2006nested} algorithms, etc. In this paper, we use {\sc{dynesty}}~\cite{speagle2020dynesty, sergey_koposov_2023_7600689}, an extensively used Python implementation of the nested sampling algorithm for GW data analysis. In this work, we have only focused on speeding up the PE algorithm by quickly evaluating the likelihood function at any point proposed by the sampling algorithm. Admittedly, another aspect of designing a fast-PE algorithm would involve optimizing the sampling algorithm itself - which is not considered in this work. It is easy to see that such improvements to the sampling algorithm can positively impact the performance of many fast-PE methods (including ours). For example, a significant improvement in computation time has been demonstrated by efficiently populating the parameter space as proposed by the {\sc{VARAHA}} sampling technique~\cite{varaha}.

\subsection{The likelihood function}
\label{sec:likelihood_chap4}

Let  $\boldsymbol{d}^{(i)}$ be the strain data recorded at the $i^{\text{th}}$ detector containing an astrophysical GW signal $\tilde{h}^{(i)}(\vec\Lambda)$. We pause to remark that complex $\tilde{h}^{(i)}(\vec \Lambda)$ will denote the frequency-domain Fourier transform (FT) of the signal $h^{(i)}(\vec \Lambda)$. Assuming uncorrelated noise among the $N_{\text{d}}$ detectors in the network, the coherent log-likelihood~\cite{lalinference} is given by 
\begin{equation}
\label{eq:multigenlikelihood_chap4}
\ln \mathcal{L}(\vec\Lambda) =  \sum_{i=1}^{N_{\text{d}}} {\langle \boldsymbol{d}^{(i)} \mid \tilde{h}^{(i)}(\vec\Lambda)\rangle} 
- \frac{1}{2} \sum_{i=1}^{N_{\text{d}}} \left [ \| \tilde{h}^{(i)}(\vec\Lambda)\|^2 + \| \boldsymbol{d}^{(i)} \|^2 \right ].
\end{equation}
For a non-precessing GW signal model, one can use the relation ${\tilde{h}_{\times} \propto -j\ \tilde{h}_{+}}$ between the two polarization states to express the signal at the $i^{\text{th}}$ detector as:
\begin{equation}
\label{eq:detframwaveform_chap4}
\tilde{h}^{(i)}(\vec\Lambda) \equiv \tilde{h}(\vec\Lambda, t^{(i)}) = \mathcal{A}^{(i)} \tilde{h}_{+}(\vec \lambda, t^{(i)}), \mathcal{A}^{(i)}\, \tilde{h}_{+}(\vec \lambda)\, e^{-j\,2\pi f t_c} \, e^{-j\,2\pi f \Delta t^{(i)}}.
\end{equation}
The complex amplitude of the signal $\mathcal{A}^{(i)}$ depends only on the extrinsic parameters $\vec\theta \in \vec \Lambda$ through the antenna pattern functions, luminosity distance, and the inclination angle:
\begin{equation}
\mathcal{A}^{(i)} = 
    \frac{1}{d_L} \left[ \frac{1+\cos^2 \iota}{2}  F^{(i)}_{+}(\alpha, \delta, \psi) \right.
            \left. - \: j \cos \iota \ F^{(i)}_{\times}(\alpha, \delta, \psi) \right]
\end{equation}
where ${F^{(i)}_{+}(\alpha, \delta, \psi)}$ and ${F^{(i)}_{\times}(\alpha, \delta, \psi)}$ are respectively the `plus' and `cross' antenna pattern functions of the $i^{\text{th}}$ detector. The antenna pattern functions of a detector describe the angular response of the detector to incoming GW signals. It arises from contracting the position and geometry dependent 'detector tensor'~\cite{detectorTensor} with the metric perturbations (GW); thereby mapping the latter to a GW strain amplitude $h^{(i)}(\vec\Lambda)$ recorded at the detector.

As indicated by the Eq.~\eqref{eq:detframwaveform_chap4}, the signal acquires an additional phase difference during its projection from the geocentric frame to the detector frame, which corresponds to a time delay denoted by $\Delta t^{(i)}$. This temporal offset originates due to the relative positioning of the $i^{\text{th}}$ detector in relation to the Earth's center, and it can be expressed explicitly as follows:
\begin{equation}
\label{eq:time_delay}
    \Delta t^{(i)} \equiv t^{(i)} - t_{c} 
    = \frac{\vec x^{(i)} \cdot \hat{N}(\alpha, \delta) }{c},
\end{equation}
where $\vec x^{(i)}$ is a vector pointing from the the Earth's centre to the location of the $i^{\text{th}}$ detector, $t^{(i)}$ is the time at the $i^{\text{th}}$ detector, and $\hat{N}(\alpha, \delta)$ is the direction of the GW propagation~\cite{PhysRevD.92.023002}. 
In our analysis, we consider the log-likelihood function marginalized over the coalescence phase~\cite{thrane_2019} parameter, which can be written as follows:
\begin{equation}
\label{eq:multiphaselikelihood1}
    \left.\ln \mathcal{L}(\vec\Lambda \mid \boldsymbol{d}^{(i)})\right|_{\phi_c} 
    = \ln I_{0}\left[\left|\sum_{i=1}^{N_{d}}\langle \boldsymbol{d}^{(i)}\mid \tilde{h}^{(i)}(\vec \Lambda)\rangle\right|\right]
    - \frac{1}{2}\sum_{i=1}^{N_{d}}\left[ \| \tilde{h}^{(i)}(\vec\Lambda)\|^2 + \| \boldsymbol{d}^{(i)} \|^2 \right],
\end{equation}
where $I_0(\cdot)$ is the modified Bessel function of the first kind. By marginalizing over extrinsic parameters, we effectively reduce the dimensionality of the problem, resulting in accelerated likelihood calculations and enhanced sampling convergence. Substituting Eq.~\eqref{eq:detframwaveform_chap4} in the above equation, we get
\begin{multline}
\label{eq:multiphaselikelihood2}
    \left.\ln \mathcal{L}(\vec\Lambda \mid \boldsymbol{d}^{(i)})\right|_{\phi_c} 
    = \ln I_{0}\left[\left|\sum_{i=1}^{N_{d}}{{\mathcal{A}}^{(i)}}^{*}\, \langle \boldsymbol{d}^{(i)} \mid \tilde{h}_{+} (\vec\lambda, t^{(i)}) \rangle \right|\right] \\
    - \frac{1}{2}\sum_{i=1}^{N_{d}}\left[ \left|\mathcal{A}^{(i)}\right|^2 \sigma^2(\vec \lambda)^{(i)} + \| \boldsymbol{d}^{(i)} \|^2 \right];
\end{multline}
where, 
$\langle \boldsymbol{d}^{(i)} \mid \tilde{h}_{+}(\vec \lambda, t^{(i)}) \rangle$ is the complex overlap integral, while ${\sigma^2(\vec\lambda)^{(i)} \equiv \langle \tilde{h}_{+}(\vec \lambda, t^{(i)}) \mid \tilde{h}_{+}(\vec \lambda, t^{(i)}) \rangle}$ is the squared norm of the template $\: \tilde{h}_{+} (\vec\lambda)$. ${\sigma^2(\vec\lambda)^{(i)}}$ depends on the noise power spectral density (PSD) of the $i^{\text{th}}$ detector. 
The squared norm of the data vector, $\| \boldsymbol{d}^{(i)} \|^2$, remains constant during the PE analysis and, therefore, doesn't impact the overall `shape' of the likelihood. As such, it can be excluded from the subsequent analysis.

\section{Meshfree likelihood interpolation}
\label{sec:meshfree_chap4}
\begin{figure*}
\centering
\includegraphics[width=\textwidth, clip=True] {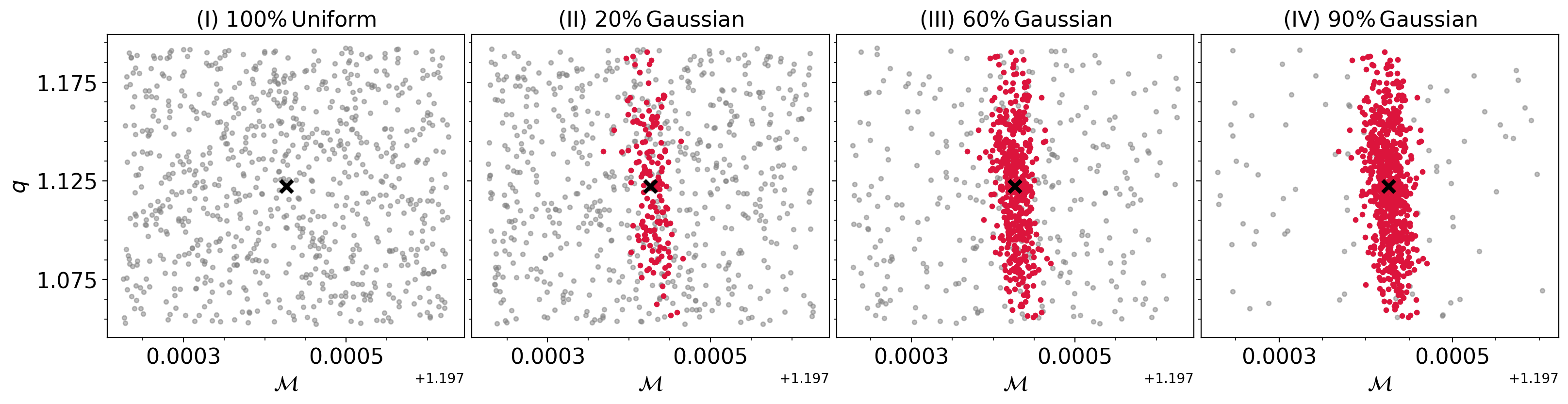}
\caption{
The figure displays randomly selected RBF interpolation nodes in the intrinsic parameter space with varying fractions from Gaussian and uniform distributions (2D slice in $\mathcal{M}$ and $q$ coordinates). Randomly placed nodes drawn from a Gaussian distribution are denoted in red, while those from a uniform distribution are shown in gray. The covariance matrix of the Gaussian distribution can be estimated semi-analytically given the signal model. Sub-figure-(I) represents a choice with $100 \%$ (all) nodes drawn from a uniform distribution, and the other sub-figures exhibit mixtures of different percentages of nodes from the uniform and Gaussian distributions. The center (marked as a ``black cross") of the sample space is determined through an optimization routine based on the highest network signal-to-noise ratio (SNR). This search is guided by the best-matched trigger obtained from the upstream search pipeline.
}
\label{fig:nodes_distr}
\end{figure*}

As discussed in our previous work~\cite{PhysRevD.108.064055}, the meshfree method comprises two distinct stages:
\begin{enumerate}
    \item Start-up Stage: During this phase, we create radial basis function (RBF) interpolants for the pertinent quantities.
    \item Online Stage: In this stage, we swiftly compute the likelihood function at query points proposed by the sampling algorithm.
\end{enumerate}
Next, we will delve into the various stages that constitute meshfree interpolation.
%

\subsection{Start-up stage} 
\label{subsec:startup-stage_chap4}
During the start-up stage, we generate RBF interpolants for the relevant quantities, which enable rapid likelihood calculations in the online stage. The start-up stage can be further divided into the following parts: 
\begin{enumerate}
    \item[i.] Placement of RBF nodes over the intrinsic parameter space, $\lambda$. These discrete nodes are denoted as ${\vec \lambda^n: n = 1, 2, ..., N.}$ 
    
    \item[ii.] Generation of complex time-series ${\vec z^{(i)}(\vec \lambda^{n}) \equiv  \langle \boldsymbol{d}^{(i)}\mid \tilde{h}_{+}(\vec \lambda^{n}, t^{(i)})\rangle}$ and template norm square ${\sigma^2(\vec \lambda^{n})^{(i)}}$ at each of the RBF nodes ${\vec \lambda^{n}}$, for each detector. In the start-up stage, we calculate the $\vec z^{(i)}(\vec \lambda^{n})$ at discrete values of circular time shifts $t_c$ (with respect to some reference geocentric coalescence time)  uniformly spaced over a specified range.  The range of time shifts $|t_c|$ should be larger than the maximum light travel time between two detectors (corresponding to the longest baseline). For the analysis presented here we choose the time shifts in the range $\pm 150 \, \text{ms}$ around the geocentric time of coalescence as reported by an upstream search pipeline.
    Note that we set $\Delta t^{(i)} = 0$ for all the detectors while calculating ${\vec z^{(i)}(\vec \lambda^{n})}$. Additional time delays $\Delta t^{(i)}$ for the signal to reach the $i^\text{th}$ detector depending on the source's location in the sky   
    can be taken care of during the online stage while sampling the posterior distribution over the sky location parameters as explained later. 

    \item[iii.] Singular value decomposition (SVD) of the complex time-series matrix ${ \mathcal{Z}^{(i)}}$ at the $i^\text{th}$ detector, where the $n^{th}$ row is defined as ${\vec z^{(i)}(\vec \lambda^{n})}$, and 

    \item[iv.] Generation of RBF interpolating functions for the SVD coefficients and the template norm square $\sigma^{2}$ for each detector. 
\end{enumerate}
We now describe some of these steps in greater detail.

\subsubsection{RBF nodes placement}
\label{subsec:nodesplacement_chap4}
\begin{figure}[!hbt]
\centering
\includegraphics[width=0.55\textwidth, clip=True]{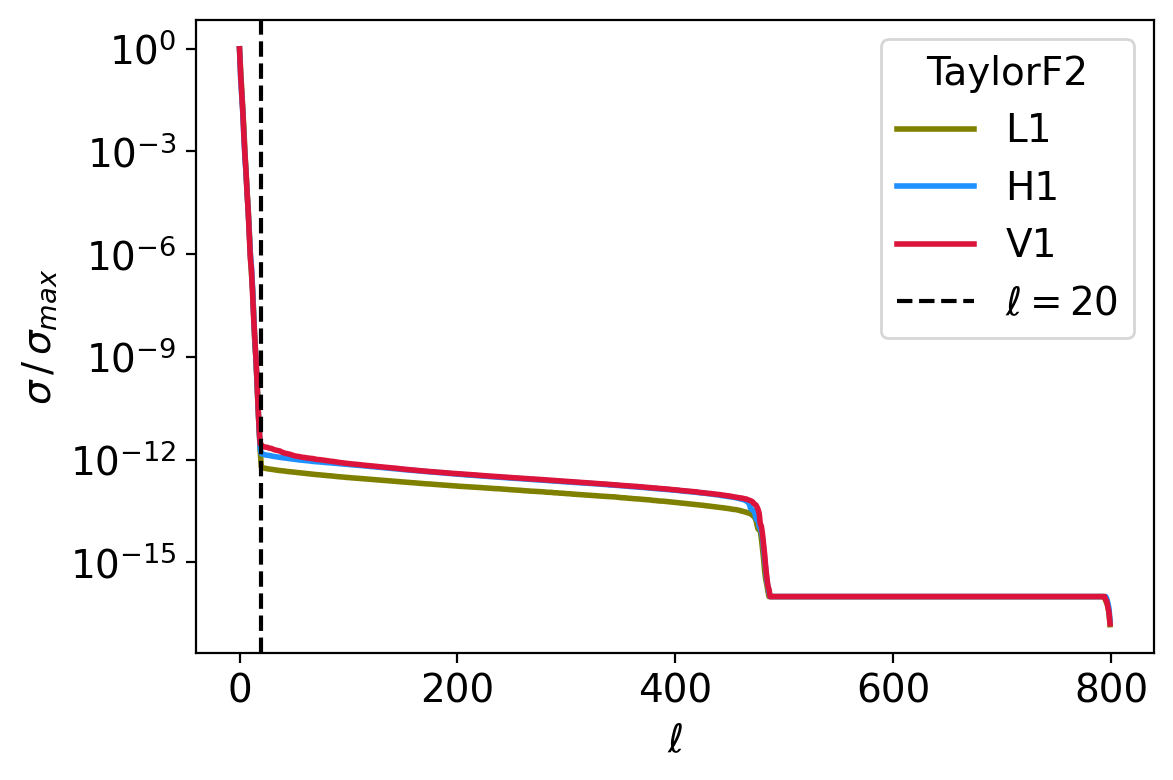}
\caption{
The figure shows the profile of the singular values normalized by the highest singular value. Notably, there is a sharp decline in the singular values (indicated by a black dashed line) due to the large correlation between the time-series $\vec z^{(i)}(\vec \lambda^n)$ at different RBF interpolation nodes, revealing that only the linear combination of the top-few singular vectors (around 20) is sufficient for effectively reconstructing the time series $\vec z^{(i)}(\vec \lambda^q)$ with minimal reconstruction error. The 3.5~PN \texttt{TaylorF2} post-Newtonian signal model is assumed.
}
\label{fig:sing_values}
\end{figure}
The accurate construction of meshfree RBF interpolants relies significantly on the strategy of randomly distributing the RBF nodes, denoted as $\vec \lambda^{n}$, across the sample space. However, the task of finding an By ooptimal set of nodes poses a considerable challenge. Typically, a hypercube in the sample space is selected around a reference point during the sampling of the posterior distribution. This reference point commonly known as trigger or best-matched template $\vec \lambda^{*}$ and trigger time (geocentric) $t_{\text{trig}}$ are  derived from an upstream search pipeline~\cite{GstLAL_2010, usman2016pycbc}.

In our previous work, we opted for a uniform distribution of nodes within the hypercube centered around the reference point (injection in the case of a simulated BNS event). This approach works well for events with low SNRs since the corresponding likelihood function exhibited a relatively flatter profile across the parameter space. However, for high SNR events, where the likelihood profile exhibits a sharp peak in certain regions of the parameter space, uniformly distributing nodes within the hypercube may not adequately cover the support of the posterior distribution, particularly in areas where we seek higher accuracy in interpolated likelihood estimates. As a result, we adopted a combination of nodes drawn from both a multivariate Gaussian distribution and a uniform distribution across the sample space within the hypercube. The multivariate Gaussian distribution is defined by a mean vector $\vec \mu$ and a covariance matrix $\boldsymbol{\Sigma}$, and its expression is as follows:
\begin{equation}
    p(\vec \lambda) \propto \exp\left[-(\vec \lambda - \vec \mu)^T \: \boldsymbol{\Sigma}^{-1} \: (\vec \lambda - \vec \mu)\right]
\end{equation}
where $\mu$ can be selected as the trigger point $\vec \lambda^{*}$, and $\boldsymbol{\Sigma}$ can be computed using the inverse of the Fisher matrix $\Gamma$ evaluated at $\vec \lambda^{*}$. While the distribution of nodes generated using the Fisher matrix may not perfectly match the actual posterior distribution across the sample space, it does exhibit significant overlap with the true distribution, especially in the vicinity of the likelihood peak.

Since the trigger originates from a search pipeline that uses a discrete set of templates over a grid (also known as a template bank)~\cite{temp_bank_soumen, PhysRevD.103.084047}, we can select a reference point, denoted as $\vec \lambda_{\text{ref}}$, in close proximity to the trigger point ($\vec \lambda^{*}$), which may have a higher network SNR. We employ an optimization strategy~\cite{2020SciPy-NMeth} aimed at maximizing the network SNR in order to identify this reference point. Subsequently, a portion of the interpolation nodes are generated from a multivariate Gaussian distribution, $\mathcal{N}(\vec \lambda_{\text{ref}}, \boldsymbol{\Sigma})$, where $\boldsymbol{\Sigma}$ is computed from the inverse of the Fisher matrix evaluated at $\vec \lambda_{\text{ref}}$, to adequately cover the region around the peak. For regions of the sample space where Gaussian nodes are sparse or possibly nonexistent, we spray a uniform distribution of samples across the hypercube. This approach ensures high accuracy in interpolated likelihood estimates throughout the entire sample space.

For a visual representation illustrating the various combinations of Gaussian and uniform nodes, please refer to Figure~\ref{fig:nodes_distr}. In particular, note that choosing a small fraction of random RBF nodes from the multivariate Gaussian distribution leads to more accurate posterior distributions over certain parameters (see Appendix~\ref{appendix:effect_of_nGauss}).

\subsubsection{SVD of ${\mathcal{Z}^{(i)}}$ and Interpolants generation}
\label{subsec:svdinterpolantsgeneration_chap4}
After obtaining the appropriate RBF nodes $\vec \lambda^{n}$, we can efficiently compute the time-series ${\vec z^{(i)}(\vec \lambda^{n}) \equiv z^{(i)}(\vec \lambda^{n}, t_c)}$ using Fast Fourier Transform (FFT) based circular correlations where $t_c$ are the uniformly spaced discrete-time shifts taken from a specified range around a reference coalescence time $t_{\text{trig}}$. Given that the predominant portion of the posterior support originates from the vicinity of the likelihood peak, we opt for ${t_{c} \in [t_{\text{trig}} \pm 0.15\; \text{s} ]}$ as the interval for sampling $t_{c}$.

Similarly, we compute the template norm square ${\sigma^2(\vec \lambda^{n})^{(i)}}$ at all the RBF nodes $\vec \lambda^{n}$.

Our objective is to identify the basis vectors capable of spanning the space of $\vec z^{(i)}(\vec \lambda^{n})$. This can be achieved by stacking (row-wise) the time-series vectors $\vec z^{(i)}(\vec \lambda^{n})$ for each $\vec \lambda^{n}$, where $n = 1, 2, \cdots, N$. Subsequently, performing the Singular Value Decomposition (SVD) on the resultant matrix ${\mathcal{Z}^{(i)}}$ yields our desired basis vectors ${\vec u_\mu}$. It can be shown that any vector $\vec z^{(i)}(\vec \lambda^{n})$ within the specified parameter range can be expressed as a linear combination of these basis vectors ${\vec u_\mu}$,
\begin{equation}
\label{eq:svdbasistimeseries_chap4}
\vec z^{(i)}(\vec \lambda^{n}) = \sum_{\mu = 1}^N\, C^{n (i)}_{\mu}\ \vec u^{(i)}_{\mu}
\end{equation}
where, ${C^{n (i)}_{\mu} \equiv C^{(i)}_{\mu}(\vec \lambda^{n})}, \ \mu=1, 2, \cdots N$ are the $N$ SVD coefficients. These coefficients $C^{n (i)}_{\mu}$ are characterized as smooth functions (see Appendix~\ref{appendix:svd_coeff_surf}) over $\vec \lambda^{n}$ within a sufficiently narrow boundary encompassing the posterior support.
Therefore, each of these $N$ coefficients can be treated as a scalar-valued, smooth function over the $d$-dimensional intrinsic parameter space and expressed as a linear combination of the radial basis functions~\cite{doi:10.1142/6437} (RBFs) $\phi$ centered at the interpolation nodes:
\begin{equation}
\label{eq:rbfcoeff_chap4}
C^{q (i)}_{\mu} = \sum_{n=1}^N\, a^{(i)}_{n}\, \phi(\|\vec \lambda^q - \vec \lambda^{n}\|_2) + \sum_{j = 1}^{M}\, b^{(i)}_{j}\, p_j(\vec \lambda^q)
\end{equation} 
where $\phi$ is the RBF kernel centered at ${\vec \lambda^{n} \in \mathbb{R}^d}$, and $\left \{  p_{j} \right \}$ denotes the monomials that span the space of polynomials with a predetermined degree $\nu$ in $d$-dimensions. 
The inclusion of these monomial terms has been shown to enhance the accuracy of RBF interpolation~\cite{2016JCoPh}. 

Similarly, we can also express $\sigma^2(\vec \lambda^{q})^{(i)}$ in terms of the RBF functions and monomials by treating it as smoothly varying scalar-valued function over the interpolation domain.

From Eq.~\eqref{eq:rbfcoeff_chap4}, we find that there are a total of $(N + M)$ coefficients to be solved for each interpolating function. 
The SVD coefficients $C_{\mu}^{q (i)}$ and $\sigma^2(\vec \lambda^{q})^{(i)}$ are known at each of the $N$ RBF nodes $\vec \lambda^n$. These provide $N$ interpolation conditions. To uniquely determine the coefficients $a_{n}$ and $b_j$, we impose $M$ additional conditions of the form ${\sum_{j=1}^M a^{(i)}_j p_j(\vec \lambda^q) = 0}$. This leads to the following system of $N + M$ equations:
\begin{equation}
	\begin{bmatrix}
		\boldsymbol{\Phi} & \boldsymbol{P} \\
		\boldsymbol{P}^{T} & \boldsymbol{O} 
	\end{bmatrix} \
		\begin{bmatrix}
		\boldsymbol{a^{(i)}} \\ 
		\boldsymbol{b^{(i)}}
	\end{bmatrix} \ 
	=  
	\begin{bmatrix}
		C^{n (i)}_{\mu} \\ 
		\boldsymbol{0} 
	\end{bmatrix}
\label{Eq:rbf-sle}
\end{equation}
where the matrices $\boldsymbol{\Phi}$ and $\boldsymbol{P}$ have components ${\Phi_{ij} = \phi(\norm{\vec\lambda^{i}-\vec\lambda^{j}}_2)}$ and ${P_{ij} = p_j(\vec\lambda^{i})}$ respectively; ${\boldsymbol{O}_{M\times M}}$ is a zero-matrix and ${\boldsymbol{0}_{M\times 1}}$ is a zero-vector.

The additional $M$ conditions arise due to the fact that $\boldsymbol{\Phi}$ needs to be a real, symmetric, and conditionally positive definite matrix of order $\nu + 1$, and the following theorem (reproduced here from \cite{doi:10.1142/6437} for completeness) guarantees the uniqueness of the solutions:
\begin{theorem}
\label{thm:rbf_solve}
\textit{Let $\boldsymbol{A}$ be a real symmetric $N \times N$ matrix that is conditionally positive definite of order one, and let $\boldsymbol{B} = [1,...,1]^T$ be an $N \times 1$ matrix (column vector). Then the system of linear equations}
\begin{equation*}
	\begin{bmatrix}
		\boldsymbol{A} & \boldsymbol{B} \\
		\boldsymbol{B}^T & \boldsymbol{O} 
	\end{bmatrix} \
		\begin{bmatrix}
		\boldsymbol{c} \\ 
		\boldsymbol{d}
	\end{bmatrix} \ 
	=  
	\begin{bmatrix}
		\boldsymbol{y} \\ 
		\boldsymbol{0} 
	\end{bmatrix}
\label{Eq:theorem_sle}
\end{equation*}
\textit{is uniquely solvable.}
\end{theorem}
Although Theorem~\ref{thm:rbf_solve} is applicable for cases where $\nu = 0$, it can be generalized to polynomials of higher degrees (For a proof, refer to Theorem $7.2$ in~\cite{doi:10.1142/6437}). 

Now, the set of equations Eq.~\eqref{Eq:rbf-sle} can be uniquely solved for the unknown coefficients $\boldsymbol{a}^{(i)}$ and $\boldsymbol{b}^{(i)}$ to determine the meshfree interpolating functions  corresponding to the coefficients $C_{\mu}^{(i)}$. 
From the sharply decreasing spectrum of eigenvalues shown in Fig.~\ref{fig:sing_values}, it is clear that only the ``top-few'' basis vectors are sufficient to reconstruct $\vec z^{(i)}(\vec \lambda^q)$ at minimal reconstruction error, we generate only top-$\ell$ meshfree interpolants of $C_{\mu}^{q(i)}$ where $\mu = 1,....,\ell$. 
Note that $\ell$ is chosen based on the spectrum of eigenvalues of the matrix ${\mathcal{Z}^{(i)}}$. We have chosen $\ell=20$ where the normalized eigenvalues fall~(see Fig.~\ref{fig:sing_values}) to $\sim 10^{-12}$. However, this is a subjective choice.

In a similar way, we generate the meshfree interpolants for the scalar-valued function $\sigma^2(\vec \lambda^q)^{(i)}$ as well. 

Note that to optimize computational efficiency, we need to generate a minimum number of RBF nodes $\vec \lambda^{n}$ given the degree ($\nu$) of the polynomial and dimension ($d$) of the intrinsic parameter space and it is given by $N_{\text{min}} = \binom{\nu + d}{\nu}$. Increasing the number of nodes increases not only the computational cost of generating the interpolants but also their evaluation cost during the online stage, which is evident from Eq.~\eqref{eq:rbfcoeff_chap4} as the number of coefficients $a_n^{(i)}$ increase linearly with the number of RBF nodes $N$. In our study, we empirically choose at least twice (or more) the minimum number of required nodes.   

At the end of the start-up stage, we thus have uniquely determined $(\ell +1)$ interpolating functions

\subsection{Online stage}
\label{subsec:online-stage_chap4}
Once we have the meshfree interpolants ready, we can evaluate the meshfree interpolated values of $C^{q (i)}_{\mu}$ and $\sigma^2(\vec \lambda^q)^{(i)}$ at any query point $\vec \lambda^q$ rapidly. We then combine the interpolated values of the coefficients $C^{q (i)}_{\mu}$ with the corresponding basis vectors $\vec u_{\mu}$ according to Eq.~\eqref{eq:svdbasistimeseries_chap4}, restricting the summation to the first $\ell$ terms.

In the startup stage~\ref{subsec:startup-stage_chap4}, we set ${\Delta t^{(i)} = 0}$ while calculating ${\vec z^{(i)}(\vec \lambda^n)}$ and didn't include the time-delay relative to the Earth's center, which depends on the sky location. Now, we incorporate it while reconstructing the interpolated time-series ${\vec z^{(i)}(\vec \lambda^q)}$. Moreover, we focus on constructing only ${\vec z^{(i)}(\vec \lambda^q)}$ at approximately 10 time-samples centered around query time $t^{q (i)}$ (Eq.~\eqref{eq:time_delay}). To ensure sub-sample accuracy, we fit these samples with a cubic spline, which incurs negligible computational cost. Subsequently, we calculate ${\vec z^{(i)}(\vec \lambda^q)}$ at the query $t^{q (i)}$ using the cubic-spline interpolant. Similarly, we evaluate the interpolated value of $\sigma^2(\vec \lambda^q)^{(i)}$. Finally, combining ${\vec z^{(i)}(\vec \lambda^q)}$ and ${\sigma^2(\vec \lambda^q)^{(i)}}$ with coefficients dependent on extrinsic parameters $\vec \theta$, as exemplified in Eq.~\eqref{eq:multiphaselikelihood2}, enables the computation of $\ln \mathcal{L}_{\text{RBF}}$.

\section{Numerical experiments}
\label{sec:numerical_chap4}

\subsection{GW170817 event}
\label{subsec:gw17_chap4}

\begin{figure*}[t]
\begin{subfigure}{0.49\linewidth}
    \includegraphics[width=\linewidth]{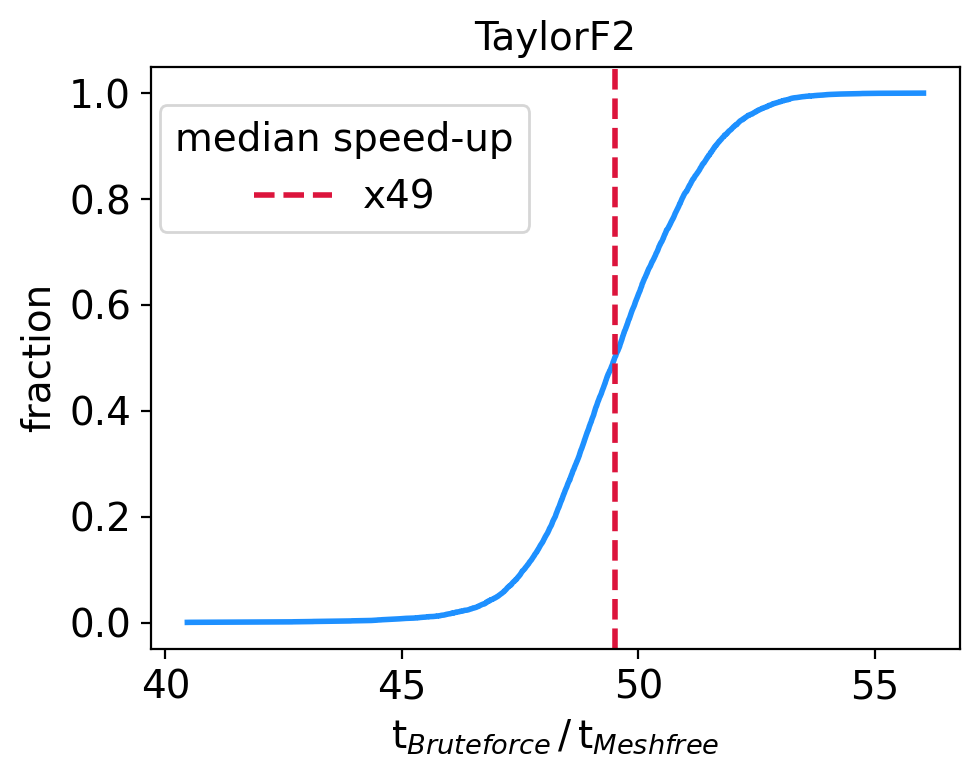}
    \caption{\texttt{TaylorF2} signal model}
    \label{figspeedup:TF2}
\end{subfigure}\hfill
\begin{subfigure}{0.49\linewidth}
    \includegraphics[width=\linewidth]{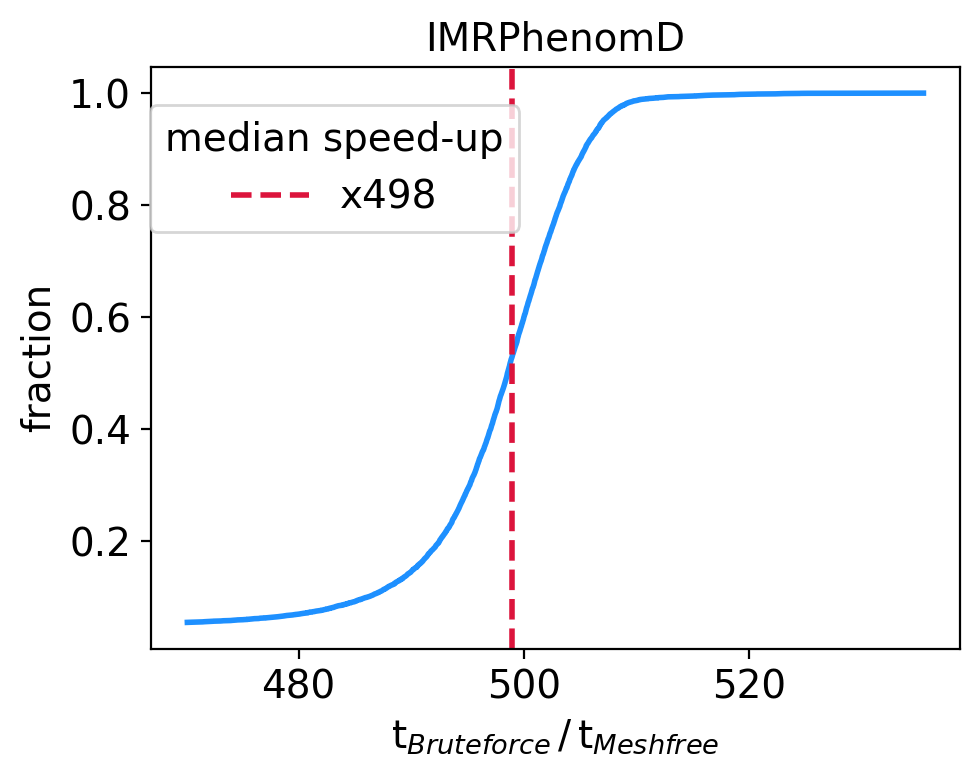}
    \caption{\texttt{IMRPhenomD} signal model}
    \label{figspeedup:IMR}
\end{subfigure}
\caption{
The plots compare the speed-up of the meshfree likelihood evaluation with bruteforce likelihood for two waveform models, \texttt{TaylorF2} and \texttt{IMRPhenomD}. The meshfree method is around $49$ times faster for \texttt{TaylorF2} and approximately $498$ times faster for \texttt{IMRPhenomD} (compared to the standard calculation). The median absolute error is approximately $\mathcal{O}(10^{-2})$ for both cases.
In the proposed meshfree method, one can trade-off speed in favor of better accuracy by retaining a larger number of SVD basis vectors (see Eq.~\eqref{eq:svdbasistimeseries_chap4}) and vice-versa. The optimal subjective choice is made from the spectrum of eigenvalues (Fig.~\ref{fig:sing_values}).
}
\label{fig:speedup}
\end{figure*}

We conducted a Bayesian PE study on the GW170817 binary neutron star (BNS) event to test our method. The strain data from the two advanced-LIGO and Virgo detectors were obtained from the open archival data sets available on GWOSC~\cite{GW170817data, RICHABBOTT2021100658}. We used a $360$ seconds data segment that was cleaned with a $4$th-order high-pass filter, with a cut-off frequency of $18$ Hz. The seismic cutoff frequency was set to $20$ Hz. The \texttt{TaylorF2} waveform model~\cite{PhysRevD.49.1707, PhysRevD.59.124016, Faye_2012, PhysRevLett.74.3515} was employed to recover the source parameters. The noise PSD was estimated from the strain data using overlapping segments of length $2$ seconds, using the \texttt{median-mean} PSD estimation method by applying a \texttt{Hann} window as implemented in \pycbc. Subsequently, we generated $N = 800$ RBF nodes using the node placement algorithm described in section~\ref{subsec:startup-stage_chap4}.

Out of total RBF nodes, ${\ngauss = 200}$ nodes (number of Gaussian nodes) are selected from a multivariate Gaussian distribution, $\mathcal{N}(\vec \lambda_{\text{ref}}, \boldsymbol{\Sigma})$ where the mean is set as $\vec \lambda_{\text{ref}}$, and the covariance matrix, $\boldsymbol{\Sigma}$ is estimated by evaluating inverse of the Fisher matrix, $\boldsymbol{\Gamma}$~\cite{vallisneri2008use} at the reference point $\vec \lambda_{\text{ref}}$ (see Section ~\ref{subsec:nodesplacement_chap4}). We utilize the \texttt{gwfast}\cite{Iacovelli_2022} Python package to calculate the covariance matrix. The remaining nodes ${\nunif = 600}$ are uniformly sampled from the ranges specified in Table~\ref{tab:priordistr_chap4}. For generating the RBF interpolants, we employ the publicly available \texttt{RBF} python package~\cite{RBF_github}. A Gaussian RBF kernel ${\phi = \exp(-\varepsilon r^2)}$ is used as the basis function, where $\varepsilon$ serves as the shape parameter. We set $\varepsilon = 10$ through trial and error, but it can be optimized using the leave-one-out cross-validation~\cite{ROCHA20091573} method. The degree of the monomials is chosen as $\nu = 7$, and we retain the top $20$ basis vectors ($\ell=20$) for the reconstruction of ${\vec z^{(i)}(\vec \lambda^{q})}$.

\begin{figure}[!hbt]
\centering
\includegraphics[width=0.55\textwidth, clip=True]{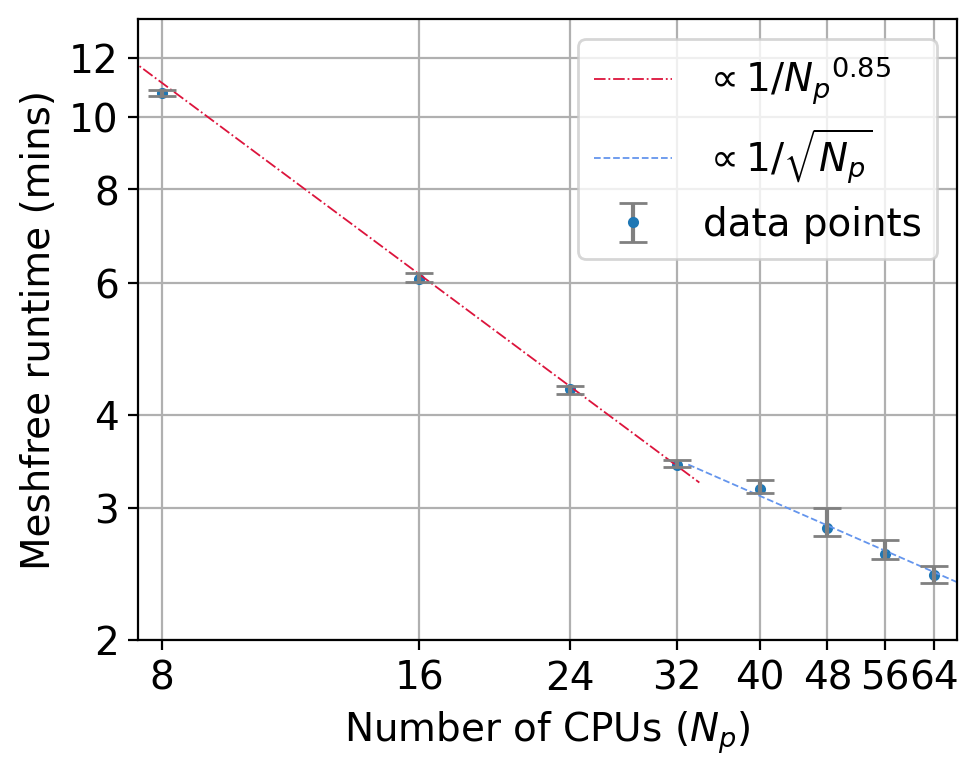}
\caption{The runtimes of the meshfree method have been plotted with varying the number of cores. We choose the GW170817 event for this specific example, and \texttt{TaylorF2} waveform model is taken for obtaining the posteriors. The plot shows a notable decrease in meshfree runtimes as the number of cores increased. We anticipate significant reductions in runtimes can be achieved with an even higher number of cores, though the gains will eventually saturate as we increase the number of cores.
}
\label{fig:timing_vs_cores}
\end{figure}
\begin{table}[hbt]
\centering
\begin{tabular}{c|c|rc}
\hline\hline
Parameters     &   Range    &   Prior distribution \\
\hline
$\mathcal{M}$  &   $[\mathcal{M}_{\text{cent}} \pm 0.0002]$ &   $\propto \mathcal{M}$ \\
$q$            &   $[q_{\text{cent}} \pm 0.07]$             & $ \propto \left [ (1 + q)/q^3 \right ]^{2/5}$     \\
$\chi_{1z, 2z}$   &   $[\chi_{1z \,\text{cent}} \pm 0.0025]$   & Uniform  \\
$d_L$          & $[10, 60]$                & Uniform in volume\\
$t_c$          & $t_{\text{trig}} \pm 0.12$& Uniform\\
$\alpha$       & $[0, 2\pi]$               & Uniform\\
$\delta$       & $\pm \pi/2$       & $\sin^{-1} \left [ {\text{Uniform}}[-1,1]\right ]$\\
$\iota$        & $[0, \pi]$                & Uniform in $\cos \iota$\\
$\psi$         & $[0, 2\pi]$               & Uniform angle\\
\end{tabular}
\caption{Prior parameter space over the ten-dimensional parameter space $\vec \Lambda$.}
\label{tab:priordistr_chap4}
\end{table}
To sample the ten-dimensional parameter space $\vec \Lambda$, we utilize nested sampling implemented in the \texttt{dynesty} python package. The likelihood function used for the analysis is $\ln \mathcal{L}_{\text{RBF}}$. The prior distributions and their boundaries for all sampling parameters are provided in Table~\ref{tab:priordistr_chap4}. The sampler configuration is outlined as follows: \texttt{nlive} $=500$, \texttt{walks} $=100$, \texttt{sample} = ``rwalk'', and \texttt{dlogz} $=0.1$. These parameters crucially determine the accuracy and time taken by the nested sampling algorithm. Here, \texttt{nlive} represents the number of live points. Opting for a larger number of \texttt{nlive} points results in a more finely sampled posterior (and consequently, evidence), albeit at the expense of requiring more iterations for convergence. \texttt{walks} denotes the minimum number of points needed before proposing a new live point, \texttt{sample} indicates the chosen sampling approach for generating samples, and \texttt{dlogz} signifies the proportion of the remaining prior volume's contribution to the total evidence, which functions as a stopping criterion for terminating the sampling process. Further details on the dynesty's nested sampling algorithm and its implementation can be found in ~\cite{speagle2020dynesty, sergey_koposov_2023_7600689}. For comparison, we also employ the bruteforce marginalized phase likelihood implemented in \pycbc with the dynesty sampler, using the same sampler configuration mentioned above.

The PE run with the meshfree likelihood (for \texttt{TaylorF2}) completed in $\sim 3.4$ minutes (including the startup and sampling stages) when run on 32 cores. In a similar analysis with the \texttt{IMRPhenomD}~\cite{khan2016frequency} waveform model, we found good agreement between the posterior distributions obtained from the meshfree and \pycbc ~likelihood methods, and these results are broadly consistent with the LVK analysis of the same event. In this case, the meshfree method generated the posterior in $\sim 3.6$ minutes utilizing $32$ cores.
\begin{figure*}[htpb]
\label{fig:corner_plot_tay}
\centering
\includegraphics[width=\linewidth]{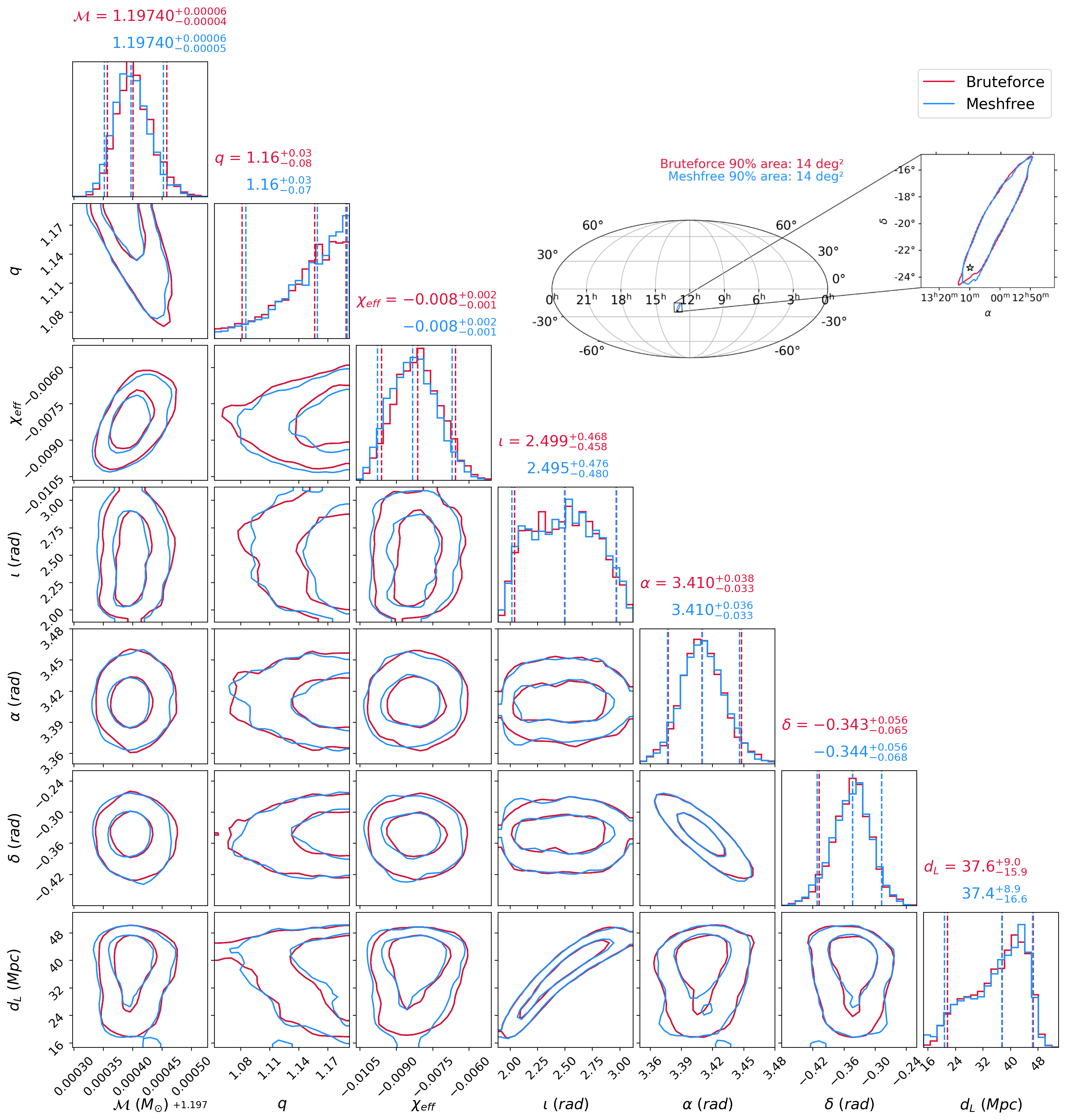}
\caption{\label{fig:cornerplot_20Hz} The corner plot shows the posterior distributions over a ten-dimensional parameter space using both the meshfree method and the standard bruteforce likelihood. The seismic cutoff frequency is set to $20$ Hz, assuming the \texttt{TaylorF2} waveform model. The estimated distributions from both methods exhibit good agreement, as indicated by the median values in the title of the 1-D marginalized posterior plots along the diagonal. The meshfree likelihood-based PE process was completed in $\sim 3.4$ minutes when performed on $32$ cores. 
 A skymap (inset) is also generated for both methods where the star denotes the actual location of the galaxy from which GW170817 is believed to have originated and is in agreement with~\cite{PhysRevX.9.011001}}. 
\end{figure*}
As shown in Fig.~\ref{fig:cornerplot_20Hz}, the posterior distributions estimated using the meshfree method and bruteforce likelihood exhibit good agreement. Additionally, sky maps for GW170817 were generated using the \texttt{ligo-sky-map}~\cite{ligo_sky_map} utility for two methods, further demonstrating the effectiveness of the meshfree method in reconstructing source parameters. For the same analysis using $64$ cores, the number of likelihood evaluations for meshfree PE (standard \pycbc~PE) is $1396388$ $(1404588)$ whereas the number of posterior samples collected during sampling is $3633$ $(3546)$.

To estimate the speed-up achieved by using the meshfree likelihood compared to the bruteforce likelihood, we generated $10^{4}$ random query points from the prior distribution specified in Table~\ref{tab:priordistr_chap4}. Subsequently, we calculated the log-likelihood at these points using both the meshfree and bruteforce likelihood methods. The results show a speed-up of $\sim 49$ over standard bruteforce likelihood, as shown in Fig.~\ref{fig:speedup}. In the case of \texttt{IMRPhenomD}, we observed a relative speed-up of  $\sim 498$ over standard bruteforce likelihood. This difference in speed-up can be attributed to the difference in time taken to generate the template in the two different waveform models: {\texttt{TaylorF2}} waveforms are far less computationally expensive to generate than  \texttt{IMRPhenomD} model waveforms. The corresponding error plots are shown in Appendix~\ref{appendix:likehood_errs}. 

Furthermore, we investigated the scaling behavior of PE run times (startup + online) using the meshfree method with respect to the number of cores employed. For the \texttt{TaylorF2} analysis of the GW170817 event, we performed the ten-dimensional PE analysis with varying core counts ranging from $8$ to $64$. Our study indicates that the analysis can be completed in $\sim 2.4$ minutes using $64$ cores. Furthermore, we found that the current implementation scales linearly with the number of live points (keeping \texttt{dlogz} fixed).

It is worth noting that further reductions in estimation times are expected if the number of cores is increased, as the scaling does not saturate at 64 cores. However, as depicted in Fig.~\ref{fig:timing_vs_cores}, the rate at which the estimation times decrease diminishes as the number of cores continues to increase, which may eventually result in run time saturation beyond a certain number of cores. The total time shown in Fig.~\ref{fig:timing_vs_cores} includes the time taken by both the start-up and online stages of the meshfree algorithm, in addition to the time required for the dynamic nested sampling algorithm to compute the evidence integral up to a desired level of accuracy. To clarify further:
(a) The start-up stage of the meshfree method benefits from the availability of many CPU cores as it can be performed in parallel. (b) Conversely, the online stage is not designed for parallel processing. This is because estimating the likelihood is a relatively fast operation, typically taking $\sim \mathcal{O}(1)$ ms, achieved by evaluating the interpolating functions. This part of the computational cost remains nearly constant regardless of the specific sample point being processed.
As a result, the runtime scaling, as depicted in Fig.~\ref{fig:timing_vs_cores}, effectively combines the time consumed by the Dynesty sampler and the start-up stage of our method.

\subsection{Simulated Data}
\label{subsec:simulated Data_chap4}
To assess the performance of our method across a range of SNRs, we generated simulated data sets with different source parameters that mimic binary neutron star (BNS) systems with network-matched filter SNRs spanning from $10$ to $100$. In our approach, the number of Gaussian nodes, denoted as $\ngauss$, for a given source is intricately tied to its SNR. Specifically, the likelihood profile for events with lower SNRs exhibits a relatively broader spread over the parameter space compared to the more sharply peaked likelihood profiles observed for high SNR events. Consequently, for higher SNR events, we select a larger number of nodes from the Gaussian distribution in contrast to lower SNR events. Thus, we opt for different values of $\ngauss$ based on the SNR of each source. We will now detail the distribution of source parameters.  

The component masses were uniformly drawn from the interval ${[1.2, \, 2.35] M_{\odot}}$, while the magnitude of the dimensionless component spins (aligned with the orbital angular momentum) was uniformly drawn from the interval ${[-0.05, \, 0.05]}$. The sources were uniformly distributed in sky location and followed a uniform distribution with $d_{L}$ ranging from $10$ to $200$ Mpc. The inclination angle $\iota$ of the binary's plane with respect to the line of sight was chosen to be uniformly distributed in $\cos \iota$. The tidal deformability parameter $\Lambda_{\text{tide}}$ of the neutron stars was set to zero for all signals. These signals were simulated using the {\texttt{TaylorF2}}  waveform approximant, and the noise PSD was chosen from {\texttt{aLIGOZeroDetHighPower}}~\cite{aLIGO_ZDHP} for the Livingston and Hanford detectors, and \texttt{AdvVirgo}~\cite{AdVirgo} for the Virgo detector, as implemented in \pycbc.

It's important to recognize that the distribution of injection parameters we've chosen may not hold any astrophysical relevance. These choices have been made solely for the purpose of testing our method across various SNR scenarios, as our node placement algorithm is intricately linked to the SNRs of these events.

\begin{figure}[!hbt]
\centering
\includegraphics[width=0.65\textwidth, clip=True]{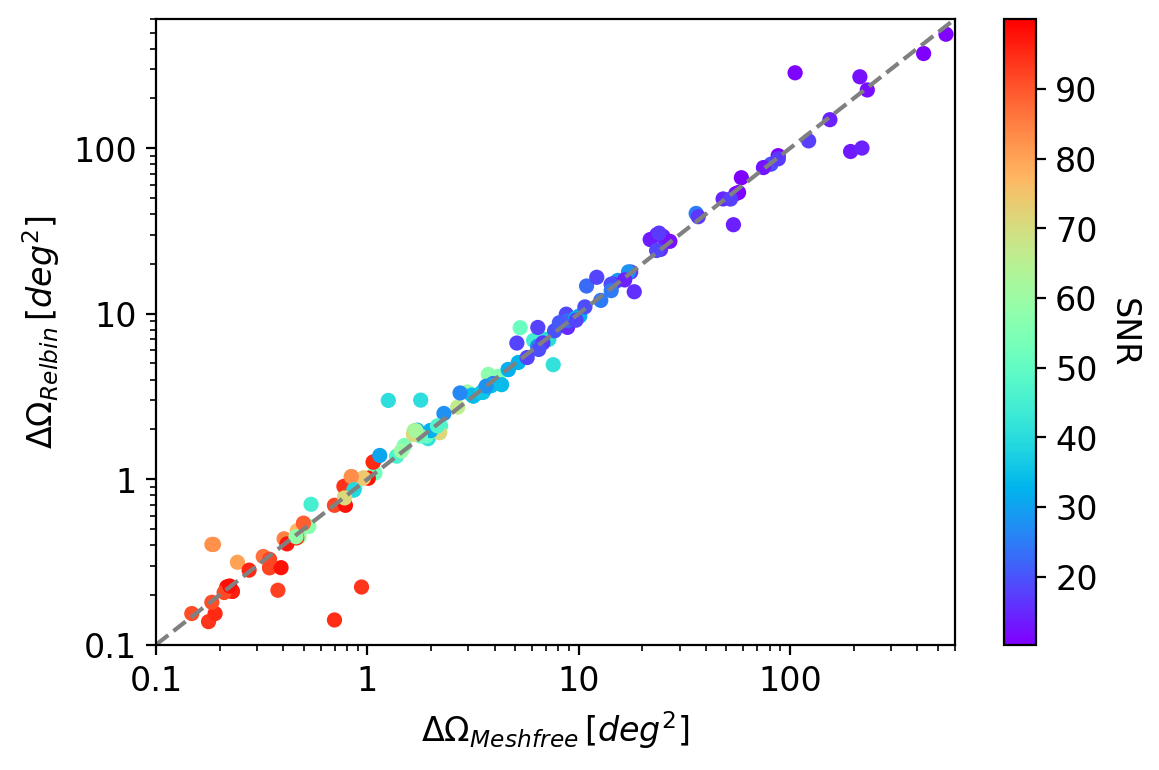}
    \caption{$90\%$ credible interval (CI) sky area as a function of SNR are shown. We compared the accuracy of meshfree with the \relbin method. The scattered points lie along the positive slope (dashed straight line), implying that the meshfree method has similar accuracy as the \relbin in estimating these parameters, i.e., sky location. It is observed that estimating these parameters with high SNR events is more accurate than with low SNR events. Note that there are some points that are away from the straight line. This is due to the fact that posteriors corresponding to those points are multimodal, giving rise to higher areas for one method in comparison to the other method.
}
\label{fig:simulated_error}
\end{figure}
\begin{figure}[!hbt]
\centering
\includegraphics[width=0.55\textwidth, clip=True]{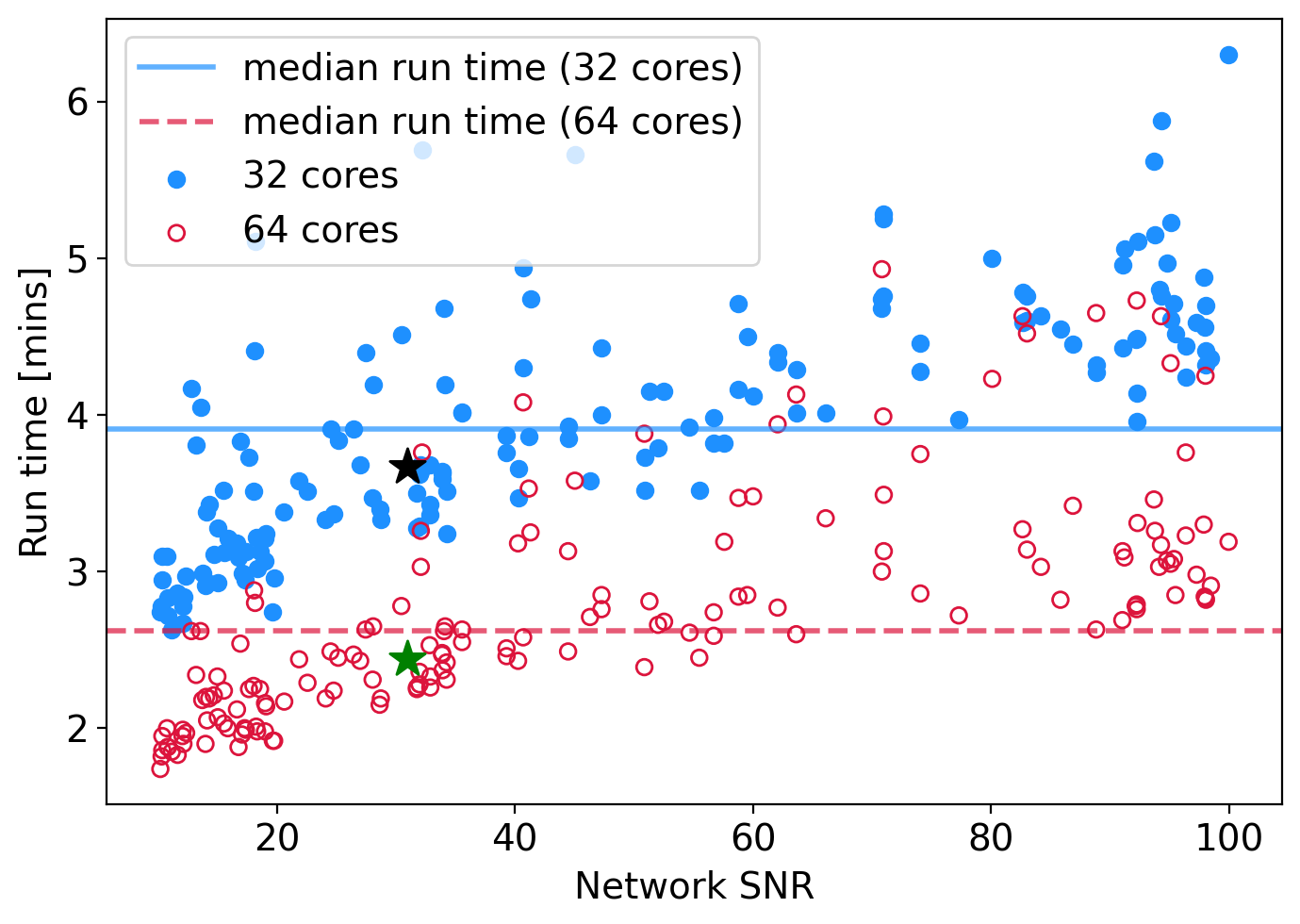}
\caption{
We present the run times for reconstructing a number of simulated BNS events injected in Gaussian noise (Sec.~\ref{subsec:simulated Data_chap4}) using the meshfree method. Higher SNR injections took longer to reconstruct due to their narrowly peaked likelihood. The analysis utilized 32(64) cores, represented by blue dots (red circles). The black(green) star shows the GW170817 event's run time on 32(64) cores for reference. With 64 cores, the median run time decreased to around $\sim 2.6$ min compared to $\sim 3.9$ min with 32 cores. For \relbin~method, using $\epsilon = 0.000198$, the median run time is $\sim 22$ minutes on $32$ cores. The \relbin~PE run times crucially depend on the value $\epsilon$, and taking higher values of $\epsilon$ can result in lower run times.
}
\label{fig:simulated_timings}
\end{figure}
At this point, we employ both the meshfree technique and the relative-binning method (\relbin) for estimating the BNS source parameters. In an ideal scenario, we would select the brute force likelihood method for comparison, but opting for such an approach might result in extended PE run times. Given that \relbin ~ is a well-established and widely used fast PE algorithm within the GW community, recognized for its combination of speed and accuracy, we have opted for this alternative over the standard approach. 

For all the simulated events, we chose $N = 800$ RBF nodes. However, $\ngauss$ was empirically chosen depending on the SNR of the events. Heuristically we found that for SNR $\geq 50$, $\ngauss \in [0.4, 0.6]$ was a suitable choice. While for events with SNR $\leq 50$, $\ngauss$ was chosen from $[0, 0.2]$. There were few exceptions to this prescription as well.

For the ~\relbin  method, we set $\epsilon = 0.000198$, which determines the number of frequency bins in the \relbin ~ method (For more details, refer to ~\cite{Venumadhav2018, cornish2021heterodyned}). This parameter determines the accuracy and run-time of the relative binning method. We choose the reference parameters (or ``fiducial point'' in the case of ~\relbin) to be the injection parameters of each simulated event, even though in a realistic scenario, it would have been obtained by an optimization process starting from the best-matched search template.

The prior distributions over the extrinsic parameters ${(\alpha, \delta, \iota, \psi, d_L)}$ used in the PE are the same as those used for the GW170817 event in the earlier section. As for the intrinsic parameters, we select them according to the following distributions: (i) Uniform distribution in component masses $(m_1, m_2)$ such that ${\mathcal{M} \in \mathcal{M}_{\text{ref}} \pm 0.0002}$ and ${q \in q_{\text{ref}} \pm 0.07}$\footnote{In cases where $q_{\text{ref}} - 0.07 < 1$, we sample from $q \in [1, 1.14]$.} (ii) Uniformly in dimensionless component spins from the interval ${\chi_{1z/2z} \in {\chi_{1z/2z}}_{\text{ref}} \pm 0.0025}$ where $\mathcal{M}_{\text{ref}}, q_{\text{ref}}, {\chi_{1z/2z}}_{\text{ref}}$ are chirp mass, mass-ratio, and aligned spins components respectively, corresponding to the reference point found by the optimizing the network-SNR starting from the best-matched template. The prior over the time of coalescence $t_c$ is chosen to be uniform from the interval ${t_{\text{ref}} \pm 0.12}$ where $t_{\text{ref}}$ is the geocentric time of the trigger. 

As can be seen from the $90\%$ CI sky-area distribution in Fig.~\ref{fig:simulated_error}, the meshfree method accurately recovers the sky location with a similar level of accuracy as ~\relbin. The timing results for the simulated signals, shown in Fig.~\ref{fig:simulated_timings}, illustrate that the run-time increases with the SNR since signals with higher SNRs exhibit a more sharply peaked likelihood function that requires a longer time to accumulate a sufficient number of posterior samples to evaluate the evidence integral upto the desired accuracy.

All the tests were performed on AMD EPYC 7542 CPU@2.90GHz processors.

\section{Conclusion and Outlook}
\label{sec:concl_outl_chap4}
This work extends our previously prescribed meshfree method \cite{PhysRevD.108.064055} to perform parameter estimation on a ten-dimensional parameter space for aligned spin waveform models using data from a multi-detector network framework. In the previous version of our method, we demonstrated its effectiveness on a simulated BNS event, as seen in a single detector, where we directly chose the injection parameters as our center for placing uniformly distributed RBF nodes. In this work, we start from the best-matched template parameters and optimize the network-SNR to reach a nearby point with a higher network-SNR value, which acts as a guide for placing RBF nodes. This node placement algorithm incorporates a blend of nodes from both multivariate Gaussian and uniform distributions. It enhances the accuracy of likelihood reconstruction (see Appendix~\ref{appendix:effect_of_nGauss} for a study of marginalized posterior profiles on different fractions of random interpolation nodes drawn from a multivariate Gaussian distribution). 

We tested our method for several simulated BNS events with varying SNRs and found good agreement between the meshfree and the relative binning methods. Further, we have demonstrated that our meshfree method can rapidly estimate the posteriors of the GW170817 binary neutron star (BNS) event and efficiently locate the electromagnetic (EM) counterpart in the sky within ${\sim}$2.4~(3.4) minutes after detection using 64~(32) CPU cores. 

These run times quoted above need to be put in the context of other fast-PE algorithms for which we chose the relative-binning method. Note that the time taken by the relative binning algorithm crucially depends on the choice of the $\epsilon$ parameter, which sets the frequency-bin resolution in the heterodyning process. We set $\epsilon=0.000198$ for which the time taken by the relative-binning method was found to take $\sim 20$ minutes on 32 cores to analyze the GW170817 event using data from a 3-detector (HLV) network. This choice of $\epsilon$ was taken to match the PE run times claimed by Finstad and Brown (2020)~\cite{Finstad_2020}. The sampler parameters were kept identical for both methods. Since we do not have the hardware (nor know the parameters) used by Finstad and Brown, this ad-hoc procedure was used to arrive at this reference time. As such, it won’t be prudent to arrive at any definite comparative conclusion based on these timing results. Nevertheless, we will continue to tune the meshfree method to explore further optimizations for speeding up the PE analysis. 

Note that the run time quoted here doesn't include the time spent in finding the center using optimization, which, when run on a single core, took $\sim 32$ seconds. However, it can be sped up by running in parallel over multiple CPU cores using multiprocessor versions of the optimization algorithm. For example, see the parallel version of the \texttt{scipy.optimize}~\cite{RJ-2019-030, florian_gerber_2020_3888570} Python routine. 

We also used our method for the NSBH event GW200115~\cite{NSBH_GW20015} (See Appendix~\ref{appendix:GW200115_app}) and obtained a reasonably good match of posterior samples between our method and bruteforce likelihood. Our meshfree framework has the potential to contribute to future LIGO low-latency PE efforts. However, it is also essential to acknowledge some limitations of the current implementation of our method. 
Firstly, the RBF meshfree interpolating functions are seen to be accurate only over a relatively narrow domain in the intrinsic parameter space, which leads to using narrow boundaries to carry out the PE analysis. This constraint restricts the flexibility of the meshfree PE analysis compared to standard approaches. A potential solution to overcome this issue could be to divide the intrinsic parameter space into smaller ``patches'' and generate interpolants for each patch independently~\cite{Borne2019DomainDM}. 

In the present implementation, the prior bounds (albeit narrow) were obtained from the best-matched template as explained in Sec.~\ref{subsec:nodesplacement_chap4}. Since the PE pipeline is expected to be triggered by an upstream search pipeline, we have used the most significant trigger parameters ($\vec \lambda^\ast$) as the starting point. Using a suitable optimization algorithm,  this point ultimately leads us to a nearby reference point ($\vec \lambda_{\text{ref}}$), which serves as the center of the hyper-rectangle from which the samples are drawn. The dimensions of this hyper-rectangle along the different directions are heuristically set to get good accuracy in the likelihood reconstruction (i.e. we do not use the LVK results apriori to determine the prior bounds).
In future, we would like to explore an automated way of determining the dimensions of this hyper-rectangle from the size of the $90\%$ overlap ellipsoid calculated at $\vec \lambda_{\text{ref}}$. Such an idea has been mooted in Pankow et al.~\cite{Pankow_2015}. However, since the axes of the overlap ellipsoid do not necessarily align with the eigen-directions of the data-driven covariance matrix, such an implementation would require careful thought. Secondly, it is worth noting that despite improvements in the placement of nodes compared to our previous work, it still needs investigation for a more generic prescription that can be applied to any CBC system. Incorporating an adaptive strategy~\cite{Zhang:2017aa} to identify significant nodes could enhance interpolation accuracy and strengthen the node placement strategy. 

In this analysis, we have focused exclusively on non-precessing signal models (\texttt{TaylorF2} and \texttt{IMRPhenomD}), with a particular emphasis on the dominant harmonic of the GW signal. These models do not incorporate tidal terms~\cite{DelPozzo:2013ala, Dietrich_2019}. Since the tidal deformability parameters corresponding to the binary components are intrinsic in nature, we need to include them as interpolating parameters in our method, which increases the dimensionality of the intrinsic parameter space to six. Additionally, a greater number of nodes are required to ensure an accurate interpolation of the SVD coefficients and template norm square. We will explore these ideas in future follow-ups of this work (see Appendix~\ref{appendix:tidal_deform_meshfree} for preliminary results).

Although we anticipate that extending our method to include higher-order harmonics should be relatively straightforward, it may result in increased start-up costs. In the case of precessing-spin waveform models, a meshfree framework is particularly suitable since the number of RBF nodes does not exponentially increase with the dimensionality of the parameter space. In this regard, the relative binning algorithm has been recently implemented for gravitational-wave parameter estimation with higher-order modes and precession~\cite{narola2023relative}. In the future, we would like to overcome these limitations by enhancing the node placement algorithm further, expanding the boundaries of the sample space, and incorporating models that account for in-plane spins~\cite{IMRPhenomXPHM}. 

%
%

In the next chapter, we apply this coherent PE meshfree method for a large-scale simulation study of the impact of including LIGO-Aundha (A1) in the International Gravitational Wave Detector Network (IGWN) on the localization of BNS sources.

\section*{Acknowledgements}
We thank Vaibhav Tiwari for carefully going through our manuscript and giving helpful suggestions. We also thank Abhishek Sharma and Sachin Shukla for the helpful discussions at various stages of this work. We especially thank the anonymous referee for their careful review and helpful suggestions. 

L.~P. is supported by the Research Scholarship Program of Tata Consultancy Services (TCS). S.~M. acknowledges INSPIRE fellowship by the Department of Science and Technology (DST), India. A.~R is supported by the research program of the Netherlands Organisation for Scientific Research (NWO). A.~S. gratefully acknowledges the generous grant provided by the Department of Science and Technology, India, through the DST-ICPS (Interdisciplinary Cyber Physical Systems) cluster project funding. We thank the HPC support staff at IIT Gandhinagar for their help and cooperation. 

Code availability: Codes used in this analysis are publicly available in the following Github repository~\cite{meshfree_github}.
%
\clearpage
\section{Appendix}
\subsection{Effect of different choices of Gaussian distributed nodes on the posteriors}
\label{appendix:effect_of_nGauss}
The distribution of input nodes plays an important role in the meshfree PE algorithm as described in Section~\ref{subsec:nodesplacement_chap4}. One would like to place nodes intelligently so as to provide good coverage near the support of the log-likelihood function, which falls very steeply around its mode. Here, we show the effect of node placement on the posteriors obtained. We choose a certain fraction of RBF nodes ($\ngauss$) from a normal (Gaussian) distribution described by the covariance matrix. The remaining nodes are distributed uniformly over the sample space. We find that the overall accuracy increases as we choose a higher fraction of Gaussian distributed nodes (particularly on the $\iota$ and $d_L$ parameters). Although, the estimations of $\alpha$, $\delta$, and $\mathcal{M}$ are quite robust against different choices of $\ngauss$.
\begin{figure*}[!hbt]
\centering
\includegraphics[width=0.9\linewidth]{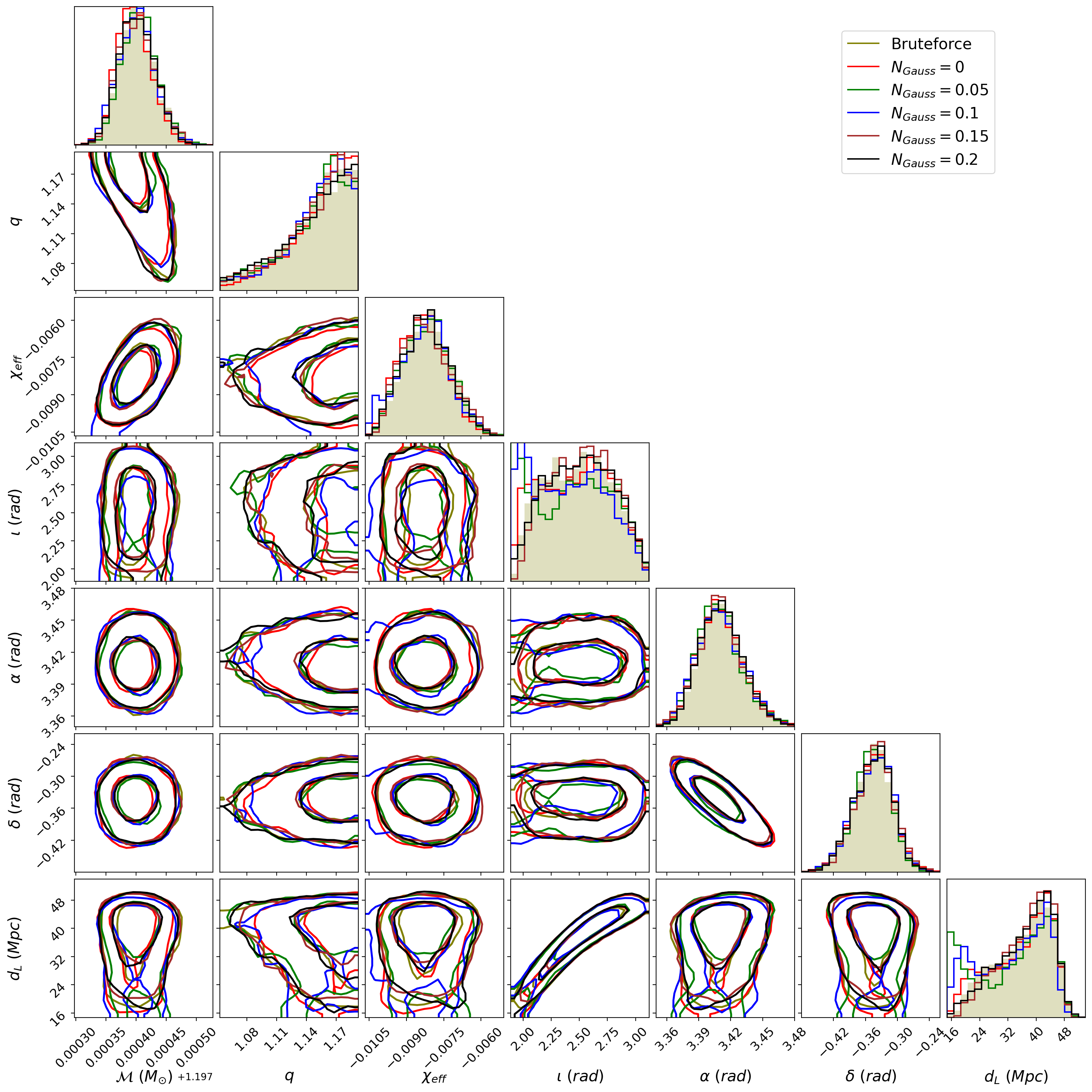}
\caption{The effect of choosing different values of $\ngauss$ (fraction of Gaussian nodes) on the reconstructed parameters.}
\label{fig:corner_plot_Ngauss}
\end{figure*}
%
\subsection{Likelihood errors}
\label{appendix:likehood_errs}
To assess the accuracy of the meshfree method to approximate the likelihood, we generated $10^4$ random query points from the prior distribution specified in Table~\ref{tab:priordistr_chap4}. Then we calculated the likelihood using both the meshfree and bruteforce likelihood methods. The median absolute error for both \texttt{TaylorF2} and \texttt{IMRPhenomD} waveform models were found to be $\mathcal{O}(10^{-2})$.
\begin{figure*}[hbt!]
\begin{subfigure}{0.49\linewidth}
    \includegraphics[width=\linewidth]{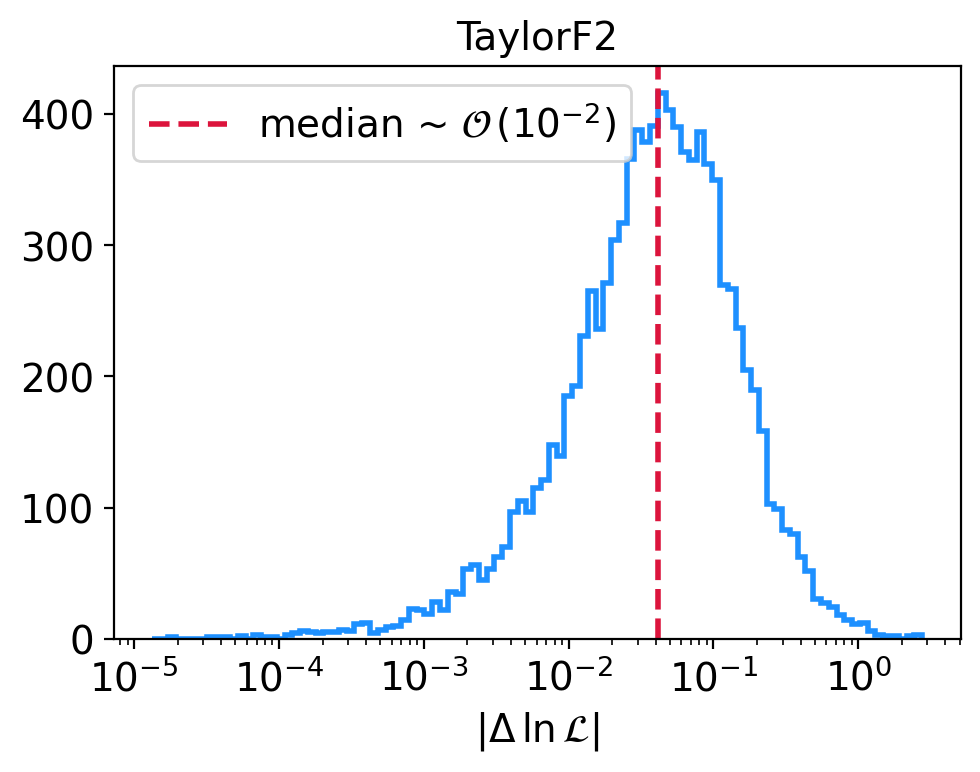}
    \caption{\texttt{TaylorF2} signal model}
    \label{figerror:TF2}
\end{subfigure}\hfill
\begin{subfigure}{0.49\linewidth}
    \includegraphics[width=\linewidth]{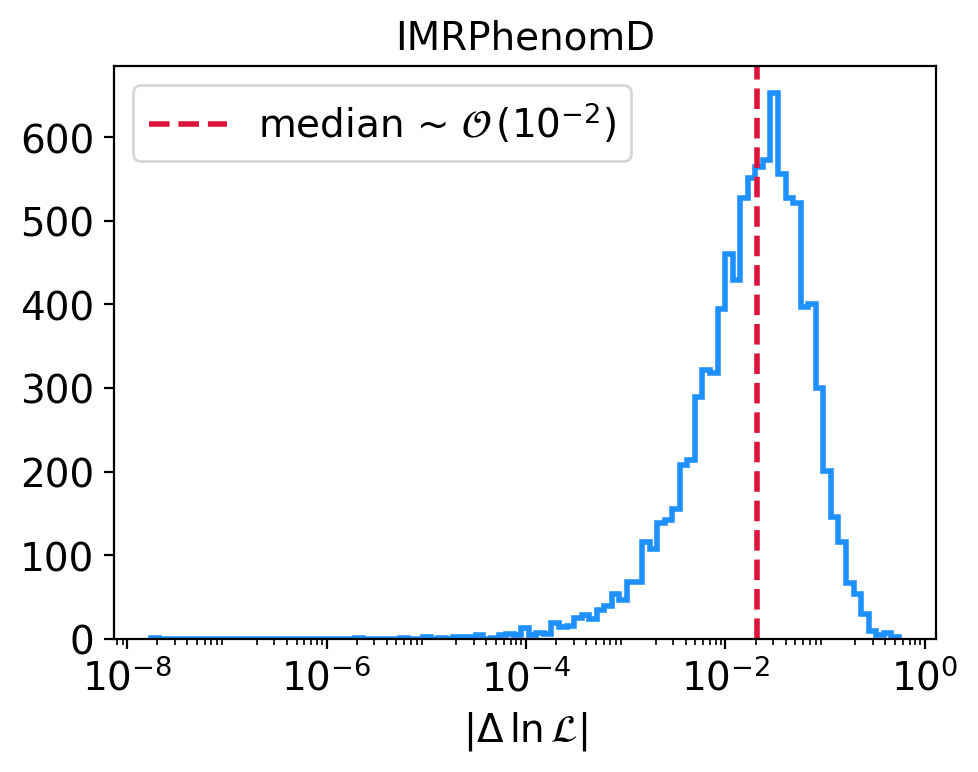}
    \caption{\texttt{IMRPhenomD} signal model}
    \label{figerror:IMR}
\end{subfigure}
\caption{The likelihood errors ($\Delta \ln \mathcal{L} \equiv \ln \mathcal{L}_{\text{PyCBC}} - \ln \mathcal{L}_{\text{RBF}}$) are shown for both \texttt{TaylorF2} and \texttt{IMRPhenomD} waveform models (on GW170817 event). The red-dashed line represents the median absolute error ($|\Delta \ln \mathcal{L}|$) that are $\sim \mathcal{O} \, (10^{-2})$. The are very few points with high errors ($|\cdot| \sim 4$($2$) for \texttt{TaylorF2}(\texttt{IMRPhenomD})) which are not shown here.}
\label{fig:error}
\end{figure*}
\subsection{Surface of SVD coefficients and template norm square}
\label{appendix:svd_coeff_surf}
\begin{figure*}[!hbt]
\centering
\includegraphics[width=\linewidth]{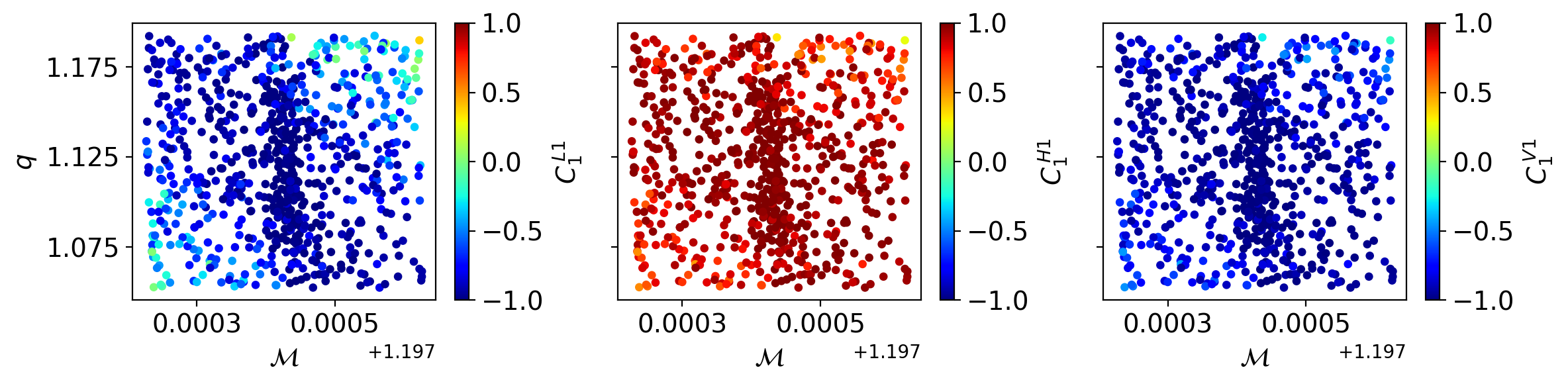}
\caption{Surface of the SVD coefficients corresponding to the first basis for three detectors (L1, H1, and V1). Note that they are smoothly varying functions of $\mathcal{M}$ and $\eta$, thereby making them suitable for interpolation.}
\label{fig:svd_coeff_surf}
\end{figure*}

As mentioned in Sec.~\ref{subsec:svdinterpolantsgeneration_chap4}, the SVD coefficients need to smooth over the intrinsic parameter space in order to be interpolated over it. Fig.~\ref{fig:svd_coeff_surf} shows the smooth variation of the SVD coefficients ($C^{q (i)}_{\mu}$) with $\mathcal{M}$ and $\eta$. Similarly, the surface of the template norm square ($\sigma^2(\vec \lambda^q)^{(i)}$) shows smooth variation over the sample space so that we can interpolate it using RBFs.
\clearpage
\subsection{Addition of tidal deformability in the meshfree framework}
\label{appendix:tidal_deform_meshfree}
To demonstrate meshfree PE for a case where we also include tidal parameters ($\Lambda_1,\, \Lambda_2$), we use the example of the GW170817 event with the $\text{IMRPhenomD\_NRTidalv2}$~\cite{PhysRevD.99.024029} waveform model. We choose a uniform prior over $\Lambda_1$ and $\Lambda_2$ lying between $[100 - 400]$ and spray $2500$ nodes, and retain the top-$20$ basis vectors to be used for snr reconstruction. For comparison, we also run a bruteforce PE (using PyCBC standard likelihood). The meshfree PE took $\sim 5.2$ minutes (including start-up), whereas the bruteforce PE took $\sim 3$ hours $25$ minutes. Both methods utilized $64$ CPU cores. Here are the corner plots of the resulting posterior distributions:
\begin{figure*}[!hbt]
    \centering
    \includegraphics[width=0.75\linewidth]{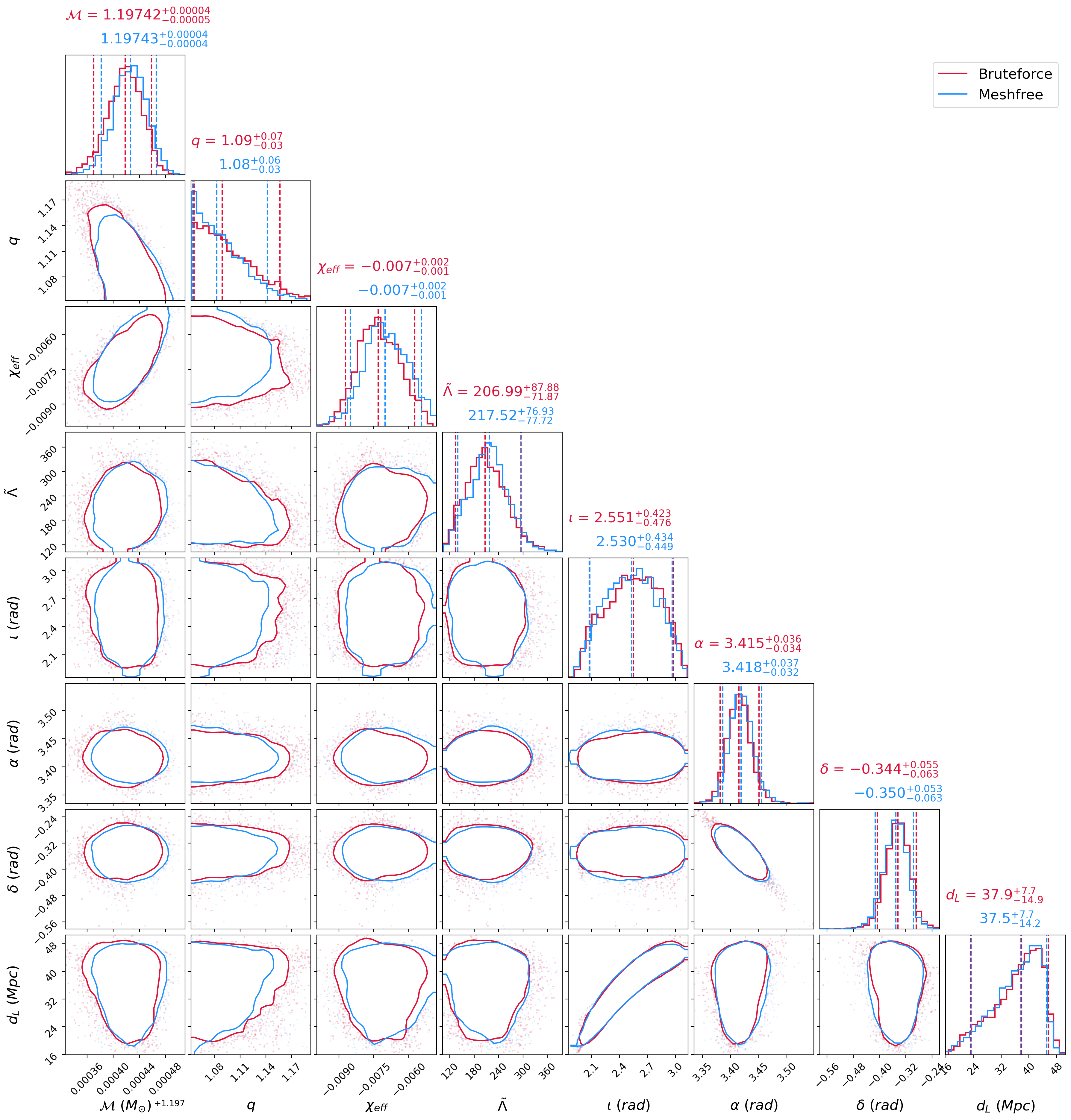}
    \caption{Posterior distribution of various binary parameters, including the tidal deformability parameters. Note that we have shown the posterior over $\tilde{\Lambda}$~\cite{PhysRevX.9.011001, PhysRevD.89.103012} (a weighted linear combination of two tidal parameters, {$\tilde{\Lambda} = \frac{16}{13} \frac{(m_1 + 12 m_2)m_1^4\Lambda_1 + (m_2 + 12 m_1)m_2^4\Lambda_2}{(m_1 + m_2)^5}$}), which is a leading tidal contribution to the GW phase evolution.}
    \label{fig:gw170817_tidal_meshfree}
\end{figure*}
\clearpage

\subsection{GW200115: An NSBH event}
\label{appendix:GW200115_app}
\begin{figure*}[!hbt]
    \centering
    \includegraphics[width=0.75\linewidth]{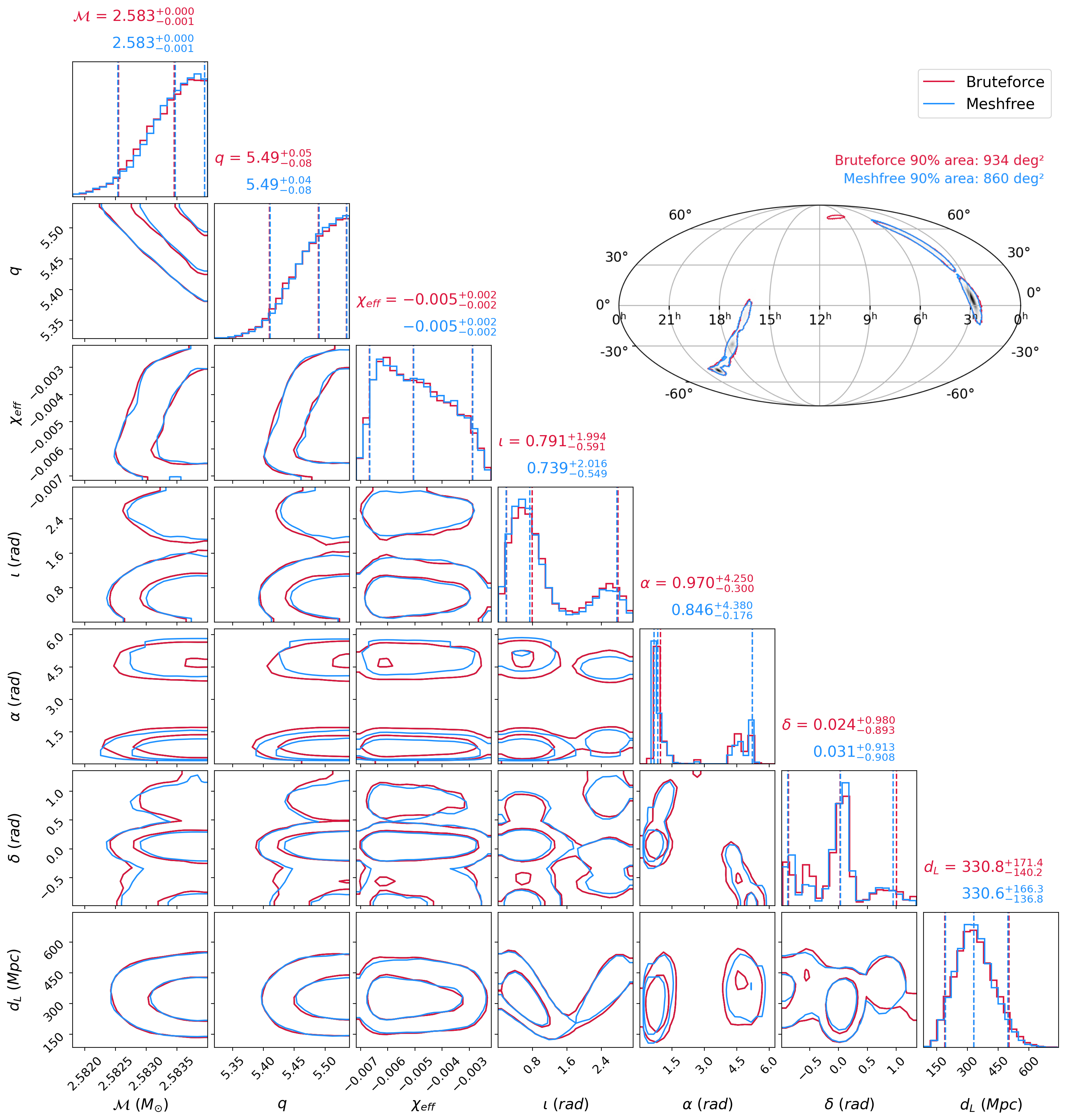}
    \caption{The corner plot shows the posterior distributions of ten-dimensional parameters using two different methods: the meshfree method and the bruteforce ~ likelihood. The estimated distributions from both methods exhibit good agreement, as indicated by the median values in the marginalized posterior titles. The \textit{inset} shows the corresponding skymaps and are consistent with the results published by the LIGO, Virgo, and Kagra collaborations~\cite{Abbott_2021}.}
    \label{fig:corner_skymap_plot_GW200115}
\end{figure*}

We present the results of reconstructing the GW200115 NSBH event using the meshfree and the bruteforce ~ likelihood (for reference). 
The seismic cutoff frequency is set to $20$ Hz, assuming the \texttt{IMRPhenomD} waveform model. We used a $64$ seconds data segment around the GW200115 NSBH event, and PSD was generated using the same data with \texttt{median-mean} estimation. The center of the sample space was chosen as the MAP values of the PE samples obtained by previous LIGO analysis~\cite{abbott2021gwtc-3}. The parameters for RBF interpolation were the same as the one used for the GW170817 BNS event earlier, except $\nu = 10$ and $\epsilon = 20$. The sampling parameters for \texttt{dynesty} nested sampler were the following: (i) \texttt{nlive} $= 2500$, (ii) \texttt{walks} $= 350$, (iii) \texttt{sample} = `rwalk'', and (iv) \texttt{dlogz} $ = 0.1$. The meshfree likelihood-based PE process (see Fig.~\ref{fig:corner_skymap_plot_GW200115}) was completed in $\sim 14$ minutes when performed on $64$ cores. 
%
\chapter{Pinpointing coalescing binary neutron star sources with the IGWN, including LIGO-Aundha}
\label{chap:chapter_5}
\paragraph{\textbf{Abstract}}
LIGO-Aundha (A1), the Indian gravitational wave detector, is expected to join the International Gravitational-Wave Observatory Network (IGWN) and begin operations in the early 2030s. We study the impact of this additional detector on the accuracy of determining the direction of incoming transient signals from coalescing binary neutron star (BNS) sources with moderately high signal-to-noise ratios. 
It is conceivable that A1's sensitivity, effective bandwidth, and duty cycle will improve incrementally through multiple detector commissioning rounds to achieve the desired `LIGO-A+' design sensitivity. For this purpose, we examine A1 under two distinct noise power spectral densities. One mirrors the conditions during the fourth science run (O4) of the LIGO Hanford and Livingston detectors, simulating an early commissioning stage, while the other represents the A+ design sensitivity. We consider various duty cycles of A1 at the sensitivities mentioned above for a comprehensive analysis. 
We show that even at the O4 sensitivity with a modest $20\%$ duty cycle, A1's addition to the IGWN leads to a $15\%$ reduction in median sky-localization errors ($\Delta \Omega_{90\%}$) to $5.6$~sq.~deg. At its design sensitivity and $80\%$ duty cycle, this error shrinks further to $2.4$~sq.~deg, with 84\% sources localized within a nominal error box of $10$~sq.~deg! This remarkable level of accuracy in pinpointing sources will have a positive impact on Gravitational Wave (GW) astronomy and cosmology. Even in the worst-case scenario, where signals are sub-threshold in A1, we demonstrate its critical role in reducing the localization uncertainties of the BNS source. 
Our results are obtained from a large Bayesian parameter estimation study using simulated signals injected in a heterogeneous network of detectors using the recently developed meshfree approximation aided rapid Bayesian inference pipeline. We consider a seismic cut-off frequency of 10 Hz for all the detectors.
We also present hypothetical improvements in sky localization for a few Gravitational-Wave Transient Catalog-like events injected in real data after including a hypothetical A1 detector to the sub-network in which such events were originally detected. We also demonstrate that A1's inclusion could resolve the degeneracy between the luminosity distance and inclination angle parameters, even in scenarios where A1 does not directly contribute to improving the network signal-to-noise ratio for the event.

\section{Introduction}
\label{sec:intro}

\label{subsec: GW170817 and multimessenger astronomy}
This chapter is based on the publication \textit{Pinpointing Binary Neutron Star sources with a global network of GW Detectors, including LIGO-Aundha}, \href{https://link.aps.org/doi/10.1103/PhysRevD.109.044051}{Phys. Rev. D \textbf{109}, 044051 (2024)}.\newline

The detection and prompt localization of GW170817 event~\cite{PhysRevLett.119.161101} can be regarded as 
a monumental discovery that led to follow-up observations over the entire electromagnetic (EM) spectrum. This discovery resulted in the first extensive multi-messenger astronomical observing campaign~\cite{Abbott_2017} undertaken to follow up post-merger emissions from compact binary coalescence. The concurrent observations of such events via electromagnetic, neutrino, and gravitational wave (GW) detectors facilitate complementary measurements of phenomena that would otherwise remain inaccessible when observed independently. For example, the standard siren measurement of Hubble-Lema\^{i}tre constant $H_0$ from the gravitational wave data is independent of the cosmic distance-ladder method as opposed to that in astronomy with EM radiation. This provides a complementary measurement of $H_0$, which can be used to elucidate the Hubble-Lema\^{i}tre tension associated with the disparity between the measurements of Hubble constant~\cite{Riess_2019,2017, Abbott_2021} in the early and late universe.

\label{Different bands of EM emissions & related info}
Early observations of post-merger emissions of binary neutron star (BNS) sources 
can be used to constrain the physical models behind the internal mechanisms of these emission processes. 
For instance, there are different models proposed explaining the origin of the early blue emission of the kilonova associated with GW170817. Even though the model proposing radioactive decay of heavy elements in low-opacity ejecta being a theoretically motivated candidate~\cite{Arcavi_2018, Grossman_2014, Roberts_2011, Metzger_2010} fits the rise time curve well enough, there are also other models (for example: by cooling of shock-heated ejecta) which fit the decline of the emission equally well. In the case of GW1701817, the associated kilonova was detected ${\sim 11}$ hours after the merger, limiting the information about the rise time of the kilonova, particularly in the ultraviolet band~\cite{Arcavi_2018}. Capturing the rise time of the emission light curves by early detection may provide a useful measure to differentiate between the kilonova origin models. The gamma-ray burst (GRB) GRB 170817A was detected independently $\sim 1.7$ seconds after the trigger time of GW170817, with studies later confirming its association with the BNS merger~\cite{Abbott_2017_GRB}. This association confirmed BNS mergers as the progenitors of at least certain short GRBs~\cite{Abbott_2017_GRB}. 
The simultaneous detection of GWs and GRB may provide remarkable insights into the central engine of short gamma-ray bursts (SGRBs). The time delay between the GW and GRB events ($\sim 1.7$ secs as in the case of GW170817) offers valuable information into the underlying physics and intrinsic processes within the core, including the formation of a remnant object and the subsequent jet. It is expected that the EM waves and GW must have identical propagation speeds. The time delay between the GW and GRB events can be used to put constraints on the deviation of the speed of gravity waves from the speed of light, hence allowing for the tests of fundamental laws of physics~\cite{Abbott_2017_GRB}. Radio emissions enable tracing the fast-moving ejecta from BNS coalescence, providing insights into explosion energetics, ejecta geometry, and the merger environment~\cite{Geng_2018, Ghirlanda_2019}. Meanwhile, X-ray observations are crucial for determining the energy outflow geometry and system orientation relative to the observer's line of sight~\cite{Troja_2017}. The post-merger transients possess observational time scales ranging from a few seconds to weeks, owing to their potential to harness radiation across the entire EM spectrum. Detecting EM counterparts to $\sim 50$ BNS mergers may enable determining $H_0$ with $2\%$ fractional uncertainty~\cite{Chen_2018}, sufficient to verify the presence of any systematic errors in the local measurements of $H_0$. Even in the absence of potential EM counterparts, the accurate localization of compact binary coalescence (CBC) sources allows for dark siren measurements of $H_0$~\cite{1986Natur.323..310S, Finke_2021, Soares_Santos_2019}. For instance, observations from more than $\sim 50$ BNS dark sirens may be required to obtain $H_0$ measurements with $6\%$ fractional error~\cite{Chen_2018}. Thus, pinpointing these mergers is key in providing promising probes of fundamental physics, astrophysics, and cosmology. 

\label{Rapid PE for EM follow-up & Intro to LIGO India}
However, in order to accurately locate the EM counterparts and supplement regular follow-ups, more accurate and rapid 3D-localization of the sources using GW observations is of utmost importance. With imminent improvements in the detector sensitivities for future observing runs, there is an inevitable need for rapid Parameter Estimation (PE) methods. This arises from the fact that an increased bandwidth of detectors towards the lower cut-off frequencies would result in a momentous increase in the computational cost of Bayesian PE, hence affecting the prompt localization of the GW source. 
Currently, LIGO-Virgo-KAGRA Collaboration (LVK)~\cite{Abbott_2020_Prospects} uses a Bayesian, non-MCMC-based rapid sky localization tool, known as {\texttt{BAYESTAR}}~\cite{Singer_2016}, to locate (posterior distributions over sky location parameters, $\alpha$, and $\delta$) CBC sources within tens of seconds following the detection of the corresponding GW signal. However, as shown by Finstad et al.~\cite{Finstad_2020}, a full Bayesian analysis including both intrinsic and extrinsic parameters can significantly increase the accuracy of sky-localization (by ${\sim {14 \,\text{deg}^2}}$ in their analysis) of the CBC sources. This is of primary significance in reducing the telescope survey area \& time for locating the EM counterparts, making rapid Bayesian PE-based methods a preferable as well an evident necessity. Nevertheless, since \texttt{BAYESTAR} can construct skymaps in the order of a few tens of seconds, it would be an interesting case to test if the \texttt{BAYESTAR} skymap samples can be used as priors for sky location estimates in rapid Bayesian PE methods. This scheme might enable a more accurate localization measurement of the source. Future improvements in detector noise sensitivities shall lead to an increase in the detector ``sensemon range''\footnote{\textit{``sensemon range"} is defined as the radius of a sphere of volume in which a GW detector could detect a source at a fixed SNR threshold, averaged over all sky locations and orientations~\cite{Virgo_status_research_gate}. For lower redshifts ($z \lesssim 1$), the \textit{sensemon range} is approximately equal to the horizon distance divided by $2.264$.~\cite{Chen_2021}}~\cite{lsc2010sensitivity, Chen_2021, PhysRevD.47.2198}. 
Hence, higher BNS detections are expected from distances further away than current ranges~\cite{Pankow_2020}.
%
It is projected that $\sim 180$ BNS events could be detected in future O5 run~\cite{kiendrebeogo2023updated}. Hence, it is judicious to start the EM follow-up right from the merger epoch, which would require early warning alerts~\cite{Sachdev_2020, Magee_2021} for the EM telescopes in future runs. 
Given the finite observational resources of EM facilities allotted for the GW localized regions, it is important to prioritize the follow-up based on the chirp-mass estimates of GW events, as suggested by Margalit et al.~\cite{Margalit_2019}. This can be facilitated by a rapid Bayesian PE analysis, which, in addition to the sky localization, more importantly, provides a significantly accurate estimation of chirp mass for these compact binary systems. However, there are studies~\cite{Biscoveanu:2019ugx} which show that the chirp mass can be estimated accurately (with uncertainty no larger than $\sim 10^{-3}\, M_{\odot}$)  in low-latency searches. Although the mass ratio and effective aligned spin estimates may be severely biased in this case. With the addition of new detectors in the ground-based detector network, it becomes important to observe improvements expected in the sky localization and source parameter estimations of these compact binary sources.

LIGO-Aundha~\cite{LIGO_India, LIGO_India_location, UNNIKRISHNAN_2013, unnikrishnan2023ligoindia} (hereafter, `A1') is set to join the network of ground-based GW detectors in the early years of the next decade. Based on our experience with currently operational interferometric GW detectors, we expect A1 to go through multiple commissioning rounds, resulting in incremental improvement to its noise sensitivity, effective bandwidth, and duty cycle. We expect that it will eventually attain the target \texttt{aLIGO} A+ design sensitivity~\cite{Abbott_2020_Prospects, Noise_curves_used_for_Simulations_in_the_update_of_the_Observing_Scenarios_Paper} (sometimes referred to as the `O5' sensitivity in this paper) and acquire higher up-times, resulting in duty cycles that are commensurate with the stable operations of other detectors in the IGWN.
To model this progression, we consider A1 to have two distinct noise sensitivities: one at conditions that mirror the fourth science run (O4) of the LIGO Hanford and Livingston detectors, simulating an early commissioning stage, and the other at the A+ design sensitivity of the detector. Additionally, we consider various duty cycles of the A1 detector at the sensitivities mentioned above for a comprehensive analysis. 

By the time A1 begins its maiden science run, the current detectors namely LIGO-Hanford (H1) and LIGO-Livingston (L1)~\cite{Advanced_LIGO_2015}, are likely to reach their A+ design sensitivities or beyond; meanwhile Virgo (V1)~\cite{Acernese_2014} and KAGRA (K1)~\cite{akutsu2020overview} are also expected to reach their target design sensitivities. A1 would add significantly longer baselines to the existing GW network and also contribute to increasing the network SNR~\cite{Schutz_2011, Saleem_2021}, leading to better sky localizations of GW sources along with improvements in the estimations of source parameters. A number of studies done previously have addressed the localization capabilities of GW networks which include analytical studies~\cite{Fairhurst_2011, Wen_2010} as well as simulations using \texttt{BAYESTAR}, \texttt{LalInference} or other localization algorithms ~\cite{Pankow_2018, Chen_2017, Pankow_2020, Singer_2014, Rodriguez_2014}. 
%
\begin{figure*}[!hbt]
    \centering
    \hspace{-8mm}
    \includegraphics[width=0.55\textwidth, height=0.45\textwidth, clip=True]{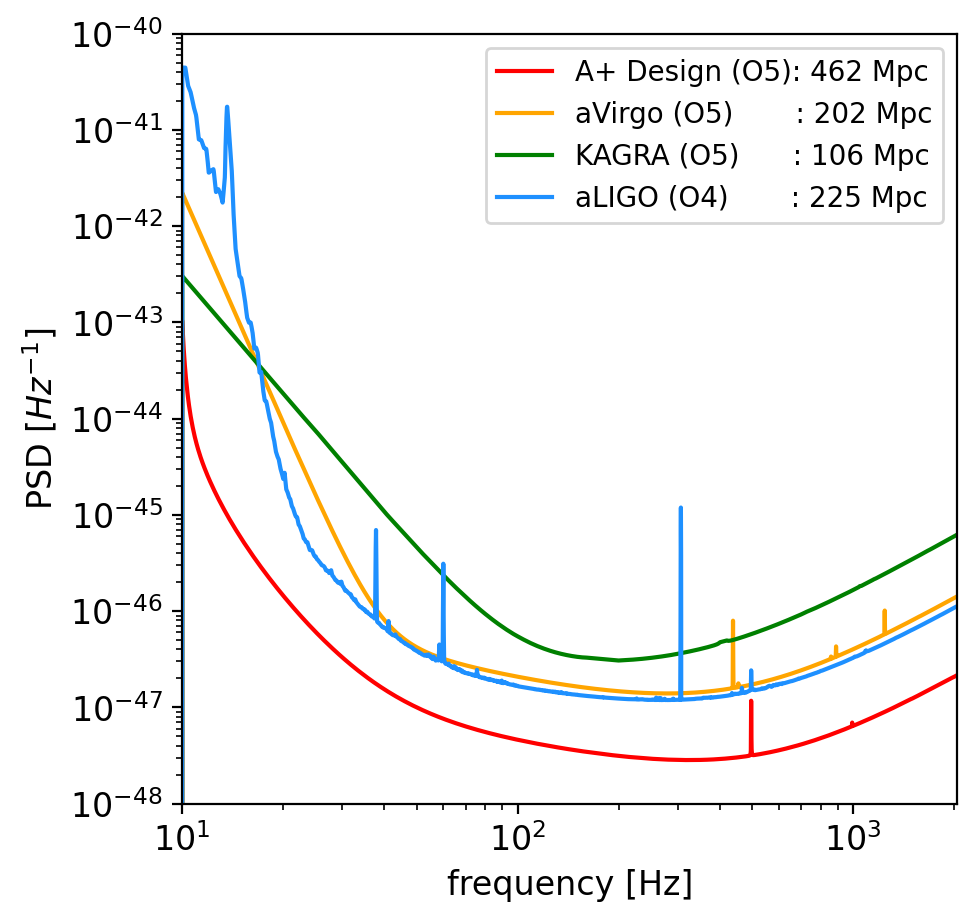}
    \caption{The noise Power Spectral Density (PSD)  for GW detectors. L1 and H1 are at \texttt{aLIGO A+ Design Sensitivity}(O5). V1 and K1 are at at \texttt{aVirgo}(O5) and \texttt{KAGRA(80 Mpc)} (O5) sensitivities respectively. We study A1 at two different sensitivities: first at \texttt{aLIGO} (O4) and then at \texttt{aLIGO A+ Design Sensitivity} (O5), respectively. The distances quoted here are the ``sensemon'' ranges for the respective detectors for a $1.4M_\odot+1.4M_\odot$ BNS at $\rho_{\text{th}}=6$. Note that the sensemon ranges for these PSD estimates presented in~\cite{Abbott_2020_Prospects, Noise_curves_used_for_Simulations_in_the_update_of_the_Observing_Scenarios_Paper} correspond to a detection signal-to-noise ratio threshold of $8$ in a single detector.}
    \label{fig:PSD}
\end{figure*}
\label{Our study & related info}
In this article, we aim to provide an illustration of the contribution A1 shall make to the current network of terrestrial GW detectors with a focus on improvements in sky localization capabilities for BNS mergers. 
We focus on the scenarios where the localization uncertainties are of the orders that enable a potential EM follow-up. This is possible when the source is localized by three or more detectors. 
We perform a full Bayesian Parameter Estimation using a rapid PE method developed by Pathak et al.~\cite{pathak2022rapid, pathak2023prompt}. The method enables us to perform rapid Bayesian PE for BNS systems from a lower seismic cut-off frequency $f_{\text{low}} = 10$ Hz, hence allowing for the bandwidths (especially at lower frequencies) that would be typical of the future (\text{O5} \& beyond) observing runs. Analyzing CBCs from ${f_{\text{low}} = 10}$ Hz increases the number of cycles in the frequency band and leads to an improvement in signal-to-noise ratio (SNR) accumulation and information content of the event. 

The study of the localization capabilities of a GW detector network can be broadly characterized by the intrinsic source parameters (e.g. component masses)~\cite{Pankow_2018, Pankow_2020} of binary systems, detector sensitivities, and duty cycles of individual detectors~\cite{Schutz_2011, Pankow_2020}. We hereby discuss these aspects in relation to our work. 

\subsection{Detector Networks}
\label{describing the Heterogeneous network}
The ground-based GW detector network is comprised of detectors with different sensitivities. The heterogeneous nature of the noise Power Spectral Density (PSD) curves of detectors plays an important role in deciding the localization abilities of the network. Hence, we assume the ground-based GW detectors to be at different noise sensitivities for our analysis. The first two LIGO detectors L1 and H1 are configured to \texttt{aLIGO A+ Design Sensitivity}
PSDs~\cite{Advanced_LIGO_2015, Abbott_2020_Prospects}. The detector V1 is taken to be at its projected \texttt{aVirgo} PSD~\cite{Acernese_2014}. Meanwhile, the K1 detector is assumed to be at \texttt{KAGRA (80 Mpc)} O5 design sensitivity~\cite{akutsu2020overview, PhysRevD.88.043007, Somiya_2012}. 

To study the improvements in localization capabilities of the network with the addition of A1, we analyze the A1 detector at two different sensitivities. We first consider the case where A1 is in its initial operating phase at \texttt{aLIGO} O4 sensitivity. In the second case, we take A1 to be at \texttt{aLIGO A+ Design Sensitivity} configuration, marking the target sensitivity it shall achieve after undergoing staged commissioning over time. All the above-mentioned PSDs can be found in~\cite{Noise_curves_used_for_Simulations_in_the_update_of_the_Observing_Scenarios_Paper}. The aforementioned noise sensitivity curves are shown in Fig.~\ref{fig:PSD}. These sensitivity configurations, in conjunction with the duty cycles taken into account, shall provide a more comprehensive approach in presenting a science case for the addition of A1 to the current GW detector network. 

\subsection{Duty Cycles}
\label{describing the duty cycles}
The duty cycle of a detector/network is defined as the fraction of time for which it successfully collects data of scientific significance during an observing run~\cite{Saleem_2021}. The duty cycle for a detector depends on the specific phase in which the detector is relative to its intended target operating configuration. Additionally, environmental effects also play a role in affecting the detector duty cycle during an observation run. No detector can practically acquire science-quality data at all times. This translates to detectors working at different duty cycles depending on the commissioning of the detectors.

To understand the impact of duty cycles on the localization of BNS sources by a network, we {\it{assume}} the following three cases as suggested by Pankow et al.~\cite{Pankow_2020} representing different stages of a detector's operation:

\begin{enumerate}
    \item [1)] 20\% duty cycle: A representative of the early stages of commissioning and engineering runs, resulting in reduced operational time.
    
    \item [2)] 50\% duty cycle: A representative of unresolved technical issues with the instrumental setup as well as signaling challenges like suboptimal environmental conditions.
    
    \item[3)] 80\% duty cycle: A representative of a detector operating near the possible target operating point. 
\end{enumerate}

Consider a GW network with $N$ detectors. Over the course of an observing run, there can be a $k$ number of detectors ($N_{\text{min}} \leq k\leq N$) participating in data collection, depending on their duty cycles. Here, $N_{\text{min}}$ represents the minimum number of detectors assumed to be participating in the observation of a GW event. These $k$ participating detectors can comprise different subnetworks of distinct detectors. For instance, a GW network with $N=5$ detectors may have only $k=4$ detectors in operation. Subject to which detector is out of operation, there can be $\binom{N}{k}=5$ different subnetworks of distinct detectors. Depending on the duty cycles of individual detectors, the effective duty cycle of a subnetwork can be evaluated. Out of a total of $N$ detectors in the network, we assume a set $\boldsymbol{m}$ composed of all the detectors participating in data collection and a set $\boldsymbol{n}$ comprising detectors that are out of operation (possibly due to maintenance) during this observation period. We represent the probability ($p$) of being in operation defined by a given duty cycle (i.e. $p=0.5$ for $50\%$ duty cycle). The probability representing the effective duty cycle of a subnetwork ($p_{\text{eff}}$) is given as
\begin{equation}
\label{equation_duty_cycle}
    p_{\text{eff}} = \prod_{m_i} \ \prod_{n_j} p_{m_{i}} (1 - p_{n_{j}}),
\end{equation}
where $p_{m_{i}}$ and $p_{n_{j}}$ are the duty cycles of the $i$th detector in set $\boldsymbol{m}$ and $j$th detector in set $\boldsymbol{n}$ respectively. For instance, consider a network of $N=4$ detectors, namely L1, H1, V1, and K1. The probability of being in operation for individual detectors is given by $p_{\text{L1}}$, $p_{\text{H1}}$, $p_{\text{V1}}$, and $p_{\text{K1}}$ representing their respective duty cycles. There may be a scenario where any one of these four detectors may get out of operation due to maintenance or environmental causes. Depending on which detector is not in observation mode, there can be $\binom{4}{3}=4$ subnetworks, comprising of $k=3$ detectors participating in data acquisition. 
To evaluate the probability of being in operation $p_{\text{eff}}$ for one of the subnetworks consisting of say H1, V1, and K1 detectors (this is the case when \texttt{L1} is not in operation) we have
\begin{equation*}
    p_{\text{eff}}\Big|_{\text{H1V1K1}} = p_{\text{H1}} \cdot p_{\text{V1}} \cdot p_{\text{K1}} \cdot (1 - p_{\text{L1}})
\end{equation*}
where, H1, V1, K1 $\in \textbf{m}$ and L1 $\in \textbf{n}$ respectively. The effective duty cycle for $N$-detector network is evaluated as the probability obtained by adding the probabilities of being in operation for all subnetworks over $k \leq N$ number of participating detectors. 

In this study, we consider the L1, H1, V1, and K1 detectors operating near their target operating point at 80\% duty cycle each and are fixed during the analysis. As A1 is expected to join this network by the early 2030s, we aim to show how the addition of A1 improves the localization capabilities of the GW network. We vary duty cycles for the A1 detector, simulating the cases for various phases of its configuration relative to its target operating point. 
Single detectors are nearly omnidirectional due to the structure of the antenna pattern functions - leading to poor source localization. For a two-detector network, solving for the direction in the sky corresponding to a fixed time-delay between the coalescence times recorded at the detectors leads to a `ring` like pattern on the sky, and as such, events localized with two detectors are generally not useful enough for EM follow-ups. Hence, we do not include the cases with $k \leq 2 $ in our analysis. For the purpose of this study, we assume $3 \leq k \leq N$ (i.e. $N_{\text{min}} = 3$) and evaluate the network duty cycles accordingly.  
The impact of A1 as an addition to the second-generation GW network consisting of L1, H1, V1, and K1 detectors is presented by the implementation of varying duty cycles (20\%, 50\%, 80\%) in conjunction with different detector sensitivities (\texttt{aLIGO} O4 and \texttt{aLIGO A+ Design Sensitivity}) for A1 detector. 

\subsection{Injected BNS sources}
\label{motivating for SNR range choice & injection parameters}

The remarkably high signal-to-noise ratio (SNR) due to the fortunate proximity of the GW170817 event enabled its effective localization and multi-messenger efficacies. Since the number of such `golden events' is expected to be low even in future observing runs~\cite{Schutz_2011, Baibhav_2019}, it becomes increasingly important to study a network's ability to localize such events. 
Hence, we aim to focus on studying the impact of the addition of the LIGO-Aundha detector in localizing moderately high SNR events, which may lead to potential multi-messenger observations. 
Taking this into consideration, we choose to generate 500 BNS events having an optimal network SNR in the range of $20$ to $25$ in the GW network comprising L1, H1, V1, and K1 detectors. We use these events for the purpose of our investigation. We would like to highlight that this is not a population study but focuses on possible improvements to the localization capabilities of a GW network with the addition of the LIGO-Aundha detector in accurately locating such `golden events'.
An event is considered to be detected if the individual detector optimal SNR is greater than a threshold value of 6 $(\rho_{\text{th}}> 6)$ in at least two detectors. From all the generated BNS sources, an event must follow the detection criteria to be considered detected by a subnetwork/network. The effectiveness of a network with an additional A1 detector is studied against the four-detector network with L1, H1, V1, and K1  detectors. 

\label{mass and spin choice}
The intrinsic parameters, like component masses, spins, etc., affect the localization of CBC sources. The effective bandwidth, as defined in~\cite{Fairhurst_2011}, measures the frequency content of the signal. Effective bandwidth is one of the important factors affecting the localization of CBC sources~\cite{Pankow_2018}. The signals from BNS sources mostly span through the entire bandwidth of the ground-based detectors owing to their relatively small component mass values in comparison to binary black holes or neutron star-black hole sources. The mass ranges for the BNS are also narrow, leading to small variations in effective bandwidths. In fact, it has also been shown by Pankow et al.~\cite{Pankow_2020} that the sky localization uncertainties for BNS systems are effectively independent of the population model of their component masses and spins. Hence, in order to simplify our simulations, we work with a particular choice of component source masses and spins. The tidal parameters are also expected to have a negligible effect on the source localization of these systems~\cite{Pankow_2018} and, therefore, are not included as source parameters.

The source-frame intrinsic parameters are chosen to have the maximum a posteriori (MAP) values of the posterior samples obtained from the LIGO PE analysis of GW170817~\cite{Romero_Shaw_2020} using \texttt{BILBY}~\cite{Ashton_2019} Python package. The source frame component masses are ${m^{\text{src}}_{1} = 1.387 \:M_{\odot}}$, ${m^{\text{src}}_{2} = 1.326\:M_{\odot}}$, while the dimensionless aligned spin component parameters are ${\chi_{1z} \approx 1.29 \times 10^{-4}}$,  ${\chi_{2z} \approx 3.54 \times 10^{-5}}$. We choose the inclination angle $\iota=\pi/6$ \texttt{radians}.
We set the polarization angle arbitrarily to $\psi=0$ \texttt{radians}. The sources are distributed uniformly in sky directions. We distribute the sources in luminosity distances corresponding to the redshifts following a uniform in comoving volume distribution up to a redshift of $\sim 0.14$, which is greater than the detection range of a detector at \texttt{aLIGO A+ Design Sensitivity} (O5) for a BNS with component masses $m^{\text{src}}_1$ and $m^{\text{src}}_2$. In addition to the sources that are to be detected with certainty, this limit also allows for events that are barely near the threshold in some detectors. We simulate uncorrelated Gaussian noise in each detector characterized by their associated PSDs respectively. The injection sources are generated with the \texttt{IMRPhenomD}~\cite{khan2016frequency} waveform model, and the source parameters are recovered using the \texttt{TaylorF2}~\cite{PhysRevD.49.1707, PhysRevD.59.124016, Faye_2012, PhysRevLett.74.3515} waveform model for the Bayesian PE analysis. Since BNS mergers are mostly inspiral-dominated in the LIGO-Virgo-KAGRA (LVK) detector band, the use of the \texttt{TaylorF2} waveform model sufficiently extracts the required information from strain data.

\subsection{Bayesian inference}
\label{Bayesian PE}
In order to estimate the parameters of GW sources, we use a Bayesian framework, where for a given waveform model $\boldsymbol{h}$ and data $\boldsymbol{d}$ from the detectors, the posterior distribution of the source parameters $\vec \Lambda$ can be estimated via Bayes' theorem:
\begin{equation}
    p(\vec \Lambda \mid \boldsymbol{d}) = \frac{\mathcal{L}(\boldsymbol{d}\mid \vec \Lambda)\: p(\vec \Lambda)}{p(\boldsymbol{d})}
\end{equation}
where ${\mathcal{L}(\boldsymbol{d}\mid \vec \Lambda)}$ is the likelihood function, $p(\vec \Lambda)$ is the prior over the source parameters ${\param \equiv \{ \vec \lambda, \vec \theta, t_{c} \}}$, and $p(\boldsymbol{d})$ is called evidence, which describes the probability of data given the model. Here $\vec \lambda$ represents the intrinsic parameters, whereas $\vec \theta$ denotes the extrinsic parameters. In principle, ${p(\vec \Lambda \mid \boldsymbol{d})}$ can be estimated by placing a grid over the parameter space $\vec \Lambda$, which for a typical compact binary coalescence (CBC) source described by a ${\sim 15}$ dimensional parameter space, would become practically intractable. In the case of a BNS system, it increases to $17$ dimensional space due to the addition of two tidal deformability parameters. Instead, stochastic sampling methods such as Markov chain Monte Carlo (MCMC)~\cite{Foreman_Mackey_2013} and Nested Sampling~\cite{skilling2006nested} are employed to generate representative samples from the posterior distribution ${p(\vec \Lambda\mid \boldsymbol{d})}$. However, this process still requires evaluating the likelihood function, which involves a computationally expensive step of generating model (template) waveforms at the proposed points by the sampler and calculating the overlap between these waveforms and the data. This computational cost is notably significant, especially for low-mass systems such as binary neutron star (BNS) events with lower cutoff frequency decreased to $10$ Hz. The situation is exacerbated by the enhanced sensitivity of detectors, resulting in a large number of in-band waveform cycles. Furthermore, incorporating additional physical effects can further escalate the computational burden of waveform generation. These factors have significant implications for the feasibility of promptly following up on EM counterparts of corresponding BNS systems.
As mentioned earlier, a high number of BNS detections are expected in the O5 runs; it would be prudent to prioritize the EM follow-ups given the limited observational resources. This underscores the importance of the development of rapid PE methods, which can efficiently estimate both intrinsic and extrinsic parameters. Various rapid PE methods have been proposed in the recent past. They broadly come under two categories: (i) ``likelihood-based'' approaches such as Reduced order models~\cite{canizares2013gravitational, canizares2015accelerated, smith2016fast, Morisaki_2020, morisaki2023rapid}, Heterodyning (or Relative Binning)~\cite{Venumadhav2018, cornish2021heterodyned, Finstad_2020, islam2022factorized}, and other techniques such as RIFT~\cite{https://doi.org/10.48550/arxiv.1805.10457}, simple-pe~\cite{fairhurst2023fast}, multibanding~\cite{Morisaki:2021ngj} (ii) ``likelihood-free'' approaches which aim to directly learn the posteriors employing Machine-learning techniques such as deep learning, normalizing flows, and variational inference as well ~\cite{chua2019reduced, chua2020learning, Green_2020, green2020complete, gabbard2022bayesian}. 

In this work, we use a likelihood-based rapid PE method developed by Pathak et al.~\cite{pathak2022rapid, pathak2023prompt}, which combines dimensionality reduction techniques and meshfree approximations to swiftly calculate the likelihood at the proposed query points by the sampler. This algorithm is interfaced with {\sc{dynesty}}~\cite{speagle2020dynesty, sergey_koposov_2023_7600689}, a Python implementation of the Nested sampling algorithm to quickly estimate the posteriors distribution over the source parameters. In the forthcoming sections, we will first define the likelihood function and subsequently provide a concise overview of how the meshfree method expeditiously computes the likelihood at the sampler's proposed query points.

\subsubsection{Likelihood function}
Given a stream of data $d^{(i)}$ from the $i^{\text{th}}$ detector and a template $\tilde{h}^{(i)}(\vec \Lambda)$, under an assumption of uncorrelated noise across the detectors, the coherent network log-likelihood is given by 
\begin{equation}
\label{eq:multigenlikelihood}
\ln \mathcal{L}(\vec\Lambda) =  \sum_{i=1}^{N_{\text{d}}} {\langle \boldsymbol{d}^{(i)} \mid \tilde{h}^{(i)}(\vec\Lambda)\rangle} - \frac{1}{2} \sum_{i=1}^{N_{\text{d}}} \left [ \| \tilde{h}^{(i)}(\vec\Lambda)\|^2 + \| \boldsymbol{d}^{(i)} \|^2 \right ]
\end{equation} 
where $\tilde{h}^{(i)}(\vec \Lambda)$ represents the frequency domain Fourier Transform (FT) of the signal $h^{(i)}(\vec \Lambda)$ and $N_d$ is the number of detectors. Here, the inner product is defined as 
\begin{equation}
    \langle x \mid y\rangle = 4\,\text{Re}\int_{0}^{\infty} \, df\,\frac{{\tilde{x}}(f)^{*}\,\,\tilde{y}(f)}{S_h(f)}
    \label{eq:inner_prod_def}
\end{equation}
In this paper, we focus on the non-precessing GW signal model, which can be decomposed into factors dependent on only intrinsic and extrinsic parameters as follows:
\begin{equation}
\label{eq:detframwaveform}
\begin{split}
\tilde{h}^{(i)}(\vec\Lambda) \equiv \tilde{h}(\vec\Lambda, t^{(i)}) 
    &= \mathcal{A}^{(i)} \tilde{h}_{+}(\vec \lambda, t^{(i)}), \\
    &= \mathcal{A}^{(i)}\, \tilde{h}_{+}(\vec \lambda)\, e^{-j\,2\pi f t_c} \, e^{-j\,2\pi f \Delta t^{(i)}}
\end{split}
\end{equation}
where $\mathcal{A}^{(i)}$, the complex magnitude of the signal depends only on the extrinsic parameters $\vec\theta \in \vec \Lambda$ through the antenna pattern functions, luminosity distance $d_L$, and the inclination angle $\iota$, and can be expressed as the following:
\begin{equation}
\mathcal{A}^{(i)} = 
    \frac{1}{d_L} \left[ \frac{1+\cos^2 \iota}{2}  F^{(i)}_{+}(\alpha, \delta, \psi).
            - \: j \cos \iota \ F^{(i)}_{\times}(\alpha, \delta, \psi) \right],
\end{equation}
$\Delta t^{(i)}$ corresponds to the time-delay introduced due to the relative positioning of the $i^{\text{th}}$ detector in relation to the Earth's center~\cite{pankow2015novel}, the ${F^{(i)}_{+}(\alpha, \delta, \psi)}$ and ${F^{(i)}_{\times}(\alpha, \delta, \psi)}$ are respectively the `plus' and `cross' antenna pattern functions of the $i^{\text{th}}$ detector, which are functions of right-ascension $\alpha$, declination $\delta$, and polarization angle $\psi$. The antenna pattern functions describe the angular response of the detector to incoming GW signals~\cite{detectorTensor}. 

In our analysis, we opt for the log-likelihood function marginalized over the coalescence phase~\cite{thrane_2019}. With $\tilde{h}^{(i)}(\vec \Lambda)$ given by Eq.~\eqref{eq:detframwaveform}, the expression of the marginalized phase likelihood is given by 
\begin{multline}
\label{eq:multiphaselikelihood}
    \left.\ln \mathcal{L}(\vec\Lambda \mid \boldsymbol{d}^{(i)})\right|_{\phi_c} 
    = \ln I_{0}\left[\left|\sum_{i=1}^{N_{d}}{{\mathcal{A}}^{(i)}}^{*}\, \langle \boldsymbol{d}^{(i)} \mid \tilde{h}_{+} (\vec\lambda, t^{(i)}) \rangle \right|\right] \\
    - \frac{1}{2}\sum_{i=1}^{N_{d}}\left[ \left|\mathcal{A}^{(i)}\right|^2 \sigma^2(\vec \lambda)^{(i)} + \| \boldsymbol{d}^{(i)} \|^2 \right];
\end{multline}
where $I_0(\cdot)$ is the modified Bessel function of the first kind and ${\vec z^{(i)}(\vec \lambda^n) \equiv \langle \boldsymbol{d}^{(i)} \mid \tilde{h}_{+}(\vec \lambda, t^{(i)}) \rangle}$ is the complex overlap integral, while ${\sigma^2(\vec\lambda)^{(i)} \equiv \langle \tilde{h}_{+}(\vec \lambda, t^{(i)}) \mid \tilde{h}_{+}(\vec \lambda, t^{(i)}) \rangle}$ is the squared norm of the template $\: \tilde{h}_{+} (\vec\lambda)$. ${\sigma^2(\vec\lambda)^{(i)}}$ depends on the noise power spectral density (PSD) of the $i^{\text{th}}$ detector. 
The squared norm of the data vector, $\| \boldsymbol{d}^{(i)} \|^2$, remains constant throughout the PE analysis and, hence, does not affect the overall `shape' of the likelihood. Consequently, it can be excluded in the subsequent analysis. Note that the marginalized phase likelihood will not be an appropriate choice for systems with high precession and systems containing significant power in subdominant modes~\cite{Pratten_2021}.

\subsubsection{Meshfree likelihood interpolation}
The meshfree likelihood interpolation, as outlined in~\cite{pathak2022rapid, pathak2023prompt}, comprises two stages: (i) Start-up stage, where we generate radial basis functions (RBF) interpolants of the relevant quantities and (ii) Online-stage, where the likelihood is calculated by evaluating the interpolants at the query points proposed by the sampler. Let's briefly discuss both stages.

\begin{itemize}
    \item \label{start-up stage}\textbf{Start-up stage}: First, we generate $N$ RBF interpolation nodes in the intrinsic parameter space ($\mathcal{M}$, $q$, $\chi_{1z}$, and $\chi_{2z}$ in this context). The center $\vec \lambda^{\text{cent}}$ around which these interpolation nodes are positioned is determined by optimizing the network-matched filter SNR, starting from the best-matched template or trigger $\vec \lambda^{\text{trig}}$ and $t_{\text{trig}}$ identified by the upstream search pipelines~\cite{GstLAL_2010, cannon2021gstlal, usman2016pycbc}. For simulated systems, the injection parameter is taken as the central point for node placement. We employ a combination of Gaussian and uniform nodes, where the Gaussian nodes are sampled from a multivariate Gaussian distribution (MVN) with a mean of $\vec \lambda^{\text{cent}}$ and a covariance matrix calculated using the inverse of the Fisher matrix evaluated at $\vec \lambda^{\text{cent}}$. A hybrid node placement approach ensures that nodes are positioned near the peak of the posterior, where higher accuracy in likelihood reconstruction is necessary. Once the nodes $\vec \lambda^{n}$ are generated, we efficiently compute the time-series ${\vec z^{(i)}(\vec \lambda^{n}) \equiv z^{(i)}(\vec \lambda^{n}, t_c)}$ using the Fast Fourier Transform (FFT) circular correlations, with $t_c$ being uniformly spaced discrete-time shifts within a specified range ($ \pm 150$ ms\footnote{This range should be larger than the maximum light travel time between two detectors.}) around a reference coalescence time $t_{\text{trig}}$. During this calculation, we set $\Delta t^{(i)} = 0$ for overlap time series, handling extra time offsets introduced due to sampling in the sky location parameters during the online stage.
    Similarly, we compute the template norm square $\sigma^2(\vec \lambda^n)^{(i)}$ at the RBF nodes $\vec \lambda^n$. We then stack the time series (row-wise) and perform Singular Value Decomposition (SVD) of the resulting matrix, producing a set of basis vectors spanning the space of $\vec z^{(i)}(\vec \lambda^n)$:
    \begin{equation}
    \label{eq:svdbasistimeseries}
    \vec z^{(i)}(\vec \lambda^{n}) = \sum_{\mu = 1}^N\, C^{n (i)}_{\mu}\ \vec u^{(i)}_{\mu}
    \end{equation}
    where the SVD coefficients $C^{n (i)}_{\mu}$, smooth functions of $\vec \lambda^{n}$ within the sufficiently narrow boundaries encompassing the posterior support, can be interpolated over the $\vec \lambda$ using a linear combination of RBFs and monomials~\cite{doi:10.1142/6437}: 
    \begin{equation}
        \label{eq:rbfcoeff}
        C^{q (i)}_{\mu} = \sum_{n=1}^N\, a^{(i)}_{n}\, \phi(\|\vec \lambda^q - \vec \lambda^{n}\|_2) + \sum_{j = 1}^{M}\, b^{(i)}_{j}\, p_j(\vec \lambda^q)
    \end{equation}
    where $\phi$ is the RBF kernel centered at ${\vec \lambda^{n} \in \mathcal{R}^d}$, and $\left \{  p_{j} \right \}$ denotes the monomials that span the space of polynomials with a predetermined degree $\nu$ in $d$-dimensions. Since the coefficients are only known at $N$ RBF nodes $\vec \lambda^n$, we impose $M$ additional conditions of the form ${\sum_{j=1}^M a^{(i)}_j p_j(\vec \lambda^q) = 0}$ to uniquely solve for the coefficients $a_n$ and $b_j$ in the Eq.~\eqref{eq:rbfcoeff}. Furthermore, it turns out that only ``top-few'' basis vectors are sufficient to reconstruct $\vec z^{(i)}(\vec \lambda^q)$ at minimal reconstruction error. Consequently, we generate only top-$\ell$ meshfree interpolants of $C_{\mu}^{q(i)}$ where $\mu = 1,....,\ell$, where $\ell$ can be chosen based on the singular value profile. Similarly, we express $\sigma^2(\vec \lambda^q)$ in terms of RBFs and monomials, treating them as smoothly varying functions over the interpolation domain. Finally, we have uniquely constructed the $\ell + 1$ RBF interpolants, which are to be used in the online stage. 

    \item \textbf{Online stage}: In the online stage, we rapidly compute interpolated values of $C^{q (i)}_{\mu}$ and $\sigma^2(\vec \lambda^q)^{(i)}$ at any query point $\vec \lambda^q$ within the interpolation domain. Subsequently, we determine the corresponding $\vec z^{(i)}(\vec \lambda^q)$ using Eq.\eqref{eq:svdbasistimeseries}. Rather than generating the entire time series, we focus on creating $\vec z^{(i)}(\vec \lambda^q)$ with around $\mathcal{O}(10)$ time samples centered around the query time $t^{q (i)}$, which contain the additional time-offset $\Delta t^{(i)}$. We fit these samples with a cubic spline, from which we calculate $\vec z^{(i)}(\vec \lambda^q)$ at the query time $t^{q (i)}$. Similarly, we compute the interpolated value of $\sigma^2(\vec \lambda^q)^{(i)}$. Finally, we integrate these interpolated values with the factors related to extrinsic parameters, as outlined in Eq.\eqref{eq:multiphaselikelihood}, to compute the interpolated likelihood $\ln \mathcal{L}_{\text{RBF}}$.
\end{itemize}

\begin{table}[!hbt]
\centering
\def\arraystretch{1.3}
\begin{tabular}{lcl}
\hline \hline
 Parameters     &   Range    &   Prior distribution \\
\hline
 $\mathcal{M_{\text{det}}}$  &   $[\mathcal{M}^{\text{cent}}_{\text{det}} \pm 0.0001]$ &   $\propto \mathcal{M_{\text{det}}}$ \\
 $q$            &   $[q^{\text{cent}} \pm 0.07]$ & $ \propto \left [ (1 + q)/q^3 \right ]^{2/5}$     \\
 $\chi_{1z, 2z}$   &   $[\chi_{1z}^{\text{cent}} \pm 0.0025]$   & Uniform  \\
 $V_{\text{com}}$          & $[\sim16.9e3, \sim8.9e8]$                & Uniform \\
 $t_c$          & $t_{\text{trig}} \pm 0.12$& Uniform\\
 $\alpha$       & $[0, 2\pi]$               & Uniform\\
 $\delta$       & $\pm \pi/2$       & $\sin^{-1} \left [ {\text{Uniform}}[-1,1]\right ]$\\
 $\iota$        & $[0, \pi]$                & Uniform in $\cos \iota$\\
 $\psi$         & $[0, 2\pi]$               & Uniform angle\\
 \hline \hline
\end{tabular}
\caption{Prior parameter space over the ten-dimensional parameter space $\vec \Lambda$.}
\label{tab:priordistr_chap5}
\end{table}
%

\section{Analysis of Simulated Events}
\label{analysis of simulated events}
As discussed previously in Section~\ref{motivating for SNR range choice & injection parameters}, we create injections with fixed source-frame masses and dimensionless aligned spin component parameters. However, the detector-frame parameters (masses) for these events vary according to their associated redshifts. We define the intrinsic detector-frame parameters by ${\vec\lambda = (\mathcal{M}_{\text{det}}, q, \chi_{1z}, \chi_{2z})}$. Similarly, the injected intrinsic parameters in the detector frame are denoted as $\vec\lambda^{\text{cent}}$. To perform Bayesian PE for each event, we first generate $N_{\text{nodes}} = 800$ RBF nodes as described in Section~\ref{start-up stage}. We sample $20\%$ of the total RBF nodes ($N_{\text{Gauss}} = 160$) from a multivariate Gaussian distribution $\mathcal{N}(\vec \lambda^{\text{cent}}, \mathbf{\Sigma})$, where $\vec\lambda^{\text{cent}}$ is the mean and $\mathbf{\Sigma}$ is the covariance matrix obtained from inverse of the Fisher matrix $\mathbf{\Gamma}$ around the center $\vec\lambda^{\text{cent}}$ using the \texttt{gwfast}~\cite{Iacovelli_2022} python package. The remaining $80\%$ of the total RBF nodes ($N_{\text{Unif}} = 640$) are sampled uniformly from the ranges provided in Table \ref{tab:priordistr_chap5}. We choose $\phi = \exp(-(\epsilon r)^2)$ as the Gaussian RBF kernel in our analysis, with $\epsilon$ being the shape parameter. For the purpose of this analysis, we use $\epsilon=10$, monomial terms with degree $\nu=7$ and $l=20$ top basis vectors for reconstructing the time-series in Eq.~(\ref{eq:svdbasistimeseries}). After the successful generation of interpolants, the likelihood function can be evaluated using $\ln \mathcal{L}_{\text{RBF}}$ by sampling the ten-dimensional parameter space $\vec \lambda$  using the \texttt{dynesty} sampler. The sampler configuration is outlined as follows: \texttt{nlive} $=500$, \texttt{walks} $=100$, \texttt{sample} = ``rwalk'', and \texttt{dlogz} $=0.1$. These parameters play a critical role in determining both the accuracy and the time required for the nested sampling algorithm to converge. In this context, the parameter \texttt{nlive} represents the number of live points. Opting for a larger value of \texttt{nlive} leads to a more finely sampled posterior distribution (and consequently, the evidence), but it comes at the cost of requiring more iterations to achieve convergence. The parameter \texttt{walks} specifies the minimum number of points necessary before proposing a new live point, \texttt{sample} indicates the chosen approach for generating samples, and \texttt{dlogz} represents the proportion of the remaining prior volume's contribution to the total evidence. In this analysis, \texttt{dlogz} $=0.1$ serves as a stopping criterion for terminating the sampling process. For a more comprehensive understanding of dynesty's nested sampling algorithm and its practical implementation, one can refer to the following references ~\cite{speagle2020dynesty, sergey_koposov_2023_7600689}.

The prior distribution for $\vec \lambda$, along with the associated parameter space boundaries, are presented in Table \ref{tab:priordistr_chap5}. The prior distributions for the extrinsic parameters ($\alpha, \delta, V_{\text{com}}, \iota, \psi, t_c$) and their respective parameter space boundaries are also presented in Table~\ref{tab:priordistr_chap5}. To evaluate the Bayesian posteriors of source parameters, we sample over the entire ten-dimensional parameter space involving four intrinsic and six extrinsic source parameters. This ensures accounting for the correlations between parameters. However, the focus of this study lies in discussing the sky localization uncertainties obtained from the posteriors over $\alpha$ and $\delta$ parameters.

In accordance with the previous discussion in Section~\ref{Bayesian PE}, we perform PE for the simulated events with different subnetworks of a GW network to take into account the effect of duty cycles. For instance, in the case of a network with L1, H1, V1, K1, and A1, there can be $10$ different subnetworks consisting of three distinct detectors ($k=3$), and $5$ different subnetworks of four distinct detectors ($k=4$) taking observations depending on the duty cycles. In addition to these,  there is a subnetwork consisting of all the five detectors for $k=5$ case.

Bayesian PE analyses are performed for the events detected in each of these subnetworks. The total number of subnetworks for all $3\leq k \leq 5$ is 16 for the five detector networks comprising L1, H1, V1, K1, and A1 detectors. The exercise is repeated for two cases: 
\begin{enumerate}
    \item [(i)] Keeping A1 at \texttt{aLIGO} O4 noise sensitivity in the GW network. Here, the A1 sensitivity is close to \texttt{aVirgo} (O5) sensitivity (Refer Fig.~\ref{fig:PSD}).

    \item [(ii)] Setting A1  at \texttt{aLIGO A+ Design Sensitivity} (O5) in the GW network. In this case, the A1 detector would be at the same sensitivity as the other two LIGO detectors.
\end{enumerate}
We represent the network with $N=5$ detectors as the L1H1V1K1A1 network and, similarly, the network with $N=4$ detectors as the L1H1V1K1 network. Using the \texttt{ligo-skymap}~\cite{Singer_2016} utility, we compute the $90\%$ credible sky localization areas $\Delta \Omega_{90\%}$ (in sq. deg) from the posterior samples over right ascension ($\alpha$) and declination ($\delta$) obtained from Bayesian PE.

\section{Sky Localization Results}
\label{sky localization results}
In order to take into account the effect of duty cycles in the sky localization of our simulated BNS events, we first evaluate the probabilities associated with the effective duty cycles of each subnetwork of a detector network using Eq.~\eqref{equation_duty_cycle}. Each subnetwork is assigned a fixed number of events depending on their observation probabilities. Taking this into account and integrating all such cases across $3\leq k\leq N$ with area samples of corresponding events gives the localization $\Delta \Omega_{90\%}$ distribution related to the network duty cycle for a given GW network. 
%
\begin{figure*}[!hbt]
    \centering
    \begin{subfigure}{0.49\linewidth}
        \centering
        \includegraphics[height=2.3in, width=\linewidth]{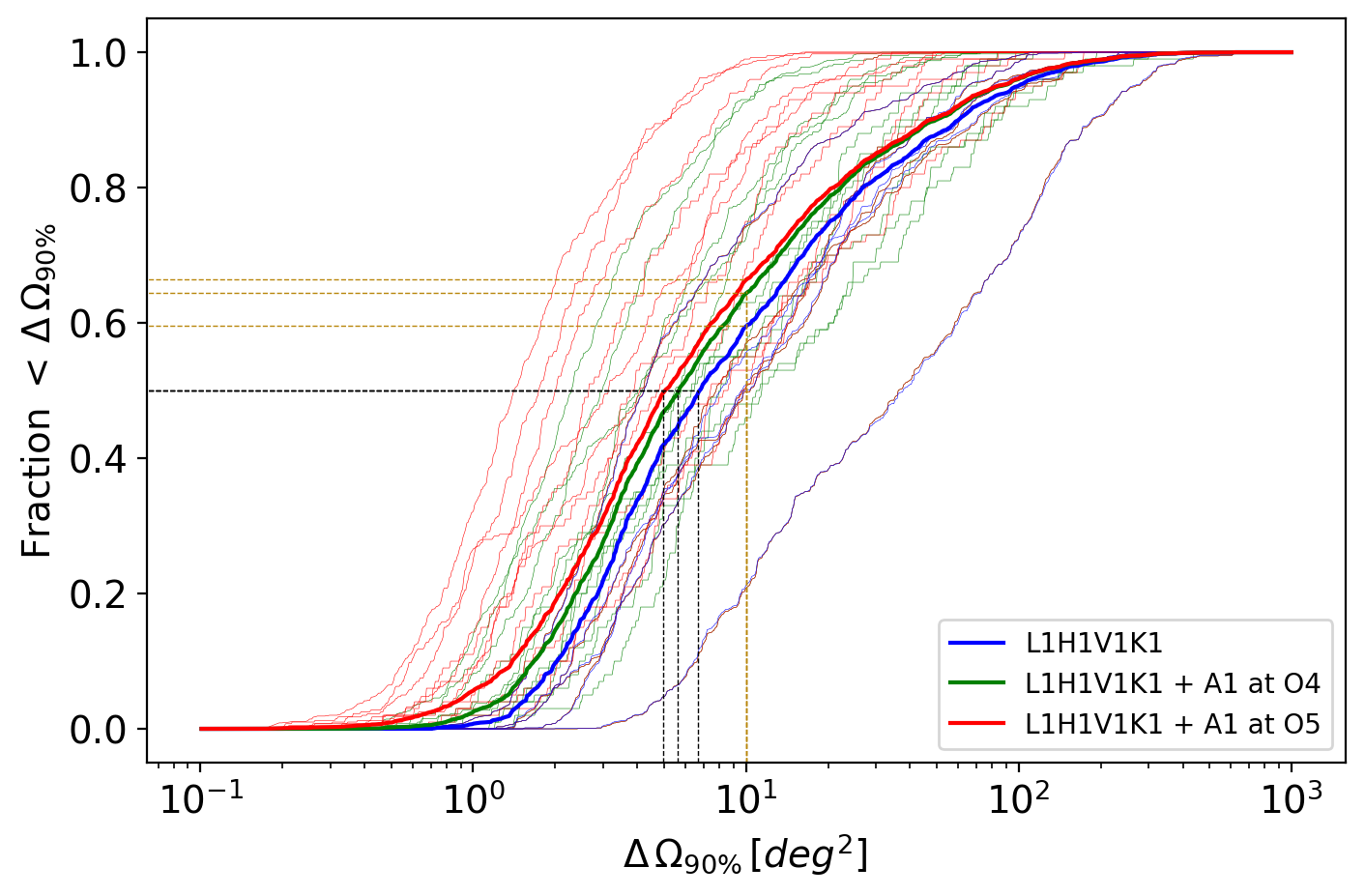} 
        \caption{A1 at 20\% duty cycle}
        \label{fig: duty cycle plot1}
    \end{subfigure}%
    \hfill
    \begin{subfigure}{0.49\linewidth}
        \centering
        \includegraphics[height=2.3in, width=\linewidth]{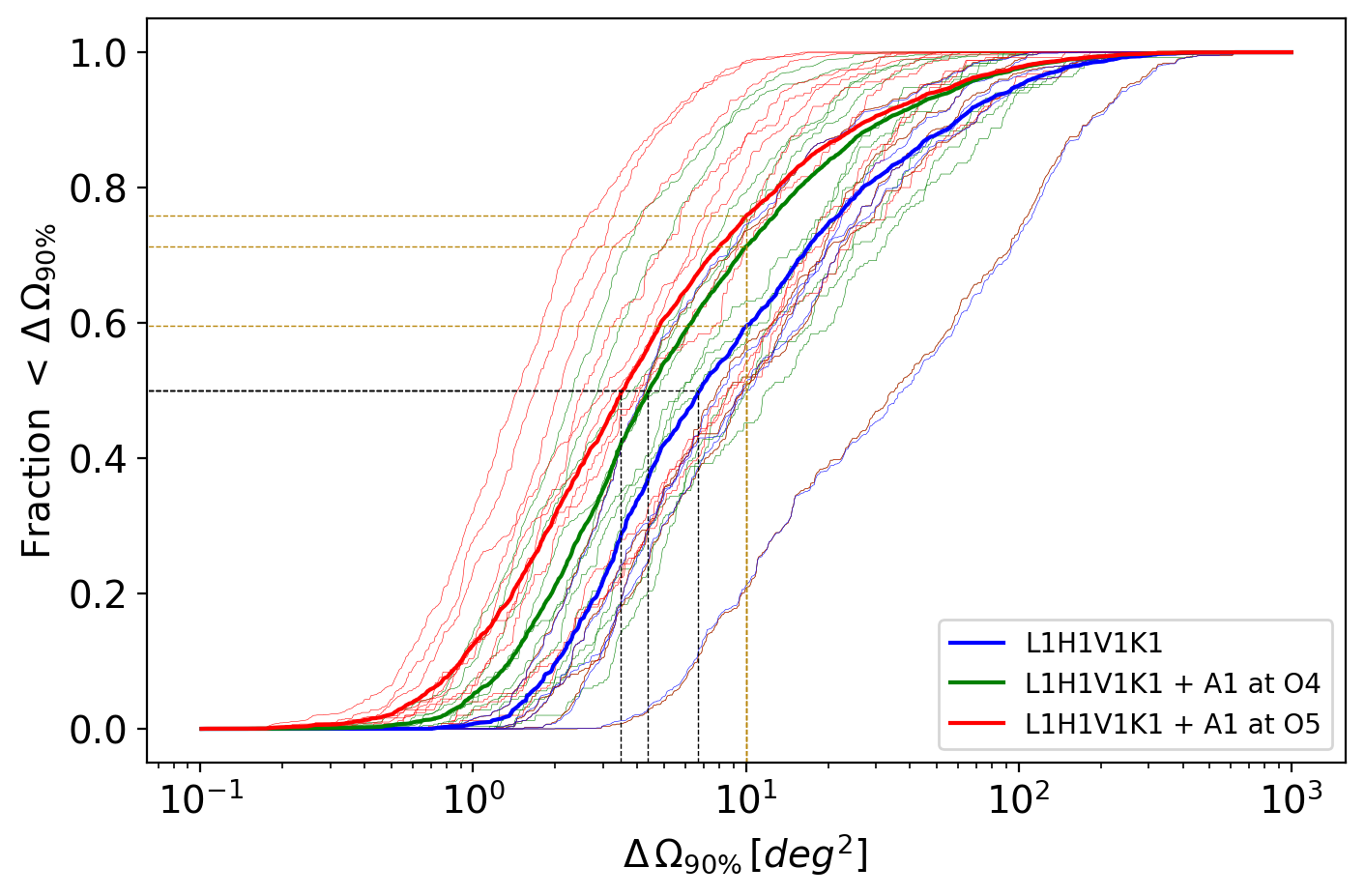} 
        \caption{A1 at 50\% duty cycle}
        \label{fig:duty cycle plot2}
    \end{subfigure}
    \hfill
    \\
    \begin{subfigure}{0.49\linewidth}
    \vspace{5mm}
        \centering
        \includegraphics[height=2.3in, width=\linewidth]{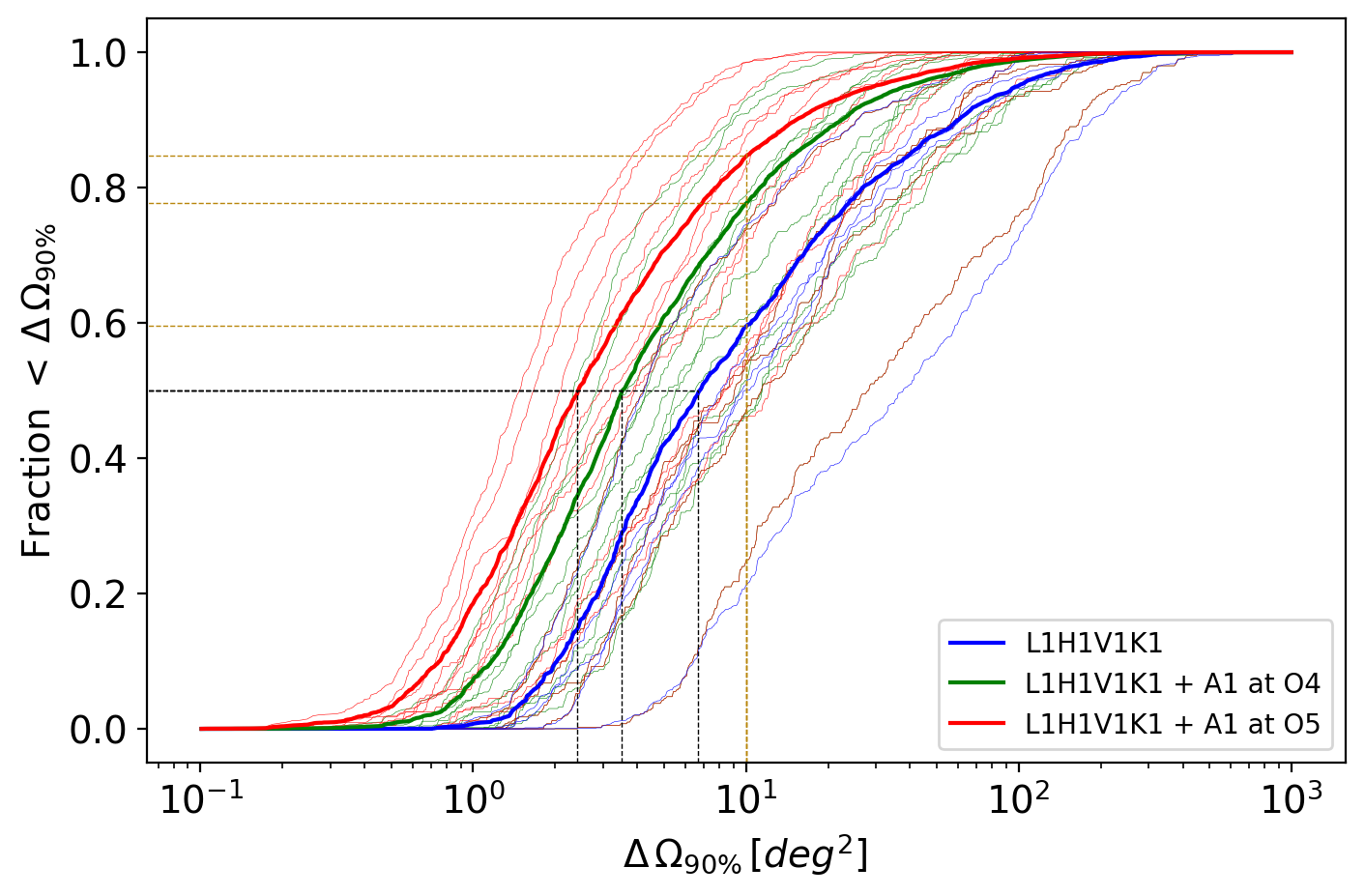} 
        \caption{A1 at 80\% duty cycle}
        \label{fig:duty cycle plot3}
    \end{subfigure}
    \caption{Cumulative Distributions (CDF) of $90\%$ localization area of simulated BNS events obtained by different networks are shown. As mentioned in \ref{describing the duty cycles}, duty cycles for each of the L1, H1, V1, and K1 detectors are set to $80\%$. Three different cases considering A1 at duty cycles of $20\%$,  $50\%$, and $80\%$ are shown, along with taking into account two different noise sensitivities (A1 at O4 sensitivity and A1 at O5 sensitivity) and presented in Fig.~\ref{fig: duty cycle plot1}, \ref{fig:duty cycle plot2} and \ref{fig:duty cycle plot3} respectively. The blue solid lines represent the CDF obtained by the L1H1V1K1 network with the aforementioned duty cycle. The green solid lines represent the CDF constructed by the L1H1V1K1A1 network, where A1 is at \texttt{aLIGO} (O4) sensitivity. The red solid lines are the CDF constructed with L1H1V1K1A1 network, where A1 is at \texttt{A+ Design Sensitivity} (O5) sensitivity. In each of the subplots, the A1 detector is at a duty cycle, as mentioned in the captions of Fig.~\ref{fig: duty cycle plot1}, \ref{fig:duty cycle plot2} and \ref{fig:duty cycle plot3} respectively. The lighter curves represent the CDFs of $\Delta \Omega_{90\%}$ areas from different subnetworks over all possible $k$ for the corresponding GW network. The vertical black color dashed lines mark the median localization areas obtained by each network. The horizontal golden color dashed lines represent the fraction of events (in percentage) recovered under 10 sq. deg. by each detector network.}
    \label{fig:CDF plots O4, O5}
\end{figure*}
\clearpage
\begin{table*}[!hbt]
\centering
\begin{tabular}{c c c c}
\hline \hline
\multicolumn{1}{c}{Network} & \multicolumn{3}{c}{Fraction within $ < 10$ sq. deg.(in \%) \& Median $\Delta \Omega_{90\%}$ Area (sq. deg.)}\\
\hline
L1H1V1K1 &  & $59$\%  &  \\
(Median $\Delta \Omega_{90\%}$)  & & $6.6$  & \\
\cline{1-4}
&  A1 at $20\%$ duty cycle &A1 at $50\%$ duty cycle& A1 at $80\%$ duty cycle \\
\cline{2-4}

\hspace{4mm}L1H1V1K1+A1 (O4) &  64\% & 71\% & 77\%\\
(Median $\Delta \Omega_{90\%}$)  & 5.6 & 4.3 & 3.5 \\

\cline{1-4}
\hspace{4mm}L1H1V1K1+A1 (O5) &  66\% & 75\% & 84\%\\
(Median $\Delta \Omega_{90\%}$) & 4.9  & 3.4 & 2.4  \\
\hline \hline
\end{tabular}
\caption{We present the percentage of detected events localized within $10$ sq. deg sky area and the median $\Delta \Omega_{90\%}$ for different detector networks. The median $\Delta \Omega_{90\%}$ for detector networks are shown separately in the rows below the values representing the percentage of detected events localized within $10$ sq. deg sky area. 
The noise sensitivity of A1, along with the duty cycles associated with A1, are mentioned in the relevant table sections. Duty cycles for L1, H1, V1, and K1 detectors are fixed at $80\%$.}
\label{table 10 sq deg + median area}
\end{table*}

The results of our simulations are presented in Fig.~\ref{fig:CDF plots O4, O5}. The Cumulative Distribution Function (CDF) plots in Fig.~\ref{fig:CDF plots O4, O5} are constructed from $\Delta \Omega_{90\%}$  sky area samples obtained by inverse sampling from the localization distributions for each subnetwork.
We find that with the L1H1V1K1 network, the median $90\%$ localization area $\Delta \Omega_{90\%}$ is $6.6$ sq. deg., meanwhile $59\%$ of the BNS sources are localized within less than $10$ sq. deg. area in the sky.
With the addition of an A1 detector to this network, we find significant improvements in the localization capabilities of the terrestrial detector network. Similar improvements were also reported by~\cite{Schutz_2011, Saleem_2021}. It is also evident from Fig.~\ref{fig:CDF plots O4, O5} that duty cycles and detector noise sensitivities play a vital role in the effective localization of sources. We shall discuss these in detail as follows:\\

\subsection{A1 at \texttt{aLIGO}-O4 sensitivity}
\label{Explain CDF A1 at O4}
As a part of the five-detector network, the A1 detector is initially set to \texttt{aLIGO} O4 noise sensitivity. The median $\Delta \Omega_{90\%}$ area in the decreasing order are found to be $5.6$, $4.3$, and $3.5$  in sq. deg. when A1 is set to $20\%$, $50\%$, and $80\%$ duty cycle respectively. We find that $64\%$, $71\%$, and $77\%$ of the events are localized with less than $10$ sq. deg in sky area, given that A1 is at $20\%$, $50\%$, and $80\%$ duty cycles respectively. Our results suggest that even when A1 is at $20\%$ duty cycle, which can be interpreted as the early commissioning phase of the detector, the five-detector network reduces the median $\Delta \Omega_{90\%}$ localization uncertainty to $5.6$ sq. deg. in comparison to $6.6$ sq. deg. obtained by the four detectors L1H1V1K1 network. This reduction in the sky localization area plays a crucial role in the `tiled mode' search for EM counterparts undertaken by the EM facilities such as the GROWTH India Telescope~\cite{Kumar_2022} with a field of view of $~0.38$ sq. deg. in area, to tile the GW localization regions.
As the A1 detector is upgraded to be at $80\%$ duty cycle, the median localization area $\Delta \Omega_{90\%}$ remarkably reduces by approximately a factor of two in comparison to that achieved by the four detector L1H1V1K1 network. We reiterate that the L1, H1, V1, and K1 detectors are taken to be operating at $80\%$ duty cycles.\\

\subsection{A1 at \texttt{A+ Design} (O5) sensitivity}
\label{Explain CDF A1 at O5}
By upgrading the A1 configuration to \texttt{aLIGO A+ Design Sensitivity} (O5), the improvement in the localization capabilities of the five-detector network relative to the four-detector network as well as the five-detector network with A1 at O4 sensitivity is considerable. 
The median $\Delta \Omega_{90\%}$ localization uncertainties in the decreasing order are found to be $4.9$, $3.4$, and $2.4$ sq. deg. in area, when A1 is set to $20\%$, $50\%$ and $80\%$ duty cycle respectively. We find that $66\%$, $75\%$, and $84\%$ of the events are localized with less than $10$ sq. deg in sky area, given that A1 is at $20\%$, $50\%$ and $80\%$ duty cycles respectively, where A1 is set at \texttt{A+} sensitivity. We observe that with the A1 detector operating at $50\%$ duty cycle, the median localization area $\Delta \Omega_{90\%}$ reduces by a factor of two with respect to the values obtained by the L1H1V1K1 network. As A1 reaches its target operating point with $80\%$ duty cycle, we find the median $\Delta \Omega_{90\%}$ to reduce by a factor of three against the median localization achieved with the four-detector network. As mentioned previously, for A1 operating at O4 sensitivity and $80\%$ duty cycle, the median $\Delta\Omega_{90\%}$ is $3.5$ sq. deg., whereas this reduces to $2.4$ sq. deg. when A1 is set to $80\%$ duty cycle and at O5 sensitivity.

\begin{figure*}[!hbt]
    \centering
    \includegraphics[width=0.9\textwidth, height=0.55\textwidth, clip=True]{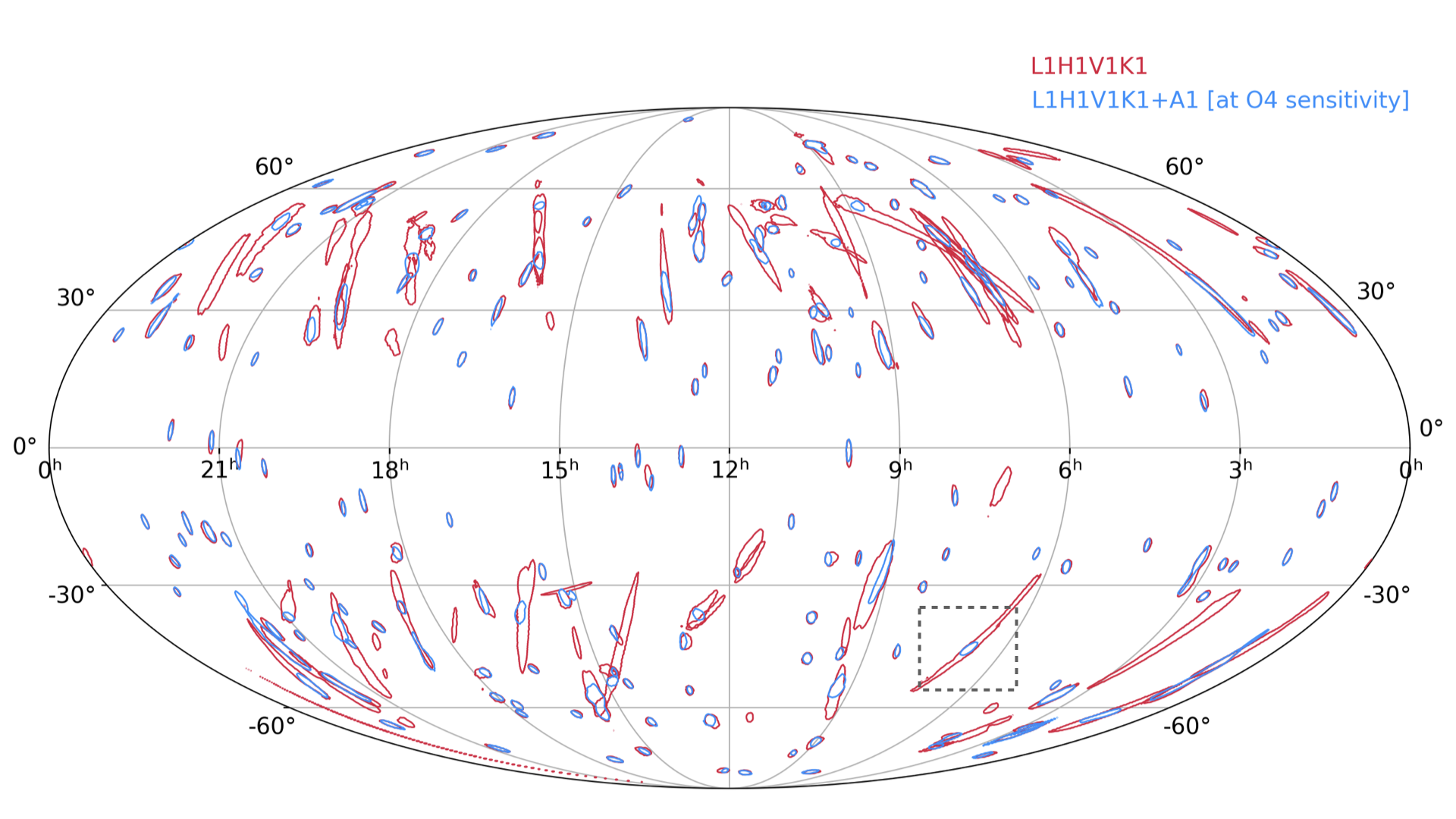}
    \caption{The crimson and blue probability contours correspond to 90\% credible sky area ($\Delta \Omega_{90\%}$) localization of the events by L1H1V1K1 and L1H1V1K1A1 networks, respectively for 191 events out of our 500 simulated BNS events. Here, A1 is at O4 sensitivity, and the events are subthreshold in A1. The contours marked in a dashed square box are discussed in Section~\ref{Subthreshold Events in LIGO-Aundha}. This skymap plot exhibits the improvement achieved in the localization of BNS events by the inclusion of an A1 detector, even when it does not contribute to the detection. Note that for some sky locations, the red contours appear in the background of the blue contours and may not visible.} 
    \label{fig: A1 subthreshold O4 skymap}
\end{figure*}

The A1 detector, with an upgraded O5 sensitivity, leads to a two-fold impact on the network capabilities. On the one hand, it leads to an increase in the number of detections (events satisfying the SNR threshold $\rho_{\text{th}} \geq 6 $) in the subnetworks, including A1. It also results in an increase in network SNR, which is one of the important factors contributing to the effective localization of sources. By adding A1 to the GW network, there is an increase in the observation probability of three or more detectors by $7\%$ (for A1 operating at the early $50\%$ duty cycle) in comparison to the four detector networks. We summarize a few important remarks about the localization $\Delta \Omega_{90\%}$ for the different network configurations in Table \ref{table 10 sq deg + median area}. Note that the CDF plots in Fig.~\ref{fig:CDF plots O4, O5} may indicate slightly lower median $\Delta \Omega_{90\%}$ localization values than those obtained in related studies~\cite{Pankow_2020, Pankow_2018}. One of the reasons is that the events are analyzed from $f_{\text{low}} = 10$ Hz, which increases the effective bandwidths, as well as due to the exclusion of cases with two detectors subnetworks or single detectors participation in source localization.
 
The focus of this study is to explore the localization capabilities of the GW network with A1, with possibilities leading to potential EM follow-ups and providing better ground for astrophysical and cosmological investigations.

\section{Localization of simulated BNS events subthreshold in A1}
\label{Subthreshold Events in LIGO-Aundha}
During the observation of GW170817, the event was detected in L1 and H1 detectors but was below the detection threshold in the V1 detector. Yet, the presence of V1 contributed to localizing the source to a few tens of sq. deg. Out of the $500$ simulated BNS events described in our previous discussion, a total of $191$ events detected in the five-detector network were found to be subthreshold ($\rho_{\text{th}}<6$) in A1 is at \texttt{aLIGO} O4 sensitivity. We compare the sky-localizations for these events obtained from a four-detector L1H1V1K1 network to those achieved by the L1H1V1K1A1 network. Since these events are subthreshold in A1, the contribution of A1 in improving the network SNR is negligible. 
Yet, the presence of an A1 detector leads to an improvement in reducing the localization uncertainties of these events. This is shown in Fig.~\ref{fig: A1 subthreshold O4 skymap}. 
We find that even in the case where these events are subthreshold in A1 (at \texttt{aLIGO} O4 sensitivity), the percentage of events localized with less than $10$ sq. deg. in the sky increase from $72\%$ to $89\%$ in comparison to the L1H1V1K1 network. Even though the CDFs (for $\Delta \Omega_{90\%}$) used for the estimation of these improvements are not very smooth due to a lesser number of such events (191 in this case), they summarize the essence of overall nature of the improvement well enough.

As the noise sensitivity configuration of A1 is upgraded to \texttt{aLIGO A+ Design Sensitivity}, there is an increase in the number of detections in the A1 detector.
In this case, the number of events that are subthreshold in the A1 detector reduces from 191 to just 44 out of all the 500 simulated BNS sources. Due to the improved sensitivity, further improvements in the localization of such events are achieved in comparison to the localizations obtained relative to the four-detector L1H1V1K1 network and L1H1V1K1A1 network with A1 at O4 sensitivity. Note that in the context of this section, we do not consider the duty cycles for these networks. Therefore, we make a direct comparison between the localization results achieved for such events with the L1H1V1K1 network and the L1H1V1K1A1 network. For instance, the event marked in Fig.~\ref{fig: A1 subthreshold O4 skymap} is localized to $44$ sq. deg. with L1H1V1K1 network. The same event when detected by the L1H1V1K1A1 network with A1 is at \texttt{aLIGO} O4 sensitivity, is recorded at an optimal SNR value $\rho_{A1} = 3.1$ in A1 detector ($\rho_{A1} < \rho_{\text{th}}$) and is localized to $\sim 6$ sq. deg. Meanwhile, when A1 is set to \texttt{aLIGO A+ Design Sensitivity}, this event is localized to $\sim 3.5$ sq. deg. area in the sky. The baselines added to the network with the addition of the A1 detector, and its antenna patterns are some of the factors leading to better localization of such events. An improved noise PSD results in an increase in effective bandwidth and hence leads to the reduction in localization uncertainties.

\section{Experiments with GWTC-like events in real noise}
\label{GWTC-Events}
In the preceding section, we showed that even if a detector does not detect an event, it nevertheless adds a valuable contribution to the network in localizing the source. In this section, we provide an illustration of how the incorporation of an additional detector could have facilitated the source localization of events from GWTC for compact binary mergers. The two BNS events, GW170817 and GW190425, and an NSBH event, GW200115, are chosen as examples from GWTC for this purpose. In our analysis, we consider A1 as a supplementary detector. We simulate the aforementioned events and inject them into real detector noise to account for a realistic scenario. The noise strains from the L1, H1, and V1 detectors were acquired by using the \texttt{GWpy}~\cite{gwpy} python package, which allows the extraction of noise strain timeseries from the datasets publicly available on GWOSC~\cite{2022AAS...24034809B}. The noise strain for A1 is taken to be that of the detector, which recorded the least SNR during the observation of these events. In fact, among all the detectors observing these events, the lowest individual SNR was recorded in the Virgo detector. The noise strain data for all the detectors is chosen hundreds of seconds away from the trigger times of the GW events in consideration. Even though the A1 noise strain is taken from V1 data, the noise strain data for both detectors belong to different stretches of data. The PE analysis for these events is performed from a lower seismic frequency of $20$ Hz.

\subsection{Noise Strain and PSDs} 
A noise strain data of a fixed segment length (360 seconds for GW170817 and GW190425; 64 seconds for GW200115) is used in our analysis, which is cleaned by a high-pass filter of $4$th order and setting the frequency cut-off at $18$ Hz. We analyze the event from a lower seismic cutoff frequency of $20$ Hz, which is illustrative of the observing runs associated with their detections. The estimation of noise PSD uses $2$ seconds overlapping segments of the strain data with the implementation of the median-mean PSD estimation method from \textsc{PyCBC}~\cite{usman2016pycbc}. The noise PSD for A1 is constructed from the strain data of the V1 detector.
    
\subsection{Choice of injection parameters:}
    \subsubsection{GW170817-like event} 
        The intrinsic detector-frame parameters $(m_1^{\text{det}}, m_2^{\text{det}}, \chi_{1z}, \chi_{2z})$ and extrinsic parameters like inclination ($\iota$) take values chosen by evaluating the MAP values from the posterior samples of detector-frame parameters in the \texttt{Original BILBY results file}~\cite{Romero_Shaw_2020} for GW170817. We assume the BNS system with spins aligned in the direction of orbital angular momentum. The sky location coordinates of NGC $4993$-the potential host galaxy of the GW170817 event, are taken as the injection values for $(\alpha, \delta)$ sky position parameters~\cite{Ebrova:2018gtz}. The luminosity distance takes the value $d_L = 40.4$ Mpc~\cite{Hjorth_2017} for our simulated BNS system. The polarization angle is taken to be zero ($\psi=0$) since the associated posterior samples in  \texttt{Original BILBY results file} are found to be degenerate. As mentioned previously, the simulated signal is injected in uncorrelated real noise in the detectors. The simulated signal is generated using the \texttt{IMRPhenomD} waveform model. The source parameters are recovered using the \texttt{TaylorF2} waveform model (see Fig.~\ref{fig:GW170817_like} in Appendix~\ref{sec:append_gwtc_corner} for the corner plot and skymap). 
        
    \subsubsection{GW190425-like event} 
        The intrinsic (detector-frame) and extrinsic parameter values are chosen by evaluating the MAP values
        of the posterior samples for parameters obtained from 
        \texttt{C01:IMRPhenomPv2\_NRTidal:LowSpin} LIGO analysis file of GW190425 event~\cite{ligo_scientific_collaboration_and_virgo_2022_6513631}. The polarization angle is chosen to be $\psi=0$ as the posterior samples for $\psi$  from the LIGO analysis follow a uniform distribution. We generate the simulated signal using the \texttt{IMRPhenomD} waveform model. The source parameters are recovered using the \texttt{TaylorF2} waveform model for the PE analysis (see Fig.~\ref{fig:GW190425_like} in Appendix~\ref{sec:append_gwtc_corner} for the corner plot and skymap).

    \subsubsection{GW200115-like event} 
        For simulating the NSBH event, we choose the intrinsic (detector-frame) and extrinsic parameter values by evaluating the MAP values of the posterior samples for parameters from \texttt{C01:IMRPhenomNSBH:LowSpin} file of GW200115 LIGO analysis~\cite{ligo_scientific_collaboration_and_virgo_2021_5546663}. We take the polarization angle $\psi=0$. We generate the simulated signal using the \texttt{IMRPhenomD} waveform model. For recovering the source parameters during the PE analysis, we again use the \texttt{IMRPhenomD} waveform model, as it also accounts for the post-inspiral regime, which occurs within the LIGO-Virgo band for the NSBH system (see Fig.~\ref{fig:GW200115_like} in Appendix~\ref{sec:append_gwtc_corner} for the corner plot and skymap).

\subsection{Analysis and Configurations} 
The prior distributions and prior boundaries for ($\mathcal{M_{\text{det}}}, q, \chi_{1z}, \chi_{2z}, V_{\text{com}} $) parameters, chosen for the three simulated events are presented in Table \ref{Hybrid table for GWTC events}. The priors for the parameters ($t_c$, $\alpha$, $\delta$, $\iota$, $\psi$)  are same as that shown in Table \ref{tab:priordistr_chap5} and hence are not seperately mentioned here. The Bayesian PE analysis follows a similar methodology of generating interpolants for the likelihood function, as discussed previously. The analysis involves the generation of RBF nodes.  We specify the total number of RBF nodes ($N_{\text{nodes}}$) by mentioning the number of nodes sampled from a multivariate Gaussian $\mathcal{N}(\vec \lambda^{\text{cent}}, \mathbf{\Sigma})$, represented by $N_\text{Gauss}$; meanwhile the number of nodes uniformly sampled around $\vec\lambda^{\text{cent}}$ are represented as $N_\text{Unif}$ for each event. The \texttt{dynesty} sampler configurations are also mentioned in Table \ref{Hybrid table for GWTC events} for the three simulated events. 

\begin{table}[!hbt]
\def\arraystretch{1}
\centering
\begin{tabular}{l c c c c l}
\hline \hline 
\multirow{1}{*}{}&\multicolumn{1}{c|}{GW170817-like} & \multicolumn{1}{c|}{\centering \hspace{-2mm}GW190425-like}& \multicolumn{1}{c}{\hspace{-5mm}GW200115-like}& \multirow{1}{*}{} \\
\cline{2-4}
Parameter & Prior Range & Prior Range & Prior Range & Prior Distribution\\
\hline
$\mathcal{M_{\text{det}}}$  & 
$[\mathcal{M}^{\text{cent}}_{\text{det}} \pm 0.0002]$& $[\mathcal{M}^{\text{cent}}_{\text{det}} \pm 0.0005]$& $[\mathcal{M}^{\text{cent}}_{\text{det}} \pm 0.0011]$ & 
$\propto \mathcal{M_{\text{det}}}$ \\

$q$  & $[1, 1.14]$ & $[1, 1.28]$  & $[q^{\text{cent}} \pm 0.1]$ & $\propto \left [ (1 + q)/q^3 \right ]^{2/5}$  \\

$\chi_{1z, 2z}$ & 
$[\chi_{1z, 2z}^{\text{cent}} \pm 0.0025]$   & 
$[\chi_{1z, 2z}^{\text{cent}} \pm 0.0025]$   &
$[\chi_{1z, 2z}^{\text{cent}} \pm 0.0025]$   &
Uniform  \\

$V_{\text{com}}$ & $[5e3, 1e8]$ & $[5e3, 4e8]$ & $[1e6, 3e8]$ & Uniform \\
\hline
\multirow{2}{*}{No. of RBF Nodes} & $N_\text{nodes}=800$ & $N_\text{nodes}=800$ & $N_\text{nodes}=1100$ & \\ 

& ($N_\text{Gauss}=20\%$) & ($N_\text{Unif}= 100\%$) & ($N_\text{Gauss} = 10\%$) & \\

\hline

RBF Parameters & $\epsilon=10, \nu=7$ & $\epsilon=10, \nu=7$ & $\epsilon=30, \nu=10$ & $\phi = \exp(-(\epsilon r)^2)$ \\

\hline

\multirow{2}{*}{Sampler Configurations} & \texttt{nLive}=500 & \texttt{nLive}=1500 & \texttt{nLive}=500 & \\

& \texttt{nwalks}=100 & \texttt{nwalks}=500 & \texttt{nwalks}=100 & \\

& \texttt{sample}=``rwalk" & \texttt{sample=}=``rwalk" & \texttt{sample}=``rwalk" & \\

& \texttt{dlogz} $=0.1$ & \texttt{dlogz} $=0.1$ & \texttt{dlogz} $=0.1$ & \\
\hline \hline 
\end{tabular}
\caption{The prior parameter space is presented in the top part of the table. The hybrid placement for the RBF nodes is mentioned by specifying $N_\text{Gauss}$ and $ N_\text{Unif}$ for each event, followed by the Gaussian RBF kernel parameters. The \texttt{dynesty} sampler configurations chosen for the PE analysis of each event are also presented in the end. The priors for the extrinsic source parameters ($t_c$, $\alpha$, $\delta$, $\iota$, $\psi$) are same as that shown in Table \ref{tab:priordistr_chap5} and hence are not mentioned here.}
\label{Hybrid table for GWTC events}
\end{table}

\begin{figure*}[!hbt] 
    \centering
    \hspace{-13mm}
    \includegraphics[width=0.55\textwidth, height=0.55\textwidth, clip=True]{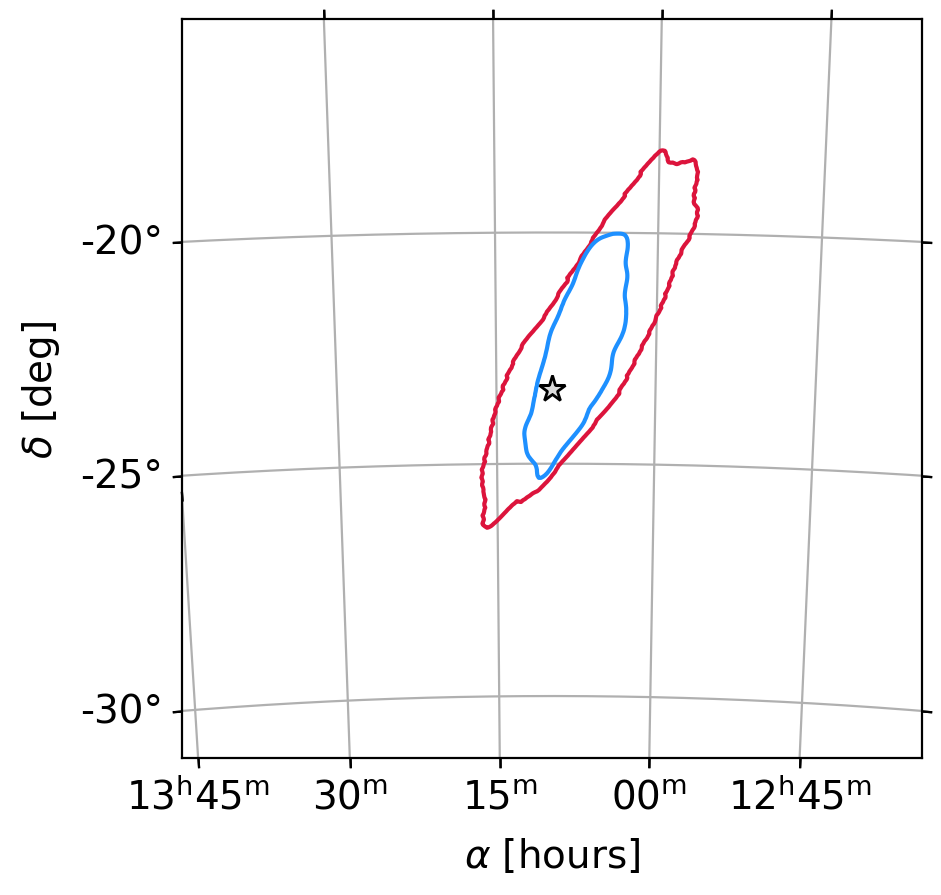}
    \caption{The crimson and blue probability contours represent $90$\% credible localization area ($\Delta \Omega_{90\%}$) for the GW170817-like event by L1H1V1 and L1H1V1A1 networks respectively. Noisy real strain data from V1 are used as a surrogate for A1 in this exercise. The star marks the true location $(\alpha, \delta)$ for the source. Though A1 does not contribute to the network SNR, its presence leads to a significant improvement in source localization.}
    \label{fig: GW170817 with A1}
\end{figure*}

The network comprising the L1, H1, and V1 detectors detected the GW170817 event. As discussed earlier, we simulate a GW170817-like signal in the non-Gaussian real noise and find the source localization uncertainty
in the presence of an A1 detector added to the L1H1V1 network. The matched-filter SNR in L1, H1, V1 and A1 are $22.6$, $18.6$, $5.4$ and $6.3$ respectively for the given noise realization. Even though, in this case, the addition of A1 to the L1H1V1 network does not lead to any considerable improvements in the network matched-filter SNR for the event, a significant reduction in $90\%$ credible localization area is observed. We find the localization uncertainty $\Delta \Omega_{90\%}$ to be $15$ sq. deg for the L1H1V1 network, whereas the localization area $\Delta \Omega_{90\%}$ reduces to $6$ sq. deg. with L1H1V1A1 network. Hence, the localization uncertainty is reduced by a factor of more than two in this case. The localization probability contours representing $\Delta \Omega_{90\%}$ obtained from the two different networks for GW170817-like event is presented in Fig.~\ref{fig: GW170817 with A1}. For a GW190425-like event, we compare the sky localization with the then-observing network of the L1V1 network to the L1V1A1 network. The sky localization uncertainty ($\Delta \Omega_{90\%}$) reduces from 9350 sq. deg. with the L1V1 network to a sky region of area 212 sq. deg with the L1V1A1 network. The matched filter SNR in L1, V1, and A1 are $10.1$, $5.1$, and $5.3$, respectively, for this case. It is evident that the event in V1 and A1 is at subthreshold SNR for the given noise realization. Yet, there is a contribution in reducing the sky localization areas. For the case of the GW200115-like event, the source localization with the L1H1V1 network, which was the observing network during the real event, is compared to that with the L1H1V1A1 network. The source localization error ($\Delta \Omega_{90\%}$) is reduced from $662$ sq. deg. obtained with L1H1V1 to $87$ sq. deg. achieved with L1H1V1A1. Here, we observe that the majority of the SNR is accumulated by the initial two LIGO detectors. Meanwhile, V1 and A1 contribute negligibly to improving the network SNR. This is because both V1 and A1 are at similar noise sensitivities for the aforementioned events. 

Note that these results vary with different realizations of the detector noise. Nevertheless, the antenna patterns and baselines added to a network by incorporating an additional detector (here, A1) may lead to an enhancement in the localization abilities of the network, even if the signal is subthreshold in one of the detectors.

\subsection{Degeneracy between luminosity distance and inclination angle}
\label{dL_iota_degeneracy_sec}
The GW190425-like event, when observed by the L1V1 network, shows a degeneracy between luminosity distance and inclination angle parameters, which was also observed for the real event. 
\begin{figure*}[!hbt]
    \centering
    \includegraphics[width=0.75\textwidth, height=0.75\textwidth, clip=True]{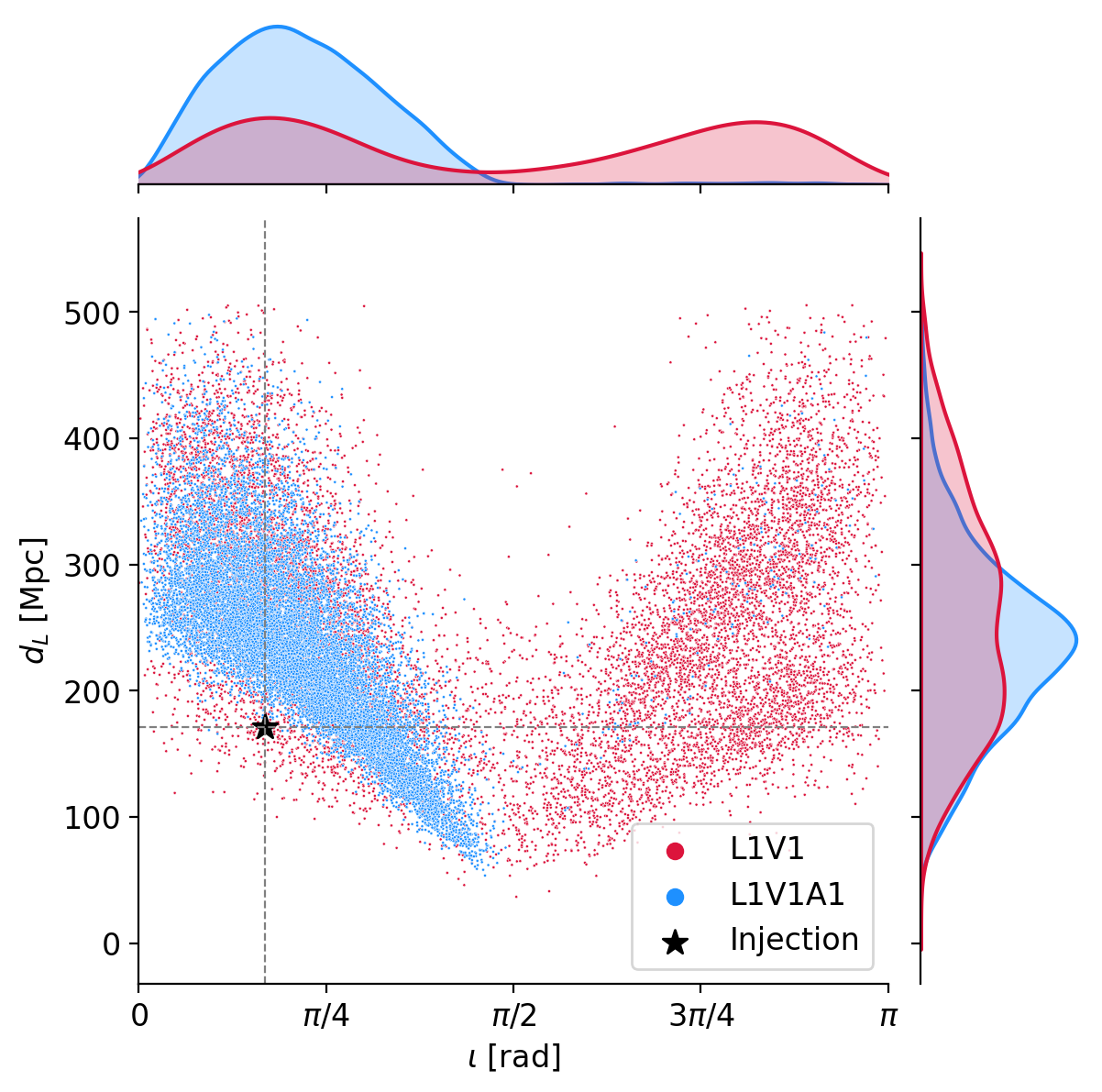}
    \caption{The crimson and blue points mark the posterior samples over $d_L$ and $\iota$ for the two-detector L1V1 and three-detector L1V1A1 networks respectively. The posteriors are constructed over 28 different noise realizations. This signifies that a three-detector network (L1V1A1) with a third detector contributing negligibly to the network SNR (in this case, A1) is able to play an important role in resolving the $\iota$--$d_L$ degeneracy relative to a two detector network (L1V1).}
    \label{fig:dL_vs_iota}
\end{figure*}
As the number of detectors in the network increases from L1V1 to L1V1A1, we observe a resolution of the distance-inclination angle degeneracy. For further investigation, we present the case for GW190425-like events with $28$ different real non-Gaussian noise realizations. The noise strains and PSD are obtained as mentioned at the beginning of Section~\ref{GWTC-Events}, where different noise strains correspond to different segments of detector strain data. The events for which the injected chirp-mass ($\mathcal{M}_c$) is within the $90\%$ credible interval of the posterior samples are chosen.

The results are summarized in Fig.~\ref{fig:dL_vs_iota}. The degeneracy between the luminosity distance ($d_L$) and inclination angle ($\iota$) parameters is resolved with an additional detector (here A1), even when the contribution of A1 in increasing the network SNR is not appreciable relative to the two detector network (L1V1).
Note that, here, GW190425-like events are generated with a waveform model (\texttt{IMRPhenomD}), which does not include higher-order modes. Also, both the compact objects (in this case: BNS) are of approximately equal masses i.e. $q\approx 1$. Hence, we can safely assume that the higher-order modes do not play a role in the resolution of degeneracy between the parameters. We obtain similar results on relaxing the condition over $\mathcal{M}$ and performing a similar analysis for $60$ different noise realizations. An investigation addressing the luminosity distance and inclination angle degeneracy for BNS systems has also been done in~\cite{Rodriguez_2014}. It is not clear that a better measurement of both the polarizations ($h_+$ \& $h_{\times}$) in a larger network leads to a more precise measurement of the inclination - especially for face-on systems ($\iota< 45$ deg.). In our study, we show the result as an empirical observation for a GW190425-like event. An extensive study constraining the inclination angle with a network of GW detectors has been performed by Usman et al.~\cite{Usman_2019}.

The improvement in the measurements of luminosity distance has direct implications in cosmology, as mentioned in Section~\ref{sec:intro}. The accurate measurements of inclination angle may lead to improvements in the constraints on the models for gamma-ray bursts and X-ray emissions from BNS mergers~\cite{Finstad_2018}. Similar improvements in the measurements of luminosity distance and inclination angles for binary black hole mergers by a three-detector network relative to a two-detector network have been obtained in~\cite{Saleem_2021}. 

\section{Conclusion}
\label{Conclusion_section}
The addition of A1 to the GW network is observed to improve the overall localization capabilities of the global detector network, even when A1 is in its early commissioning stages. To estimate the source parameters, we performed a full Bayesian PE from a lower cut-off frequency  $f_{\text{low}}= 10$ Hz, which is representative of the future LVK Collaboration analysis of GW sources. We find that addition of A1 detector (at \texttt{aLIGO} O4 sensitivity) to the GW network leads to a reduction of the median $\Delta \Omega_{90\%}$ area to $5.6$, $4.3$, and $3.5$ sq. deg. for cases where A1 is operating at $20\%$, $50\%$, and $80\%$ duty cycles respectively, in comparison to the median $\Delta \Omega_{90\%}$ area of $6.6$ sq. deg. obtained with the four detector L1H1V1K1 network for BNS sources with potential for multi-messenger follow-ups.

Our results suggest that an expanded GW detector with at an early phase A1 operating at a $20\%$ duty cycle and operating at a weaker sensitivity (\texttt{aLIGO} O4) as compared to the other LIGO detectors (\texttt{aLIGO} A+ Design Sensitivity) is capable of localizing $64\%$ of these BNS sources under $10$ sq. deg in comparison to $56\%$ by the four detector network. With the imminent improvement in the duty cycle and noise PSD of the A1 detector, an apparent enhancement in the localization capabilities of the GW network is observed (Refer Table~\ref{table 10 sq deg + median area}). With the addition of an A1 detector to the GW network, the observation probability for the sub-networks of $k\geq3$ detectors increases, leading to a decrease in localization uncertainties in the sky area. This allows for an optimized ``tiled mode'' search for post-merger emissions by telescopes such as the GROWTH India facility with a field of view of the order of $\sim0.4$ sq. deg. in sky area. We show that improvements in duty cycles and noise sensitivity for A1 detector play a crucial role in enhancing the localization capabilities of the GW network. Hence, in order to get the maximal payoff from the addition of the A1 detector, efforts should be made towards maximizing the operational duty cycle and improving the noise sensitivity as soon as the detector becomes operational. 

Furthermore, we show that even for BNS sources that are sub-threshold in A1, the sky-localization uncertainties with the five detector L1H1V1K1A1 network are reduced in comparison to that obtained from the four detector L1H1V1K1 network. Thus, even in a situation where A1 does not detect the BNS event independently, it plays a crucial role in pinpointing the sources that enable a fast and efficient electromagnetic follow-up by ground and space-based telescopes.

Taking the examples of two BNS events and one NSBH event from GWTC, we show the possible source localization improvements with A1 as an additional detector in the network with real noise. For this exercise, the real noisy strain data from Virgo is used as surrogate noise in A1 detector - to simulate a scenario where the Indian detector is observing the event but has not achieved its design sensitivity. 

We reaffirm the role of an additional detector (A1 in our case) in resolving the degeneracy between luminosity distance and inclination angle parameters relative to a two-detector network for a GW190425-like BNS source. This is shown by reconstructing the source parameters for GW190425-like BNS events in real, non-Gaussian noise, with the L1V1 and L1V1A1 detector networks, respectively, where data samples from V1 are used as surrogates for A1. 

\section{Discussion}
\label{Discussion}
In order to maximize the incentives from the GW detection of BNS sources, the EM follow-up of these events is of utmost importance. A1, joining the network of terrestrial GW detectors in the early 2030s, will enhance the localization capabilities of the network. We studied the impact of the addition of A1 in the detector network in the localization of BNS sources with moderately high signal-to-noise ratios. 

The observation of an event with three or more detectors working in conjunction is fundamental for achieving localization uncertainties small enough so as to allocate telescope time for subsequent electromagnetic follow-ups. Our results presented in Fig.~\ref{fig:CDF plots O4, O5} from Section~\ref{sky localization results} can be considered optimistic, owing to the assumption of a BNS event being observed by more than two detectors at any given time. 
Including sub-networks of two detectors will lead to the broadening of the distribution of localization uncertainties, causing a slight shift to the right in the Cumulative Distribution Functions (CDFs) shown in Fig.~\ref{fig:CDF plots O4, O5}.
However, this is beyond the scope of this work, and a more realistic study taking the case of two detector subnetworks into account can be performed in the future. Along with this, considering the case where one or more detectors turn out to be at duty cycles that are lower than expected, as is the case for Virgo and KAGRA during the O4 run, can provide a more realistic account depicting the localization capabilities of a GW network. For instance, in the context of our study, the median $\Delta \Omega_{90\%}$ area of $\sim 13.5$ sq. deg. is obtained with the four detector L1H1V1K1 network, where L1 and H1 are at $80\%$ duty cycle and both V1 and K1 detectors are operating at a lower $20\%$ duty cycle. With the addition of A1 detector to this network, where A1 is set to \texttt{aLIGO} O4 sensitivity and operates at $20\%$ duty cycle (same as that of V1 and K1), there is a significant reduction in median $\Delta \Omega_{90\%}$ area to $\sim 8$ sq. deg.
A case study of the localization capabilities considering only the three LIGO detectors (L1, H1, and A1) is presented in~\cite{Saleem_2021}, where all three are considered to be at A+ sensitivity. 

For this study, we generate the BNS events in Section~\ref{motivating for SNR range choice & injection parameters} using the \texttt{IMRPhenomD} waveform model and reconstruct the source parameters using the \texttt{TaylorF2} model template waveforms. Using a waveform model that includes tidal deformability parameters, higher-order modes, and other physical effects captured by additional intrinsic parameters in the analysis can make the study more comprehensive. We aim to 
incorporate the tidal parameters and higher-order modes within the meshfree framework in the future. This extension will enable us to achieve a more comprehensive and rigorous analysis. We have also fixed the values of $\iota$ and $\psi$ as shown in Section~\ref{motivating for SNR range choice & injection parameters}. This may also have an effect on the localization results. For a more general treatment, the events under consideration should be generated such that all the parameters should be allowed to vary in parameter space. This shall allow for a more exhaustive assessment of the localization capabilities of different GW networks. We simulate uncorrelated Gaussian noise in the detectors for our analysis. In this context, it has been shown by Berry et al.~\cite{Berry_2015} that no appreciable impact is observed in the localization results for the case of simulated signal injected in real detector noise. 

Another aspect that might affect the sky localization area evaluated from the Bayesian posterior samples is the narrow prior boundaries taken over the intrinsic parameters. For consistency, we evaluate the sky localization areas from the posterior samples with wide boundaries over intrinsic parameters using another rapid PE method (relative binning in this case) and compared the results with the meshfree framework adopted here. We find that the difference between the localization areas obtained from these two approaches is not significant enough to affect the localization results, at least for a network involving three or more detectors. The results showcased in this work serve as a demonstration of what can be accomplished by adding A1 as a new detector to the gravitational wave (GW) network. The primary emphasis is on evaluating the GW network's ability to pinpoint the source of gravitational waves, particularly in the context of potential electromagnetic follow-up observations.

\section*{Acknowledgements}
We would like to thank Varun Bhalerao, Gaurav Waratkar, Aditya Vijaykumar, Sanjit Mitra, and Abhishek Sharma for useful suggestions and comments. We especially thank the anonymous referee for their careful review and helpful suggestions. 

S.~S. is supported by IIT Gandhinagar. L.~P. is supported by the Research Scholarship Program of Tata Consultancy Services (TCS). A.~S. gratefully acknowledges the generous grant provided by the Department of Science and Technology, India, through the DST-ICPS cluster project funding. We thank the HPC support staff at IIT Gandhinagar for their help and cooperation. 
\clearpage
\section{Appendix}
\subsection{GWTC-like events corner plots}
\label{sec:append_gwtc_corner}
Here are the corner plots of the three GWTC-like events, which are injected into the real noise representing O2 sensitivities. We want to study how the addition of LIGO-Aundha (A1) to the GW network observing a given GW-like event might have impacted their sky localization. 
\vspace{2mm}
\begin{figure*}[!hbt]
    \centering
    \includegraphics[width=\textwidth, clip=True]{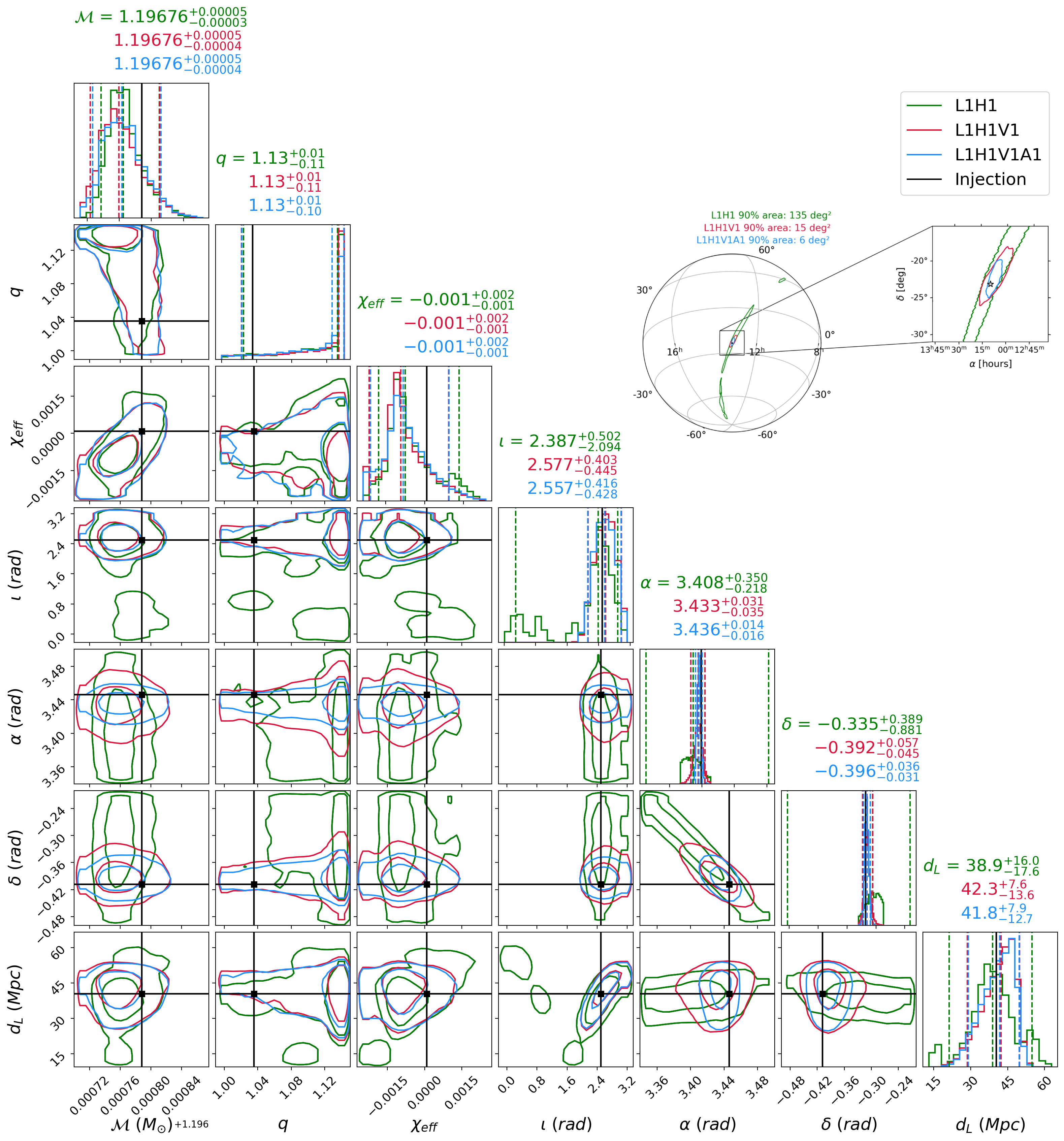}
    \caption{Corner plot showing the posterior distributions (with $90\%$ CI) of various binary parameters of GW170817-like event. The black vertical line represents the corresponding injection parameters. As evident from the sky-map (in \textit{inset}), the addition of A1 decreases the $90\%$ credible sky localization area by ${\sim 60 \%}$ (in comparison to L1H1V1 network) while the estimation of other parameters broadly remains similar.}
    \label{fig:GW170817_like}
\end{figure*}

\begin{figure*}[!hbt]
    \centering
    \includegraphics[width=\textwidth, clip=True]{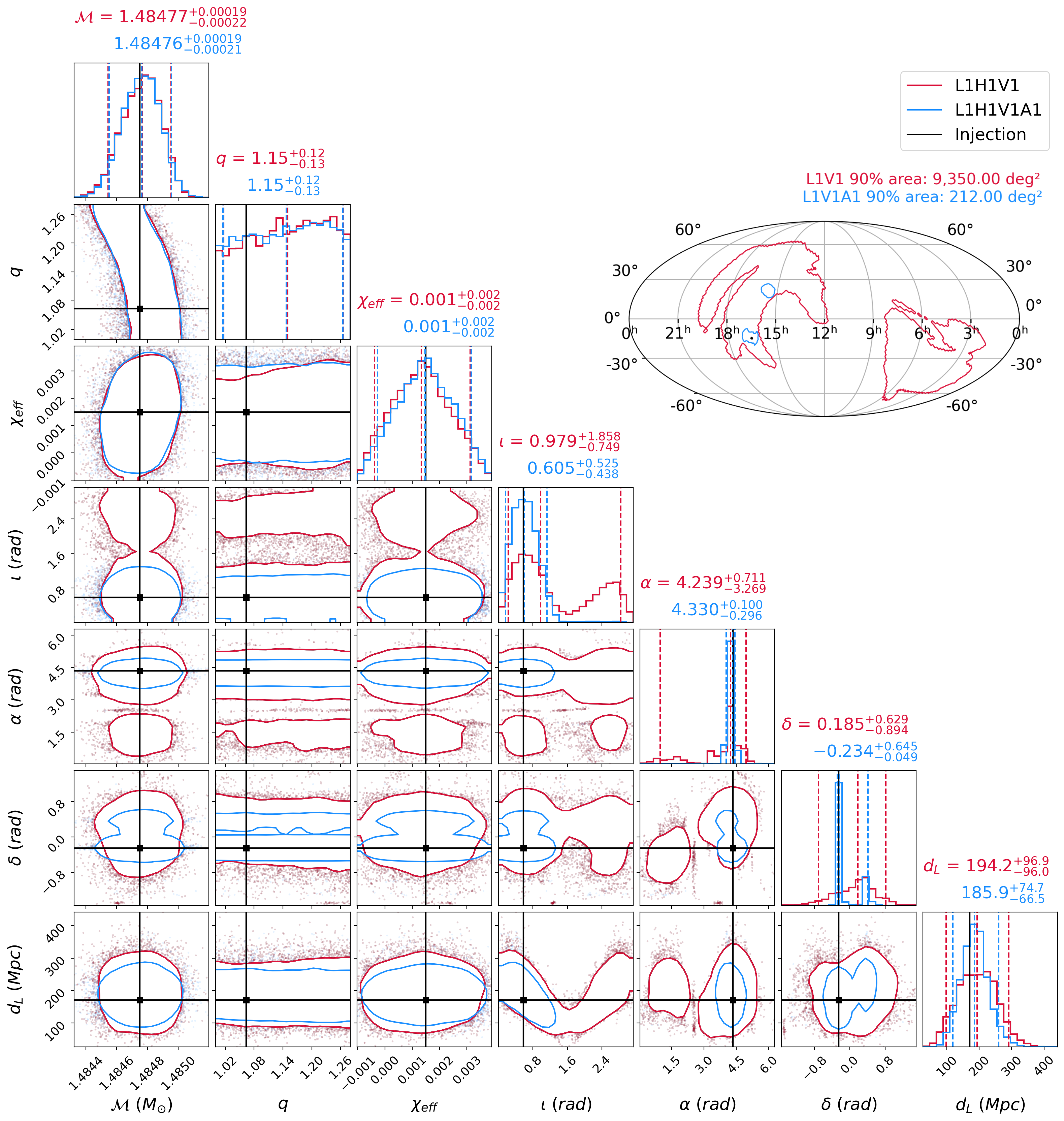}
    \caption{Corner plot showing the posterior distributions (with $90\%$ CI) of various binary parameters of GW190425-like event, injected in real noise representing O2 sensitivity. The black line represents the corresponding injection parameters. As evident from the sky-map (in \textit{inset}), the addition of A1 significantly decreases the $90\%$ credible sky localization area by ${\sim 98 \%}$. In addition, the degeneracy between distance and inclination is also broken, which helps constrain the distance and inclination effectively. Similar to the GW170817-like case, the estimation of other parameters broadly agrees across the two cases.}
    \label{fig:GW190425_like}
\end{figure*}

\begin{figure*}[!hbt]
    \centering
    \includegraphics[width=\textwidth, clip=True]{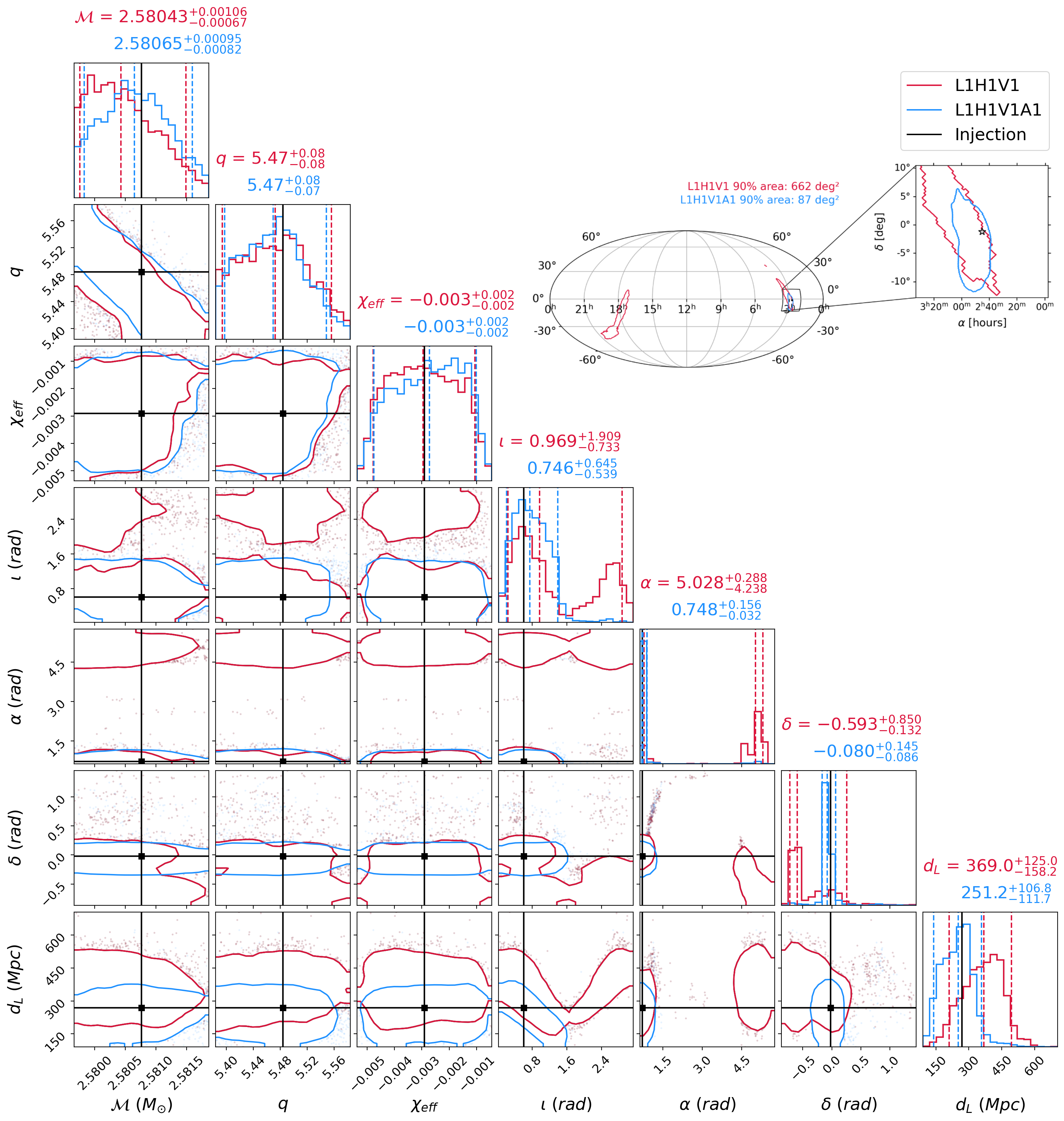}
    \caption{Corner plot showing the posterior distributions (with $90\%$ CI) of various binary parameters of GW200115-like event, injected in real noise representing O2 sensitivity. The black line represents the corresponding injection parameters. As evident from the sky map (in \textit{inset}), the addition of A1 significantly decreases the $90\%$ credible sky localization area by ${\sim 86 \%}$. In addition, the degeneracy between distance and inclination is again broken, which helps constrain the distance and inclination effectively. However, the estimation of other parameters broadly agrees across the two cases.}
    \label{fig:GW200115_like}
\end{figure*}

\subsection{Simulated BNS event}
\label{sec:append_sims_corner}
Here, we present a corner plot representing the posterior distributions of various binary parameters corresponding to one of the $500$ simulated BNS events injected in the L1, H1, V1, and K1 network operating at O5 design sensitivity. In the same plot, we also show the posterior distributions of the binary parameters for the cases when the same event is also injected in LIGO-Aundha (A1) in addition to L1, H1, V1, and K1 detectors (at O5 sensitivities), operating at O4 and O5 design sensitivity. 
\begin{figure*}[!hbt]
    \centering
    \includegraphics[width=0.85\textwidth, clip=True]{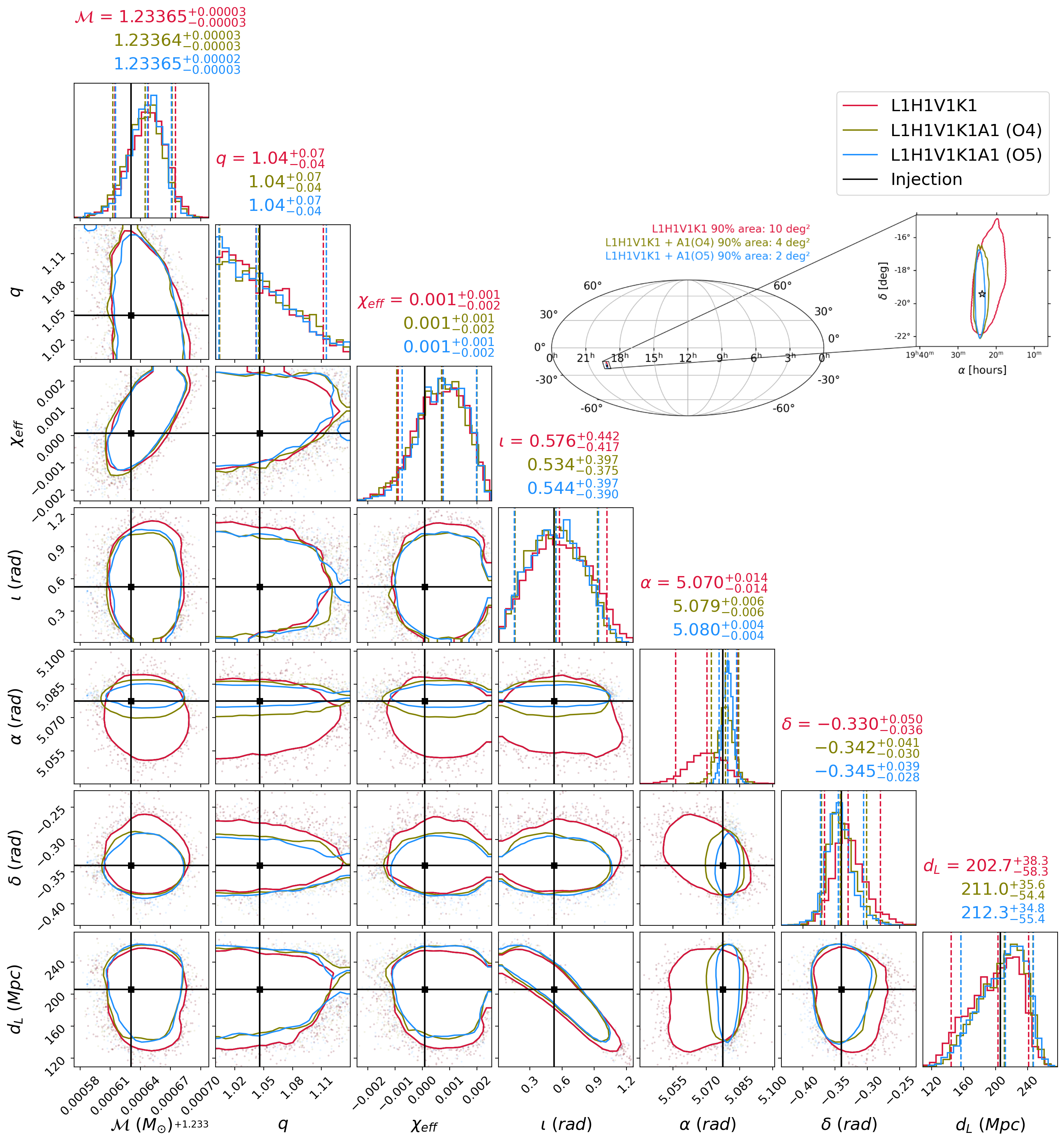}
    \caption{Corner plot showing the posterior distributions (with $90\%$ CI) of various parameters of one of $500$ simulations for three different cases. \textit{Red}: When the network consists of four GW detectors, all operating at O5 design sensitivity. \textit{Olive}: When the network consists of five GW detectors, out of which four detectors are operating at O5 design sensitivity while A1 is at O4 sensitivity. \textit{Blue}: When the network consists of five GW detectors, all operating at O5 design sensitivity. The black vertical line represents the injection parameter. The \textit{inset} shows the corresponding skymaps. Evidently, the $90\%$ CI decreases from $10$ sq. deg. to $4$ sq. deg.($2$ sq. deg.) as A1 operating at O4(O5) sensitivity is added to the L1, H1, V1, K1 network.
    }
    \label{fig:corner_sim}
\end{figure*}
\chapter{Conclusion and Future Outlook}
\label{chap:chapter_6}

In this thesis, we explored rapid parameter estimation methods by utilizing a blend of singular value decomposition (SVD), a widely recognized linear algebra approach, and meshfree approximation, commonly applied in the interpolation of scattered data. The chapter~\ref{chap:chapter_1} provides a foundational overview of Gravitational waves theory, along with a brief exploration of methods for their detection. Additionally, we delve into diverse sampling techniques commonly employed for probability distribution sampling, specifically Markov Chain Monte Carlo (MCMC) and Nested Sampling. These methodologies serve as the foundational elements for various estimation approaches utilized in the analysis of Gravitational Wave data.

The second chapter introduces a rapid method for generating waveforms that leverages SVD and meshfree approximation. This technique involves creating meshfree surrogates for GW waveform models by producing interpolations for the amplitude and phase of the waveform separately. A greedy algorithm is employed to identify an appropriate basis (for both amplitude and phase) to cover the given sample space of binary parameters (such as component masses and spins). Once these interpolations are prepared, they enable the swift generation of otherwise computationally expensive waveforms without sacrificing accuracy. To validate the reliability of our mesfhree surrogates, we apply them in reconstructing a simulated BBH source.

In chapter~\ref{chap:chapter_3}, we present a rapid parameter estimation technique that circumvents waveform generation entirely and directly interpolates the likelihood. This method consists of two stages: In the initial stage (referred to as the "start-up stage"), we employ SVD to identify a suitable basis for the space spanned by the overlap time series (between the GW strain data and templates). Subsequently, the corresponding SVD coefficients are fitted with a combination of radial basis functions (RBFs) and monomials to generate their interpolating functions. Once these functions are prepared, they are utilized to quickly evaluate the likelihood of the samples proposed by the sampler (i.e., MCMC or Nested Sampling) during the sampling process, known as the "online stage." We validated the effectiveness of our approach using a simulated BNS injected into Gaussian noise from a Power Spectral Density (PSD), emulating the design sensitivities of the LIGO-Livingston detector "L1." Our results demonstrated a substantial speed-up of over three orders of magnitude in likelihood evaluation compared to the bruteforce likelihood calculation implemented in \textsc{PyCBC}~\cite{usman2016pycbc, biwer2019pycbc} while showing good agreement between the meshfree and bruteforce likelihoods.

In chapter~\ref{chap:chapter_4}, we expand upon the meshfree framework introduced in chapter~\ref{chap:chapter_3} to enable coherent network parameter estimation (PE) by incorporating gravitational wave (GW) strain data from multiple detectors. This extension facilitates the estimation of various parameters, including sky location and inclination, crucial for follow-ups of the electromagnetic counterparts of such events. Additionally, we propose an effective strategy for node placement to distribute nodes in the intrinsic parameter space. This placement is guided by the Fisher matrix calculated at the center of the selected sample space box, enhancing node concentration near the likelihood peak for higher accuracy in meshfree likelihood approximation. To validate our method's robustness, we inject numerous simulated binary neutron star (BNS) signals into Gaussian noise, mimicking the detector sensitivities akin to the theoretical design sensitivities of ground-based detectors such as LIGO-Virgo. We estimate their source parameters using our meshfree approach and observe substantial agreement with the relative binning method~\cite{Venumadhav2018, cornish2021heterodyned}, an alternative rapid PE method employed in the GW community. Considering that BNS systems are influenced by tidal effects, we conduct a preliminary analysis within a meshfree framework, incorporating tidal parameters for both components of a BNS system. This inclusion increases the interpolating parameter space dimensionality to six, necessitating a higher number of nodes for an accurate likelihood approximation. We again observe a broad agreement between the posterior distributions of binary parameters in PE analyses conducted with both meshfree and bruteforce likelihood methods.

In chapter~\ref{chap:chapter_5}, we employ our meshfree-based rapid parameter estimation (PE) pipeline to conduct an extensive study of the potential impact of LIGO-Aundha. LIGO-Aundha is an upcoming Indian Gravitational Waves Observatory located in Hingoli, Maharashtra, expected to commence scientific operations around 2030. This observatory is poised to join the international gravitational wave detector network (IGWN). The focus of our study is on evaluating the influence of LIGO-Aundha joining the IGWN on the sky-localization capabilities of BNS systems, considering their significance as potential sources for electromagnetic (EM) follow-ups. Our findings suggest that, even with moderate sensitivity, the addition of LIGO-Aundha has the potential to significantly enhance the overall scientific objectives of second-generation gravitational wave observatories.

Finally, we highlight some of the potential extensions of the rapid PE methods presented in this thesis. In the current implementation, as outlined in chapter~\ref{chap:chapter_2}, the waveform interpolation primarily addresses aligned spin waveform models, specifically focusing on the dominant "quadrupolar" mode. In future iterations, our aim is to broaden the scope of our method to encompass precessing waveform models and models that include subdominant modes. The likelihood interpolation technique, initially introduced in chapter~\ref{chap:chapter_3} and extended for coherent network parameter estimation in chapter~\ref{chap:chapter_4}, currently requires a narrow region within the intrinsic parameter space for accurate interpolation of SVD coefficients and template norm squares. Additionally, it focuses on aligned spin waveform models featuring only the dominant mode. In the subsequent development of these works, we intend to overcome these limitations by proposing robust strategies for node placement, expanding boundaries in the intrinsic parameter space, and accommodating waveform models that incorporate both spin precession and sub-dominant modes. Some of the ongoing efforts include the incorporation of eccentricity into our meshfree framework for both waveform and likelihood interpolation. Furthermore, the techniques introduced in this thesis hold promise for application to parameter estimation challenges in scenarios involving overlapping signals~\cite{Samajdar_2021, Janquart:2023hew, alvey2023things} in third-generation detectors like Cosmic Explorer and the Einstein Telescope. Additionally, they could find utility in addressing the parameter estimation challenges in the context of the Laser Interferometer Space Antenna (LISA) where the signal duration is $\mathcal{O}(1\, \text{year})$ and a rapid PE method becomes essential for timely estimation of the source parameters.
%


\bibliographystyle{unsrt}
\bibliography{references_comb}

\end{document}